\documentclass{aa}  

\usepackage{graphicx}
\usepackage[dvipsnames, svgnames, table, xcdraw]{xcolor}
\usepackage{comment}
\usepackage{caption}
\usepackage{natbib}
\usepackage{subcaption}
\renewcommand\thesubfigure{\arabic{subfigure}}
\DeclareCaptionFormat{cont}{#1 (cont.)#2#3\par}
\usepackage{multirow}
\usepackage{multicol}
\usepackage{arydshln}
\usepackage{graphicx, graphics,amssymb,amsmath} 
\usepackage{booktabs}
\usepackage{float,capt-of}
\usepackage{natbib}
\usepackage[english]{babel}
\usepackage{morefloats}
\usepackage{sidecap}
\usepackage{epsfig,color}
\usepackage{wrapfig}
\usepackage{longtable}
\usepackage{rotating}
\usepackage{pdflscape}
\usepackage{array}

\newcommand\kms{km~s$^{-1}$}

\def\app#1#2{%
  \mathrel{%
    \setbox0=\hbox{$#1\sim$}%
    \setbox2=\hbox{%
      \rlap{\hbox{$#1\propto$}}%
      \lower1.1\ht0\box0%
    }%
    \raise0.25\ht2\box2%
  }%
}
\def\approxprop{\mathpalette\app\relax}

\definecolor{skobeloff}{rgb}{0.0, 0.48, 0.45}

\usepackage{txfonts}
\usepackage{hyperref}
\begin{document}

   \title{A sensitive APEX and ALMA CO(1--0), CO(2--1), CO(3--2), and [CI](1--0) spectral survey of 40 local (U)LIRGs}

   \author{I. Montoya Arroyave
        	\inst{1}
          \and
          C. Cicone\inst{1}
          \and E. Makroleivaditi\inst{2, 3}
          \and A. Weiss \inst{2}
          \and A. Lundgren\inst{4}
          \and P. Severgnini\inst{5}
          \and C. De Breuck\inst{4}
          \and B. Baumschlager\inst{1}
          \and A. Schimek\inst{1}
          \and S. Shen\inst{1}
          \and M. Aravena\inst{6}
          }
   \institute{Institute of Theoretical Astrophysics, University of Oslo,
   P.O. Box 1029, Blindern, 0315 Oslo, Norway \\  
              \email{isabemo@uio.no}
          \and Max-Planck-Institut fur Radioastronomie, Auf dem Hugel 69, D-53121 Bonn, Germany 
          \and Rheinische Friedrich-Wilhelms-Universitat Bonn, Germany     
          \and European Southern Observatory, Karl-Schwarzschild-Strasse 2, 85748 Garching, Germany 
          \and INAF - Osservatorio Astronomico di Brera, Via Brera 28, I-20121 Milano, Italy 
          \and Instituto de Estudios Astrof\'{\i}sicos, Facultad de Ingenier\'{\i}a y Ciencias, Universidad Diego Portales, Av. Ej\'ercito Libertador 441, Santiago, Chile 
        }

   \date{Received 2022-09-26 ; Accepted 2023-02-09 }

  \abstract
 {We present a high sensitivity, ground-based spectral line survey of low-\textit{J} Carbon Monoxide (CO($J_{\rm up}\rightarrow J_{\rm up}-1$) with $J_{\rm up}=1,2,3$) and neutral Carbon [CI] $^{3}P_{1}-^{3}P_{0}$ ([CI](1--0)) in 36 local ultra luminous infrared galaxies (ULIRGs) and 4 additional LIRGs, all with previous \textit{Herschel} OH~119 $\mu$m observations. The study is based on new single-dish observations 
 conducted with the Atacama Pathfinder Experiment (APEX), complemented by archival APEX and Atacama Large Millimeter Array (ALMA and ACA) data. Our methods are optimized for a multi-tracer study of the total molecular line emission from these ULIRGs, including any extended low-surface brightness components. We find a tight correlation between the CO and [CI] line luminosities suggesting that the emission from CO(1--0) (and CO(2--1)) arises from similar regions as the [CI](1--0), at least when averaged over galactic scales. By using [CI] to compute molecular gas masses, we estimate a median CO-to-H$_2$ conversion factor of $\langle\alpha_{\rm CO}\rangle =1.7\pm 0.5$ M$_{\odot}$ (K \kms pc$^2)^{-1}$ for ULIRGs. We derive median galaxy-integrated CO line ratios of $\langle r_{21} \rangle = L{'}_{\mathrm{CO(2-1)}}/L{'}_{\mathrm{CO(1-0)}} = 1.09$, $\langle r_{31} \rangle = L{'}_{\mathrm{CO(3-2)}}/L{'}_{\mathrm{CO(1-0)}} = 0.76$, $\langle r_{32} \rangle = L{'}_{\mathrm{CO(3-2)}}/L{'}_{\mathrm{CO(2-1)}} = 0.76$, significantly higher than normal star forming galaxies, confirming the exceptional molecular gas properties of ULIRGs. We find that the $r_{21}$ and $r_{32}$ ratios are poor tracers of CO excitation in ULIRGs, while $r_{31}$ shows a positive trend with $L_{IR}$ and SFR, and a negative trend with the H$_2$ gas depletion timescales ($\tau_{\rm dep}$). Our investigation of CO line ratios as a function of gas kinematics shows no clear trends, except for a positive relation between $r_{21}$ and $\sigma_v$, which can be explained by CO opacity effects. These ULIRGs are also characterized by high $L{'}_{\rm [CI](1-0)}/L{'}_{\rm CO(1-0)}$ ratios, with a measured median value of $\langle r_{\rm CICO} \rangle =0.18$, higher than previous interferometric studies that were affected by missing [CI] line flux. The $r_{\rm CICO}$ values do not show significant correlation with any of the galaxy properties investigated, including OH outflow velocities and equivalent widths. We find that the line widths of [CI](1--0) lines are $\sim10\%$ smaller than CO lines, and that this discrepancy becomes more significant in ULIRGs with broad lines ($\sigma_v>150$~\kms) and when considering the high-v wings of the lines. This suggests that the low optical depth of [CI] can challenge its detection in diffuse, low-surface brightness outflows, and so its use as a tracer of CO-dark H$_2$ gas in these components. Finally, we find that higher $L_{\rm AGN}$ are associated to longer $\tau_{\rm dep}$, consistent with the hypothesis that AGN feedback may reduce the efficiency of star formation. Our study highlights the need of sensitive single-dish multi-tracer H$_2$ surveys of ULIRGs that are able to recover the flux that is missed by interferometers, especially in the high-frequency lines such as [CI]. The Atacama Large Aperture Submillimeter Telescope (AtLAST) will be transformational for this field.}
 

\titlerunning{CO and [CI] observations of 40 local (U)LIRGs}
\authorrunning{Montoya Arroyave et al.}

\maketitle

\section{Introduction}\label{sec:Introduction}

In the local ($z \lesssim 0.2$) Universe, (ultra) luminous infrared galaxies (ULIRGs, $L_{\rm IR}(8-1000$  $\mu \textrm{m}) \ge 10^{12}L_{\odot}$, and LIRGs if $L_{\rm IR}(8-1000$  $\mu \textrm{m}) \ge 10^{11}L_{\odot}$) pinpoint gas-rich galaxy mergers undergoing intense starbursts (SBs) and supermassive black hole (SMBH) accretion \citep{Sanders+Mirabel96, Genzel+98, Lonsdale+06, Perez-Torres+21, U22}. These processes cooperate to deeply modify the physical and dynamical properties of the interstellar medium (ISM), likely leading to a permanent morphological transformation and quenching (e.g., \citealt{Hopkins+08}).

(Sub)millimeter interferometric observations of CO lines \citep[e.g.,][]{Downes+Solomon98, Wilson+08, Ueda+14} and dense H$_2$ gas tracers such as HCN and HCO$^+$
\citep{Aalto+12, Imanishi+14, Imanishi+19, Ledger+21}
show that the extreme star formation rates of (U)LIRGs are fueled by massive ($M_{{\rm H}_2}>10^9$~M$_{\odot}$), dense H$_2$ gas reservoirs, characterized by a high surface brightness in the central kiloparsec-scale region. However, feedback mechanisms and tidal forces can disperse ISM material outside of the nuclear regions \citep{Springel+05, Narayanan+06, Narayanan+08, Duc+Renaud13}. Hence, we may expect a portion of the ISM of (U)LIRGs to reside in diffuse, low-surface brightness structures, possibly missed by high resolution interferometric observations. 

Galactic outflows have been observed ubiquitously in (U)LIRGs since decades, in the ionized \citep{Westmoquette+12, Arribas+14} and atomic \citep{Rupke+05, Martin+05, Cazzoli+14} gas phases, as expected in sources affected by strong radiative feedback from SBs and AGN \citep[e.g.,][]{Costa+18b, Biernacki+Teyssier18}. More recent is the discovery that the outflows of (U)LIRGs can embed large amounts of molecular gas, traveling at speeds of up to $v\sim1000$~\kms. Such molecular outflows have been detected unambiguously by {\it Herschel}, via observations of P-Cygni profiles and/or blue-shifted absorption components of far-infrared (FIR) OH, H$_2$O, and OH$^{+}$ transitions \citep{Fischer+10, Sturm+11, Spoon+13, Veilleux+13, Stone+16, Gonzalez-Alfonso+17, Gonzalez-Alfonso+18}, as well as through the investigation of broad and/or high-velocity components of CO \citep{Feruglio+10, Cicone+12, Cicone+14, Feruglio+15, Pereira-Santaella+18, Lutz+20, Fluetsch+19, Lamperti+22}, HCN, HCO$^+$ \citep{Aalto+12, Aalto+15, Barcos-Munoz+18}, and CN \citep{Cicone+20} emission lines. These sensitive observations have also shown that the molecular ISM of (U)LIRGs, and especially the low surface brightness outflow components \citep[e.g.,][]{Feruglio+13a, Cicone+18, Herrera-Camus+20}, can extend by several kpc, up to the edge of the field of view of single-pointing interferometric data. 

Obtaining robust H$_2$ mass measurements of the total ISM reservoirs as well as of the gas embedded in outflows is crucial to understand the impact of gas-rich galaxy mergers - and of the collateral powerful SB and AGN feedback mechanisms - on galaxy evolution. 
Low-\textit{J} CO lines such as CO(1--0) and CO(2--1) can be used to estimate H$_2$ masses through a CO-to-H$_2$ conversion factor (hereafter $\alpha_{\textrm{CO}}$), which is however highly dependent on the physical state of the gas. The $\alpha_{\textrm{CO}}$ parameter can vary by up to a factor of $\sim10$ in different ISM environments, depending on the CO optical depth, on the metallicity of the medium, as well as on the exposure to far-UV radiation and Cosmic Rays (CRs) that can destroy CO more than H$_2$ \citep[see, e.g.,][]{Bisbas+15, Glover+15, Offner+14}. 
For the molecular ISM of disk galaxies, the conventional $\alpha_{\textrm{CO}}$ factor is $4.3$ M$_{\odot}$ (K km s$^{-1}$ pc$^{-2}$)$^{-1}$ \citep{Strong+96, Abdo+10, Bolatto+13}, while for more perturbed galaxies such as gas-rich mergers and starbursts, a lower $\alpha_{\textrm{CO}}$ factor of $\sim 0.8-1.0$ M$_{\odot}$ (K km s$^{-1}$ pc$^{-2}$)$^{-1}$ is often preferred \citep{Downes+Solomon98}.

Combining multiple molecular transitions can help constrain the physical properties of molecular gas and so derive a better estimate of the $\alpha_{\textrm{CO}}$ factor.
A particularly valuable H$_2$ gas tracer is the forbidden $^3P_1 - ^3P_0$ fine structure line of atomic Carbon, hereafter [CI](1--0), which has an excitation temperature of $T_{\textrm{ex}}=23.6$ K, and a critical density similar to that of CO(1--0), i.e., $n_{crit,\textrm{[CI]}}\sim 1000$ cm$^{-3}$.
The [CI](1--0) line is optically thin and has a simple three-level partition function, which makes it easier to interpret than low-\textit{J} CO lines (see discussion in \citealt{Papadopoulos+22}). 
Early models of photodissociation regions predicted [CI] to exist in a thin transition layer between the central region of molecular clouds (molecular gas - CO) and its envelope (ionized gas - [CII]) \citep[for the standard PDR view see][]{Tielens+85}. However, observations have shown [CI] to be well mixed with CO widely throughout the cloud \citep{Valentino+18, Saito+20, Papadopoulos+22} suggesting that both species might trace the \textit{bulk} of molecular gas mass \citep{Ojha+01, Papadopoulos+04, Kramer+08, Salak+19, Izumi+20, Jiao+17}.
Moreover, theoretical models show that CO (but not H$_2$) may be destroyed in environments dominated by cosmic rays (CRs), shocks, or intense radiation fields, leaving behind CO-\textit{dark} or CO-poor reservoirs, giving more value to [CI] as an alternative H$_2$ gas mass tracer.

Observing the atomic Carbon emission from local galaxies requires a sensitive sub-mm telescope located at a very high and dry site. Indeed, the [CI](1--0) transition, at a rest frequency of $\nu_{\rm [CI](1-0)}^{\rm rest}=492.161$~GHz (609.135~$\mu$m), in absence of a significant red-shift, can be observed only if the atmospheric opacity is low (precipitable water vapor PWV$<1$~mm). For this reason, sensitive observations of [CI](1--0) in the local Universe are still very sparse, even for bright (U)LIRGs. 
\cite{Cicone+18} used the {\it Atacama Large Millimeter/sub-millimeter Array} (ALMA) and the {\it Morita Array} (ACA) to obtain high S/N [CI](1--0) observations of NGC~6240. These data, combined with archival CO(1--0) and CO(2--1) observations, were used to study the $r_{21}\equiv L^{\prime}_{\rm CO(2-1)}/L^{\prime}_{\rm CO(1-0)}$ and the $r_{\rm CICO}\equiv L^{\prime}_{\rm [CI](1-0)}/L^{\prime}_{\rm CO(1-0)}$ line ratios (the latter can be used to estimate $\alpha_{\textrm{CO}}$) in the massive molecular outflow of NGC~6240. Through a spatially resolved analysis, \cite{Cicone+18} found that the outflowing ISM in NGC~6240 is robustly characterized by a lower $\alpha_{\textrm{CO}}$ than the non-outflowing H$_2$ medium, and that $r_{21}$ is higher for high-$\sigma_v$ outflow components, especially at large distances from the nuclei. 
The \cite{Cicone+18} analysis suggested that: (i) despite its obvious limitations, a multi-component decomposition of galaxy-integrated spectra, performed simultaneously to multiple transitions, enabling an investigation of line ratios separately for spectral components with different widths and central velocities, can deliver results that are consistent with a proper outflow/disk spatial decomposition of the ISM; (ii) the outflowing H$_2$ gas may be characterized by different physical properties from the non-outflowing ISM of NGC~6240, and in particular by a lower CO optical depth and a higher CO excitation. These results have been obtained on a single, extreme source, and further statistics is required.

Our study builds upon these previous results and aims at expanding the analysis of \cite{Cicone+18} to a sample of 36 local ULIRGs and 4 additional LIRGs with low-\textit{J} CO (up to \textit{J=3}) and [CI](1--0) line observations. In designing our survey, we paid particular attention to capturing the total flux from these sources, including possible extended low surface brightness components that may be dominated by outflows and tidal tails and may be missed by high-resolution, low S/N interferometric data. The final survey contains proprietary and archival data from the {\it Atacama Pathfinder EXperiment} telescope (APEX), ALMA, ACA and IRAM {\it Plateau de Bure Interferometer} (PdBI, now the {\it NOrthern Extended Millimeter Array}, NOEMA), which we have re-reduced and re-analyzed in a consistent and uniform way. Therefore we can rely both on a consistent data analysis and on high-quality spectra, all taken with receivers whose large instantaneous intermediate frequency (IF) bandwidth can properly sample the extremely broad emission lines of (U)LIRGs.

This paper is organized as follows. 
In Section \ref{sec:sample} we describe the sample selection. 
In Section \ref{sec:observations} we describe the observing strategy, the observations and data reduction. 
In Section \ref{sec:methodology} we explain the methodology used for the spectral fitting and the data analysis.
Our results are presented in Section~\ref{sec:results}, where they are also contextualized through a comparison with relevant results from the literature. A more general discussion is reported in Section~\ref{sec:discussion}. 
Finally, Section~\ref{sec:conclusions} summarizes the main results and presents the conclusions of our work.
Throughout this work, we adopt a $\Lambda$CDM cosmology, with $H_0=67.8$ \kms Mpc$^{-1}$, $\Omega_{\textrm{M}}=0.307$ and $\Omega_{\Lambda}=0.693$ \citep{Planck_cosmology+14}.

\section{The sample}\label{sec:sample}

In absence of additional spatial information, the only unambiguous method to assess the presence of molecular outflows trough spectroscopy, is the detection of P-Cygni profiles or blueshifted absorption components in molecular transitions, such as the OH$119\mu$m line observed by Herschel \citep{Fischer+10, Sturm+11}.
However, the absence of these features does not necessarily rule out the presence of outflows, as in the case of NGC~6240, studied in \cite{Cicone+18}. In this source, OH is detected only in emission despite the presence of an extreme molecular outflow detected in multiple tracers. For this reason, molecular emission line observations can provide valuable and unique information on the presence and properties of galactic outflows, complementary to OH data.

The \textit{Herschel} OH targets studied by \cite{Veilleux+13} and \cite{Spoon+13} represent the only conspicuous sample of (U)LIRGs that has uniform and unambiguous prior information about their molecular outflows, hence providing a robust comparison dataset for our investigation based on emission lines. Moreover, the southern targets in this sample also have plenty of ancillary data from ALMA and APEX, allowing us to capitalize on public archives, which is a main focus of this work. 
For these reasons, the targets in our sample are selected from the \cite{Spoon+13} and \cite{Veilleux+13} samples, regardless of the detection of a molecular outflow in OH.

From the parent \textit{Herschel} samples, we have included all sources with declination $\delta < 15$ deg, except IRAS 12265+0219 and IRAS 00397-1312 for which we did not have any data available\footnote{The galaxy IRAS 00397-1312 was registered to have APEX CO(2--1) archival data, however, when opened, it returns an empty file, and it was not possible to re-observe this source within our PI programmes.}. NGC~6240 satisfies our selection criteria but is excluded from our work because it was the target of the pilot study by \cite{Cicone+18}. 
Our sample includes 36 ULIRGs, whose physical properties such as redshifts, $L_{\rm IR}$, SFRs, and AGN fractions ($\alpha_{\textrm{AGN}} \equiv L_{\rm AGN} /L_{\rm bol}$) are reported in Table \ref{table:source_list} with their corresponding references. The 4 additional LIRGs reported in the Appendix \ref{sec:appendix_LIRGs} have been reduced and analyzed consistently with the rest of the sample, but they have been excluded from the main body of the paper to avoid biasing the relations given the low statistics for these low $L_{\rm IR}$ sources. Hereafter, we will use the term ``(U)LIRGs'' when referring to the entire sample, and ``ULIRGs'' when we exclude the 4 LIRGs.

The parent {\it Herschel} samples from which our targets were selected have a redshift upper limit of $z < 0.2$. As a result, our study investigates local (U)LIRGs with redshifts ranging from $z=0.00708$ (IRAS F12243-0036) to $z=0.1935$ (IRAS F05024-1941). The sample is by definition composed by high-$L_{\rm IR}$ galaxies with $L_{\rm IR}$ ranging from $10^{11}$ L$_{\odot}$ to $10^{12.8}$ L$_{\odot}$, covering luminosities within the (U)LIRG regime.
The sources span a wide range in $\alpha_{\textrm{AGN}}$ values from 0.0 up to 0.92, with $\sim 50 \%$ of the sources having $\alpha_{\textrm{AGN}} \geq 0.5$. 
The star formation rates (SFRs) are taken from the parent sample papers, with the exception of galaxies selected in \cite{Veilleux+13} for which there are no SFRs reported. In those cases, we followed the method by \cite{Sturm+11} \citep[also used in][]{Spoon+13} to obtain the SFRs using: $\textrm{SFR} = (1-\alpha_{\rm AGN})\times 10^{-10}L_{\rm IR}$, so that all values are computed uniformly. The SFRs range from a couple of solar masses a year up to $\sim$300 M$_{\odot}\textrm{yr}^{-1}$. 
We have checked that our sample of ULIRGs is representative, in terms of physical properties (e.g. SFR, $L_{\rm AGN}$, $\alpha_{\rm AGN}$) of the local ULIRG population by comparing it with the QUEST (Quasar/ULIRG Evolutionary Study) sample at $z<0.2$ \citep[see][]{Veilleux+09a}.
Most previous works studying the molecular gas in local (U)LIRGs have included both LIRGs and ULIRGs. We will compare some of our results with the works of \cite{Herrero-Illana+19} and \cite{Jiao+17}, whose samples are however heavily dominated by LIRGs as opposed to ours. As the distinction between LIRGs and ULIRGs is based on an arbitrary $L_{\rm IR}$ cut, several galaxies that are officially LIRGs (such as NGC~6240) belong to the same population as the more IR luminous ULIRGs.

Table \ref{table:source_list} lists the OH outflow velocity values for the sources (29  out of the whole sample of 40) that show an OH outflow detection according to \cite{Veilleux+13} and \cite{Spoon+13}. We also report the OH equivalent widths for the whole sample.

Our sample, which focuses on southern (U)LIRGs targeted by previous \textit{Herschel} OH observations, has naturally a large overlap with the APEX and ALMA/ACA public archives. Indeed, many of these sources have been observed in previous projects targeting different molecular tracers. In this work, we make the most out of such archives, focusing on the low-$J$ CO transitions and [CI] atomic carbon line, and we complement them with our own new proprietary high sensitivity single-dish observations with APEX. The observations and the data reduction process are described in detail in Section~\ref{sec:observations}.

\begin{table*}[]
	\centering \footnotesize
	\caption{List of galaxies analyzed in this work along with some general properties.}\label{table:source_list}
	\begin{tabular}{lcccccccccc}
		\hline
		\hline
		Galaxy name                &   z       &       RA      &      Dec     &  $\alpha_{\textrm{AGN}}$   & $\log L_{\textrm{IR}}$   & $\log L_{\textrm{AGN}}$ & SFR   &   OH$_{\rm max}^{\dagger}$  &   OH$_{\rm EQW}^{\dagger}$ & Ref.  \\
		&           &               &       &       & [$L_{\odot}$]   & [$L_{\odot}$] & [$M_{\odot}\textrm{yr}^{-1}$]    &      [\kms]   &      [\kms]   &    \\
		(1)                 & (2)       & (3)           & (4)          & (5)        & (6)              & (7)               & (8)           & (9)           & (10)     & (11)     \\ \hline
		IRAS 00188$-$0856     & 0.1284    & 00:21:26.522  & $-$08:39:25.98 &  0.51   & 12.39   & 12.16     & 120$\pm$50$^{*}$       &    $-$1781    &   $-$305   & $\gamma$  \\ 
        IRAS 01003$-$2238     & 0.1178    & 01:02:50.007  & $-$22:21:57.22 &  0.83   & 12.32   & 12.30         & 36$\pm$14      &   $-$1238  &   $-$276   &   $\gamma$   \\
        IRAS F01572+0009    & 0.1631    & 01:59:50.253  & +00:23:40.87 &  0.65   & 12.62   & 12.49         & 150$\pm$60$^{*}$           &   $-$1100::    &   0   &   $\beta$    \\
        IRAS 03521+0028     & 0.1519    & 03:54:42.219  & +00:37:03.41 &  0.06   & 12.52   & 11.39         & 309$\pm$120      &   $-$100     &   $-$26   &  $\gamma$ \\
        IRAS F05024$-$1941    & 0.1935   & 05:04:36.555  & $-$19:37:02.83 &  0.07   & 12.37   & 11.30         &  220$\pm$80$^{*}$   &  -850::   &   $-$83    &    $\beta$     \\
        IRAS F05189$-$2524    & 0.0426    & 05:21:01.392  & $-$25:21:45.36 &  0.72   & 12.16   & 12.07         & 40$\pm$16$^{*}$           &  $-$850     &   $-$10    &   $\beta$  \\
        IRAS 06035$-$7102     & 0.0795    & 06:02:54.066  & $-$71:03:10.48  &  0.60   & 12.22   & 12.06         & 70$\pm$30        &	 $-$1117  &   $-$128   &   $\gamma$ \\
		IRAS 06206$-$6315     & 0.0924    & 06:21:01.210  & $-$63:17:23.5  &  0.43   & 12.23   & 11.92         & 100$\pm$40      &   $-$750  &   $-$272    &  $\gamma$  \\
        IRAS 07251$-$0248     & 0.0876    & 07 :27:37.544  & $-$02:54:54.67 &  0.30   & 12.39   & 11.92         &  170$\pm$70$^{*}$	  &     $-$550     &   $-$56    &  $\beta$  \\
        IRAS 08311$-$2459     & 0.1005    & 08:33:20.600  & $-$25:09:33.7 &  0.79   & 12.50   & 12.46         & 70$\pm$30        & 	   &   163   &   $\gamma$ \\
		IRAS 09022$-$3615     & 0.0596    & 09:04:12.689  & $-$36:27:00.76 &  0.55   & 12.29   & 12.09         &  90$\pm$30$^{*}$  &    $-$650   &   17   &  $\beta$   \\
        IRAS 10378+1109     & 0.1363    & 10:40:29.169  & +10:53:18.29 &  0.30  & 12.31   & 11.85         & 140$\pm$60      &   $-$1300  &   $-$155   & $\gamma$  \\
		IRAS 11095$-$0238     & 0.1066    & 11:12:03.377  & $-$02:54:22.58 &  0.49   & 12.28   & 12.03         & 100$\pm$40       &		   &   107   &  $\gamma$  \\
		IRAS F12072$-$0444    & 0.1286    & 12:09:45.132  & $-$05:01:13.76 &  0.75   & 12.40    & 12.33         & 60$\pm$20$^{*}$      &    $-$1200      &   $-$51    &  $\beta$   \\
		IRAS F12112+0305    & 0.07309    & 12:13:45.978  & +02:48:40.4  &  0.18   & 12.32    & 11.63         &  170$\pm$70$^{*}$  &   $-$400    &   $-$2    &    $\beta$   \\
		IRAS 13120$-$5453     & 0.0308    & 13:15:06.358  & $-$55:09:23.23 &  0.33   & 12.24   & 11.83         & 120$\pm$50$^{*}$     &    $-$1200    &   $-$113    &   $\beta$   \\
		IRAS F13305$-$1739    & 0.1484    & 13:33:16.540  & $-$17:55:10.7  &  0.88   & 12.26   & 12.26         &  23$\pm$9$^{*}$  & 	      &     &    $\beta$   \\
		IRAS F13451+1232    & 0.1217    & 13:47:33.425  & +12:17:24.32 &  0.81   & 12.32   & 12.29         & 41$\pm$15$^{*}$        & 	       &   136   &   $\beta$   \\
		IRAS F14348$-$1447    & 0.0830    & 14:37:38.317  & $-$15:00:23.29  &  0.17   & 12.34   & 11.64         &   180$\pm$70$^{*}$   &     $-$900    &   $-$25   &    $\beta$    \\
		IRAS F14378$-$3651    & 0.068127   & 14:40:59.008  & $-$37:04:31.94 &  0.21   & 12.11   & 11.50         &  100$\pm$40$^{*}$   &  $-$1200     &   $-$119    &    $\beta$ \\
		IRAS F15462$-$0450    & 0.100283    & 15:48:56.813  & $-$04:59:33.61 &  0.61   & 12.21   & 12.05         & 60$\pm$20$^{*}$     &    $-$600:     &   80    &   $\beta$     \\
		IRAS 16090$-$0139     & 0.1336    & 16:11:40.432  & $-$01:47:06.56 &  0.43   & 12.55   & 12.25         & 200$\pm$80     &    $-$1422  &   $-$332   & $\gamma$  \\
		IRAS 17208$-$0014     & 0.0428    & 17:23:21.920  & $-$00:17:00.7  &  $\leq$ 0.05  & 12.39 & 11.15   & 230$\pm$90$^{*}$  &  &   $-$148   &    $\beta$  \\
        IRAS 19254$-$7245     & 0.06149    & 19:31:21.400  & $-$72:39:18.0 &  0.74   & 12.09   & 12.02         & 32$\pm$12      &   $-$1126  &   $-$130    &   $\gamma$ \\
        IRAS F19297$-$0406    & 0.08573    & 19:32:21.250  & $-$03:59:56.3  &  0.23   & 12.38   & 11.81         &  180$\pm$70$^{*}$  &  $-$1000  &   $-$119    &  $\beta$   \\
		IRAS 19542+1110     & 0.0624   & 19:56:35.786  &  +11:19:05.45  &  0.26   & 12.06   & 11.52         &   90$\pm$30$^{*}$   &    $-$700    &   $-$29    &    $\beta$     \\
		IRAS 20087$-$0308     & 0.1057    & 20:11:23.870  & $-$02:59:50.7 &  0.20   & 12.42   & 11.79         & 210$\pm$80       &    $-$812   &   $-$386    &  $\gamma$  \\
		IRAS 20100$-$4156     & 0.1296    & 20:13:29.540  & $-$41:47:34.9 &  0.27   & 12.67   & 12.16         & 340$\pm$130    &    $-$1609   &   $-$461   &   $\gamma$ \\
		IRAS 20414$-$1651     & 0.0871   & 20:44:18.213  & $-$16:40:16.22 &  0.00   & 12.22   & <11.46       & 170$\pm$60    & 	$-$100	  &   $-$101    &  $\gamma$  \\
        IRAS F20551$-$4250    & 0.0430    & 20:58:26.781  & $-$42:39:00.20 &  0.57   & 12.05   & 11.87         &    48$\pm$19$^{*}$   &    $-$1200    &   $-$70    &    $\beta$   \\
		IRAS F22491$-$1808    & 0.0778    & 22:51:49.264  & $-$17:52:23.46 &  0.14  & 12.84   & 12.05         & 590$\pm$220$^{*}$       & 	   &   $-$25    &    $\beta$  \\
		IRAS F23060+0505     & 0.1730    & 23:08:33.952  & +05:21:29.76 &  0.78   & 12.53   & $\cdot\cdot\cdot$         & 80$\pm$30       &		   &         &  $\delta$  \\
        IRAS F23128$-$5919    & 0.0446    & 23:15:46.749  & $-$59:03:15.55 &  0.63  & 12.03   & 11.89         &   40$\pm$15$^{*}$   & 		   &          &   $\beta$   \\
		IRAS 23230$-$6926     & 0.1066    & 23:26:03.620  & $-$69:10:18.8 &  0.32   & 12.37   & 11.93         & 160$\pm$60       &   $-$845   &   $-$55    &  $\gamma$ \\
		IRAS 23253$-$5415     & 0.1300    & 23:28:06.100  & $-$53:58:31.0 &  0.23   & 12.36   & 11.78         & 180$\pm$70       &     $-$650    &   $-$134   &  $\gamma$ \\
	    IRAS F23389+0300    & 0.1450    & 23:41:30.306  & +03:17:26.44 &  0.23   & 12.13   & 11.54         &   100$\pm$40$^{*}$   &   $-$600    &   46    &     $\beta$  \\
		\hline
        IRAS F00509+1225    & 0.0611    & 00:53:34.940  & +12:41:36.0  &  0.90   & 11.95   & 11.96         & 36$\pm$14            & 		     &   65    &    $\beta$    \\
        PG 1126$-$041         & 0.0600    & 11:29:16.729  & -04:24:07.25 &  0.89  & 11.46   & 11.47         &   3$\pm$1$^{*}$   & 		     &        &    $\beta$    \\
        IRAS F12243$-$0036    & 0.00708    & 12:26:54.620  & $-$00:52:39.40 &  0.56   & 11.00   & 10.81         & 15$\pm$6          &         &   $-$63    &    $\beta$   \\
		PG 2130+099         & 0.0630    & 21:32:27.813  & +10:08:19.46 &  0.92   & 11.71   & 11.76         &   4$\pm$2$^{*}$   & 	     &      &      $\beta$  \\
		\hline
	\end{tabular}

	\tablefoot{(1) Source name. (2) Redshift. (3) Right ascension. (4) Declination. (5) Reported fractional contribution of the AGN to the bolometric luminosity in the reference papers ($\alpha_{\textrm{AGN}} = L_{\rm AGN} /L_{\rm bol} $). (6) Infrared luminosity $(8-1000$  $\mu \textrm{m})$, $^{*}$computed using $L_{\rm IR} = L_{\rm bol}/1.15$ for sources retrieved from the reference paper $\beta$. (7) Reported AGN luminosity in the reference papers, derived using method 6 by \cite{Veilleux+09b} which uses the 15 to 30 $\mu$m continuum ratio ($f_{30}/f_{15}$) to infer $\alpha_{\rm AGN}$, resulting in uncertainties of $\sim$20\% on average for the $L_{\rm AGN}$ for sources retrieved from the reference paper $\beta$, similar to the method used in the reference paper $\gamma$. (8) Star formation rate, $^{*}$computed using $\textrm{SFR} = (1-\alpha)\times 10^{-10}L_{\rm IR}$ . (9) In case of a detection of OH outflow through P-Cygni profile, the maximum velocity of the absorption feature is reported according to the reference paper. (10) The equivalent width of the OH 119 $\mu$m doublet as reported by the reference papers; a negative equivalent width implies that the absorption component is stronger than the emission component. (11) Reference papers: $\beta$: \citet{Veilleux+13}, $\gamma$: \citet{Spoon+13}.\\
    $^{\dagger}$ For sources taken from $\gamma$, the OH velocities have uncertainties of $\pm 200$ \kms. Sources taken from $\beta$ have uncertainties typically of $50$ \kms, unless the value is followed by a colon, meaning uncertainties from $50$ to $150$ \kms, or a double colon, meaning uncertainties larger than $150$ \kms.}

\end{table*}

\section{Observations}\label{sec:observations}

\subsection{Observing strategy and data reduction}\label{sec:datareduction}

We want to study simultaneously the total integrated line emission from the three lowest-$J$ transitions of CO and from [CI](1--0) in our sample of 40 (U)LIRGs. To do so, we combine proprietary and archival single dish (APEX) and interferometric (ACA, ALMA, and IRAM PdBI) observations. The final, reduced spectra employed in our analysis are all shown in Appendix~\ref{sec:appendix_spectra}, Figures~\ref{fig:spectra1} to \ref{fig:spectra6}.

In Table \ref{table:data_available} we report all the data sets considered in this paper with their respective project IDs. In those cases where multiple spectra are available for the same source and transition, we report at the top of the corresponding row in Table~\ref{table:data_available} the dataset that was used for our main analysis, followed by the one(s) that are not employed in the analysis. In such cases of duplication, we assign higher priority to datasets with the highest sensitivity to large-scale structures, i.e.: 1) APEX PI data, 2) APEX archival data, 3) ACA archival data and, lastly, 4) ALMA archival data. 
In this way, we prioritize single-dish data that better trace the total flux, including possible extended emission that can be filtered out by interferometric observations. If, for a given transition, single-dish data exist but are of poor quality, i.e., have a low S/N or are affected by instrumental issues, we prefer the ACA or ALMA data when available for the same transition, after carefully checking that there is no significant flux loss. 
The ALMA/ACA archival data used here are not tailored to the aim of our study, and therefore, the angular resolutions and maximum recoverable scales (MAS) of the interferometric observations vary over a wide range, from a fraction of arcsec (in the most extended ALMA antenna configurations), up to $\sim 40$ arcsec (in the ACA antenna configurations) for the angular resolution, and a few arcsec up to $\sim 90$ arcsec for the MAS. We do, however, pay special attention to define an aperture for extracting the total flux that is equal to or smaller than the MAS of the observation.
The duplicated spectra that were discarded from our main analysis, have been nevertheless reduced and are shown in Appendix~\ref{sec:appendix_duplication_spectra} (Figures~\ref{fig:spectra_dupli_co21}, ~\ref{fig:spectra_dupli_co32}, and ~\ref{fig:spectra_dupli_ci10}). 

As summarized in Table \ref{table:data_available}, we have CO(1--0) line spectral data for 22 galaxies (20 ULIRGs and 2 LIRGs), where 21 datasets are obtained from the ALMA/ACA data archive, and one from the IRAM PdBI (analysed by \citet{Cicone+14}). CO(2--1) line observations are available for the whole sample, i.e., for 40 sources (36 ULIRGs and 4 LIRGs); of these, 32 galaxies were observed with APEX through PI observations, 7 have archival APEX data, and 6 have ALMA/ACA archival data. 
As many as 31 galaxies have CO(3--2) line coverage (30 ULIRGs and only one LIRG): 18 sources have APEX PI data, 16 have APEX archival data and 12 have ALMA/ACA archival data.
Lastly, we have APEX PI observations of the [CI](1--0) line for 17 galaxies of the sample (14 ULIRGs and 3 LIRGs), one of which resulted in a non-detection (the LIRG PG2130+099). For 7 of these sources there are also ACA archival [CI](1--0) data.
Summarizing, we cover all three CO transitions for 18 galaxies (45\% of the sample), two CO transitions for 17 galaxies (42\% of the sample), and a single CO transition (CO(2--1)) for the remaining 5 galaxies (13\% of the sample). Additionally, we probe the [CI](1--0) emission line for 16 galaxies (40\% of the sample), and have an [CI](1--0) upper limit for 1 additional target. In Appendix~\ref{sec:appendix_duplication_spectra} we discuss specific instances where additional archival data were available but have been discarded in our analysis because of poor quality and unreliable fluxes. 

In the following, we describe the data reduction and analysis procedure in more detail, separately for the single-dish and interferometric data.

\subsection{APEX}

\subsubsection{Observations}\label{sec:APEXdata}

The APEX PI CO(2--1) observations for 32 sources of our sample (see Table \ref{table:data_available}) were conducted  between August and December 2019 (project ID \texttt{E-0104.B-0672}, PI: C. Cicone). Our observing strategy was to reach a line peak-to-rms ratio of S/N $> 5$ on the expected CO(2--1) peak flux density in velocity channels $\delta v \sim 5-50$ \kms.
The observations of the CO(2--1) line ($\nu^{rest}_{\textrm{CO(2-1)}}$ = 230.538 GHz) were performed with the receivers SEPIA180 and PI230 (similar frequency coverage as the ALMA Band 5 and 6 receivers), depending on the target's redshift. Both PI230 and SEPIA180 are frontend heterodyne with dual-polarization sideband-separating (2SB) receivers. The instrument PI230 can be tuned within a frequency range of $195-270$ GHz with an IF coverage of 8 GHz per sideband and with 8 GHz gap between the sidebands. The backends are fourth-Generation Fast Fourier Transform Spectrometers (FFTS4G) that consists of two sidebands, upper and lower, of 4 GHz (2x4 GHz bandwidth), which lead to the total bandwidth of $\Delta \nu = 8 \textrm{ GHz}$ \footnote{\href{https://www.eso.org/sci/facilities/apex/cfp/cfp104/recent-changes.html}{https://www.eso.org/sci/facilities/apex/cfp/cfp104/recent-changes.html}}.
The instrument SEPIA180 covers a frequency range from $159-211$ GHz. For this instrument, the backends are the eXtended bandwidth Fast Fourier Transform Spectrometers (XFFTS), which also consist of two sidebands, upper and lower, each covering 4-8 GHz, for a total of $\Delta \nu = 16 \textrm{ GHz}$ IF bandwidth. Both receivers have an average noise temperature ($T_{\textrm{rx}}$) of $\sim 55$ K \citep{Belitsky+18}.

Each sideband spectral window covered 4 GHz and was divided into 65536 (64k) channels, resulting in a resolution of $\sim 61\textrm{kHz}$ which corresponds to $\sim 80-95 \textrm{ m s}^{-1}$ in velocity units at the range of redshifts covered by our sample. The CO(2--1) emission line was placed in the lower sideband (LSB) and the telescope was tuned to the expected CO(2–1) observed frequency for each source computed by using previously known optical redshifts (see Table \ref{table:source_list}).
All our PI observations were performed in the wobbler-switching symmetric mode with $60''$ chopping amplitude and a chopping rate of $R=0.5 \textrm{ Hz}$. The data were calibrated using standard methods. The on-source integration times (without overheads) varied from source to source and were calculated using the APEX Observing Time Calculator tool\footnote{\href{https://www.apex-telescope.org/heterodyne/calculator/}{https://www.apex-telescope.org/heterodyne/calculator/}}, and are reported in Table \ref{tab:properties_observations}. During the observing runs, the PWV varied from $0.8 < {\rm PWV}\textrm{ [mm]} < 3$.

APEX PI CO(3--2) observations were obtained between October and December 2020 (project ID \texttt{E-0106.B-0674}, PI: I. Montoya Arroyave). In this case, our observing strategy was to reach a line peak-to-rms ratio of S/N $\sim 7$ on the expected CO(3--2) peak flux density in velocity channels $\delta v \sim 50-100$ \kms. These CO(3--2) observations ($\nu^{rest}_{\textrm{CO(3-2)}}$ = 345.796 GHz) were carried out with the SEPIA345 receiver, which has a frequency coverage similar to ALMA Band 7. The instrument SEPIA345, similar to SEPIA180, is a frontend heterodyne with dual-polarization 2SB receiver and works with an XFFTS backend. It can be tuned within a frequency range of $272-376$ GHz and it has two IF outputs per polarization (two sidebands: USB and LSB), each covering 4-12 GHz leading up to a total of up to $\Delta \nu = 32$ GHz IF bandwidth \citep{Meledin+22}. 
Each sideband spectral window covered 8 GHz and was divided into 4096 channels. Initially, the requested resolution was for 65536 (64k) channels ($\sim 122$ kHz per channel), corresponding to $\sim107-126 \textrm{ m s}^{-1}$, however, due to the necessity of performing remote operations during the pandemic and to the limited band for transferring data, we applied a spectral binning at the acquisition stage so that the data could be transferred quickly to Europe after acquisition. This however did not affect the scientific output, since these extragalactic targets are characterized by broad emission lines and the new spectral resolution of of $\sim 1953\textrm{ kHz}$ ($\sim 1.7 -  2 \textrm{ km s}^{-1}$) was still very high for our science goals. 
The CO(3--2) emission line was placed in the LSB, and the tuning frequency was the expected CO(3--2) observed frequency plus 2 GHz: by doing so, we centered the line at IF = 8 GHz (center of sideband), rather than at IF = 6 GHz (center of backend unit) in order to have better sampling of the baselines on both sides of the line. Observations were performed in the wobbler-switching symmetric mode with $100''$ chopping amplitude and a chopping rate of $R=0.5 \textrm{ Hz}$, and we adopted standard calibration. The on-source integration times (calculated similarly as for the CO(2--1) observations) are reported in Table \ref{tab:properties_observations}. During the observing runs, the PWV varied from $0.7 < {\rm PWV}\textrm{ [mm]} < 2.5$ .

The [CI](1--0) APEX PI observations were obtained between October 2020 and June 2021 (project ID \texttt{E-0104.B-0672}, PI: C. Cicone). Our observing strategy for the atomic carbon line was to reach a S/N peak-to-rms of $\sim 5$ on the expected [CI](1--0) peak flux density in velocity channels $\delta v \sim 25-100$ \kms. The expected [CI](1--0) line flux was conservatively estimated by assuming $L^{'}_{\rm [CI](1-0)}/L^{'}_{\rm CO(1-0)} = 0.2$, i.e., the lowest value observed in NGC~6240 by \cite{Cicone+18}, which turned out to be a reasonable assumption.
The [CI](1--0) line observations ($\nu^{rest}_{\textrm{[CI](1-0)}}$ = 492.161 GHz) were carried out with nFLASH460, which covers a similar frequency range as the ALMA Band 8 receiver. The instrument nFLASH460 is a frontend heterodyne with dual-polarization 2SB receiver with instantaneous coverage in 2 bands (USB and LSB) of 4 GHz each, where the separation between the center of the two sidebands is 12 GHz. It covers the frequency window between 378 and 508 GHz, and works with a FFTS backend in each sideband. 
For our observations, each sideband spectral window covered 4 GHz and were divided into 65536 (64k) channels ($\sim 61$ kHz per channel), corresponding to a resolution of $\sim 37-44$ m s$^{-1}$ in velocity units at the range of redshifts covered by our sample. The telescope was tuned to the expected [CI](1--0) observed frequency for each source, with wobbler-switching symmetric mode with $60''$ chopping amplitude and a chopping rate of $R=0.5 \textrm{ Hz}$. The data were calibrated using standard procedures. The observing times, computed similarly as for CO(2--1) and CO(3--2), are reported in Table \ref{tab:properties_observations}.
During the observing runs, the PWV varied from $0.3 < {\rm PWV}\textrm{ [mm]} < 1.0$.

Additionally, we used APEX archival CO(2--1) and CO(3--2) data for part of the sample, from different projects with observing dates ranging from 2010 to 2017, using the SHeFI and nFLASH receivers. All project codes of the archival datasets used throughout this work are also reported in Table \ref{table:data_available}. 
The APEX archival CO(2--1) observations used in this paper have an average S/N $\sim$ 5, while APEX CO(3--2) archival observations reach an average S/N $\sim 3-5$ (computed peak-to-rms with $\delta v$ $\sim$ 50 \kms channels).

For all single-dish data, PI and archival, we adopted the same reduction and analysis steps, which are described below in Section \ref{subsubsec:datareduction}. 

\subsubsection{Data reduction}\label{subsubsec:datareduction}
For the reduction of the APEX datasets we used the \texttt{GILDAS/CLASS} software package\footnote{https://www.iram.fr/IRAMFR/GILDAS/}, and applied the following steps to all science targets:
\begin{enumerate}
	\item After collecting all spectral scans of interest for a given target and transition (which could have different observing dates), we checked all scans individually, and discarded those affected by baseline instabilities and instrumental features following a similar procedure to \citet{Cicone+17}. At the time of our APEX PI CO(2--1) observations, we found that the receiver SEPIA180 had slightly more stable baselines than PI230. For PI230, we verified that one of the polarization windows was heavily affected by standing waves, which led to discarding an average of 40\% of subscans in that polarization. In the worst cases, this polarization window had to be discarded completely, therefore effectively cutting the integration time by half. For the SEPIA180 instrument, the average percentage of discarded scans was $\sim20$\%. For the PI SEPIA345 CO(3--2) observations, the average percentage of discarded scans was 25\% (an additional flagging was performed at the edge of the window to compensate for platforming issues). The nFLASH460 [CI](1--0) observations were heavily affected by instrumental features and/or sky-lines (more common at these high frequencies) and therefore the average fraction of discarded scans was $\sim 50\%$ for most sources.
	\item We then collected all the selected subscans corresponding to a given source and transition, and fitted and subtracted a linear baseline from each subscan after masking the central $v \in (-500, 500)$ \kms~in order to avoid the expected line emission. We then averaged together all baseline-subtracted scans to produce a high S/N spectrum for each source.
	\item We smoothed the combined spectrum to a common velocity bin of $\delta v \approx 50$ \kms. We fitted and subtracted a final linear baseline by using the same masking of the central $v \in (-500, 500)$ \kms. We fitted a single Gaussian function to the spectrum and, based on the result, we refined the central masking adjusting it to the width of the detected line emission. For the non-detections, we kept the initial mask ($v \in (-500, 500)$ \kms). These spectra were binned to $\delta v \approx 50$ \kms resolution to compute the rms values reported in Table \ref{tab:properties_observations}.
	\item Lastly, we produced the final spectrum to be used in the spectral analysis. For all the galaxies, the spectrum was extracted with a resolution of $\delta v \approx 5$ \kms, in order to allow for further smoothing, if needed, in the spectral fitting stage. The final spectrum was imported into Python\footnote{https://www.python.org}, where we performed the remaining analysis.
\end{enumerate}

The output signal from the APEX telescope corresponds to the antenna temperature corrected for atmospheric losses, $T_{A}^{*}$ in [K], and it must be multiplied by a calibration factor (or telescope efficiency) in order to obtain the flux density in units of Jansky [Jy]. Both for PI230 and SEPIA180 the average Kelvin to Jansky conversion factor measured during our observation period is $\sim 36 \pm 3$ $\mathrm{Jy~K^{-1}}$, for SEPIA345 it is $\sim 37.5 \pm 3$ $\mathrm{Jy~K^{-1}}$ and for nFLASH460 it is $\sim 58 \pm 5$ $\mathrm{Jy~K^{-1}}$. For the archival data, the calibration factor used for CO(2--1) observations was $\sim 39 \pm 5$ $\mathrm{Jy~K^{-1}}$ and $\sim 46.5 \pm 7$ $\mathrm{Jy~K^{-1}}$ for CO(3--2) observations.

\subsection{ALMA and ACA}\label{subsubsec:ALMAdata}
The archival data used here have observing dates ranging from 2012 to 2018. They use different antenna configurations of the 12-m (ALMA) and 7-m (ACA) arrays, corresponding to different angular resolutions and different maximum recoverable scales (MAS). All the project IDs used in this work are reported in Table \ref{table:data_available}.
We performed calibration and imaging using the \emph{Common Astronomy Software Applications} package\footnote{https://casa.nrao.edu} (CASA, hereafter). 
The calibrated measurement sets (MS) of datasets older than 2018 were provided by the ESO ALMA helpdesk. For the newer data obtained after 2018, we retrieved the MS by running the CASA pipeline (version \texttt{5.6.1}) and executing the calibration script provided with each corresponding dataset.

We then analysed the MS within \texttt{CASA 5.6.1} and separated the spectral windows including the lines of interest (i.e., CO(1--0), CO(2--1), CO(3--2) or [CI](1--0), depending on the data set), using the task \texttt{split}. 
A first deconvolution and cleaning were performed in interactive mode with the task \texttt{tclean}, by adapting the mask to the source size. This first clean provided us with an initial datacube that we used to identify the line-free channels for continuum subtraction, and to optimize the parameters for the final clean.
We ran the cleaning process until reaching uniform residuals, and produced an image of the source in which we measured the noise level. 
We then performed the continuum subtraction using the task \texttt{uvcontsub}, by fitting a first-order polynomial and estimating the continuum emission in the line-free frequency ranges previously identified.
We produced the final data cube using once again the task \texttt{tclean} on the continuum-subtracted MS file. 
We constructed all image cubes with the highest spectral resolution available, ranging from $\Delta v\sim$1 to $\sim$10 \kms, depending on the dataset. 
Final cubes are obtained using Briggs weighting with robust parameter equal to 0.5 and primary-beam corrected. 
We extracted the final spectrum from a circular aperture that is size-matched to maximize the recovered flux. The apertures used for spectra extraction are reported in Table \ref{tab:properties_observations}. 
For the sources where only high-resolution ALMA observations were available (IRAS F01572+0009 and IRAS F12072-0444), we applied an {\it uv} tapering to enhance the sensitivity to extended structures. This however does not overcome the possible issue of missing flux from faint extended structures due to poor sampling of short {\it uv} baselines.

The spectra are exported from {\it CASA} in flux density [Jy] units, extracted in suitable format and imported into Python for further analysis. The quoted errors refer to the systematic errors on the absolute flux calibration of ALMA/ACA data (estimated to be 5\% for Band 3 data and 10\% for Bands 6-8 data, and typically the dominant source of error in the data used in this study), added in quadrature to the statistical RMS of the spectra.

\section{Methodology}\label{sec:methodology}

\subsection{Spectral line fitting}\label{sec:spectralfitting}

Our analysis is aimed at deriving source-averaged line ratios for different kinematic components of the molecular and atomic ISM in local (U)LIRGs. We also want to investigate possible  statistical trends between the molecular (CO and [CI]) line ratios as a function of the central velocity $v$ and line width $\sigma_v$ of the different components, to understand whether broader/higher-$v$ H$_2$ gas components are the origins for the extremely high {\it global} CO excitation previously found in (U)LIRGs \citep[e.g.,][]{Papadopoulos+12a}, which \cite{Cicone+18} suggested based on their pilot study on NGC~6240.

We base the analysis reported in this paper exclusively on total (i.e., galaxy-integrated) molecular line spectra, and on the results of a multi-Gaussian spectral fitting of the CO and [CI](1--0) lines. 
We acknowledge that, without spatially-resolved information, it is not possible to link in a  straightforward way the different (e.g., broad/narrow) spectral line components to outflowing/non-outflowing gas. However, performing such classification is not needed in our case, because we can rely on a large, statistically significant sample. Indeed, we are interested in studying in a statistical sense any trends observed between molecular line ratios and the central velocity ($v$) and/or velocity dispersion ($\sigma_v$) of the different spectral components. Once the presence of statistical correlations is assessed (independently of any arbitrary classification of such components in terms of outflow or disk), we can interpret the results by assuming that, in this sample of (U)LIRGs, the line luminosities of high-$\sigma_v$ and/or high-$v$ spectral components are more likely to be dominated by molecular gas embedded in outflows compared to the low-$\sigma_v$ and low-$v$ components. Such assumption would be supported by (i) the results of the pilot study on NGC~6240 see \citep[see][]{Cicone+18}, (ii) the unambiguous detection of OH$119$ $\mu$m outflows in most targets \citep{Sturm+11,Veilleux+13,Spoon+13}, and (iii) the widespread evidence for high-velocity molecular outflows in local (U)LIRGs reported in the recent literature (see review by \cite{Veilleux+20}).

After having clarified our strategy, we now describe our spectral fitting procedure. We used \texttt{mpfit} for Python\footnote{Open-source algorithm adapted from the IDL version, see \href{https://github.com/segasai/astrolibpy/blob/master/mpfit/}{https://github.com/segasai/astrolibpy/blob/master/mpfit/}}.
This tool uses the Levenberg-Marquardt technique to solve the least-squares problem in order to fit a user-supplied function (the \textit{model}) to the user-supplied data points (the \textit{data}). 
Spectral lines were modeled using single or multiple Gaussian profiles characterized by an amplitude, a peak position ($v_{cen}$), and velocity dispersion ($\sigma_v$, or, equivalently, FWHM). 
Having multiple CO transitions allows us to partially break any degeneracy in the spectral line decomposition with Gaussian functions. We therefore fitted simultaneously all CO transitions available for each source, by constraining the $v_{cen}$ and $\sigma_v$ of the Gaussian components to be equal in all CO transitions, allowing only their amplitudes to vary freely in the fit.
We allowed the fit to use up to a maximum of three Gaussian functions to reproduce the observed global line profiles, so that line asymmetries and broad wings are properly captured with separate spectral components when the S/N is high enough (see for example IRAS~13120-5453 in Figure~\ref{fig:spectra2}).
For all the fits, we verified using a reduced $\chi^2$ criterion that a fourth Gaussian component was not required for any of the sources, with the data at hand. In those cases where the statistical criterion (reduced $\chi^2$) does not indicate a clear preference between a fit performed with 1, 2, or 3 Gaussians, we performed a visual inspection of the fit. For the low S/N spectra (e.g. S/N $\sim$ 3) without clear asymmetries, we used only 1 Gaussian component, in order not to over-fit the data.
 
The fitting procedure for the [CI](1--0) line spectra is carried out separately from the CO lines due to their overall lower S/N. Furthermore, by doing so, we can avoid assuming a priori that the atomic Carbon and CO lines trace the same gas clouds and share the same kinematics. This assumption will be tested and discussed in Section~\ref{sec:sigmas}.

The final fits for all sources and transitions are shown in Figures~\ref{fig:spectra1} to \ref{fig:spectra6}. We color-coded the spectral transitions with \textit{red} for CO(1--0), \textit{green} for CO(2--1), \textit{blue} for CO(3--2), and \textit{magenta} for [CI](1--0).
Based on the fit results, we computed velocity-integrated line fluxes for both the individual Gaussian components and the entire line profiles, and the latter are reported in Table~\ref{tab:fluxlumino}. As a sanity check, we verified that the total line fluxes measured through the fit (by adding up the individual Gaussians) are consistent with the total line fluxes calculated by directly integrating the spectra within $v \in (-1000, 1000)$ \kms, after setting a threshold of $>2\sigma$ for each channel. We find the values to be consistent within the errors, and therefore using either value will not affect the analysis performed throughout our work.

\subsection{Line luminosities and ratios}\label{sec:linelumrat}

We calculated the CO and [CI] line luminosities from the integrated line fluxes following the definition of \citet{Solomon+97}:
\begin{equation}
	L{'}_{\rm line}\,\mathrm{ [K\,km\,s^{-1}\,pc^{2}]} = 3.25 \times 10^7\frac{D^{2}_{L}}{\nu^{2}_{\mathrm{obs}}(1+z)^3}\, \int{S_{v}\, \mathrm{dv}}\, ,
\end{equation}
where $D_{L}$ is the luminosity distance measured in [Mpc], $\nu_{\textrm{obs}}$ is the corresponding observed frequency in [GHz], and $\int S_{v}\,\mathrm{dv}$ is the total integrated line flux in [Jy~\kms]. In Table \ref{tab:fluxlumino} we report the total integrated fluxes and respective luminosities calculated for the different transitions.
The CO line ratios are defined as
\begin{equation}
	\begin{split}
		r_{21}=&L{'}_{\mathrm{CO(2-1)}}/L{'}_{\mathrm{CO(1-0)}}\\
		r_{31}=&L{'}_{\mathrm{CO(3-2)}}/L{'}_{\mathrm{CO(1-0)}}\\
		r_{32}=&L{'}_{\mathrm{CO(3-2)}}/L{'}_{\mathrm{CO(2-1)}}.
	\end{split}
\end{equation}
In our analysis we will use both the global CO line luminosity ratios as well as those computed for individual Gaussian components. 
Additionally, we calculated global [CI](1--0)/CO(1--0) line luminosity ratios, as:
\begin{equation}
	r_{\mathrm{CICO}}= L{'}_{\mathrm{[CI](1-0)}}/L{'}_{\mathrm{CO(1-0)}}.
\end{equation}
Line ratios were computed for all combinations of lines and sources where the individual line luminosity measurements pass a loose criterion of S/N>1, in order not to penalize cases where one of the two lines is constrained at very high significance. As a result, some line ratios have very large error-bars, which are taken into account in our analysis. 

\section{Results}\label{sec:results}

\subsection{Atomic carbon as an alternative gas tracer}

\subsubsection{A tight relation between CO(1--0) and [CI](1--0) luminosities}\label{sec:linear_relation_lum}

\begin{figure}[tbp]
	\centering
	\includegraphics[width=0.45\textwidth]{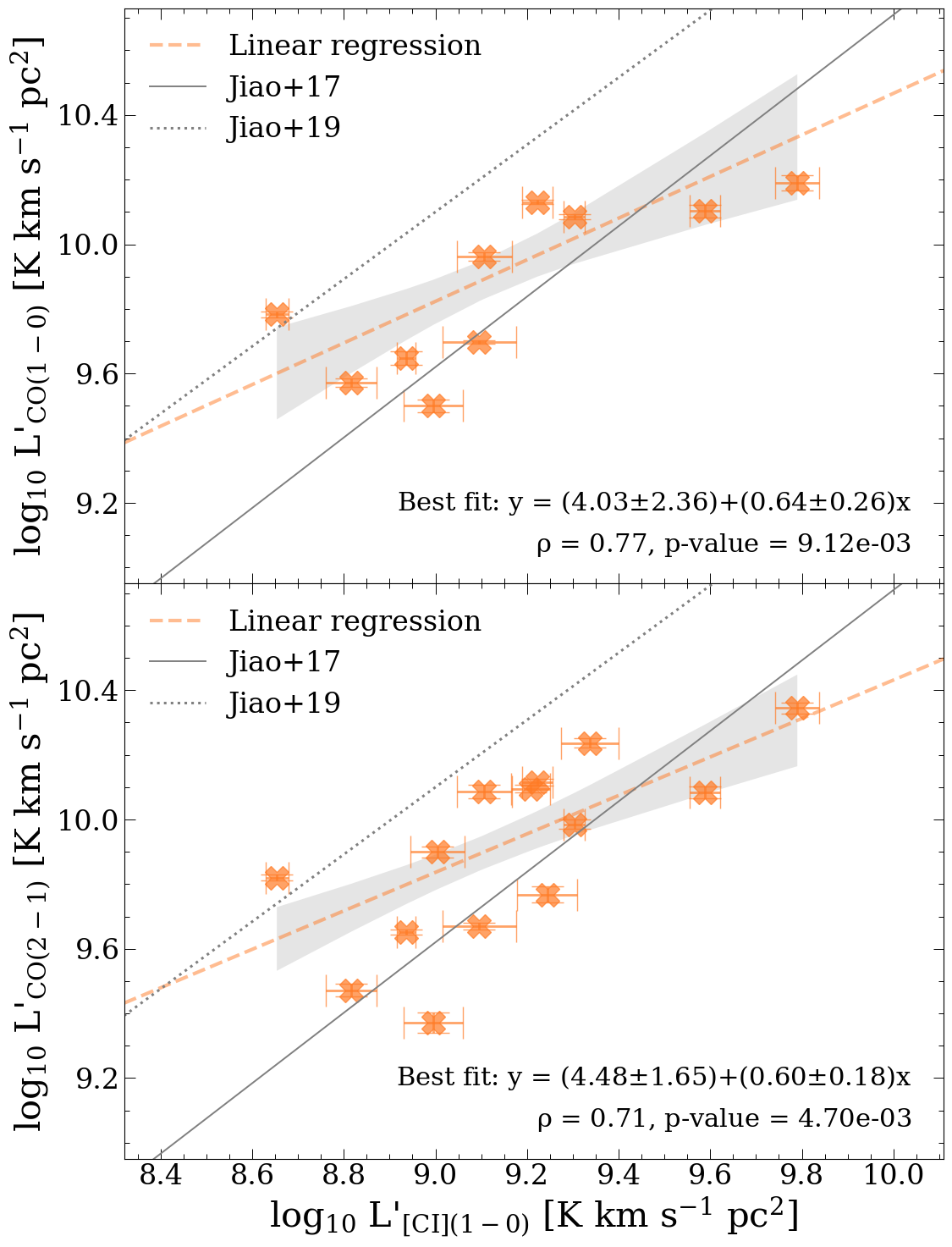}
	\caption{CO(1--0) vs. [CI](1--0) luminosity (\textit{top}) and CO(2--1) vs. [CI](1--0) luminosity (\textit{bottom}) for the ULIRGs in our sample. The best fit relations are shown as \textit{orange dashed} lines. The best-fit parameters are reported at the bottom right corner of the plots. We also display the Pearson correlation coefficients ($\rho$) and their associated {\it p}-values. The \textit{solid black lines} in both panels represent the corresponding relations reported by \citet{Jiao+17} for a sample of 71 (U)LIRGs, and the \textit{dotted lines} represent the relations of \citet{Jiao+19} for a sample of 15 nearby spiral galaxies, between $L'_{\textrm{CO(1-0)}}$ and $L'_{\textrm{[CI](1-0)}}$.}
	\label{fig:COlumCIlum}
\end{figure}

In Figure~\ref{fig:COlumCIlum} we plotted the measured total CO(1--0) and CO(2--1) line luminosities as a function of [CI](1--0) line luminosity, for the ULIRGs in our sample. 
The relation with CO(3--2) was not studied as this line starts to trace denser and more excited H$_2$ gas rather than the global molecular reservoir, while we are interested in exploring the potential of [CI](1--0) to probe similar regions as the CO J$_{\textrm{up}}=2,1$ transitions.

The plots in Figure \ref{fig:COlumCIlum} show that $L'_{\textrm{CO(1--0)}}$ and $L'_{\textrm{CO(2--1)}}$ are both tightly correlated with $L'_{\rm [CI](1-0)}$, showing Pearson correlation coefficients ($\rho$) equal to 0.77 and 0.71, and {\it p}-values of $9.1\times10^{-3}$ and $4.7\times10^{-3}$, respectively for $L'_{\textrm{CO(1--0)}}$ and $L'_{\textrm{CO(2--1)}}$. We performed a fit to the relation between $L'_{\textrm{CO(1--0)}}$ and $L'_{\textrm{[CI](1--0)}}$ using least squares, which gives:
\begin{equation}
	\textrm{log}_{10}\,L'_{\textrm{CO(1--0)}} = (4.03\pm2.36)+(0.64\pm0.26)\, \textrm{log}_{10}\,L'_{\textrm{[CI](1--0)}}.
\end{equation}
We performed a similar fit with the CO(2--1) line, whose results are reported on the corresponding plot.
We find that $L'_{\textrm{CO(1--0)}}$ and $L'_{\textrm{CO(2--1)}}$ follow very similar relations as a function of $L'_{\textrm{[CI](1--0)}}$, with variations in the best-fit parameters within one standard deviation.

In Figure~\ref{fig:COlumCIlum}, we also report (solid black line) the relation found by \cite{Jiao+17} in a study of unresolved neutral carbon emission in a sample of 71 (U)LIRGs based on {\it Herschel} observations, for which they derive a best-fit relation of log$_{10}\,L'_{\textrm{CO(1--0)}} = (-0.19\pm1.26)+(1.09\pm0.15)$ log$_{10}\,L'_{\textrm{[CI](1--0)}}$. In a similar study performed on 15 nearby spiral galaxies with spatially-resolved {\it Herschel} data, \cite{Jiao+19} obtain: log$_{10}\,L'_{\textrm{CO(1--0)}} = (0.74\pm0.12)+(1.04\pm0.02)$ log$_{10}\,L'_{\textrm{[CI](1--0)}}$ (reported with a dotted black line).
Our best-fit $L'_{\textrm{[CI](1--0)}}$ vs $L'_{\textrm{CO(1--0)}}$ relation has a flatter slope than the one found by \cite{Jiao+17}, likely due to our sample only covering a narrower dynamic range in luminosities than the sample in \cite{Jiao+17}, while our sources are exclusively ULIRGs, The \cite{Jiao+17} sample is heavily dominated by LIRGs (62 LIRGs and only 9 ULIRGs). When we explore the same relations in our extended sample (including the LIRGs), shown in Figure \ref{fig:COlumCIlum_LIRG}, we find that the best-fit relation between $L'_{\textrm{[CI](1--0)}}$ and $L'_{\textrm{CO(1--0)}}$ ($\textrm{log}_{10}\,L'_{\textrm{CO(1--0)}} = (-0.45\pm0.73)+(1.13\pm0.08)\, \textrm{log}_{10}\,L'_{\textrm{[CI](1--0)}} $)
is almost linear and well in agreement with that obtained by \citet{Jiao+17} and clearly shifted to higher $L'_{\textrm{[CI](1--0)}}/L'_{\textrm{CO(1--0)}}$ ratios with respect to the \cite{Jiao+19} fit performed on non-IR luminous local galaxies. Such difference in $L'_{\textrm{[CI](1--0)}}/L'_{\textrm{CO(1--0)}}$ ratios between (U)LIRGs and other galaxies will be further explored in Section~\ref{sec:tot_CICOratios}.

The tight correlations in Fig.~\ref{fig:COlumCIlum} suggest that the CO(1--0) and CO(2--1) lines arise from similar regions as the [CI](1--0) emission, at least when averaged over galactic scales, and strengthen the hypothesis (see, e.g., \citealt{Papadopoulos+04}) that the [CI](1--0) line is an excellent molecular gas tracer, and a valid alternative to low-\textit{J} CO line emission.

\subsubsection{Comparison between CO and [CI] line widths}\label{sec:sigmas}

\begin{figure}[tbp]
	\centering
	\includegraphics[width=0.42\textwidth]{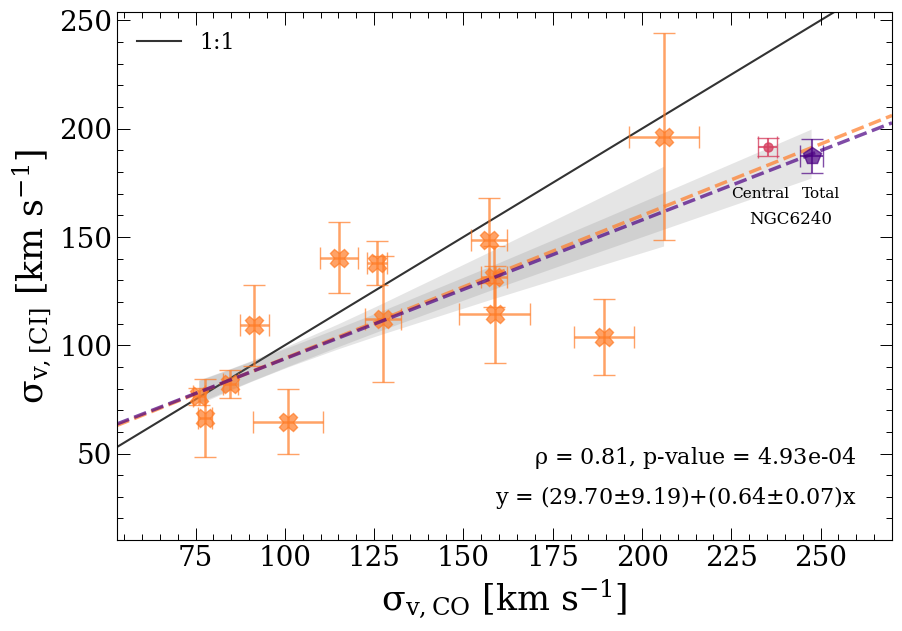}\\
	\includegraphics[width=0.42\textwidth]{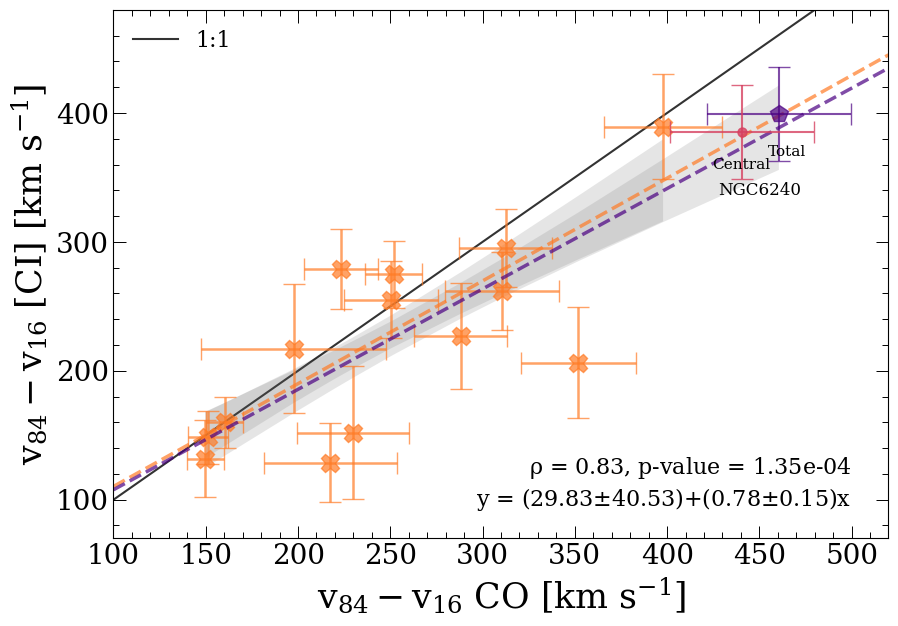}\\
    \includegraphics[width=0.42\textwidth]{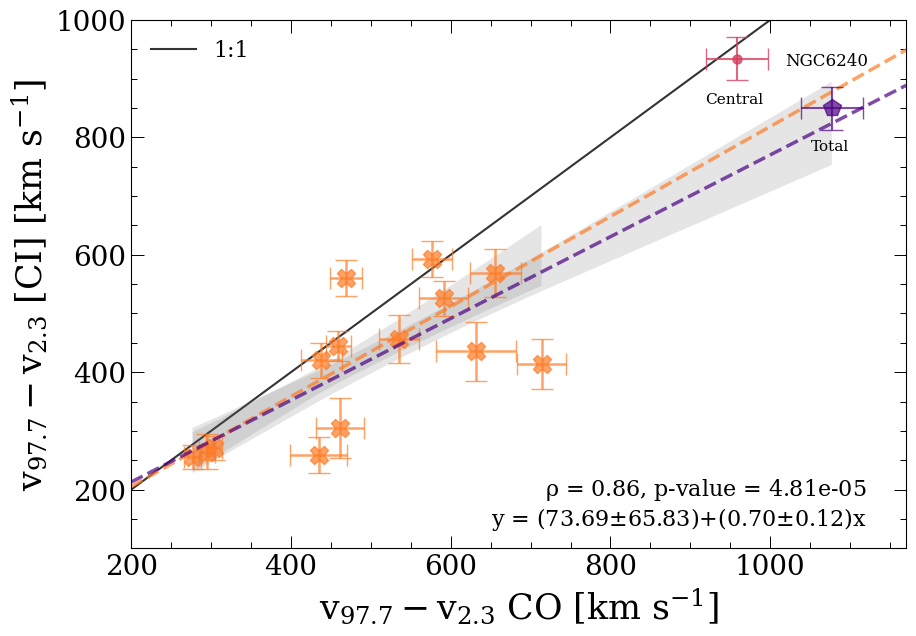}
	\caption{[CI](1--0) line width as a function of the CO(2--1) line width. \textit{Top panel:} velocity dispersion ($\sigma_{\rm v}$) obtained via a single Gaussian spectral fit to the [CI](1--0) and CO(2--1) emission lines. \textit{Middle} and \textit{bottom panels:} 16th-84th (${\rm v_{84}}-{\rm v_{16}}$) and 2.3nd-97.7th (${\rm v_{97.7}}-{\rm v_{2.3}}$) percentile velocity intervals for CO and [CI], respectively, derived from the best-fit function obtained from the multi-Gaussian fit.
    Orange data points correspond to the ULIRGs sample analyzed in this work. The purple pentagon and the pink dot represent the values obtained for NGC~6240, respectively for the total region and for the inner $\sim2''$ \citep[see][and the text in Section \ref{sec:sigmas} for a more detailed explanation of the apertures used for spectra extraction]{Cicone+18}.
	The \textit{black solid} line indicates the 1:1 relation. The \textit{orange dashed} line shows the best-fit relation obtained using only our sample, while the \textit{purple dashed} line is the best-fit relation obtained for our sample plus NGC6240 (total). The shaded gray areas corresponds to the $1\sigma$ confidence interval of the two fits.
	On the bottom right corner we report the Pearson correlation coefficient ($\rho$) and the {\it p}-values, as well as the best-fit coefficients of the purple fit. 
    }
	\label{fig:COwid_CIwid}
\end{figure}

In the previous section we have found a tight relation between the [CI](1--0) and CO total line luminosities for ULIRGs (which becomes almost linear when expanding the dynamic range in luminosity values by including the LIRGs). Here we test whether the two tracers share the same kinematics. 
We compared the line widths using the CO(2--1) and the [CI](1--0) lines. We preferred CO(2--1) over CO(1--0) to maximize the sample size, since CO(2--1) spectra are available for all 16 sources with a [CI](1--0) detection. 

We used two different approaches to study the line widths of the two tracers. Firstly, we performed a dedicated, single Gaussian spectral fit, run independently for each line.
Secondly, we computed the 16th-84th (${\rm v_{84}}-{\rm v_{16}}$) and 2.3nd-97.7th (${\rm v_{97.7}}-{\rm v_{2.3}}$) percentile velocity intervals, derived from the analytical form of the overall best-fit line profile obtained by a multi-Gaussian fit. The latter approach is likely a more robust method when dealing with complex line profiles, as is the case for some of the sources in our sample, e.g. IRAS 19254--7245.

The results are shown in Figure \ref{fig:COwid_CIwid}. The plot of $\sigma_{\textrm{v,[CI]}}$ versus $\sigma_{\textrm{v,CO}}$ obtained through a single Gaussian fit is plotted on the \textit{top-panel} of Fig.~\ref{fig:COwid_CIwid}, while the percentile velocity plots obtained using the second approach are shown in the bottom panels of Fig.~\ref{fig:COwid_CIwid}.
We find that the sources characterized by broader profiles sit preferentially below the 1:1 relation (black solid line). The best-fit relations obtained through a least squares regression analysis, indicated by the orange dashed lines in all three plots have a slope below unity. 

As an additional test, we have over-plotted on Fig.~\ref{fig:COwid_CIwid} the values obtained for NGC~6240, which is a source characterized by an extremely turbulent ISM, strongly affected by outflows. 
The purple pentagon represents the total measurement available for NGC~6240, computed from spectra extracted from a $12''\times6''$ rectangular aperture encompassing the nuclei and the molecular outflow, while the pink dot represents the central $2''\times2''$ region \citep[for a more in depth explanation of how the apertures are defined see][]{Cicone+18}.
Quite strikingly, both NGC~6240 data points sit on the best-fit relation obtained for our sample when probing the core of the lines via $\sigma_{\rm v}$ and ${\rm v_{84}}-{\rm v_{16}}$. Instead,
as we probe more towards the high-velocity wings of the line by using, e.g., the ${\rm v_{97.7}}-{\rm v_{2.3}}$ values, the nuclear spectrum results to be more consistent with the 1:1 relation between [CI] and CO line width, while the total spectrum of NGC~6240, including the extended outflows, sits on the best-fit relation obtained from the analysis of the other sources.
The fact that the total spectrum includes more of the extended outflow than the nuclear one \citep[see][]{Cicone+18}, and is also the one that departs more from the 1:1 relation, may indicate that this deviation (i.e. a narrower width of [CI] with respect to CO) is accentuated by the inclusion of diffuse outflowing gas. 

The fit that includes the total emission from NGC~6240, displayed in the top panel of Fig.~\ref{fig:COwid_CIwid} as a purple dashed-line and consistent with the one obtained from our sample alone, is:
\begin{equation}
    \sigma_{\mathrm{v,[CI]}} = (30\pm9) + (0.64\pm0.07)~\sigma_{\mathrm{v,CO}}~,
\end{equation}
and the corresponding fit for the ${\rm v_{97.7}}-{\rm v_{2.3}}$ velocity percentiles (shown in the bottom panel) is:
\begin{equation}
    {\rm v_{97.7}}-{\rm v_{2.3}}\,{\rm [CI]} = (74\pm66) + (0.70\pm0.12)\,{\rm v_{97.7}}-{\rm v_{2.3}}\,{\rm CO}~.
\end{equation}
Therefore, our data indicate that the [CI](1--0) line is narrower than CO(2--1). 
The average line width ratio is $\langle\sigma_{\mathrm{v,[CI]}}/\sigma_{\mathrm{v,CO}} \rangle = \langle r_{\sigma}\rangle = 0.91 \pm 0.07$, computed using all sources. The ratios below unity are driven by targets with $\sigma_{\rm v,CO}>150$~\kms{}, while those with $\sigma_{\rm v,CO} \lesssim 150$ \kms, which represents the majority of our sample, are consistent with the 1:1 relation.

Few comparisons of CO and [CI] line widths can be found in the literature, and most of these previous studies report a 1:1 correspondence between the line widths. \cite{Michiyama+21}, by comparing ACA CO(4--3) and [CI](1--0) observations of a sample of 36 local (U)LIRGs, found a 1:1 relation between the FWHMs of the two transitions. However, their analysis excludes sources with complex profiles (e.g., double peak emission), which we did not do.
Similarly, \cite{Bothwell+17} analyzed ALMA [CI](1--0) emission line observations in a sample of strongly lensed dusty star-forming galaxies (DSFGs) spanning a wide redshift range of $2<z<5$, and for 11 of such sources they compared [CI] and CO(2--1) line widths, using literature CO data. Their results are consistent with a 1:1 relation.
In Section~\ref{sec:discussion} we will discuss possible explanations for the difference in CO and [CI] line widths observed in our sample and specifically in the high-$\sigma_{\rm v}$ (U)LIRGs.

\subsection{Molecular gas mass estimates and the CO-to-H$_2$ factor}\label{sec:alphaCO}

Building upon Section~\ref{sec:linear_relation_lum}, we use the [CI](1--0) and CO(1--0) emission lines to derive independent estimates of the molecular gas mass ($M_{\textrm{mol}}$) of the ULIRGs of our sample. We also use the [CI]-based $M_{\textrm{mol}}$ to derive an average value for the $\alpha_{\textrm{CO}}$ factor, similarly to \cite{Cicone+18}. 

Both tracers rely on calibration factors in order to compute $M_{\textrm{mol}}$.
For [CI](1--0)-based estimates, we need to assume the optically thin condition (which applies to most extragalactic environments), a value for the [CI] abundance with respect to H$_2$ (X$_{\textrm{CI}}=[\textrm{C}/\textrm{H}_2]$), and a value for the parameter Q$_{10}$, i.e., the [CI] excitation factor. The molecular hydrogen gas mass can then be computed, following \citet{Dunne+21}, as:
\begin{equation}\label{eq:mol_masses_CI}
		M_{\rm {mol,[CI]}} [{\rm M}_{\odot}] = 1.36~\frac{(9.51\times 10^{-5})}{\rm X_{CI}~Q_{10}}~L^{\prime}_{\rm [CI](1-0)}.
\end{equation}
For CO-based mass measurements, we need to assume an $\alpha_{\textrm{CO}}$ factor:
\begin{equation}\label{eq:mol_masses_CO}
	M_{\rm mol,CO} [{\rm M}_{\odot}] = \alpha_{\rm CO}~L^{\prime}_{\rm CO(1-0)}.
\end{equation}
Both Eq~\ref{eq:mol_masses_CI} and \ref{eq:mol_masses_CO} include the Helium contribution to the molecular gas mass through a multiplicative factor of  1.36. 

All tracers of H$_2$ are affected by uncertainties. Using CO and an arbitrary $\alpha_{\textrm{CO}}$ value can introduce large errors in the computed $M_{\mathrm{mol}}$ due to its sensitivity to metallicity in a non-linear fashion, and to the turbulence and kinematics of the CO-emitting clouds that affect the global optical depth of galaxy-averaged CO measurements. In the case of [CI], different combinations of X$_{\textrm{CI}}$ and Q$_{10}$ may yield different results; X$_{\textrm{CI}}$ may be the easiest parameter to model in terms of the ISM conditions if in fact the C$^0$ abundance is determined by cosmic rays \citep[e.g.,][]{Bisbas+15, Dunne+21}.
In the following Sections~\ref{sec:CIbased} and \ref{sec:CObased} we discuss separately the [CI]-based and CO-based $M_{\rm mol}$ estimates.

\subsubsection{[CI]-based $M_{\rm mol}$ estimates}\label{sec:CIbased}

\begin{figure}[tb]
	\centering
	\includegraphics[width=0.475\textwidth]{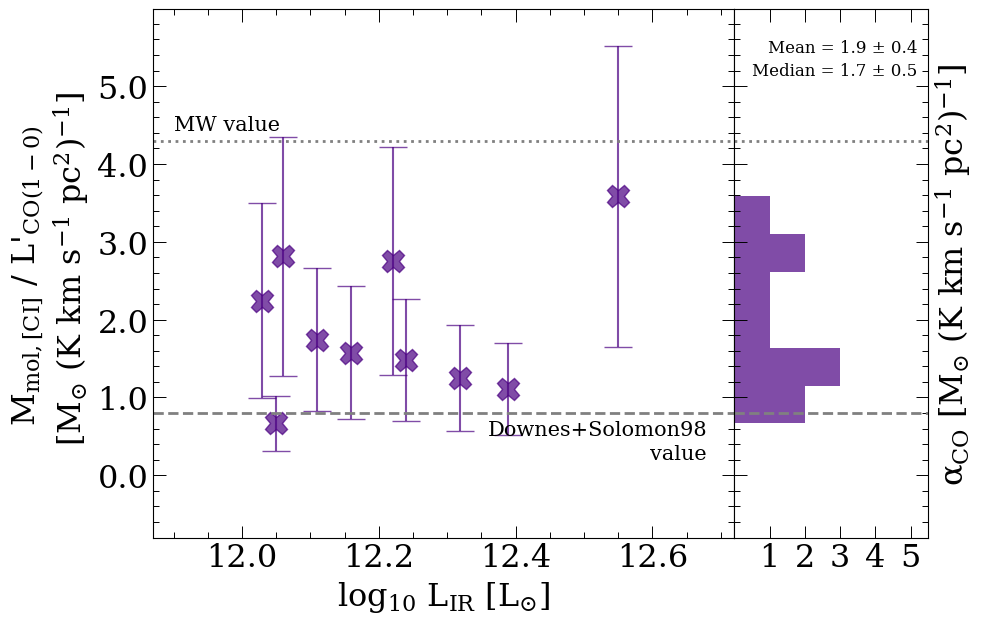}
	\caption{Ratio between [CI]-based molecular gas mass estimates and CO(1--0) line luminosity, computed for the ULIRGs in our sample that have both lines available, and plotted as a function of the infrared luminosity. This ratio is interpreted as the $\alpha_{\textrm{CO}}$ factor (see Eq.~\ref{eq:alphaCO_fromCI}). The right part of the plot shows the resulting distribution of [CI]-based $\alpha_{\textrm{CO}}$ values. The {\it dotted} line indicates the CO-to-H$_2$ conversion factor for the Milky Way galaxy, and the {\it dashed} line corresponds to the value commonly used in the literature for (U)LIRGs \citep{Downes+Solomon98}. 
	The mean and median values obtained for our sample (reported on the top right corner) are $\alpha_{\textrm{CO}}=1.9\pm 0.4$~M$_{\odot}$ and $1.7\pm 0.5$~M$_{\odot}$ (K km s$^{-1}$ pc$^2$)$^{-1}$, respectively.}
	\label{fig:aCO_histogram}
\end{figure}

We compute $M_{\mathrm{mol,[CI]}}$ using the [CI](1--0) luminosities reported in Table \ref{tab:fluxlumino} and Equation~\ref{eq:mol_masses_CI}.
We adopt a Carbon abundance of X$_{\textrm{CI}}=(3.0\pm 1.5)\times 10^{-5}$, which is an appropriate value for local star forming galaxies and has been used by several previous studies \citep[e.g.,][]{Weiss+05, Papadopoulos+04, Walter+11, Jiao+17, Cicone+18}.
For the [CI] excitation factor Q$_{10}$, we adopt a value of 0.48 (with <16\% variation), following the prescriptions by \cite{Papadopoulos+22}, who found that, for the most expected average ISM conditions in galaxies ($n_{\textrm{H}_2} = [300-10^4]$ cm$^{-3}$ and T$_{\textrm{kin}}=[25-80]$ K), the [CI] lines are \textit{globally} sub-thermally excited.

By defining a parameter $\alpha_{\textrm{[CI]}}$ analogous to $\alpha_{\textrm{CO}}$ to represent a [CI](1--0)-to-H$_2$ conversion factor, we can re-write Equation~\ref{eq:mol_masses_CI} as:
\begin{equation}
	\begin{split}
		M_{\textrm{mol,[CI]}}=&\alpha_{\textrm{[CI]}}\,L'_{\textrm{[CI](1-0)}}\\
		&\textrm{with}\, \, \alpha_{\textrm{[CI]}}=1.293\times 10^{-4} (\textrm{X}_{\textrm{CI}}\textrm{Q}_{10})^{-1}.
	\end{split}
\end{equation}
And thus, plugging in our assumptions for the X$_{\textrm{CI}}$ and Q$_{10}$ values, we obtain $\alpha_{\textrm{[CI]}}=9.0$ M$_{\odot}$ (K km s$^{-1}$ pc$^2$)$^{-1}$.

The [CI]-based $M_{\textrm{mol}}$ values can  then be used to infer $\alpha_{\textrm{CO}}$ as follows:
\begin{equation}\label{eq:alphaCO_fromCI}
	\alpha_{\textrm{CO}} = M_{\textrm{mol,[CI]}}/L'_{\textrm{CO(1-0)}}\,.
\end{equation}
The ratio $M_{\textrm{mol,[CI]}}/L'_{\textrm{CO(1-0)}}$, which in practice represents the $\alpha_{\textrm{CO}}$ factor required to force agreement between [CI]- and CO-based H$_2$ mass estimates, is plotted in Fig.~\ref{fig:aCO_histogram} against the infrared luminosity, for the 10 ULIRGs with available CO(1--0) and [CI](1--0) detections.
The distribution of the resulting $\alpha_{\textrm{CO}}$ values is also shown on the right of Fig.~\ref{fig:aCO_histogram}.

Figure~\ref{fig:aCO_histogram} demonstrates that 9 out of 10 targets require an $\alpha_{\textrm{CO}}$ value higher than the one commonly assumed for (U)LIRGs of 0.8~$\textrm{M}_{\odot}\,(\textrm{K km s}^{-1}\textrm{ pc}^{2})^{-1}$ \citep[see][]{Downes+Solomon98}. 
The mean value measured for our sample is $1.9\pm 0.4$~M$_{\odot}$, and the median value is $1.7$~M$_{\odot}$ (K km s$^{-1}$ pc$^2$)$^{-1}$, with 16$^{\rm th}$–84$^{\rm th}$ percentile equal to $1.2$ – $2.8$ (K km s$^{-1}$ pc$^2$)$^{-1}$. Our results are consistent with the dust-based $\alpha_{\textrm{CO}}$ estimate equal to $1.8^{+1.3}_{-0.8}$~M$_{\odot}$ (K km s$^{-1}$ pc$^2$)$^{-1}$) derived by \citet{Herrero-Illana+19} for 55 (U)LIRGs from the Great Observatories All-sky LIRG survey (GOALS). These authors estimated first the dust mass ($M_{\rm dust}$) from a FIR Spectral Energy Distribution (SED) fit using {\it Herschel} data, following the strategy proposed by \cite{Scoville+16} of fixing $T_{\rm dust}=25$~K for every source, and then estimated $\alpha_{\rm CO}$ by requiring that the gas-to-dust mass ratio of (U)LIRGs matches the one of local star forming spirals. Similarly, a study performed by \cite{Kawana+22} on a nearby LIRG (NGC 3110) estimated an $\alpha_{\textrm{CO}}$ value of $1.7\pm0.5$~M$_{\odot}$ (K km s$^{-1}$ pc$^2$)$^{-1}$ based on thermal dust continuum emission and assuming that the dust and the rotational temperature of the CO molecule are equal.

In the Appendix \ref{sec:appendix_LIRGs} we extend the calculations including the LIRGs in our sample with available [CI](1--0) and CO(1--0) data, i.e. 2 sources (see Figure \ref{fig:aCO_histogram_LIRGs}). The resulting mean and median values are perfectly consistent with the ones obtained in Fig. \ref{fig:aCO_histogram}.

\subsubsection{CO-based $M_{\textrm{mol}}$ estimates}\label{sec:CObased}

\begin{figure*}[tbh]
	\centering
	\includegraphics[scale=0.238]{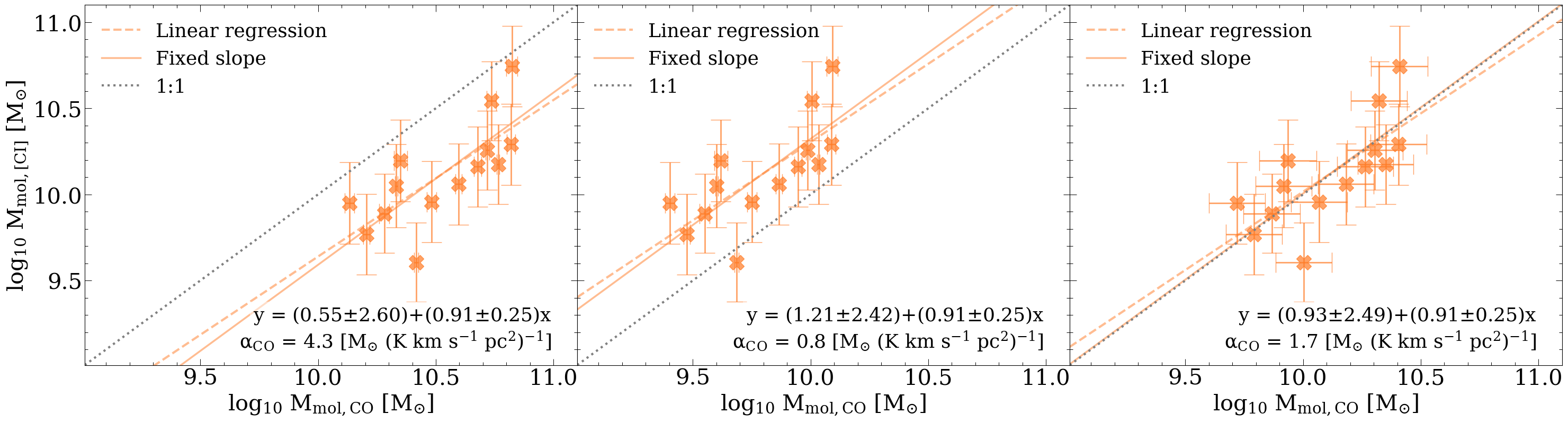}
	\caption{[CI]-based $M_{\rm mol}$ vs. CO-based $M_{\rm mol}$ values obtained with different $\alpha_{\textrm{CO}}$ assumptions ($0.8, 4.3$ and $1.7$~$\textrm{M}_{\odot}\,(\textrm{K km s}^{-1}\textrm{ pc}^{2})^{-1}$).
	The best-fit linear relation is shown as an \textit{orange dashed} line, and the best-fit parameters are reported at the bottom right corner of each plot. The \textit{solid orange} line shows the best-fit relation obtained by fixing the slope to unity, and the 1:1 relation is reported as a dotted black line.}
	\label{fig:MCI_vs_MCO_all}
\end{figure*}

We then use the lowest-\textit{J} CO transition available for each source to estimate a CO-based total molecular gas mass.
For sources without CO(1--0) data, we first estimate $L'_{\textrm{CO(1--0)}}$ based on the $L'_{\textrm{CO(2--1)}}$ transition, which is available for the whole sample. We assume a line ratio of $r_{21}=1.1\pm0.4$, which is the median value computed in this work based on the data available in our sample (see Section~\ref{sec:tot_COratios} and Figure~\ref{fig:COratios_histo}). We then proceed to compute $M_{\textrm{mol,CO}}$ using Equation~~\ref{eq:mol_masses_CO}.
We compute  $M_{\textrm{mol,CO}}$ using three different $\alpha_{\textrm{CO}}$ values, namely, $4.3$, $0.8$ and $1.7$~$\textrm{M}_{\odot}\,(\textrm{K km s}^{-1}\textrm{ pc}^{2})^{-1}$, corresponding to the value measured for the Milky Way galaxy \citep{Bolatto+13}, the value commonly employed in the past for (U)LIRGs \citep{Downes+Solomon98}, and the median value obtained for our sample using the [CI]-based method described previously (see Section \ref{sec:CIbased}). 

In Figure \ref{fig:MCI_vs_MCO_all} we show the comparison between the [CI]-based $M_{\textrm{mol}}$ values, and the CO-based $M_{\textrm{mol}}$ estimates obtained using different conversion factors. These plots report the same result as Figure~\ref{fig:aCO_histogram} (i.e., the $\alpha_{\rm CO}$ value), but visualized in a different way and including the uncertainties. For each relation we perform two linear fits, one with a free-varying slope (shown as a \textit{dashed orange} line) and another with a slope fixed to unity (\textit{solid orange} line). The 1:1 relation is also over-plotted using a \textit{dotted black} line. The plots in Figure~\ref{fig:MCI_vs_MCO_all} show clearly that an $\alpha_{\textrm{CO}}=4.3$ $\textrm{M}_{\odot}\,(\textrm{K km s}^{-1}\textrm{ pc}^{2})^{-1}$  overestimates the $M_{\textrm{mol}}$,  while an $\alpha_{\textrm{CO}}=0.8$ $\textrm{M}_{\odot}\,(\textrm{K km s}^{-1}\textrm{ pc}^{2})^{-1}$ underestimates it, for all ULIRGs of the sample. Instead, as is obvious from the definition, the value of $\alpha_{\textrm{CO}}=1.7$ $\textrm{M}_{\odot}\,(\textrm{K km s}^{-1}\textrm{ pc}^{2})^{-1}$, which is the median of the individual [CI]-based $\alpha_{\textrm{CO}}$ values estimated in the previous section, brings all data points closer to the 1:1 relation. 
We note, however, that this is an average value for the CO-to-H$_2$ conversion factor, and it is most likely to vary from galaxy to galaxy (as shown by a few outliers visible in Fig.~\ref{fig:MCI_vs_MCO_all}), as well as a function of aperture size and spatial scales probed. This variability averages out on galaxy-averaged measurements leading to a typical value corresponding to the dominant radiation-emitting regions within the galaxy. We note that large uncertainties on empirically-estimated $\alpha_{\textrm{CO}}$ values are still expected, as there are many factors that may impact on its value, e.g., density, temperature, metallicity, optical depth, among others. These results are however reassuring and indicate that the adoption of $\alpha_{\textrm{CO}}\simeq1.7$ $\textrm{M}_{\odot}\,(\textrm{K km s}^{-1}\textrm{ pc}^{2})^{-1}$ for local (U)LIRGs is reasonable.  

\subsection{Total line luminosities as a function of galaxy properties}\label{sec:gen_properties}

\begin{figure*}[tbp]
	\centering
	\includegraphics[width=.31\textwidth]{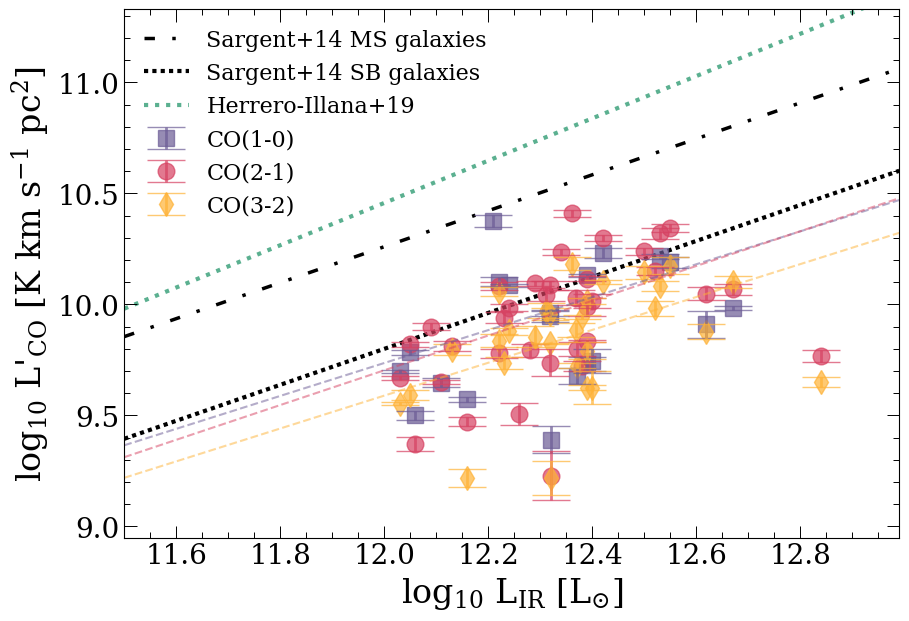}\quad
	\includegraphics[width=.31\textwidth]{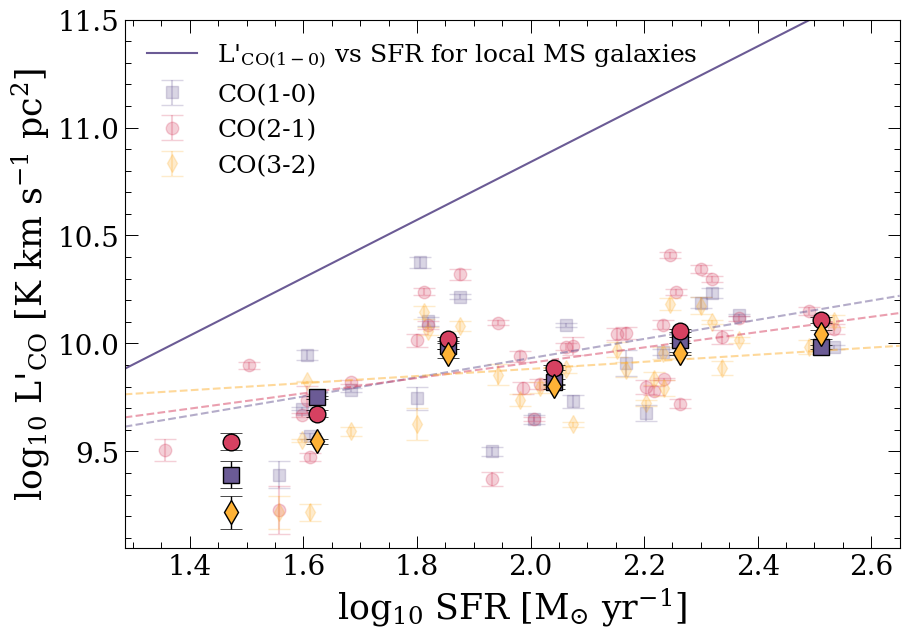}\quad
	\includegraphics[width=.31\textwidth]{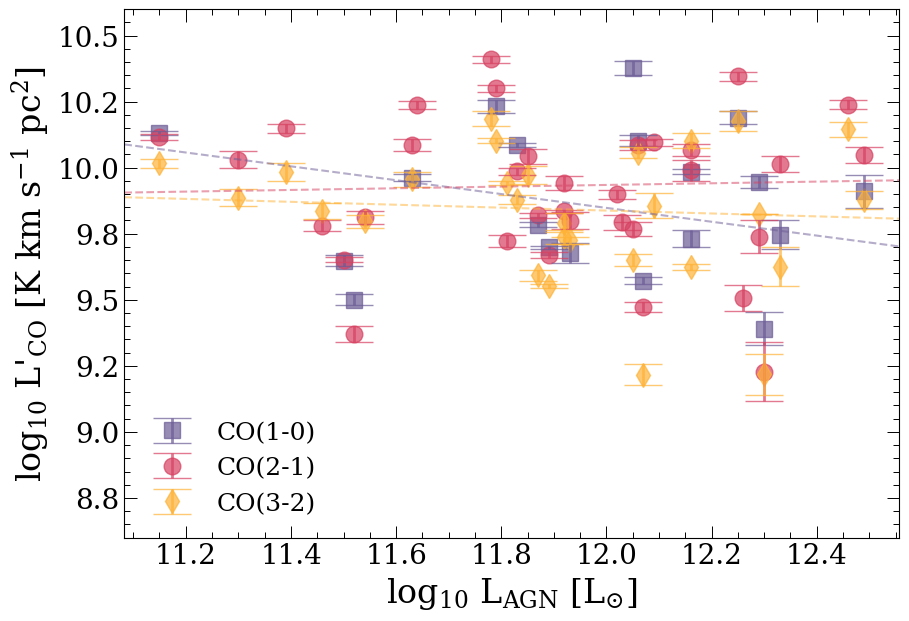}\\
	\includegraphics[width=.31\textwidth]{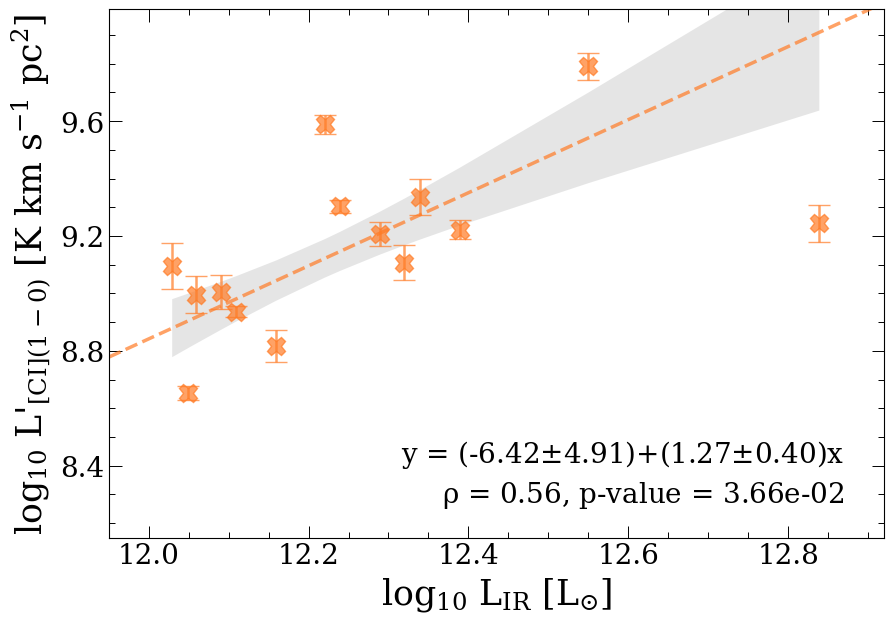}\quad
	\includegraphics[width=.31\textwidth]{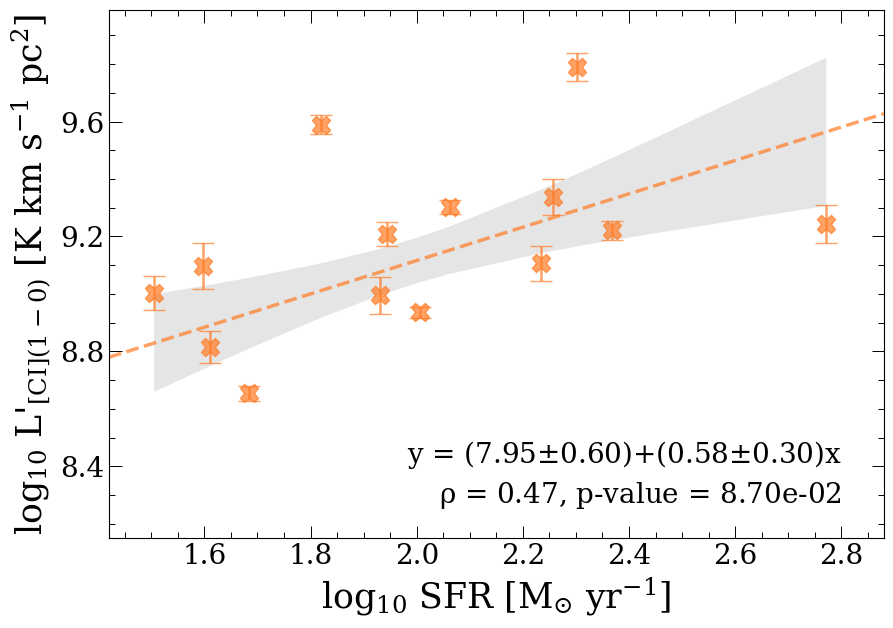}\quad
	\includegraphics[width=.31\textwidth]{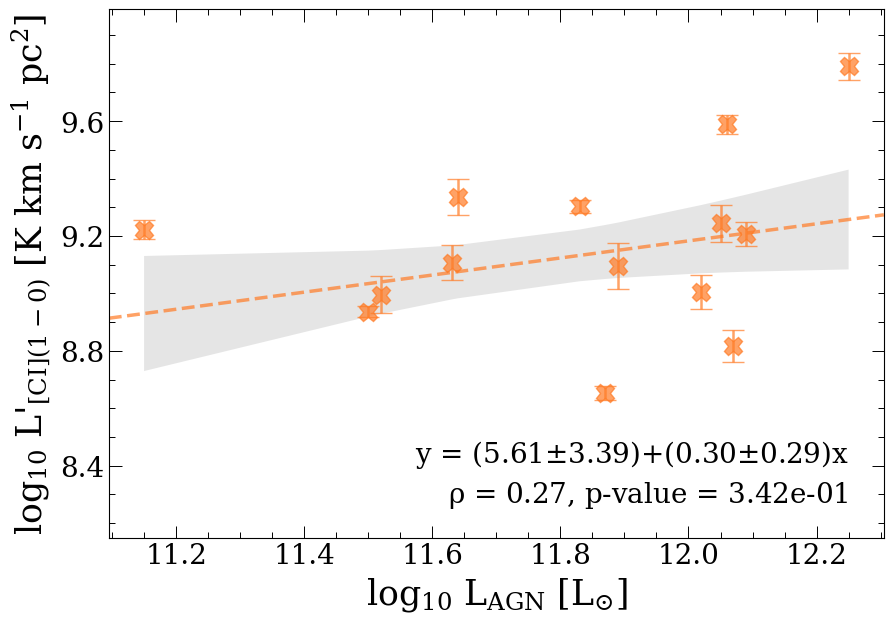}\\
	\caption{Global CO(1--0), CO(2--1), CO(3--2) (\textit{top panels}) and [CI](1--0) (\textit{bottom panels}) line luminosity plotted as a function of $L_{\textrm{IR}}$ ({\it left}), SFR ({\it middle}), and $L_{\rm AGN}$({\it right}) for our sample of ULIRGs. In each plot, the \textit{dashed} lines are the best-fit relations obtained from a least squares regression analysis conducted for each transition separately (color-coded according to the transition, see legend on the top-left panel). In the bottom panels, the shaded gray areas correspond to the $1\sigma$ confidence interval of the fit. The bottom panels report also the Pearson correlation coefficients ($\rho$) and their associated {\it p}-values.
	The {\it top-left panel} shows also the $L'_{\mathrm{CO(1-0)}}$-$L_{\textrm{IR}}$ relations found by \citet{Sargent+14} for local MS (dot-dashed black line) and starburst galaxies (\textit{dotted black} line), and the $L'_{\mathrm{CO(1-0)}}$-$L_{\textrm{TIR}}$ relation obtained by \cite{Herrero-Illana+19} using IRAM~30m CO(1--0) observations of 55 local sources in the GOALS sample including mostly LIRGs with $L_{IR}<10^{12}~L_{\odot}$ and with a $<20\%$ AGN contribution to $L_{\rm IR}$.
	In the {\it top-middle panel}, for better visualization, we over-plotted with darker colors the data in bins of SFRs. In this plot, the \textit{solid purple} line represents the $L'_{\mathrm{CO(1-0)}}$-SFR best-fit obtained by \cite{Cicone+17} for the ALLSMOG and COLDGASS samples of local star-forming galaxies, described by 
	$\textrm{log}_{10}\,L'_{\textrm{CO(1--0)}} =(8.16\pm0.04)+(1.34\pm0.07)\, \textrm{log}_{10}\,\textrm{SFR}$.}
	\label{fig:Lum_vs_properties}
\end{figure*}

Before investigating molecular line ratios and their dependencies on galaxy properties, it is worth exploring first the trends involving the line luminosities that are used to compute such ratios. We recall that our sample is, by construction, biased towards high $L_{\textrm{IR}}$. This could cause underlying galaxy scaling relations not to be properly captured by our targets, or to be detected with different slopes compared to the global star forming galaxy population (already seen, e.g., in Figure \ref{fig:COlumCIlum}), due to the limited range of intrinsic properties (e.g., SFR) probed by local (U)LIRGs \citep[see discussion on scaling relations in, e.g.,][]{Cicone+17}. It is therefore important to identify the portion of the $L^{\prime}_{\rm line}$- $L_{\rm IR}$ (or $L^{\prime}_{\rm line}$- SFR, $L^{\prime}_{\rm line}$- $L_{\rm AGN}$) parameter space occupied by our sources in order to place our results into perspective. To this aim, Figure~\ref{fig:Lum_vs_properties} shows the CO(1--0), CO(2--1), CO(3--2), [CI](1--0) line luminosities as a function of $L_{\rm IR}$, SFR, and $L_{\rm AGN}$, for all targets with corresponding line measurements available (see Table~\ref{tab:fluxlumino}).

For the $L^{\prime}_{\rm CO}$ vs $L_{\rm IR}$ relations (top-left panel of Fig.~\ref{fig:Lum_vs_properties}), we measure correlation coefficients of $\rm \rho_{L'_{CO(1-0)}-L_{IR}}=0.41$, $\rm \rho_{L'_{CO(2-1)}-L_{IR}}=0.45$, $\rm \rho_{L'_{CO(3-2)}-L_{IR}}=0.38$, with {\it p}-values = $7.6\times10^{-2}$, $6.4\times10^{-3}$, $4.1\times10^{-2}$, respectively). These relations trace, essentially, the Schmidt-Kennicutt (S-K) law \citep{Schmidt59, Kennicutt98}; the correlation coefficients, although hinting of positive relations, do not show significant relation, most likely caused by the narrow range on $L_{\rm IR}$ probed by our sample.
Such hypothesis is strengthened as the $\rho$ coefficients (and their respective \textit{p}-values) yield much tighter relations between the quantities when the LIRGs in our sample are included in the analysis (see Fig. \ref{fig:Lum_vs_properties_LIRGs}).
Our sample lies close to the $L'_{\mathrm{CO(1-0)}}$-$L_{\textrm{IR}}$ relation obtained by \cite{Sargent+14} for local starbursts ({\it dotted black} line in the top-left panel of Fig.~\ref{fig:Lum_vs_properties}), which is offset by $\simeq0.46$~dex from the main sequence galaxies' relation ({\it dot-dashed black} line). We note that our sample is instead significantly offset with respect to the \cite{Herrero-Illana+19} relation, we ascribe this discrepancy to a combination of two factors: (i) the extrapolation of a relation that is based on a lower-$L_{\rm IR}$ sample than ours; and (ii) their use of a different method for computing the $L_{\rm IR}$ based on an SED fitting, which delivers lower $L_{\rm IR}$ (by $\sim0.5$~dex) per given $L'_{\rm CO(1-0)}$. 
We verified that for the four targets in common with \cite{Herrero-Illana+19}, the CO fluxes are consistent, but the $L_{\rm IR}$ computed through their SED fitting (reported in their Table~5) are 0.45-1.0 dex lower than our $L_{\rm IR}$ values, which are instead consistent with the $L_{\rm IR}$ values reported by \cite{Armus+09} for the same sources.
The best fit $L'_{\mathrm{CO(1-0)}}$-$L_{\textrm{IR}}$ relation obtained by running a least square regression analysis on our data is $\log_{10} L'_{\rm CO(1-0)}=(0.8\pm2.2)+(0.74\pm0.18)\log_{10} L_{\rm IR}$, with $L'_{\mathrm{CO(1-0)}}/L_{\rm IR}$ ratios similar to the SB sample of \cite{Sargent+14}. This is not surprising since all of the sources in our sample of ULIRGs show enhanced star formation (see Table \ref{table:source_list}).

The bottom-left panel of Fig.~\ref{fig:Lum_vs_properties} reports $L'_{\rm [CI](1-0)}$ as a function of $L_{\rm IR}$. The [CI](1--0) luminosities span a range $\log_{10} L'_{\mathrm{[CI](1-0)}} = 8.7-9.8$ [K \kms pc$^{2}$], consistently lower than the CO(1--0) luminosities ($\log_{10} L'_{\mathrm{CO(1-0)}} = 9.4-10.4$ [K \kms pc$^{2}$]). 
Figure~\ref{fig:COlumCIlum} showed a tight relation between $L'_{\rm CO(1-0)}$ and $L'_{\rm [CI](1-0)}$, hence, a similar relation to that found between $L'_{\rm CO(1-0)}$ (or $L'_{\rm CO(2-1)}$) and $L_{\rm IR}$ is expected.
Indeed, we measure a slightly higher Pearson correlation coefficient of $\rho_{L'_{\rm [CI](1-0)}-L_{\rm IR}}=0.56$ ({\it p}-value = $0.04$).

The middle panels of Fig.~\ref{fig:Lum_vs_properties} display the CO and [CI] line luminosities as a function of SFR. These relations are not much dissimilar from those with $L_{\rm IR}$ (left panels), as expected since much of the $L_{\rm IR}$ in (U)LIRGs is powered by star formation. Since the SFRs have been computed by removing the contribution to $L_{\rm IR}$ estimated to arise from AGN-heated dust, the middle panels of Fig.~\ref{fig:Lum_vs_properties} should more truthfully trace the S-K relation. However, we struggle to retrieve a tight S-K law for this sample, probably because of selection biases due to their narrow distribution in SFRs, combined with the inevitably large uncertainties on the AGN fraction \citep{Veilleux+09b}. 
Although we measure slightly higher Pearson correlation coefficients for the $L'_{\rm CO}$ vs SFR relations ($\rho_{L'_{\rm CO(1-0)}-\rm SFR}=0.42$, $\rho_{L'_{\rm CO(2-1)}-\rm SFR}=0.46$, $\rho_{L'_{\rm CO(3-2)}-\rm SFR}=0.50$, with {\it p}-values = $6.8\times10^{-2}$, $5\times10^{-3}$ and $5.8\times10^{-3}$, respectively) than for the $L'_{\rm CO}$ vs $L_{\rm IR}$ relations, the values are still marginal at best for their correlations.
When using [CI](1--0) as a H$_2$ tracer, we obtain $\rho_{L'_{\rm [CI](1-0)}-\rm SFR}=0.47$ and {\it p}-value = $8.7\times10^{-2}$. 
The best-fit $L'_{\mathrm{CO(1-0)}}$-SFR relation ($\log_{10} L'_{\rm CO(1-0)}=(9.04\pm0.22)+(0.44\pm0.11)\log_{10} \rm SFR$) has a flatter slope than the $L'_{\mathrm{CO(1-0)}}$-$L_{\rm IR}$ one. Similarly, the $L'_{\mathrm{[CI](1-0)}}$-SFR relation (reported on the plot), also shows a considerably shallower slope than the $L'_{\mathrm{[CI](1-0)}}$-$L_{\rm IR}$ relation, with a value of $0.6\pm0.3$.
In the top-middle panels of Fig. \ref{fig:Lum_vs_properties}, we plot with a solid purple line
the best-fit $L'_{\mathrm{CO(1-0)}}$-SFR relation obtained for a much more unbiased sample of local star forming main sequence (MS) galaxies (drawn from the COLDGASS and ALLSMOG surveys, see \citealt{Cicone+17})\footnote{SFR range for ALLSMOG and COLDGASS samples: $-1.5 < {\rm log}_{10}{\rm SFR}\,\, [{\rm M}_{\odot}\, {\rm yr}^{-1}] <1.5$ .}. Our ULIRGs are characterized by significantly lower $L'_{\rm CO(1-0)}/{\rm SFR}$ ratios than MS galaxies, especially in the high-SFR regime.
The result of lower $L'_{\rm CO(1-0)}/{\rm SFR}$ ratios in (U)LIRGs than in normal MS galaxies is well known, and due to a combination of (i) a higher efficiency of star formation (i.e., higher SFE=SFR/$M_{\rm mol}$, or equivalently lower $\tau_{\rm dep}=M_{\rm mol}/{\rm SFR}$) and (ii) a lower $\alpha_{\textrm{CO}}$ compared to normal disk galaxies, as also confirmed by our analysis in Section~\ref{sec:alphaCO}.

The right panels of Fig.~\ref{fig:Lum_vs_properties} show the CO and [CI] line luminosities as a function of AGN luminosity. 
Neither of the two gas tracers, in any of the transitions, show any correlation with $L_{\rm AGN}$. 
The measured Pearson correlation coefficients and \textit{p}-values for the CO emission lines are
$\rho_{L'_{\textrm{CO(1--0)}}-L_{\rm AGN}}=-0.05$ ({\it p}-value = 0.8), 
$\rho_{L'_{\textrm{CO(2--1)}}-L_{\rm AGN}}=-0.06$ ({\it p}-value = 0.7) and 
$\rho_{L'_{\textrm{CO(3--2)}}-L_{\rm AGN}}=-0.2$ ({\it p}-value = 0.3).
These results are opposed to what has been found in the literature by, e.g., \cite{Husemann+17, Shangguan+20b}, who find signs of correlation between the CO(1--0) (or CO(2--1)) line luminosity and $L_{\rm AGN}$ on samples of tens of local quasars, although the interpretation is not entirely clear.
Indeed, while it is generally accepted that the S-K law traces a fundamental causal relation between the availability of fuel and the resulting star formation activity, the relation between $L^{\prime}_{\rm CO}$ and $L_{\rm AGN}$ can be only investigated globally, because it relates two processes that affect very different scales in galaxies. 
We note, however, the change in the relation when the LIRGs in our sample are included (see Fig. \ref{fig:Lum_vs_properties_LIRGs}), for which we obtain Pearson correlation coefficients and {\it p}-values of $\rho_{L'_{\textrm{CO(1--0)}}-L_{\rm AGN}}=0.53$ ({\it p}-value = 0.01), 
$\rho_{L'_{\textrm{CO(2--1)}}-L_{\rm AGN}}=0.37$ ({\it p}-value = 0.02), and
$\rho_{L'_{\textrm{CO(3--2)}}-L_{\rm AGN}}=0.32$ ({\it p}-value = 0.09). 
Hence the inclusion of LIRGs, with the consequent widening of the dynamic range in properties probed, produces positive trends between molecular line luminosities and $L_{\rm AGN}$, in particular, for CO(1--0) and CO(2--1) (although weak); possibly originating from an underlying scaling of both quantities with SFR and/or M$_*$ \citep[see][]{Cicone+17}.
It is, however, necessary to confirm these results with a sample including more sources in the $L_{\rm IR} < 12\,{\rm L_{\odot}}$ regime.

The relation between $L'_{\rm [CI](1-0)}$ and $L_{\rm AGN}$ in our sample of ULIRGs (bottom-right panel of Figure \ref{fig:Lum_vs_properties}) is equally weak as the relations for CO lines, with a measured correlation coefficient of $\rho_{L'_{\textrm{[CI](1--0)}}-L_{\rm AGN}}=0.27$ ({\it p}-value = $0.3$). Interestingly, when we include the LIRGs in the analysis (see Fig. \ref{fig:Lum_vs_properties_LIRGs}), [CI](1--0) shows a stronger correlation with $L_{\rm AGN}$ than CO(1--0) or CO(2--1), we measure a correlation parameter of $\rho_{L'_{\textrm{[CI](1-0)}}-L_{\rm AGN}}=0.68$ ({\it p}-value = $3.7\times 10^{-3}$), with a slope equal to $0.74\pm0.26$, a similarly tight relation to the one found for $L'_{\textrm{[CI](1-0)}}$ and SFR for the entire sample. 
Indeed, our sample poorly populates the $L_{\rm IR} < 10^{12} \, \mathrm{L_{\odot}}$ regime, and these results are solely driven by the galaxy IRAS F12243-0036 (also known as NGC~4418), a source that is an outlier in many respects.
This galaxy has also the lowest redshift ($z=0.00708$) in our sample, and was discovered by \cite{Sakamoto+10} to host an extremely compact obscured nucleus (CON). Here, $\sim10^8 \rm M_{\odot}$ of molecular gas are concentrated in a region with $<20$~pc size, pointing to extremely high column densities of $\rm N_H>10^{25}~cm^{-2}$ \citep[see][and references therein]{Falstad+21}.
If such correlation between $L'_{\textrm{[CI](1-0)}}$ and $L_{\rm AGN}$ is confirmed with larger statistics, this result could suggest different behaviors of CO and [CI] as H$_2$ gas tracers in AGN hosts.

\subsection{Galaxy-integrated CO line ratios}\label{sec:tot_COratios}

\begin{figure*}
	\centering
	\includegraphics[clip=true,trim=0.0cm 0.0cm 20cm 0.cm,width=.8\textwidth]{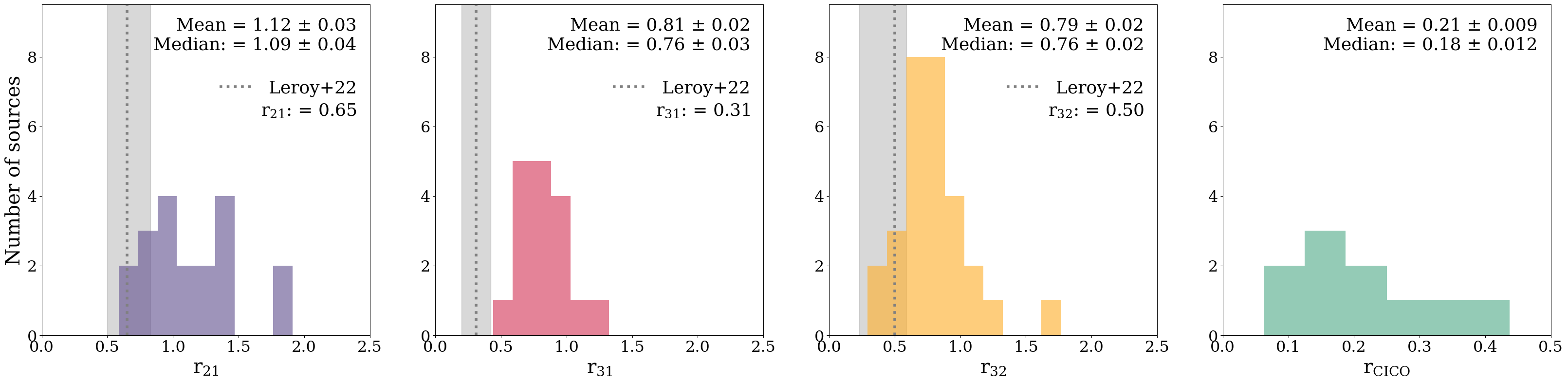}
	\caption{Distribution of galaxy-integrated CO line ratios obtained for our sample of ULIRGs, with mean and median values reported on the top-right of each plot. \textit{Left:} $r_{21}\equiv{L{'}_{\textrm{CO(2--1)}}}/{L{'}_{\textrm{CO(1--0)}}}$. \textit{Middle:} $r_{31}\equiv{L{'}_{\textrm{CO(3--2)}}}/{L{'}_{\textrm{CO(1--0)}}}$. \textit{Right:} $r_{32}\equiv{L{'}_{\textrm{CO(3--2)}}}/{L{'}_{\textrm{CO(2--1)}}}$. The mean CO line ratios measured by \cite{Leroy+22} in local massive main sequence disk galaxies are shown using vertical dotted lines, with the gray shaded regions representing their 16th-84th percentile ranges. }\label{fig:COratios_histo}
\end{figure*}

The distributions of galaxy-integrated CO line ratios ($r_{21}\equiv{L{'}_{\textrm{CO(2--1)}}}/{L{'}_{\textrm{CO(1--0)}}}$, $r_{31}\equiv{L{'}_{\textrm{CO(3--2)}}}/{L{'}_{\textrm{CO(1--0)}}}$, $r_{32}\equiv{L{'}_{\textrm{CO(3--2)}}}/{L{'}_{\textrm{CO(2--1)}}}$) for our local ULIRGs are shown in Figure~\ref{fig:COratios_histo}. Our sample spans a very wide range of values: $0.6\lesssim r_{21}\lesssim1.9$, 
$0.4\lesssim r_{31}\lesssim1.3$, and $0.4\lesssim r_{32}\lesssim1.7$. For comparison, we report in Fig.~\ref{fig:COratios_histo} also the measurements obtained by \cite{Leroy+22} across tens of local disk galaxies ($N\sim40$, 30, 20 objects with $r_{21}$, $r_{32}$, and $r_{31}$ measurements, respectively), which are based on resolved CO maps and are thus unaffected by beam mismatch. We note that the \cite{Leroy+22} sample is only representative of local massive ($\log_{10} M_* [M_{\odot}] \sim 10.2-10.8$), star forming galaxies on the main sequence ($\rm -10.5 \lesssim \log_{10} sSFR~[yr^{-1}]\lesssim -9.9$). 

Our mean and median $r_{21}$ values, respectively $1.12$ and $1.09$, are significantly higher than the average $\rm \langle r_{21}\rangle=0.79 \pm 0.03$ measured by \cite{Saintonge+17} for the xCOLD GASS sample\footnote{A stellar-mass selected ($M_*>10^9~M_{\odot}$) and molecular gas fraction-limited sample of 532 local galaxies, having single-dish CO(1--0) and CO(2--1) data from the IRAM~30m and APEX telescopes.}. Although dominated by massive main sequence galaxies, the xCOLD GASS sample, being only M$_*$-selected, includes targets above the main sequence as well as AGNs, hence it is not necessarily representative of purely star forming disks. 
Indeed, the molecular ISM of main sequence disks generally presents even lower values of $r_{21} \approx 0.6-0.7$ at $z\sim0$ \citep{denBrok+21, Leroy+22}, and perhaps also at higher redshift \citep[see results at $z\sim2$ by ][]{Aravena+14}.
Whether the $r_{21}$ ratio is higher or not in AGNs is not clear, indeed the mean $r_{21}$ measured by \cite{Husemann+17} and \cite{Shangguan+20a} in nearby unobscured quasars are $\langle r_{21}\rangle=0.9 \pm 0.3$ and $\langle r_{21}\rangle=0.6 ^{+0.15}_{-0.13}$, respectively. It is however quite well established that the central regions of star forming galaxies, and in general regions with higher $\rm \Sigma_{SFR}$, present systematically higher ratios than the outskirts (\citealt{denBrok+21, Leroy+22}).

Despite slight variations in different samples, {\it global $r_{21}$ luminosity ratios above unity} are extremely rare in the Universe, yet they are predominant in our sample of local (U)LIRGs. This finding is in agreement with \cite{Papadopoulos+12a} who, through their analysis of 70 (U)LIRGs with heterogeneous single-dish multi-{\it J} CO observations, measured $\langle r_{21} \rangle =0.91$. This is slightly lower than our mean value, probably because of their larger statistics. It is also probable that the smaller apertures of JCMT and IRAM~30m telescopes used for the CO(2--1) observations, combined with the narrower bandwidths of the heterodyne receivers used at that time, have contributed to some CO(2--1) flux loss for the most extreme (U)LIRGs of the \cite{Papadopoulos+12a} sample. In any case, the $r_{21}$ value distribution obtained by these authors is similarly broad to the one we obtain (Fig.~\ref{fig:COratios_histo}) with many sources globally characterized by $r_{21}>1$ values. 

The offset in global CO ratios from the local galaxy population is even more extreme in the $r_{31}$ values, whose mean and median of $0.81$ and $0.76$ measured in our sample of ULIRGs are significantly higher than $\langle r_{31}\rangle =0.31$, which is the mean galaxy-integrated value measured by \cite{Leroy+22}. Strikingly, Figure~\ref{fig:COratios_histo} shows that there is {\it no overlap} between our ULIRGs sample and normal massive star forming galaxies in the $r_{31}$ values. The results by \cite{Leroy+22} are consistent with literature data focusing on local massive main sequence star forming galaxies, overall confirming that the low-{\it J} CO line emission in these sources is dominated by optically thick clouds with moderately sub-thermally excited CO(2--1) and CO(3--2) transitions. As expected due to the diversity of their sample, \cite{Lamperti+20} measured higher values of $r_{31}$ compared to \cite{Leroy+22}. Specifically, \citet{Lamperti+20} obtained $\langle r_{31}\rangle =0.55 \pm 0.05$ for 25 xCOLD GASS objects with JCMT CO(3--2) data, which probe a massive ($M_*>10^{10}~M_{\odot}$), highly star forming (with $\rm \log_{10} sSFR~[yr^{-1}] >-10.5$ and with most sources being above the main sequence) portion of the parent xCOLD GASS sample presented by \cite{Saintonge+17}. The same study by \cite{Lamperti+20} included also 36 hard X-ray selected AGNs from the BASS survey with JCMT CO(3--2) and CO(2--1) observations, and, by extrapolating the CO(1--0) luminosities from the CO(2--1) data, they report consistent $r_{31}$ values compared to a non-AGN sample with matched specific star formation rate.
The average LIRG value obtained by \cite{Papadopoulos+12a} is $\langle r_{31} \rangle =0.67$\footnote{\cite{Papadopoulos+12a} use the $r_{32}$ notation to indicate the ${L{'}_{\textrm{CO(3-2)}}}/{L{'}_{\textrm{CO(1-0)}}}$ ratio, which we instead indicate as $r_{31}$.}, with a distribution displaying a significant tail including many sources with $r_{31}\gtrsim1$.

For completeness, we show in Fig.~\ref{fig:COratios_histo} also our $r_{32}$ measurements, with average and median respectively $r_{32}=0.80$ and $r_{32}=0.76$. These are also quite offset from normal galaxy disks, for which \cite{Leroy+22} report $\langle r_{32}\rangle =0.50$.

\begin{figure*}[tbp]
	\centering
	\includegraphics[width=.31\textwidth]{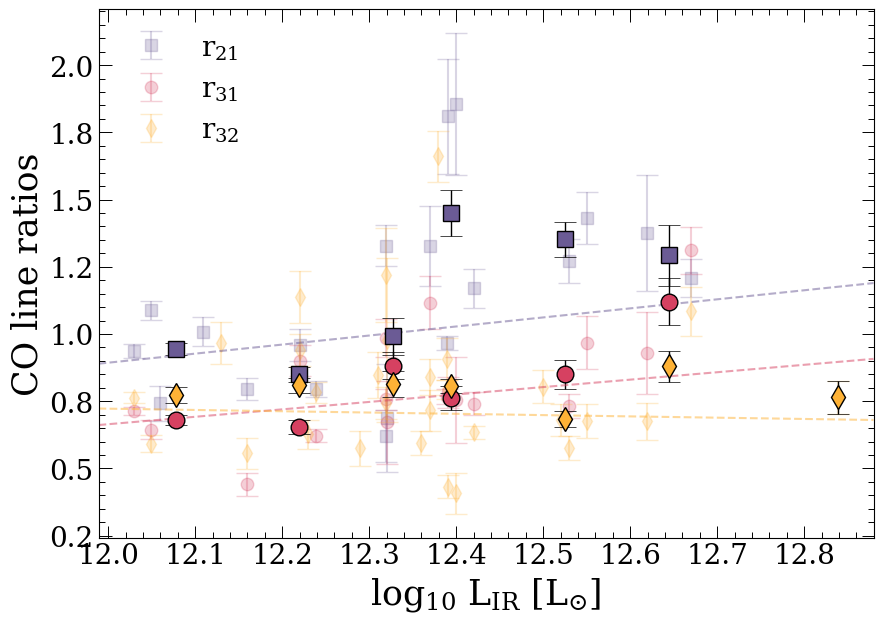}\quad
	\includegraphics[width=.31\textwidth]{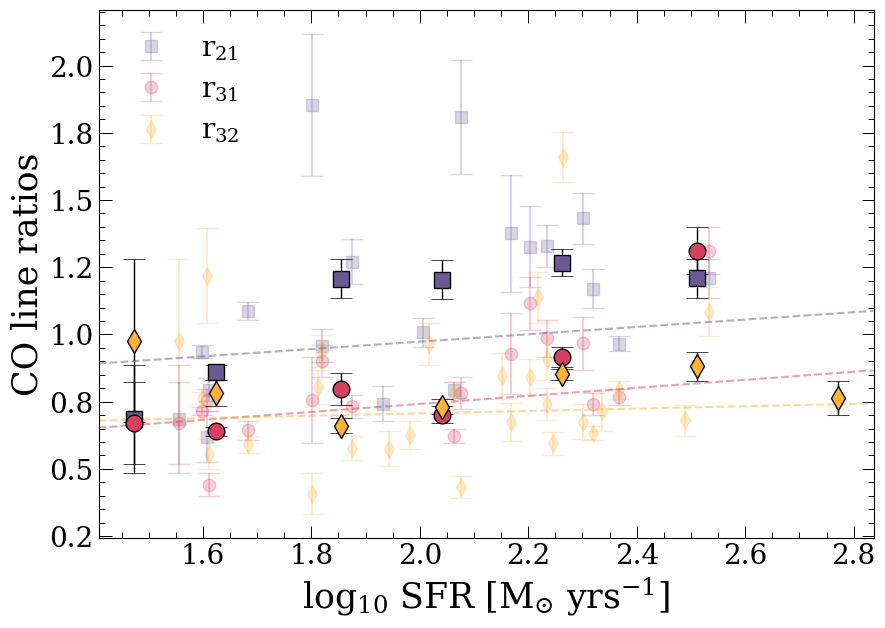}\quad
	\includegraphics[width=.31\textwidth]{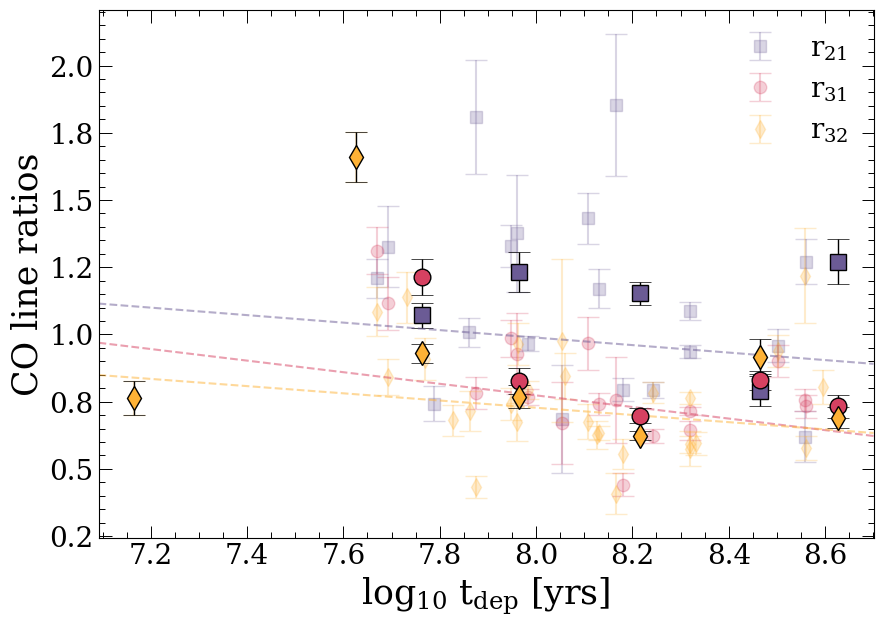}\\
	\includegraphics[width=.31\textwidth]{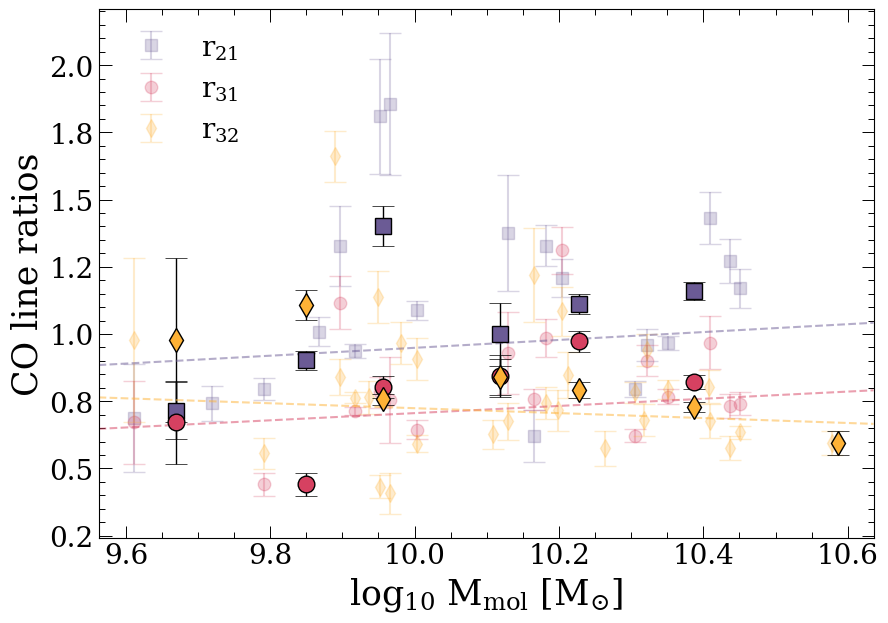}\quad
	\includegraphics[width=.31\textwidth]{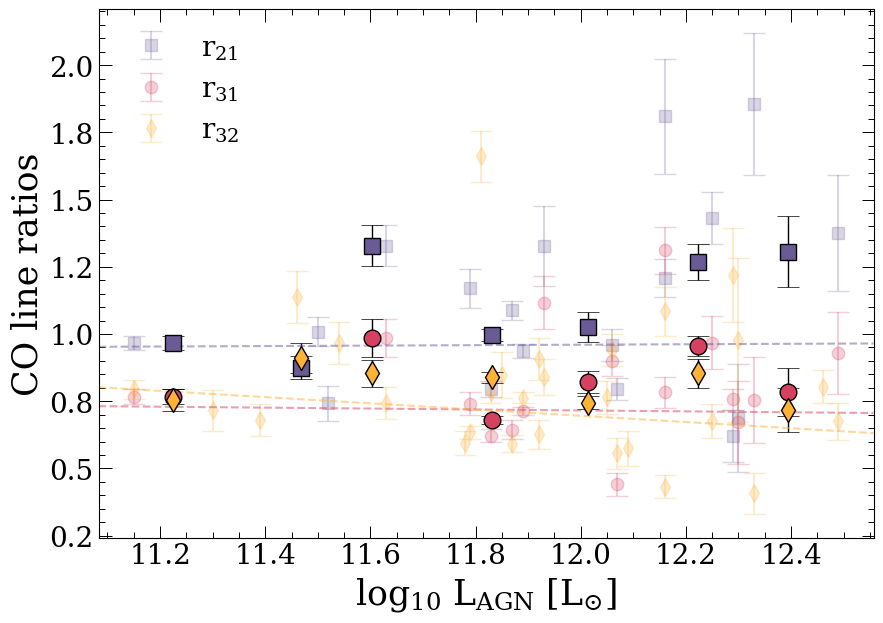}\quad
	\includegraphics[width=.31\textwidth]{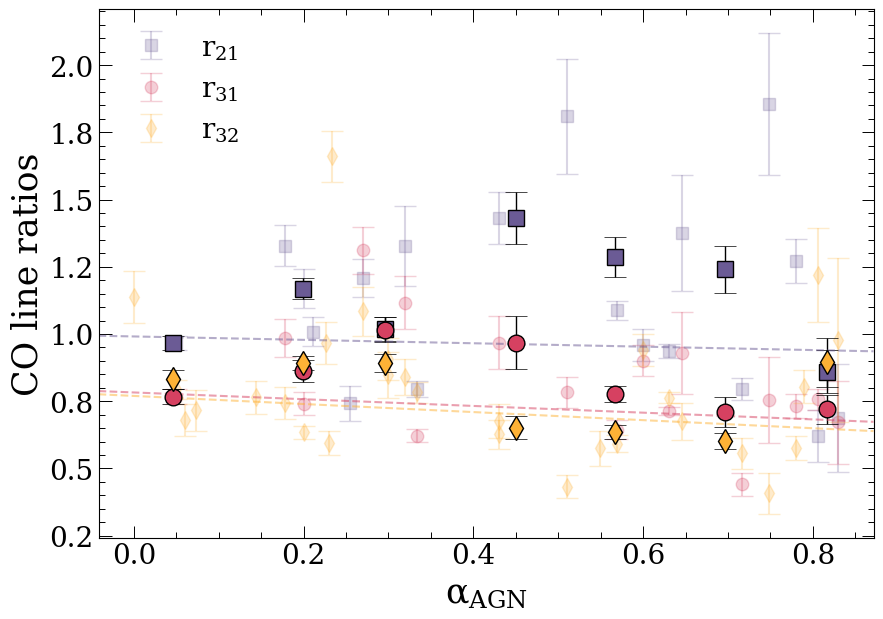}\\
	\caption{CO line ratios as a function of galaxy properties: $L_{\rm IR}$ ({\it top-left}), 
		SFR ({\it top-middle}), molecular gas depletion time-scale due to star formation $\tau_{\rm dep} \equiv M_{\rm mol}/\rm SFR$ ({\it top right}), molecular gas mass $M_{\rm mol}$ (computed by using the average $\alpha_{\rm CO}$ derived in this work) ({\it bottom-right}), $L_{\rm AGN}$ ({\it bottom-middle}), and AGN fraction ($L_{\rm AGN}/L_{\rm bol}$) ({\it bottom-right}). We show the binned values using darker colors, overplotted on the individual data points. Dashed lines indicate the best fit relations obtained from a least squares regression analysis conducted for each transition line separately. The color coding refers to different line ratios, as indicated in the legend at the top-right corner of each plot.}
	\label{fig:COratios_vs_properties}
\end{figure*}

\subsection{Global CO line ratios as a function of galaxy properties}\label{sec:COratios_galaxyprop}

In Figure~\ref{fig:COratios_vs_properties} we plotted the total CO line ratios as a function of different galaxy properties, namely: $L_{\rm IR}$ ({\it top-left}), SFR ({\it top-middle}), $L_{\mathrm{AGN}}$ ({\it top-right}), $M_{\mathrm{mol}}$ ({\it bottom-left}), molecular gas depletion timescale due to star formation, defined as $\rm \tau_{dep} \equiv  M_{\rm mol}/\rm SFR$, which is the inverse of star formation efficiency ($\rm SFE \equiv SFR/M_{mol}$) ({\it bottom-middle}), and the AGN fraction ($\alpha_{\rm AGN}\equiv L_{\rm AGN}/L_{\rm bol}$). The values of $L_{\rm IR}$, SFR, $L_{\rm AGN}$, and $\alpha_{\rm AGN}$ are taken from the literature and are listed in Table~\ref{table:source_list} together with their corresponding references. Instead, the $M_{\mathrm{mol}}$ and $\tau_{\mathrm{dep}}$ values have been computed in this work, using, for each source, the lowest-{\it J} CO transition available and adopting the 
median $\alpha_{\textrm{CO}}$ factor computed in this work, i.e., $1.7$ $\textrm{M}_{\odot}\,(\textrm{K km s}^{-1}\textrm{ pc}^{2})^{-1}$ (see Section~\ref{sec:alphaCO}). 

With the aim of better highlighting any underlying trends, we divided the ULIRG sample in bins according to the quantity on the \textit{x}-axis and calculated the mean values of the ratios in these bins, which are shown using darker symbols in Fig.~\ref{fig:COratios_vs_properties}.
The dashed lines show the best-fit relations, with color coding indicating different CO line ratios. 
An inspection of Fig.~\ref{fig:COratios_vs_properties} shows the absence of \textit{strong} correlations between the global low-{\it J} CO line ratios and the other quantities investigated here.
This is not completely surprising given the narrow dynamic range in galaxy properties spanned by our sample, which is representative of the most extremely IR-bright galaxies of the local Universe ($\log_{10} L_{\rm IR} [{\rm L}_{\odot}] \in [12.0, 12.8]$, making it a very special sample in its galaxy and ISM properties,
already evident in Figure~\ref{fig:COratios_histo}, where almost no overlap with the normal population of star-forming galaxies is seen. 
However, we retrieve (although still weak) positive correlations between some of the low-\textit{J} CO line ratios and quantities related to the strength of the star formation activity (namely, $L_{\rm IR}$, SFR and SFE$=\tau_{\rm dep}^{-1}$).

In the top-left panel of Fig.~\ref{fig:COratios_vs_properties} we investigate the relations between the line ratios and $L_{\rm IR}$. With measured Pearson correlation coefficients and \textit{p}-values of 
$\rho_{{r_{21}}-L_{\rm IR}}=0.50$ ({\it p}-value = 0.02), 
$\rho_{{r_{31}}-L_{\rm IR}}=0.62$ ({\it p}-value = 0.008) and 
$\rho_{{r_{32}}-L_{\rm IR}}=-0.02$ ({\it p}-value = 0.93),
we find that $r_{21}$ and $r_{31}$ show a positive relation with the infrared luminosity, while $r_{32}$ does not. 
In the relations with SFR (middle top panel), we find the ratio $r_{31}$ to be the only one showing a positive correlation ($\rho_{{r_{31}}-{\rm SFR}}=0.69$ and {\it p}-value = 0.002), while no significant correlation is found for $r_{21}$ ($\rho_{{r_{21}}-{\rm SFR}}=0.40$ and {\it p}-value = 0.09), and $r_{32}$ ($\rho_{{r_{32}}-{\rm SFR}}=0.10$ and {\it p}-value = 0.61).
Equivalently to the results seen in Figure \ref{fig:Lum_vs_properties}, the relations with $L_{\rm IR}$ and SFR are expected to be similar since much of the infrared luminosity in these galaxies is due to star formation.
The correlations found with $L_{\rm IR}$ are in agreement with the study performed by \cite{Rosenberg+15} on the HERCULES sample\footnote{The {\it Herschel} Comprehensive ULIRG Emission Survey.}, where a correlation was found between high-{\it J} CO ratios and $L_{\rm FIR}$ (as well as with dust color).
\cite{Lamperti+20} and \cite{Leroy+22} found, for massive main sequence galaxies with or without an AGN, positive correlations between CO line ratios and SFR, in particular for the ratio $r_{31}$.
If the CO(1--0) emission traces the total H$_2$ reservoir including the more diffuse components, and the CO(3--2) emission traces the somewhat already denser gas, then the ratio $r_{31}$ can be interpreted as a measure of the fraction of molecular gas that is in the denser star-forming regions.
This would lead to higher values of $r_{31}$ for ULIRGs and starburst galaxies, as these sources are expected to have higher fractions of dense gas,
leading also to increased star formation efficiency (and decreased depletion times).
Indeed, in our sample of ULIRGs, we compute Pearson coefficient for the CO ratios vs $\tau_{\rm dep}$ of
$\rho_{{r_{21}}-\tau_{\rm dep}}=-0.27$ ({\it p}-value = 0.27), 
$\rho_{{r_{31}}-\tau_{\rm dep}}=-0.58$ ({\it p}-value = 0.02) and 
$\rho_{{r_{32}}-\tau_{\rm dep}}=-0.26$ ({\it p}-value = 0.17), where only $r_{31}$ shows a (negative) correlation with $\tau_{\rm dep}$,  
consistent with the (opposite) trend seen by \cite{Lamperti+20} between $r_{31}$ and SFE.

We do not find any trend between these global ratios and molecular gas mass estimates ($\rho=0.19$, $\rho=0.25$ and $\rho=-0.24$, for $r_{21}$, $r_{31}$, and $r_{32}$, all having {\it p}-values $>0.05$). This result is in agreement with \cite{Yao+03}, while \citealt{Leroy+22} found a weak anti-correlation between $r_{21}$ or $r_{31}$ and the lowest-{\it J} CO transition available ($L^{\prime}_{\rm CO, low}$).
We also do not find any significant trend between CO line ratios and $L_{\rm AGN}$ ($\rho=0.29$, $\rho=0.10$ and $\rho=-0.10$, for $r_{21}$, $r_{31}$, and $r_{32}$, all having {\it p}-values $>0.05$), consistent with \cite{Lamperti+20} and \cite{Yao+03}.
The relations between CO line ratios and the AGN fraction $\alpha_{\rm AGN}$ (lower-right panel of Fig.~\ref{fig:COratios_vs_properties}), may suggest a negative trend, in particular for $r_{31}$, though still not statistically significant as measured by $\rho_{{r_{31}}-\alpha_{\rm AGN}}=-0.43$ ({\it p}-value = 0.09), while $r_{21}$ and $r_{32}$ show no correlation at all with $\alpha_{\rm AGN}$, having $\rho_{{r_{21}}-\alpha_{\rm AGN}}=-0.03$ and $\rho_{{r_{32}}-\alpha_{\rm AGN}}=-0.2$  and \textit{p}-values $>0.05$.

\subsection{Global [CI](1--0)/CO(1--0) line ratios}\label{sec:tot_CICOratios}

\begin{figure}[tbp]
	\centering
	\includegraphics[clip=true,trim=0cm 0.2cm 0cm 0.cm,width=.8\columnwidth]{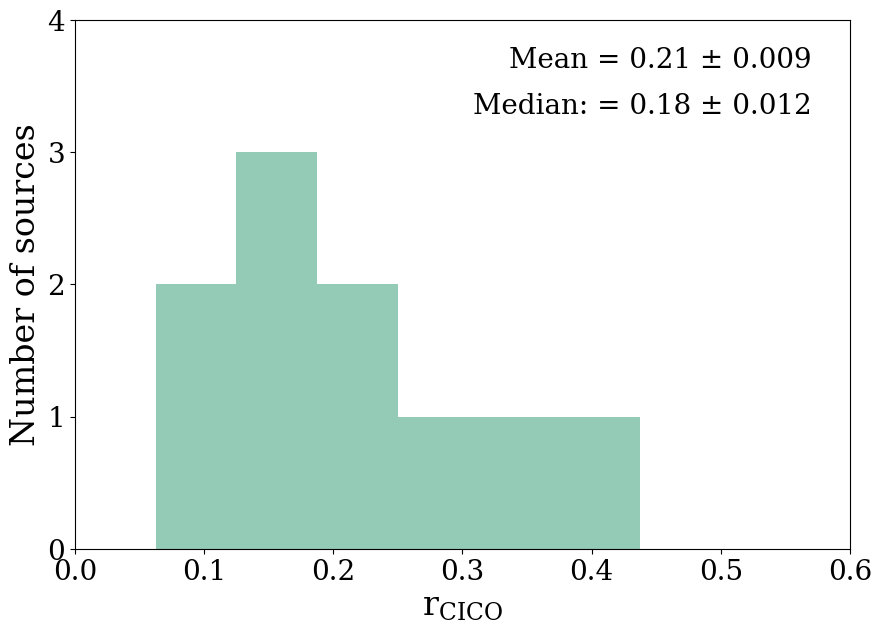}
	\caption{Distribution of galaxy-integrated $r_{\rm CICO}\equiv L'_{\rm [CI](1-0)}/L'_{\rm CO(1-0)}$ line ratios obtained for our sample of ULIRGs, with mean and median values reported on the plot. }\label{fig:CICOratio_histo}
\end{figure}

The distribution of integrated [CI](1--0)/CO(1--0) line luminosity ratios ($r_{\rm CICO}\equiv L'_{\rm [CI](1-0)}/L'_{\rm CO(1-0)}$) obtained for our sample is reported in Figure~\ref{fig:CICOratio_histo}. The measurements span the range $0.08\lesssim  r_{\textrm{CICO}}\lesssim 0.4$, with a median of $\langle r_{\textrm{CICO}} \rangle ^{\rm median}=0.18$ and a mean of 
$\langle r_{\textrm{CICO}} \rangle ^{\rm mean}=0.21$. 

Due to the paucity of [CI](1--0) detections available in the literature, it is unfortunately not possible to compare our results with statistically significant measurements performed on less extreme samples of local star forming galaxies. Furthermore, often local targets observed in [CI](1--0) do not have adequate, aperture-matched CO(1--0) line data that can be used to compute the [CI](1--0)/CO(1--0) ratio, so there are only a few works to which we can compare our results.
\cite{Jiao+19} analyzed {\it Herschel} SPIRE [CI](1--0) maps (with 1~kpc resolution) of 15 nearby galaxies, and combined them with single-dish CO(1--0) observations, convolved to the same beam, obtained with the single-dish Nobeyama 45m telescope. The 15 sources of \cite{Jiao+19} are very famous nearby galaxies: a few starbursts (such as M 82, NGC 253, M 83), one HII galaxy (NGC 891), six LINERs and three Seyferts (among which, NGC 1068). Their sample is small but it is diverse and covers almost three orders of magnitude in SFR surface density. Therefore, their median $r_{\rm CICO}$ of 0.11 (mean is 0.12), based on multiple measurements per galaxy (one per resolution element), may be considered the closest we can get to a value that is representative of typical local star forming galaxies. 
\cite{Michiyama+21} performed simultaneous CO(4--3) and [CI](1--0) observations with ACA Band~8 in 36 local (U)LIRGs, and reported single-dish CO(1--0) line luminosity measurements for 25 of their targets that have also an [CI](1--0) line detection. We used these values to compute the average and median $r_{\rm CICO}$ ratio for the \cite{Michiyama+21} sample with both CO(1--0) and [CI](1--0) measurements, both coincidentally being 0.13 with a standard deviation of 0.07. Their median value of 0.13 is lower than what we find for our sample of ULIRGs, despite our samples overlap by 7 sources (of which only 5 have a CO(1--0) value in \cite{Michiyama+21}) and span a similar range of SFRs. 
We note that only in two cases we decided to employ in our analysis the ACA Band~8 data collected by \cite{Michiyama+21} (project ID 2018.1.00994.S, see Table~\ref{table:data_available}).
For the other five overlapping targets we preferred our own APEX PI observations over the ACA archival data, because of their better quality and/or higher flux recovered (see duplicated [CI](1--0) data shown in Figure~\ref{fig:spectra_dupli_ci10}). This is consistent with the considerations \cite{Michiyama+21}, who estimate an [CI](1--0) line flux loss of $\sim30-40$~\% for their ACA data.

\begin{figure}[tbp]
	\centering
	\includegraphics[width=.95\columnwidth]{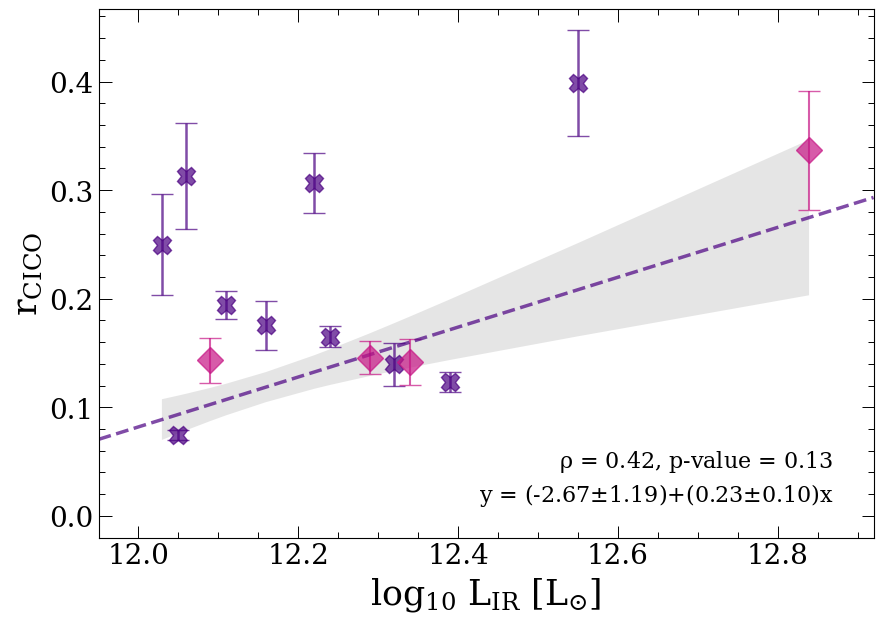}\\
	\caption{$r_{\textrm{CICO}}$ as a function of $L_{\rm IR}$ for our sample of ULIRGs. Similar relations with other galaxy properties are reported in Fig.~\ref{fig:rCICO_properties}. Sources with direct measurements of both the [CI](1--0) and CO(1--0) lines are plotted using purple crosses.
	The pink diamonds represent sources observed in [CI](1--0) but without CO(1--0) line data, for which we inferred $L'_{\textrm{CO(1-0)}}$ from the CO(2--1) line luminosity, by assuming $r_{21}=1.1$, which is the mean value computed for our sample (see Figure~\ref{fig:COratios_histo}). 
	The \textit{dashed purple} line represents the best-fit using a least squares regression analysis, and the shaded gray area corresponds to the $1\sigma$ confidence interval of the fit. The Pearson correlation coefficients ($\rho$) are reported at the upper left corner of the panel with their associated {\it p}-values.}
	\label{fig:rCICO_LIR}
\end{figure}

We then investigated the $r_{\textrm{CICO}}$ ratio as a function of the following galaxy properties: $L_{\rm IR}$, SFR, $L_{\rm AGN}$, $M_{\rm mol}$, $\tau_{\rm dep}$, and $\alpha_{\rm AGN}$. 
According to the Pearson correlation coefficients, none of these relations show a significant trend, hence, in the main body of the paper, we show only the $r_{\rm CICO}$ vs $L_{\rm IR}$ relation (Figure~\ref{fig:rCICO_LIR}), and report the other plots in the Appendix (Figure~\ref{fig:rCICO_properties}).
As explained earlier, it is hard to make a comparison between our results and previous studies of [CI] and CO in galaxies, because these sparse previous analyses have mostly considered very heterogeneous samples and datasets, with large aperture and sensitivity variations between different data. 
With these caveats in mind, we can discuss Figure~\ref{fig:rCICO_LIR} in relation to some previous findings. 
A study performed by \cite{Israel+15} in a sample of 76 local galaxies with {\it Herschel} [CI] line data and ground-based JCMT $^{13}$CO(2--1) line spectra, shows a correlation between the [CI](1--0)/$^{13}$CO(2--1) flux ratio and $L_{\rm FIR}$.
Though with a Pearson parameter indicating a non-significant correlation, Figure \ref{fig:rCICO_LIR} shows a hint of a positive trend between the [CI](1--0)/$^{12}$CO(1--0) line ratio and infrared luminosity in our sample of local ULIRGs.
There are however two important things to consider.
Firstly, the [CI](1--0)/$^{13}$CO(2--1) ratio investigated by \cite{Israel+15} may behave differently from the [CI](1--0)/$^{12}$CO(1--0) line ratio that we explore here.
Indeed the different isotopologue and rotational level of the CO transition used to compute the ratio implies a different excitation temperature, abundance, and opacity. 
Secondly, the study by \cite{Israel+15} includes sources spanning $L_{\rm IR}\sim 10^{9}~\rm L_{\odot}$ to $\gtrsim 10^{12}~\rm L_{\odot}$, while all the data points studied in Figure \ref{fig:rCICO_LIR} are in the $L_{\rm IR}>10^{12}~\rm L_{\odot}$ regime.
Performing a visual inspection of the \cite{Israel+15} [CI](1--0)/$^{13}$CO(2--1) - $L_{\rm FIR}$ relation, considering only the data points at $L_{\rm FIR}>10^{12}~\rm L_{\odot}$, we observe no clear trend.
Taking these considerations into account, our results are actually in agreement with these previous findings: our average $r_{\textrm{CICO}}$ is clearly higher than that computed in less IR luminous galaxies by \cite{Jiao+19}, as discussed above for Fig~\ref{fig:CICOratio_histo}. 
Another study we can examine is \cite{Valentino+18}. These authors investigated the [CI](1--0)/CO(2--1) luminosity ratio in a heterogeneous sample including star forming galaxies on the $z\sim1.2$ main sequence observed with ALMA, and $\sim30$ local starbursts (mostly LIRGs and AGNs) with archival {\it Herschel} [CI] and ground-based single dish low-{\it J} CO data, respectively from \cite{Liu+15} and \cite{Kamenetzky+16}. Their total sample spans $10^{10} \lesssim L_{\rm IR}~[\rm L_{\odot}] \lesssim 10^{13.5}$; they find no [CI]/CO ratio variations as a function of $L_{\rm IR}$, with an average value of $L^{\prime}_{\rm [CI](1-0)}/L^{\prime}_{\rm CO(2-1)} \simeq 0.20\pm0.02$, consistent with our results.

Similarly, for SFR and $L_{\rm AGN}$, we retrieve Pearson correlation coefficients of $\rho_{r_{\rm CICO}-{\rm SFR}} = 0.31$ and $\rho_{r_{\rm CICO}-{L_{\rm AGN}}} = 0.39$ (with \textit{p}-values of $0.28$ and $0.17$, respectively).

\section{Discussion}\label{sec:discussion}

The discussion is divided in three parts. First, in Section~\ref{sec:discus_COratios}, we discuss the possible origin of the high CO line ratios measured in our sample, by exploring also possible dependencies on gas kinematics (ISM turbulence, and presence of outflows). Then, in Section~\ref{sec:discus_rCICO} we discuss the origin of the high $r_{\rm CICO}$ ratios, and investigate their relationship with prominent molecular outflows. 
Finally, in Section~\ref{sec:discus_AGNfeedback} we discuss the role of AGN feedback in setting the molecular gas properties of (U)LIRGs.

\subsection{Physical drivers of low-J CO line ratios in (U)LIRGs}\label{sec:discus_COratios}
\subsubsection{A weak dependence on ISM excitation}
Our sample is characterized by extremely high {\it global} $r_{21}$, $r_{31}$, and $r_{32}$ values, which show little or no overlap with those measured in the local main sequence galaxy population (see Figure~\ref{fig:COratios_histo}).  
Only marginal correlations were found, mainly for $r_{31}$ with $L_{\rm IR}$, SFR, and $\tau_{\rm dep}$, while no statistical trends were found with $L_{\rm AGN}$, or $\alpha_{\rm AGN}$ (see Figure~\ref{fig:COratios_vs_properties}).
This suggests that, in these (U)LIRGs, galaxy-averaged low-{\it J} CO line ratios are unable to trace effects originated by AGN activity, while ratios depending on CO(3--2) line emission may start to reflect effects of star formation activity and gas density (seen in the relations with SFR and $\tau_{\rm dep}$).
Overall, our results agree with the conclusion of \cite{Papadopoulos+12a}, who stated that the low-excitation $J_{\rm up}\leq3$ portion of CO spectral line energy distributions (SLEDs) are ``highly degenerate to the average state of the ISM'', as we verified here for these (U)LIRGs that are not characterized by cold CO SLEDs.

We highlight the observed large variations of CO line ratios within our sample, much larger than those measured in main sequence galaxies. Does Fig.~\ref{fig:COratios_vs_properties} inform us about the drivers of such strong differences? The largest $r_{J, J-1}$ variations appear to be typical at the high-$L_{\rm IR}$ end of the sample, in particular for the global $r_{21}$ values, which become consistently $r_{21}>1$ at $\log_{10}L_{\rm IR}\,[\rm L_{\odot}]\simeq12.3$. 
However, Figure~\ref{fig:COratios_vs_properties} shows only a weak a link between large CO ratio variations and SFR, but no link to AGNs.
This may seem puzzling since AGNs are strong sources of far-UV photons and Cosmic rays, which can excite CO, but it is consistent with the hypothesis that low-{\it J} CO line ratios in these (U)LIRGs are actually degenerate for CO excitation effects. Higher-$J$ CO transitions are needed to better constrain the effects of AGN on the excitation of the ISM \citep{Gallerani+14, Rosenberg+15, Jarvis+20}.

Ratios $>1.0$ can only be obtained through highly excited gas coupled with low optical depths. More specifically, CO lines can become partially transparent (i.e., reduced line opacities) in the presence of a diffuse, warm, turbulent component characterized by a large velocity gradient. \cite{Cicone+18}, based on the analysis of NGC~6240, suggested that such turbulent/envelope H$_2$ phase may be typical of massive molecular outflows, which are widespread in (U)LIRGs. In Section~\ref{sec:disc_ratios_kinem} we explore 
the hypothesis of a connection between high CO line ratios and the gas kinematics.

\subsubsection{Effects of high-velocity and turbulent gas on the CO line ratios}\label{sec:disc_ratios_kinem}
\begin{figure*}[htb]
	\centering
	\includegraphics[width=.95\textwidth]{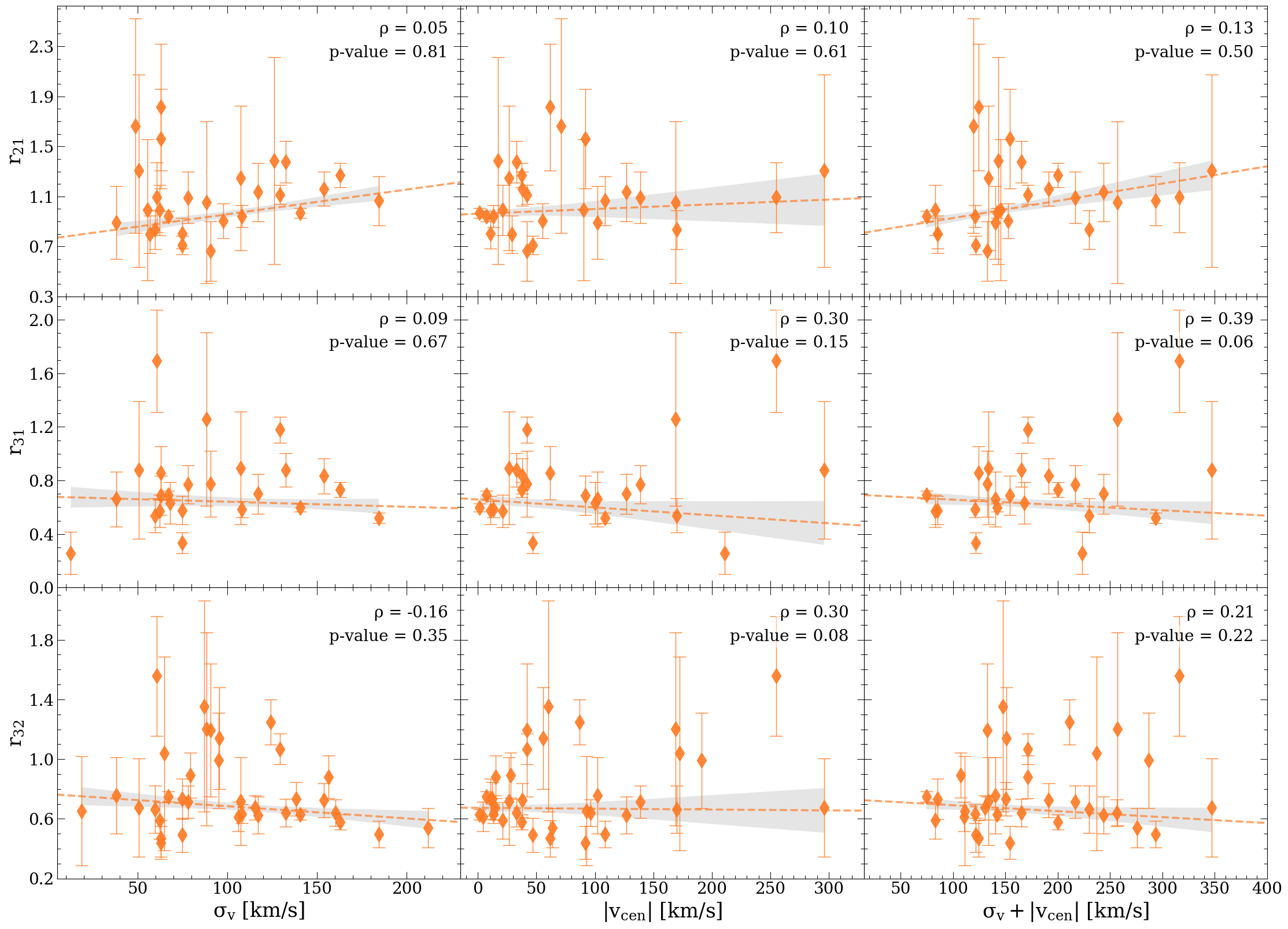}
	\caption{Individual spectral component analysis of the CO line ratios as a function of the gas kinematics for our sample of ULIRGs. We have set a detection threshold of $2\sigma$ for the flux of each Gaussian component in order to compute the ratios. 
	The top, central, and bottom rows show respectively the $r_{21}$, $r_{31}$, and $r_{32}$ plots. \textit{Left column:} line ratios as a function of the velocity dispersion $\sigma_{v}$ (line width of the spectral Gaussian component). \textit{Middle column:} line ratios as a function of the module of the central velocity of the spectral Gaussian component ($|v_{\rm cen}|$), measured with respect to the galaxy's redshift reported in Table~\ref{table:source_list}. \textit{Right column:} line ratios as a function of the $\sigma_{v} + |v_{\mathrm{cen}}|$ parameter. The dashed lines correspond to the best-fit relations obtained from a least squares regression analysis, and the shaded gray areas correspond to the $1\sigma$ confidence interval of the fit. The Pearson correlation coefficients and associated {\it p}-values are reported on the top-right corner of each plot, although they are less informative than the best-fit because they do not take the (large) error-bars into account.}
	\label{fig:ratios_kinematics}
\end{figure*}

The simultaneous multi-Gaussian component spectral fitting procedure described in Section~\ref{sec:spectralfitting} (see spectral fits in Figures~\ref{fig:spectra1} to \ref{fig:spectra6}), allows us to study CO line ratios as a function of the kinematic properties ($v_{cen}$ and $\sigma_v$) of the CO-emitting gas. 
The results of such investigation are shown in Figure~\ref{fig:ratios_kinematics}, where we plotted the $r_{21}$, $r_{31}$, and $r_{32}$ obtained for each Gaussian component employed in the fit, as a function of their kinematic parameters, namely: $\sigma_v$, $|v_{\rm cen}|$, and the combined parameter $\sigma_{v} + |v_{\mathrm{cen}}|$. We also studied the dependency on $v_{\rm cen}$, and the results were similar to the ones showed in Figure \ref{fig:ratios_kinematics}, which is why we choose to only show the absolute value.

Concerning the gas kinematics, 
the $\sigma_{v}$ values range from $\sim10$~\kms~up to $\sim230$~\kms, and $v_{\mathrm{cen}}$ from $\sim-300$~\kms~up to $\sim315$~\kms. Most Gaussian components display $|v_{\mathrm{cen}}| \lesssim100$~\kms, which are most likely tracing `disk' material with systemic velocities, though we observe components reaching up to $|v_{\mathrm{cen}}| \sim 315$ \kms~tracing blue/red-shifted velocity gas, with motions clearly deviating from ordered rotation. 
We caution against an over-interpretation of the specific best-fit kinematic properties for the individual targets, because the possibility to fit multiple components depends mainly on the S/N of the lines and on the presence of spectral sub-structures. As a result, a source with generally broad CO lines may be fitted either with one single broad component or with multiple narrow components at different velocities. We therefore stress again that the plots presented in Fig.~\ref{fig:ratios_kinematics} should be regarded for their statistical value. 

The CO line ratios measured for the individual Gaussian components span similar ranges as the global galaxy-integrated ratios (see Figure~\ref{fig:COratios_histo}), namely: $0.7 \leq r_{21} \leq 1.8$, $0.3 \leq r_{31} \leq 1.8$, and $0.5 \leq r_{32} \leq1.6$.
The Pearson correlation coefficients ($\rho$) and associated {\it p}-values (see Figure~\ref{fig:ratios_kinematics}) do not evidence any significant trend among the nine relations explored. There are clearly instances where spectral components characterized by higher $\sigma_v$ and/or higher central velocity show higher CO line ratios, but these trends are not statistically significant if considering the whole sample.

However, we note that the correlation coefficients do not take into account the error-bars, which are quite large in this case. We therefore proceed with a regression analysis, whose results are over-plotted in Figure~\ref{fig:ratios_kinematics}.
The $r_{21}$ vs $\sigma_{v}$ plot ({\it top-left panel}) shows a statistical positive trend, described by the relation:
\begin{equation}\label{eq:r21_vs_sigmav}
	r_{21} = (0.76 \pm 0.08) + (2.0 \pm 0.7)\times 10^{-3} \sigma_{v}.
\end{equation}
A similar one is found between $r_{21}$ and $\sigma_{v} + |v_{\mathrm{cen}}|$ ($r_{21} = (0.79 \pm 0.07) + (1.4 \pm 0.5)\times 10^{-3} (\sigma_{v} + |v_{\mathrm{cen}}|)$).
Since no clear trend is detected between $r_{21}$ and $|v_{\mathrm{cen}}|$, the relation found between $r_{21}$ and $\sigma_{v} + |v_{\mathrm{cen}}|$ is consistent with being entirely driven by the trend with $\sigma_v$.

Among the CO ratios explored here, $r_{21}$ is the least sensitive to ISM gas excitation, and so we can hypothesize that the scatter observed in $r_{21}$ values is dominated by optical depth effects \citep[see also][]{Zschaechner+18}. 
In particular, high $r_{21}\gtrsim1$ can be explained by a low opacity, especially in the CO(1-0) transition. The positive correlation between $r_{21}$ and $\sigma_v$ reported in Equation~\ref{eq:r21_vs_sigmav} suggests that the low CO opacity is driven by large velocity gradients, which can in turn be due to turbulence in the ISM and/or bulk motions within molecular outflows. A similar positive trend between $r_{21}$ and $\sigma_v$ was observed in NGC~6240, and in that case the high-$r_{21}$ and high-$\sigma_v$ gas was predominantly linked to the massive molecular outflow \citep{Cicone+18}.

We note that neither the $r_{31}$ nor the $r_{32}$ values show significant trends as a function of gas kinematics.
It is possible that ratios involving the CO(3--2) line transition are less sensitive to optical depth effects and more affected by gas excitation, hence the absence of trends with these ratios may support an interpretation of the $r_{21}$ vs $\sigma_v$ relation in terms of opacity of CO lines, rather than gas excitation effects.

\begin{figure*}[tbp]
	\centering
    \includegraphics[width=.42\textwidth]{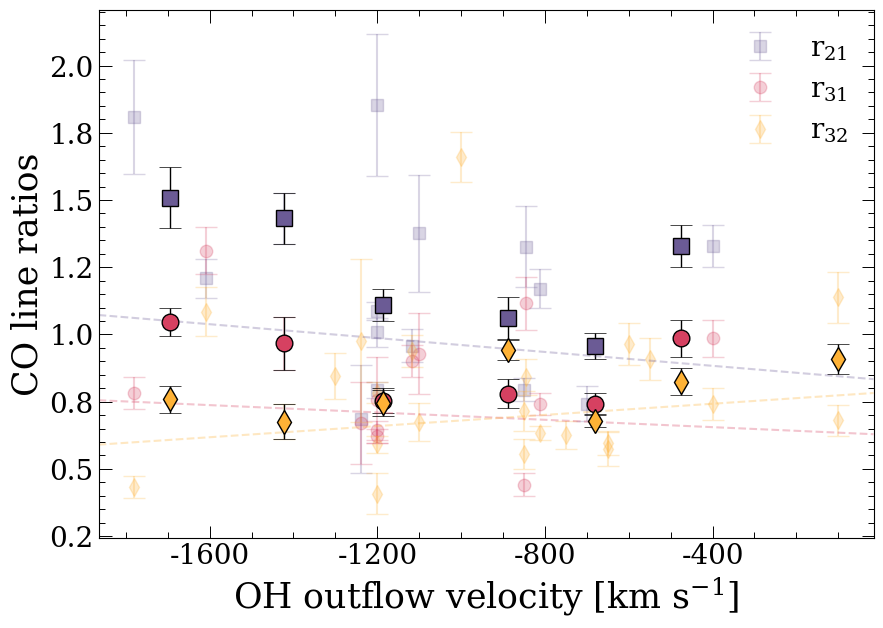}\qquad\qquad
    \includegraphics[width=.42\textwidth]{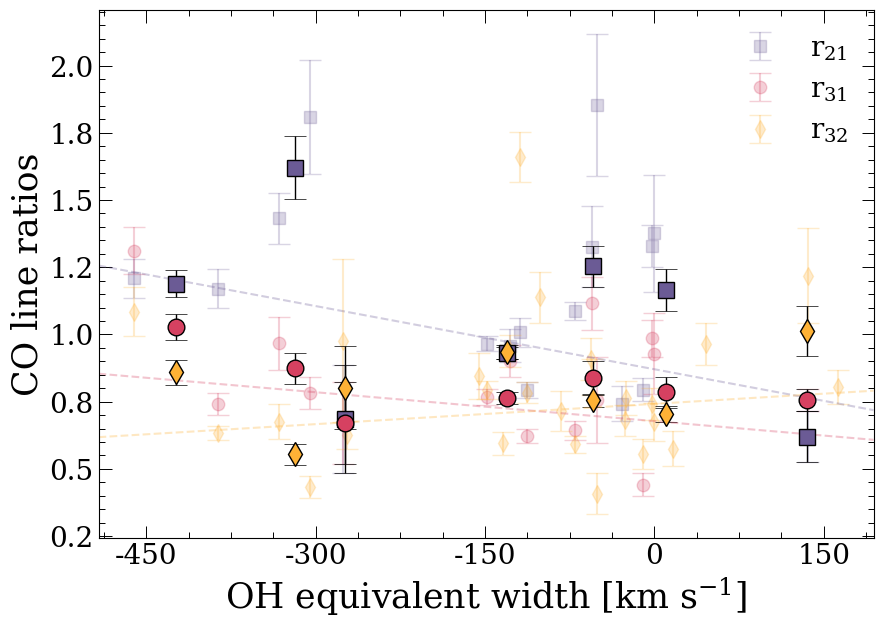}\\
	\caption{Global CO line ratios as a function of OH$_{\rm max}$ (\textit{left}) and OH$_{\rm EQW}$ (\textit{right}) for the sources with OH119 $\mu$m data as reported in the literature (see Table \ref{table:source_list}). The binned values (computed in the same way as in Section \ref{sec:tot_COratios}) are shown using darker colors and overplotted on the individual data points. Dashed lines indicate the best fit relations obtained from a least squares regression analysis conducted for each transition line separately. The color coding refers to different line ratios, as indicated in the legend at the top-right corner of each plot.}
	\label{fig:ratiosCO_OH}
\end{figure*}

As explained in Section \ref{sec:sample}, our sources are selected from the \textit{Herschel} OH parent sample, and we can make use of the literature OH119 $\mu$m data to study possible dependencies between the line ratios measured in this paper and the available OH outflow parameters; mainly the OH maximal outflow velocity (OH$_{\rm max}$) for sources showing absorption components, and the equivalent width of the OH line (OH$_{\rm EQW}$). This latter is a negative value for sources where the absorption components is stronger than the emission component of the OH119 $\mu$m doublet.
The OH parameters are also tracers of molecular gas kinematics in these galaxies and we can test whether there are correlations with the \textit{global} CO line ratios measured in our sample. In Figure \ref{fig:ratiosCO_OH} we show the global $r_{21}$, $r_{31}$ and $r_{32}$ line ratios as a function of the OH$_{\rm max}$ and OH$_{\rm EQW}$.
As we did also in Section \ref{sec:COratios_galaxyprop}, we divided the total sample in bins according to the quantity on the \textit{x}-axis and calculated the mean values of the ratios in these bins, which are shown using darker symbols. The dashed lines show the best-fit relations, with color coding indicating different CO line ratios.
We do not observe any statistically significant trend between the quantities, with Pearson correlation coefficients of
$\rho_{r_{21}-{\rm OH}_{\rm max}}=-0.33$ ({\it p}-value = 0.23), 
$\rho_{r_{31}-{\rm OH}_{\rm max}}=-0.13$ ({\it p}-value = 0.68), 
$\rho_{r_{32}-{\rm OH}_{\rm max}}=0.10$ ({\it p}-value = 0.64), 
$\rho_{r_{21}-{\rm OH}_{\rm EQW}}=-0.26$ ({\it p}-value = 0.31), 
$\rho_{r_{31}-{\rm OH}_{\rm EQW}}=-0.32$ ({\it p}-value = 0.24), 
$\rho_{r_{32}-{\rm OH}_{\rm EQW}}=0.06$ ({\it p}-value = 0.78).
Additionally, the regression analysis performed on each luminosity ratio, which takes into account the error bars, reveal best-fit parameters that are consistent with flat trends for all the variables.

Summarizing, we study any possible trends between the CO line ratios and the kinematics of the gas as traced by (i) the velocity of the individual spectral components of the CO multi-Gaussian fit, and (ii) the OH outflow properties. The former, allows us to study the ratios for each spectral component, for which we retrieve a statistically significant positive trend (via regression analysis) between $r_{21}$ and $\sigma_{v}$, likely due to a lower CO opacity in the high-$\sigma_v$ gas, which could be tracing outflows. 
On the other hand, we do not find significant trends between the global $r_{21}$ ratios and OH$119\mu$m outflow properties, showing that stronger OH outflows are not statistically associated with globally enhanced $r_{21}$ values.
These two results are only in apparent contradiction. Indeed, 
the OH$119\mu$m outflows are detected in absorption against the FIR continuum and so trace mainly the inner components of the galaxy-scale molecular outflows, while the low-\textit{J} CO lines, observed in emission and integrated over the whole extent of these galaxies, probe the totality of the molecular gas including extended and diffuse components far from the central engines.
Indeed, it would be remarkable if we were to find a trend between the global $r_{21}$ values and the OH outflow properties.
Higher-\textit{J} CO lines, tracing the medium exposed to the excitation sources, may show tighter relations with the OH outflows. This hypothesis needs to be tested with \textit{J$_{\rm up}$}$\geq 4$ CO line observations.
In conclusion, our analysis of CO line ratios as a function of gas kinematics, suggests that molecular outflows are not statistically characterized by higher CO line ratios in ULIRGs, at least for low-\textit{J} CO lines up to $J=3$. In other words, the low-\textit{J} CO line ratios do not appear sensitive to the presence or strength of massive molecular outflows.

\subsubsection{Absence of a relation between $r_{21}$ and $\alpha_{\rm CO}$}\label{sec:disc_r21_alpha}

\begin{figure}[tbp]
	\centering
	\includegraphics[width=.95\columnwidth]{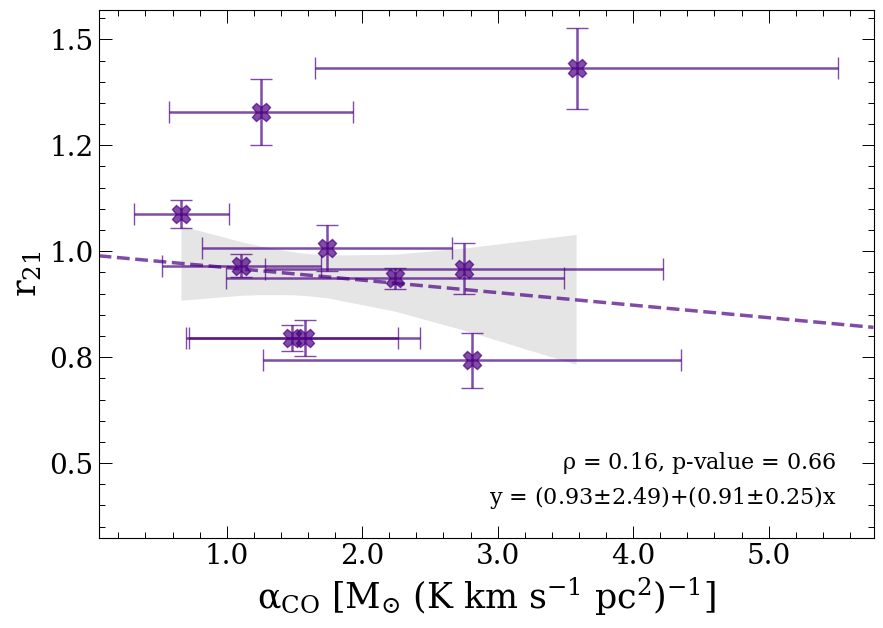}
	\caption{Global $r_{21}$ ratios plotted as a function of [CI]-derived $\alpha_{\rm CO}$ values. The purple dashed line shows the best-fit relation, with its corresponding 1$\sigma$ confidence interval represented by the gray shaded area. Neither the Pearson correlation coefficient ($\rho$) nor the regression analysis suggest any correlation between these quantities.}
	\label{fig:r21_vs_alphaCO}
\end{figure}

It has been proposed that, although the CO(2--1)/CO(1--0) luminosity ratio alone cannot be used to constrain the $\alpha_{\rm CO}$ in galaxies (which require either independent H$_2$ tracers or a much more accurate multi-{\it J} analysis of CO lines), variations in $r_{21}$ should be correlated to those observed in $\alpha_{\rm CO}$, because both these parameters are affected by the optical depth of the gas. This holds as long as other factors, such as the CO abundance, are fixed. 
In particular, \cite{Gong+20} predicted a close-to-linear anti-correlation between $r_{21}$ and $\alpha_{\rm CO}$ ($r_{\rm 21}\approxprop (1/\alpha_{\rm CO})^{1.2}$), based on their models tailored to the ISM conditions typical of galaxy disks. 
Such a trend is also strengthen by the findings in \cite{Leroy+22}: these authors, by fitting radiative transfer models to measured line ratios, find an anti-correlation between $r_{21}$ and $\alpha_{\rm CO}$. 
However, such dependency is not found in our sample. 
As shown by Figure~\ref{fig:r21_vs_alphaCO}, $r_{21}$ and  $\alpha_{\rm CO}$ values are not correlated ($\rho=0.16$, {\it p}-value = 0.66). The best-fit relation, which takes into account the large uncertainties on $\alpha_{\rm CO}$, is consistent with a flat trend.

This result further strengthens the hypothesis that the molecular ISM of (U)LIRGs, as traced by low-{\it J} CO line ratios, is extremely different from that of normal galaxy disks. The global $r_{21}>1$ measured in our sample are not even allowed by the models proposed by \cite{Gong+20}. Furthermore, even if such high $r_{21}$ values strongly suggest optically thin CO emission, the corresponding $\alpha_{\rm CO}$ estimates of $\sim1-4~\rm M_{\odot}~(K~km~s^{-1}~pc^2)^{-1}$ measured in those same sources that display $r_{21}>1$ (see Figure~\ref{fig:r21_vs_alphaCO}), are significantly higher than the optically thin $\alpha_{\rm CO}$ values measured in, e.g., disk galaxies \citep[see, e.g.,][]{Bolatto+13}. This indicates an apparent contradiction that was already observed in NGC~6240 \citep{Cicone+18}. A possible explanation can be the coexistence of a diffuse and turbulent molecular ISM phase, which is characterized by large velocity gradients (such as those typical of outflows) and dominating the low-{\it J} CO luminosity, with a denser phase, which is dominating the gas mass and driving up the [CI]-based $\alpha_{\rm CO}$ measurements. The latter phase may be preferentially traced by denser molecular gas tracers, such as CN, HCN, CS, and higher-{\it J} CO transitions.

\subsection{Effects of high-velocity and turbulent gas on the [CI]/CO line ratio}\label{sec:discus_rCICO}

\begin{figure*}[tbp]
	\centering
    \includegraphics[width=.40\textwidth]{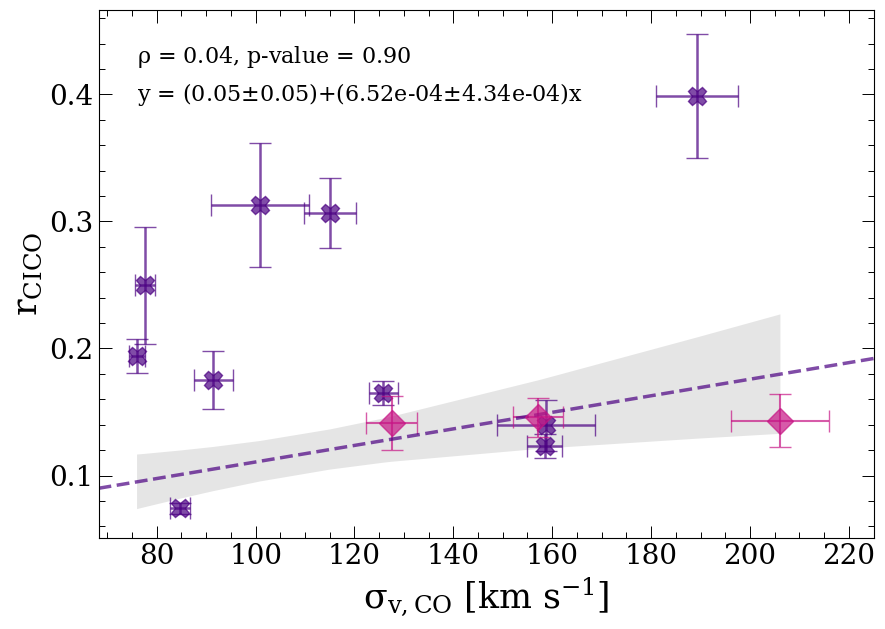}\qquad\qquad
	\includegraphics[width=.40\textwidth]{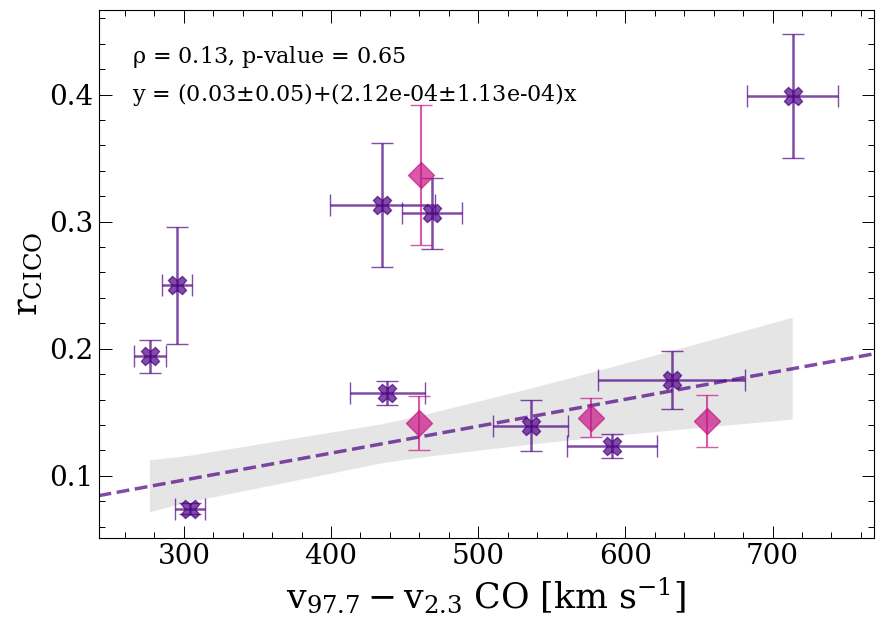}\\
    \includegraphics[width=.40\textwidth]{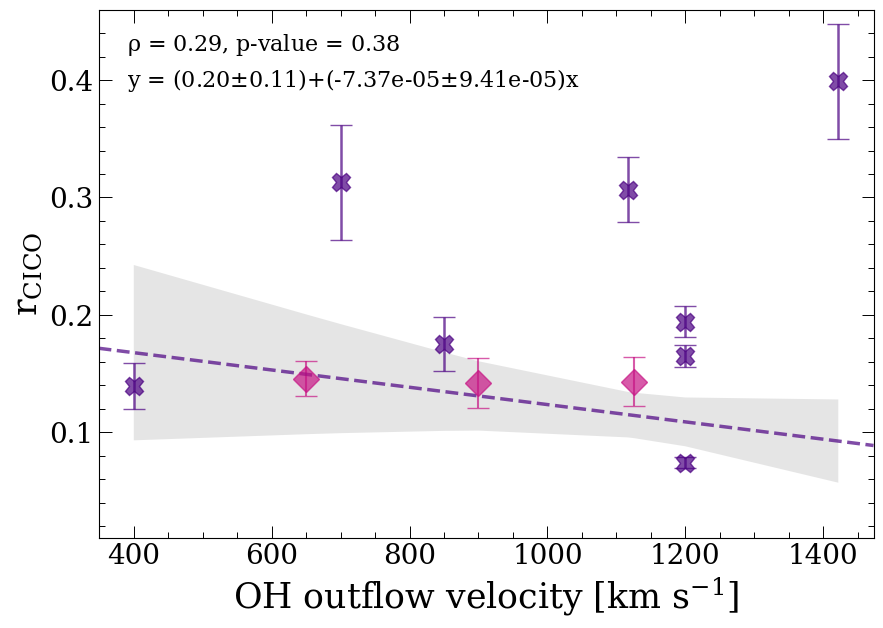}\qquad\qquad
    \includegraphics[width=.40\textwidth]{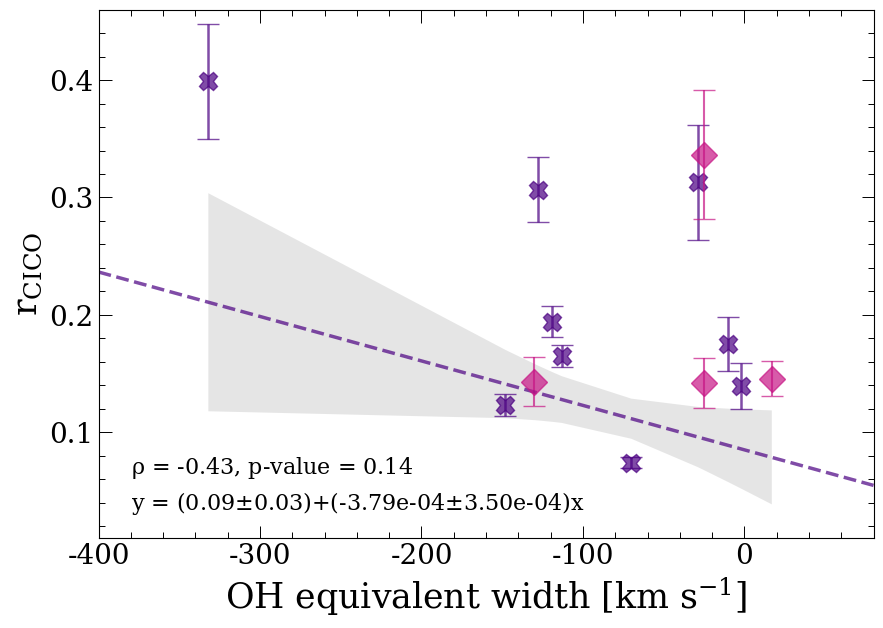}
	\caption{\textit{Top row:} [CI](1--0)/CO(1--0) luminosity ratio as a function of $\sigma_{\rm v,CO}$ from the single Gaussian fit (see Section \ref{sec:sigmas}) (\textit{left}), and as a function of ${\rm v_{97.7}}-{\rm v_{2.3}}$ velocity interval for CO (\textit{right}).
    \textit{Bottom row}: $r_{\rm CICO}$ line ratio as a function of OH$_{\rm max}$ (\textit{left}) and OH$_{\rm EQW}$ (\textit{right}).
    Sources with direct measurements of both the [CI](1--0) and CO(1--0) lines are plotted using purple crosses. For the other sources (pink diamonds), the CO(1--0) line luminosity was estimated assuming the average CO(2--1)/CO(1--0) luminosity ratio of $r_{21}=1.1$ computed for our sample of ULIRGs.
	The dashed purple line represents the best-fit using a least squares regression analysis, and the shaded gray area corresponds to the $1\sigma$ confidence interval of the fit. The Pearson correlation coefficients ($\rho$) are reported at the upper left corner of the panel with their associated {\it p}-values.
    }
	\label{fig:COwidth_rCICO}
\end{figure*}

In Section~\ref{sec:results} we showed that, in our targets, $L^{\prime}_{\rm CO(1-0)}$ (as well as $L^{\prime}_{\rm CO(2-1)}$) is tightly correlated with $L^{\prime}_{\rm [CI](1-0)}$, and that the  [CI](1--0)/CO(1--0) luminosity ratios, spanning $0.07\lesssim r_{\rm CICO}\lesssim0.4$, are significantly higher than those measured in normal star forming galaxies. Such high $r_{\rm CICO}$ values appear to be typical of IR-bright galaxies, as suggested by previous works \citep{Israel+15, Valentino+18}. 
Within our sample, we do not find any significant relation between $r_{\rm CICO}$ and intrinsic galaxy properties such as SFR, $M_{\rm mol}$, $\tau_{\rm dep}$, or $L_{\rm AGN}$. 
In this section we test the hypothesis of a link between $r_{\rm CICO}$ and gas kinematics, and in particular with molecular outflows.

As described in Section~\ref{sec:spectralfitting}, the [CI](1--0) line spectra were fitted independently from the CO lines, due to their generally lower S/N, and to the possibility that CO and [CI] lines may not have identical profiles (see also results in Sect~\ref{sec:sigmas}). Therefore, it is not possible to investigate $r_{\rm CICO}$ ratios as a function of kinematic parameters for individual Gaussian components as we did in Figure~\ref{fig:ratios_kinematics} for the CO line ratios. However, we can still investigate the global $r_{\rm CICO}$ values as a function of velocity dispersion of the gas, as traced by, e.g., $\sigma_{\rm v}$ (from the single Gaussian fit to [CI](1--0) and CO(2--1), see Section \ref{sec:sigmas}), ${\rm v_{97.7}}-{\rm v_{2.3}}$ interval for CO and [CI], and as a function of OH properties, i.e., maximal outflow velocity and equivalent width.
In Figure \ref{fig:COwidth_rCICO}, we show the relations between $r_{\rm CICO}$ and $\sigma_{\rm v,CO}$ (top-left panel), ${\rm v_{97.7}}-{\rm v_{2.3}}$ for CO (top-right panel), OH outflow maximal velocity (bottom-left panel) and OH equivalent width (bottom-right panel). 
All Pearson correlation coefficients and the respective \textit{p}-values are reported in the figures, evidencing no indication of a correlation between any of the quantities. The regression analyses, which take into account the error bars, are consistent with flat trends in all cases. 
Similarly, we do not find any trend between $r_{\rm CICO}$ and [CI] line widths, as probed by $\sigma_{\rm v,[CI]}$ and its ${\rm v_{97.7}}-{\rm v_{2.3}}$ velocity interval.
Therefore, our results do not evidence a significant impact of the gas kinematics on the global [CI]/CO luminosity ratio in (U)LIRGs.

In light of such independence of $r_{\rm CICO}$ on gas kinematics, it remains puzzling to explain the detection, shown in Fig~\ref{fig:COwid_CIwid}, of a marginal deviation of the [CI] and CO line widths from a 1:1 relation, where [CI] lines appear narrower than CO lines in sources with broad CO lines.
One hypothesis is that a slight depletion of [CI] with respect to CO in the broad wings of the lines conspires with an enhancement of [CI]/CO in the narrow core of the same lines, so that $\sigma_{\rm v,[CI]}<\sigma_{\rm v, CO}$ while the total luminosity ratio, $r_{\rm CICO}$, remains unchanged.
The reason why [CI] may be fainter in the high-velocity wings of the molecular lines could be due to its optically thin nature. This makes it challenging to detect this line in the presence of more diffuse and extended components of the medium, where the column density is much lower than in the disks. On the contrary, low-\textit{J} CO lines such as CO(1--0) and CO(2--1) remain bright and relatively easy to detect in such components.
The apparent bi-modality of the data observed in the two top panels of Figure \ref{fig:COwidth_rCICO}, where a cluster of sources with high $r_{\rm CICO}$ and narrower CO lines populate the upper-left regions of the plots, while the remaining data points show low $r_{\rm CICO}$ and high CO line widths, could strengthen that hypothesis.
We warn the reader, however, that this needs to be confirmed with larger statistics.

Moreover, the premise of the different behaviors for the two gas tracers studied here is not supported by any theoretical prediction. Actually, such effect would be at odds with the theoretical prediction that massive molecular outflows contain a significant CO-dark H$_2$ component, due to leakage of far-UV photons and CRs \citep{Papadopoulos+04, Papadopoulos+18}, which instead seems to be a promising avenue to explain the observational results by \cite{Saito+22} and \cite{Ueda+22}. In particular, \cite{Saito+22} used ALMA to explore the [CI](1--0) and CO(1--0) emission within the central 1~kpc of the Seyfert~2 galaxy NGC~1068. They measured the highest $r_{\rm CICO}$ values (median 0.72, 16th-84th range of 0.33-1.86) in a region offset by $\sim300$~pc from the AGN position, beyond the circumnuclear disk, which appears to correspond to the shocked environment between the molecular outflow (see \cite{Garcia-Burillo+14}) and the galactic disk.
These authors interpreted the high [CI]/CO ratios as arising from the shocked gas, due to efficient shock dissociation of H$_2$ gas by the shock wave, and the consequent dissociation of CO via atomic hydrogen endothermically reacting with said molecule \citep[see][]{Hollenbach+80}.
On the contrary, the starburst ring of NGC~1068, which is unaffected by the outflow, displays a much lower median $r_{\rm CICO}=0.15$ (16th-84th ranges of 0.08-0.23).
Given the limited field of view (FoV) of the [CI](1--0) observation by \cite{Saito+22}, an $r_{\rm CICO}$ map beyond the central 1~kpc, which could probe the outflow in regions not overlapping with the galactic disk (therefore not shocked), is not available.

Another example is the study of Arp~220 performed by \cite{Ueda+22}, who obtained an $r_{\rm CICO}$ map of the central $5''$ (1.9 kpc) of the galaxy. 
These authors, however, report that the archival ALMA data used in their analysis suffer from missing flux, whose fraction they estimate to be 34\% for the [CI](1--0) line through a comparison between their data and JCMT observations \citep{Papadopoulos+04}. Therefore, the results presented by \cite{Ueda+22} apply only for components smaller than the maximum recoverable scale (MRS) of their data sets, 1.75 kpc.
\cite{Ueda+22} measure $r_{\rm CICO}^{red~OF}=0.9\pm0.3$ in the collimated molecular outflow hosted by the western nucleus of Arp~220, clearly higher than the galaxy-integrated value of $r_{\rm CICO}=0.22\pm0.04$. 
They interpret such a high $r_{\rm CICO}$ value to be arising from regions with elevated [CI]/CO abundance ratio, which may be caused either by CRs or by shocks.
If compared to the results obtained by \cite{Saito+22}, the regions with high $r_{\rm CICO}$ in Arp~220 are well within its central region and as such, they could be also driven by shock waves leading to CO dissociation, as suggested for NGC 1068. However, disentangling between CRs, high radiation fields or shocks, is challenging.
Additionally, \cite{Ueda+22} also study how $r_{\rm CICO}$ relates to $r_{31}$ and find a positive trend between $r_{\rm CICO}>0.3$ measurements and high $r_{31}>1$ values, hence regions with high [CI]/CO luminosity show very high, optically thin CO(3--2)/CO(1--0) ratios. 
In our sample, although most of the sources with $r_{31}\gtrsim0.9$ have also $r_{\rm CICO}>0.2$, the correlation found between the two quantities is not statistically significant ($\rho=0.52$, {\it p}-value = 0.15, best fit relation: $r_{\rm CICO}=(-0.05\pm0.14)+(0.24\pm0.20)~r_{31}$). 
However, any kpc-scale effects of the type observed by \cite{Saito+22} and \cite{Ueda+22}, possibly triggered by collimated outflows and localized shocks, would probably be washed out in the galaxy-integrated measurements used in this work.

\subsection{Role of AGN feedback}\label{sec:discus_AGNfeedback}

The CO and [CI]/CO line ratios investigated in this work do not show any significant correlation with $L_{\rm AGN}$ or $\alpha_{\rm AGN}$, suggesting that AGN radiation has little effect on these ratios, in this sample of (U)LIRGs.

AGN feedback could manifest in different ways in galaxies. One manifestation is through powerful outflows, which expel gas from the central regions and may deplete the galaxy of its $M_{\rm mol}$ reservoir \citep{Cicone+14}. Alternatively, AGN feedback could inject turbulence in the gas (through, e.g., compact jets or outflows), making star formation inefficient even in the presence of cold and dense H$_2$ gas (see, e.g., \cite{Alatalo+15}).

\begin{figure}[tbp]
	\centering
	\includegraphics[width=.95\columnwidth]{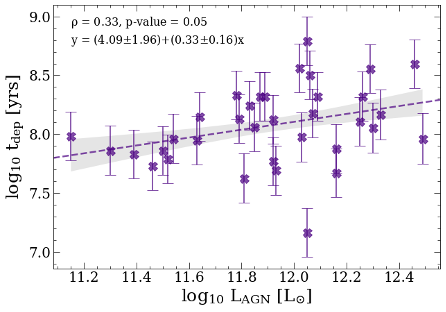}
	\caption{Depletion time-scale ($\rm \tau_{dep}\equiv M_{mol}/SFR$) plotted as a function of L$_{\textrm{AGN}}$. The dashed line shows the best-fit relation and its $1\sigma$ confidence interval plotted as a shaded gray area. The best-fit parameters and the Pearson correlation coefficient are reported at the upper left corner of the panel.}
	\label{fig:LAGN_tdep}
\end{figure}

We exploit our refined total molecular gas mass estimates, which are based on higher quality data as well as on an independent check on the $\alpha_{\rm CO}$ factor, to investigate any links between the AGN luminosity and signatures of suppressed star formation in these (U)LIRGs.
Figure~\ref{fig:LAGN_tdep} shows the molecular gas depletion timescale as a function of $L_{\rm AGN}$. We note that this is the depletion timescale due to consumption of gas by the star formation activity, without including the gas expelled via outflows (see, e.g., \cite{Cicone+14} for an investigation of the latter).

We find a weak though still statistically significant correlation ($\rho=0.33$, {\it p}-value = 0.05), implying that the ULIRGs with more luminous AGNs consume their gas slower than the ones with less luminous AGNs, or, in other words, have a less efficient star formation. 
We note, however, that the $L_{\rm AGN}$ values, taken from \cite{Veilleux+13} and \cite{Spoon+13}, were computed using method 6 by \cite{Veilleux+09b}, which is based on the 15 to 30 $\mu$m continuum ratio ($f_{30}/f_{15}$). This method carries uncertainties of the order of $\sim20$\% on average for the reported $L_{\rm AGN}$. Moreover, we have also explored the $\tau_{\rm dep}$ vs OH outflow velocity relation, and we did not find any significant trend.

\section{Summary and conclusions}\label{sec:conclusions}
We have performed a combined analysis of high S/N CO(1--0), CO(2--1), CO(3--2) and [CI](1--0) emission line spectra of 36 ULIRGs and additional 4 LIRGs at $z<0.2$, selected to have OH$119$ $\mu$m observations from {\it Herschel}. These datasets are particularly sensitive to the {\it total} line emission from the sources, including faint, low surface brightness components that may be missed by high-resolution interferometric observations. 
All data, both PI and archival, were reduced and analyzed in a uniform way, with the goal of minimizing the aperture biases that usually affect the ratios computed between lines observed at different frequencies. 
The LIRGs, which are a minority in our sample, were discussed separately from the ULIRGs to avoid that the low statistics at $L_{\rm IR} < 10^{12}~L_{\odot}$ biased our relations.
Our findings can be summarized as follows:
\begin{itemize}
	\item We find a tight relation between CO and [CI] luminosity: $\log_{10}\,L'_{\textrm{CO(1--0)}} = (4 \pm 2)+(0.64\pm0.26)\, \log_{10}\,L'_{\textrm{[CI](1--0)}}$, and a very similar one for CO(2--1). This strengthens the hypothesis that atomic Carbon is a valid alternative to CO for tracing the bulk of H$_2$ gas in galaxies. 
	\item We compared the line widths of [CI] and CO, and derived a best-fit relation  of: $\sigma_{\rm v,[CI]} = (30\pm9) + (0.64\pm0.07)~\sigma_{\rm v,CO}$, when comparing the velocity dispersion obtained via a single Gaussian fit. We have also explored the line widths computed at the wings of the lines, using the difference in percentile velocities. We obtained: ${\rm v_{97.7}}-{\rm v_{2.3}}({\rm CO}) = (74 \pm 66) + (0.70\pm0.12){\rm v_{97.7}}-{\rm v_{2.3}}({\rm [CI]})$. Hence, in our sample, [CI] lines are on average narrower than CO, especially for sources showing broader line profiles.
	\item We combined [CI](1--0) and CO(1--0) observations to derive a galaxy-averaged $\alpha_{\rm CO}$ factor. Based on the 10 sources with data available for both emission lines, we computed a median value of $\langle \alpha_{\rm CO} \rangle = 1.7 \pm 0.5$ M$_{\odot}$ (K \kms pc$^2$)$^{-1}$. This value is higher than the $\alpha_{\rm CO}$ commonly used in the literature for (U)LIRGs, and consistent with recent estimates by \cite{Herrero-Illana+19} and \cite{Kawana+22} based on the thermal dust continuum.
	\item In the ULIRGs sample, the total CO and [CI] line luminosities show weak correlations with $L_{\rm IR}$ and SFR. However, when including the LIRGs, the correlations become much stronger, though remaining weaker for SFR than $L_{\rm IR}$. Our sample of (U)LIRGs is characterized by $L'_{\rm CO}/L_{\rm IR}$ ratios that are similar or lower than the starbursts' best-fit derived by \cite{Sargent+14}, and so it is clearly not representative of the general local star forming population, but only of its most extreme (merger-driven) starbursts. 
	\item We measure extremely high galaxy-integrated CO line ratios, significantly offset from those measured in local massive main sequence galaxies. For example, more than 50\% of our targets have global $r_{21}>1$ ($\langle r_{21}\rangle^{\rm median}=1.09$). The distribution of $r_{\rm 31}$ values in our (U)LIRGs, with a median of $\langle r_{31}\rangle^{\rm median} =0.76$, does not even overlap with that of massive local galaxies. The extremely high low-\textit{J} CO ratios of local (U)LIRGs were noted by several previous studies \citep[e.g.,][]{Papadopoulos+12a}, and here we confirm this trend with better quality data.
	\item We have investigated possible drivers of such high CO line ratios. The global $r_{21}$ and $r_{31}$ values show a positive relation with $L_{\rm IR}$, and $r_{31}$ also shows a correlation with the SFR. We have studied links between CO line ratios and gas kinematics, as traced by the central velocity and $\sigma_v$ of the spectral line components simultaneously fit to all three CO transitions. We find a positive trend between $r_{21}$ and the velocity dispersion of the spectral components, probably tracing CO opacity effects, which is described by the relation: $r_{21}=(0.76\pm0.08)+(2.0\pm0.7)\times 10^{-3}~\sigma_v$. From these results we infer that, in local (U)LIRGs, low-\textit{J} CO line ratios are generally poor tracers of CO excitation, contrary to what has been found for local massive main sequence galaxies \citep[e.g.,][]{Leroy+22, Lamperti+20}. We have also explored the CO line ratios as a function of OH outflow properties, as retrieved from the literature \citep{Veilleux+13, Spoon+13} and concluded that these ratios are not sensitive to the presence or strength of molecular outflows.
	\item We measure $r_{\rm CICO} = L{'}_{\rm [CI](1-0)}/L{'}_{\rm CO(1-0)}$ values between 0.07 and 0.4, with a median of $\langle r_{\rm CICO}\rangle^{\rm median} = 0.18$. The values obtained by us, based on APEX observations are significantly higher than previous (sparse) measurements in local galaxies, including those obtained from a interferometric observations of a partially overlapping sample of ULIRGs. This strongly highlights the importance of sensitive, single dish observations that can capture the total line flux, especially at high frequencies where the interferometric flux loss becomes more severe. The $r_{\rm CICO}$ ratios do not show any significant trend with the galaxy properties explored here.
	\item We do not observe any significant correlation between $r_{\rm CICO}$ and the velocity of molecular outflows (as probed by OH absorption), or gas kinematics (as probed by the CO or [CI] velocity dispersion $\sigma_{\rm v}$, and the ${\rm v_{97.7}}-{\rm v_{2.3}}$ velocity intervals), hence we rule out that gas kinematics or turbulence has, statistically, any significant effect on the global [CI]/CO luminosity ratio. This result makes the discovery that [CI](1--0) lines are narrower than CO lines even more puzzling, because it implies that any effect causing [CI] lines to be narrower (in high-$\sigma_{\rm v}$ sources), conspires with some other effect in a way that the total luminosity ratio remains unchanged.
    \item Finally, we used our refined total $M_{\rm mol}$ estimates to investigate links between $L_{\rm AGN}$ and signatures of suppressed star formation. We find a statistically significant correlation between $\tau_{\rm dep}\equiv M_{\rm mol}/{\rm SFR}$ and $L_{\rm AGN}$, implying that (U)LIRGs with more luminous AGNs are statistically less efficient at forming stars. This could result from injected turbulence by radio jets or AGN winds (see, e.g., \cite{Alatalo+15}).
\end{itemize}

Overall, our analysis confirms the exceptional molecular gas properties of local (U)LIRGs, which are completely different from those of normal star forming galaxies in the local Universe. This study highlights the necessity of gathering large samples of high S/N observations of alternative H$_2$ gas tracers in local galaxies, especially [CI] lines, high-\textit{J} CO lines and high- density gas tracers (e.g., HCN, CN, CS, SiO), to understand the extreme mechanisms at work during accelerated phases of galaxy evolution. Due to the faintness of most of these tracers and their potentially different spatial extents in the host galaxies, such high S/N multi-tracer analyses need to be conducted with a sensitive large-aperture single dish telescope with a large instantaneous FoV. Furthermore, observations of [CI] lines and high-\textit{J} CO lines require a high atmospheric transparency that can only be obtained in a  high, dry site. Finally, the large line widths ($\sigma_{\rm v}$) of these local galaxy mergers require stable spectral baselines. The new concept for the Atacama Large Aperture Submillimeter Telescope (AtLAST\footnote{\href{https://www.atlast.uio.no}{https://www.atlast.uio.no}}) satisfies all of these conditions, and so it can be a transformational facility for this field.

\begin{acknowledgements}
	We are very grateful to the APEX and ESO staff for their unceasing efforts at preserving APEX operations and completing our projects despite the Covid19 pandemic.
    We thank the referee for their constructive and thorough reports, which significantly helped us improve the paper.
	We thank our friend and collaborator Padelis P. Papadopoulos who provided, as usual, very useful feedback and expert input on this work, and who always warns us against superficial, mainstream interpretations.
	This project has received funding from the European Union’s Horizon 2020 research and innovation programme under grant agreement No 951815 (AtLAST).
	This publication is based on data acquired with the Atacama Pathfinder Experiment (APEX) under programme IDs 0104.B-0672, 0106.B-0674, 086.F-9321, 090.B-0404, 092.F-9325, 099.F-9709, 077.F-9300, and 084.F-9306. APEX is a collaboration between the Max-Planck-Institut fur Radioastronomie, the European Southern Observatory, and the Onsala Space Observatory. 
	This paper makes use of the following ALMA data: ADS/JAO.ALMA\#2016.1.00177.S, ADS/JAO.ALMA\#2015.1.01147.S, ADS/JAO.ALMA\#2013.1.00535.S, ADS/JAO.ALMA\#2016.1.00140.S, 
	ADS/JAO.ALMA\#2016.1.00177.S, ADS/JAO.ALMA\#2015.1.00287.S, ADS/JAO.ALMA\#2013.1.00180.S, 
	ADS/JAO.ALMA\#2018.1.00503.S, ADS/JAO.ALMA\#2017.1.01235.S, ADS/JAO.ALMA\#2016.2.00006.S, 
	ADS/JAO.ALMA\#2017.1.01398.S, ADS/JAO.ALMA\#2013.1.00659.S, ADS/JAO.ALMA\#2018.1.00699.S,
	ADS/JAO.ALMA\#2012.1.00377.S, ADS/JAO.ALMA\#2017.1.00297.S, ADS/JAO.ALMA\#2015.1.00102.S,
	ADS/JAO.ALMA\#2012.1.00611.S, ADS/JAO.ALMA\#2016.2.00042.S, ADS/JAO.ALMA\#2013.1.00180.S,
	ADS/JAO.ALMA\#2018.1.00888.S, ADS/JAO.ALMA\#2018.1.00994.S. 
	ALMA is a partnership of ESO (representing its member states), NSF (USA) and NINS (Japan), together with NRC (Canada), MOST and ASIAA (Taiwan), and KASI (Republic of Korea), in cooperation with the Republic of Chile. The Joint ALMA Observatory is operated by ESO, AUI/NRAO and NAOJ.
	This work is based on observations carried out with the IRAM Plateau de Bure Interferometer. IRAM is supported by INSU/CNRS (France), MPG (Germany), and IGN (Spain).
	MA acknowledges support from FONDECYT grant 1211951, ANID+PCI+INSTITUTO MAX PLANCK DE ASTRONOMIA MPG 190030, ANID+PCI+REDES 190194 and ANID BASAL project FB210003.
    BB and SS acknowledge the support from the Research Council of Norway through NFR Young Research Talents Grant 276043. PS and CC acknowledge financial contributions from Bando Ricerca Fondamentale INAF 2022 Large Grant “Dual and binary supermassive black holes in the multi-messenger era: from galaxy mergers to gravitational waves” and from the agreement ASI-INAF n.2017-14-H.O.
\end{acknowledgements}

\bibliography{biblio}
\bibliographystyle{aa}

\begin{appendix}
\section{LIRGs}\label{sec:appendix_LIRGs}

\begin{figure}[tbp]
	\centering
	\includegraphics[width=0.45\textwidth]{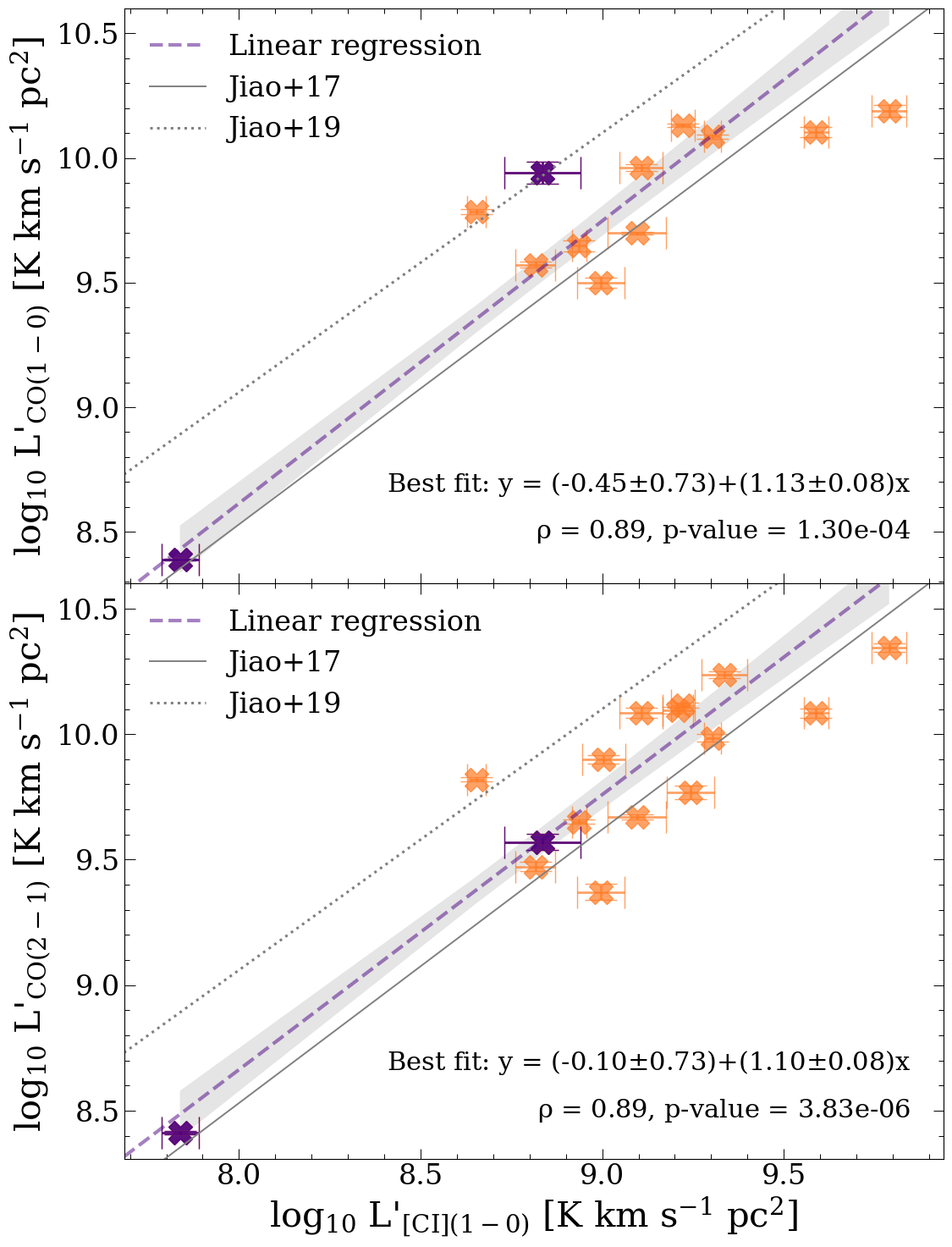}
	\caption{Same as Figure \ref{fig:COlumCIlum}, now including the two LIRGs (in purple data points) for which we have [CI](1--0) observations. The best fit relations (shown as \textit{purple dashed} lines) now include the LIRGs data points. The best-fit parameters are reported at the bottom right corner of the plots. We also display the Pearson correlation coefficients ($\rho$) and their associated {\it p}-values. The \textit{solid black lines} in both panels represent the corresponding relations reported by \citet{Jiao+17} for a sample of 71 (U)LIRGs, and the \textit{dotted lines} represent the relations of \citet{Jiao+19} for a sample of 15 nearby spiral galaxies, between $L'_{\textrm{CO(1-0)}}$ and $L'_{\textrm{[CI](1-0)}}$.}
	\label{fig:COlumCIlum_LIRG}
\end{figure}

This appendix shows several plots presented in the main body of the text, now including the 4 LIRGs in our sample. These sources populate poorly the $11.0<\log_{10} L_{\rm IR}\,{\rm L_{\odot}}<12.0$ regime, with IRAS F12243-0036 (also known as NGC~4418), with the lowest infrared luminosity and redhisft in our sample, being the main driver of the relations found for our extended sample of (U)LIRGs. 
In Fig. \ref{fig:COlumCIlum_LIRG} we plot the [CI](1--0) with respect to CO(1--0) and CO(2--1) line luminosities. The results are similar to the ones obtained in Fig. \ref{fig:COlumCIlum}, although the slope is now steeper, with a close-to-linear relation between the quantities.

\begin{figure}[tbp]
	\centering
	\includegraphics[width=0.45\textwidth]{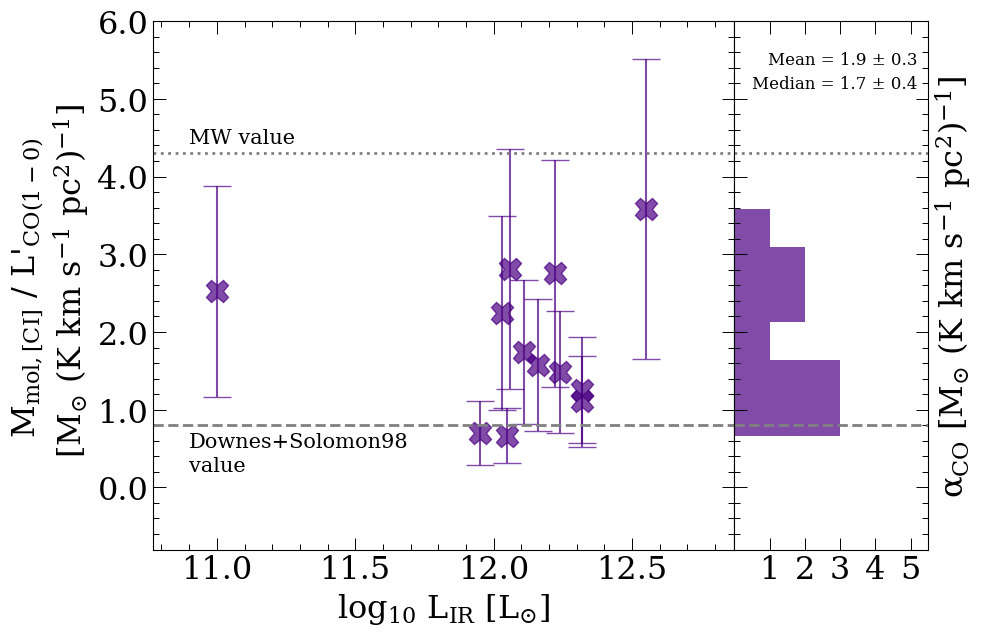}
	\caption{Same as Figure \ref{fig:aCO_histogram}, now including the two LIRGs for which we have [CI](1--0) and CO(1--0) observations.
    The right part of the plot shows the resulting distribution of [CI]-based $\alpha_{\textrm{CO}}$ values. The {\it dotted} line indicates the CO-to-H$_2$ conversion factor for the Milky Way galaxy, and the {\it dashed} line corresponds to the value commonly used in the literature for (U)LIRGs \citep{Downes+Solomon98}. 
	The mean and median values for sample of (U)LIRGs remain the same to the ones obtained solely for the ULIRGs, i.e., $\alpha_{\textrm{CO}}=1.9\pm 0.3$~M$_{\odot}$ (K km s$^{-1}$ pc$^2$)$^{-1}$ and $1.7\pm 0.4$~M$_{\odot}$ (K km s$^{-1}$ pc$^2$)$^{-1}$, for the mean and median, respectively.}
	\label{fig:aCO_histogram_LIRGs}
\end{figure}

In Figure \ref{fig:aCO_histogram_LIRGs} we explore how the CO-to-H$_2$ conversion factor changes when taking into account the 2 LIRGs in our sample for which we have both [CI](1--0) and CO(1--0) data, i.e., IRAS F12243-0036 and IRAS F00509+1225. These two sources show average $\alpha_{\textrm{CO}}$ values of $\sim 2.5$~M$_{\odot}$ (K km s$^{-1}$ pc$^2$)$^{-1}$ and $\sim 0.7$~M$_{\odot}$ (K km s$^{-1}$ pc$^2$)$^{-1}$, respectively. The overall mean and median of the $\alpha_{\textrm{CO}}$ factor, remains the same for the extended sample of (U)LIRGs.

\begin{figure*}[tbp]
	\centering
	\includegraphics[width=.31\textwidth]{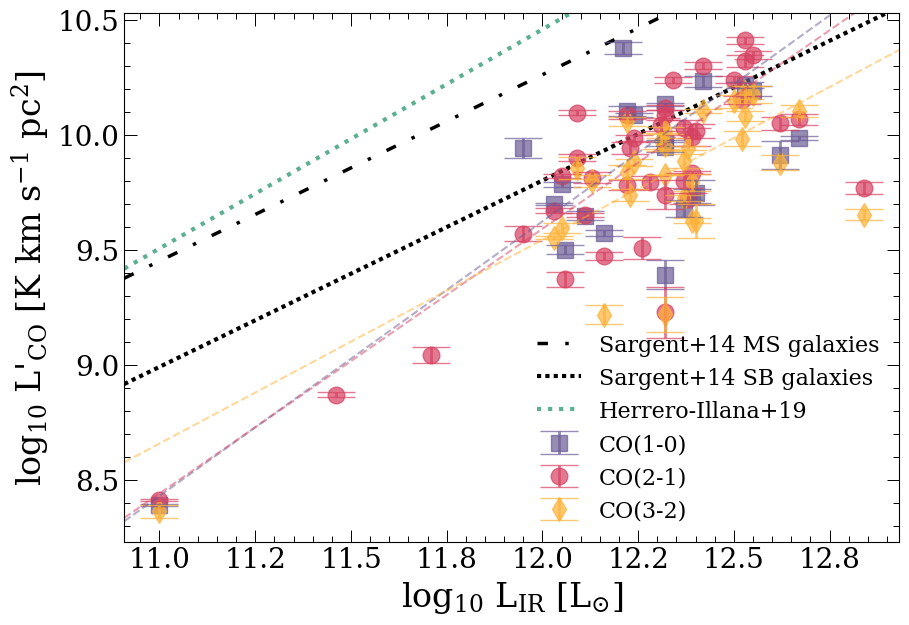}\quad
	\includegraphics[width=.31\textwidth]{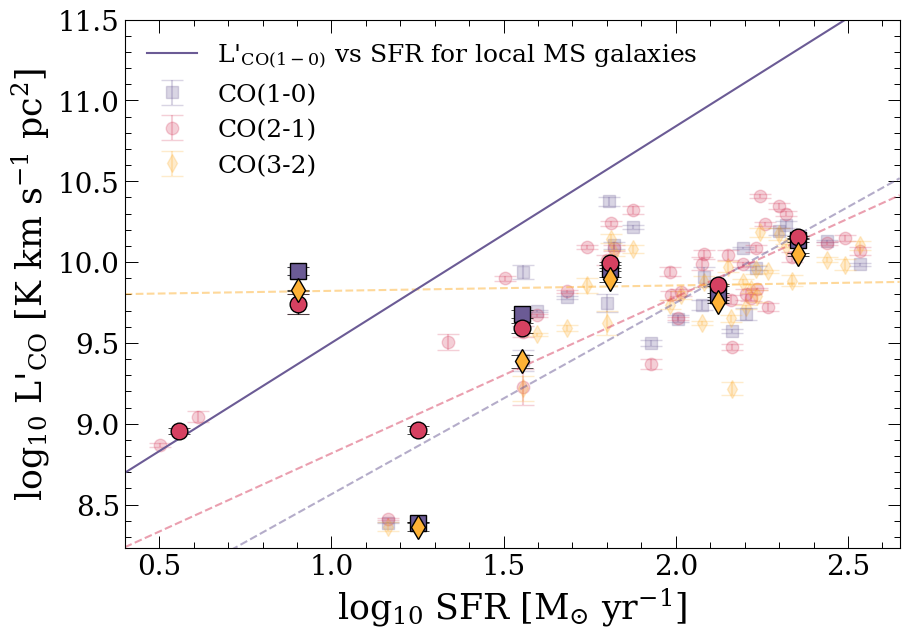}\quad
	\includegraphics[width=.31\textwidth]{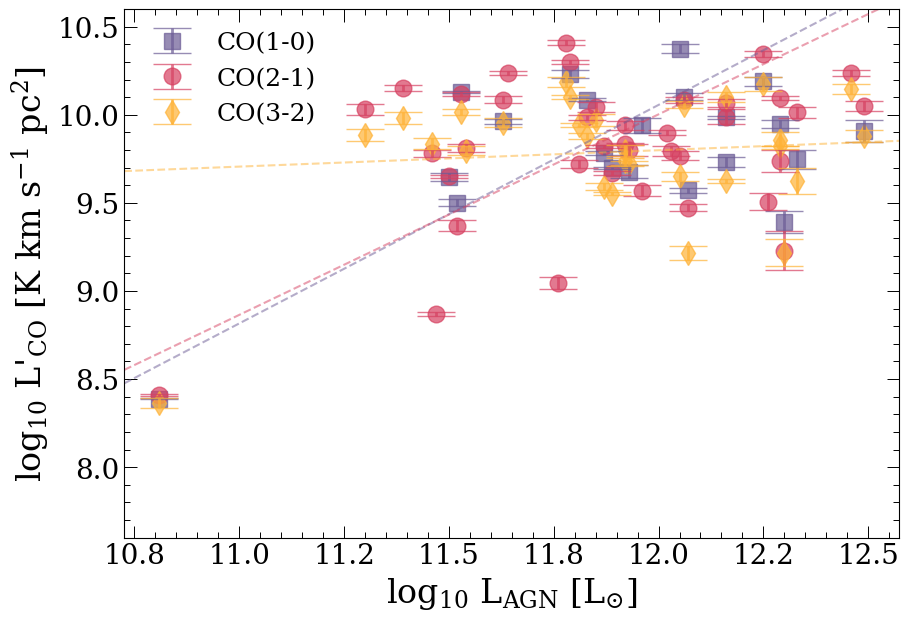}\\
	\includegraphics[width=.31\textwidth]{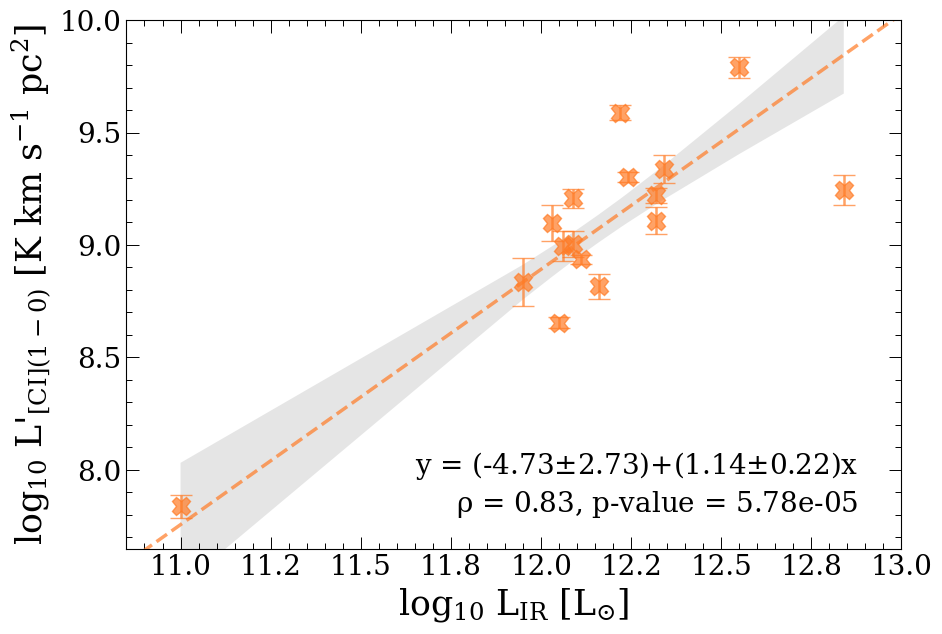}\quad
	\includegraphics[width=.31\textwidth]{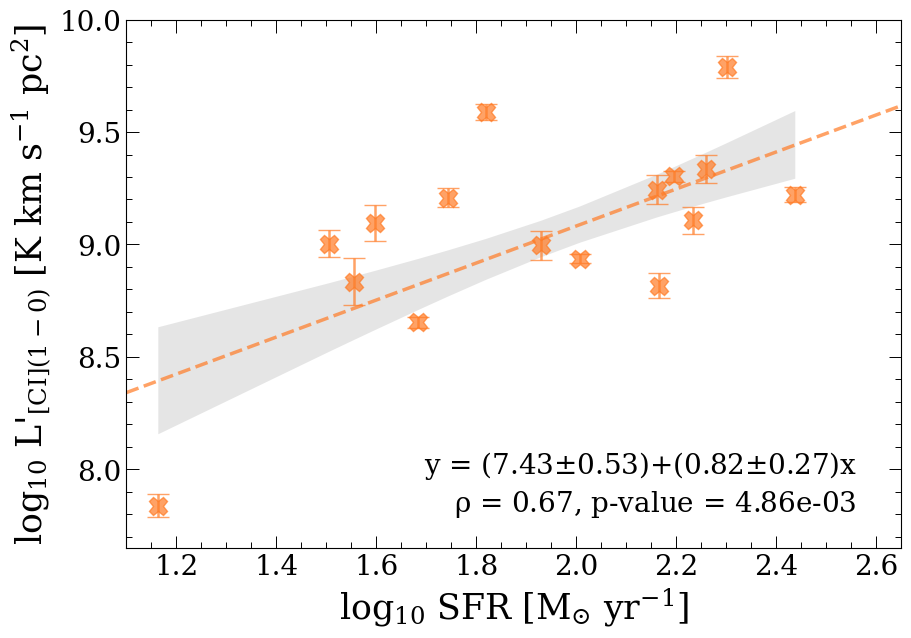}\quad
	\includegraphics[width=.31\textwidth]{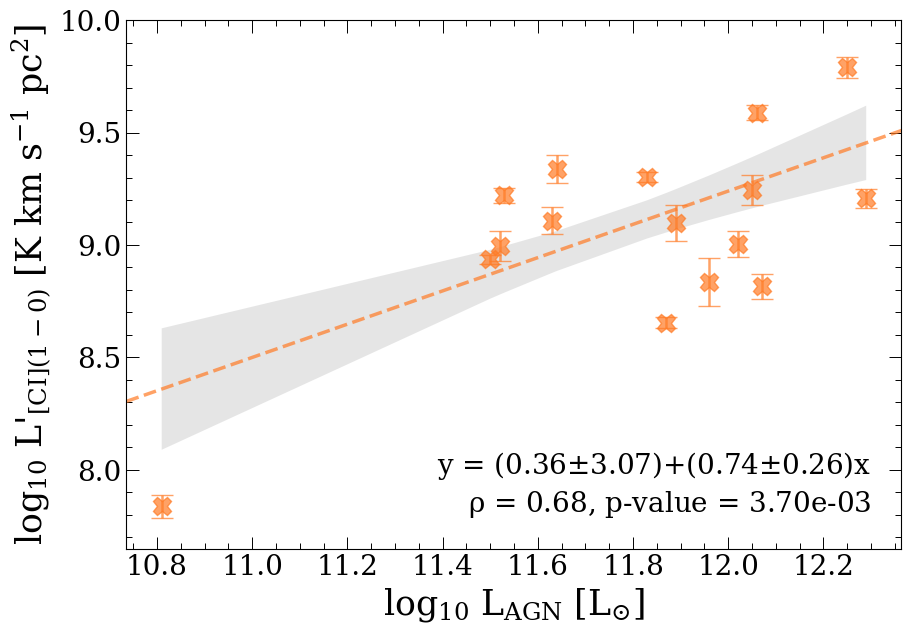}\\
	\caption{Same as Figure \ref{fig:Lum_vs_properties}, now including the four LIRGs in the extended sample.
    In each plot, the dashed lines are the best-fit relations obtained from a least squares regression analysis conducted for each transition separately (color-coded according to the transition, see legend on the top-left panel). In the bottom panels, the shaded gray areas correspond to the $1\sigma$ confidence interval of the fit. The bottom panels report also the Pearson correlation coefficients ($\rho$) and their associated {\it p}-values.
	The {\it top-left panel} shows also the $L'_{\mathrm{CO(1-0)}}$-$L_{\textrm{IR}}$ relations found by \citet{Sargent+14} for local MS (dot-dashed black line) and starburst galaxies (dotted black line), and the $L'_{\mathrm{CO(1-0)}}$-$L_{\textrm{TIR}}$ relation obtained by \cite{Herrero-Illana+19} using IRAM~30m CO(1--0) observations of 55 local sources in the GOALS sample including mostly LIRGs with $L_{IR}<10^{12}~L_{\odot}$ and with a $<20\%$ AGN contribution to $L_{\rm IR}$.
	In the {\it top-middle panel}, for better visualization, we over-plotted with darker colors the data in bins of SFRs. In this plot, the solid purple line represents the $L'_{\mathrm{CO(1-0)}}$-SFR best-fit obtained by \cite{Cicone+17} for the ALLSMOG and COLDGASS samples of local star-forming galaxies, described by 
	$\textrm{log}_{10}\,L'_{\textrm{CO(1--0)}} =(8.16\pm0.04)+(1.34\pm0.07)\, \textrm{log}_{10}\,\textrm{SFR}$.}
	\label{fig:Lum_vs_properties_LIRGs}
\end{figure*}

Figure \ref{fig:Lum_vs_properties_LIRGs} shows the relations between the line luminosities and the general galaxy properties for the extended sample of (U)LIRGs.
By including sources with $L_{\rm IR} < 12.0\, {\rm L_{\odot}}$, we now retrieve tighter relations between all CO and [CI] line luminosities and infrared luminosity (top-left and bottom-left panels). On the contrary, the relations with SFR (middle panels) become weaker for all CO lines, and the [CI] line shows now a stronger correlation to SFR.
Lastly, the relations with $L_{\rm AGN}$ become stronger for all line luninosities, with statistically significant relations for CO(1--0) ($\rho_{L'_{\textrm{CO(1-0)}}-L_{\rm AGN}}=0.53$, and {\it p}-value = 0.01), CO(2--1) ($\rho_{L'_{\textrm{CO(2-1)}}-L_{\rm AGN}}=0.37$ and {\it p}-value = 0.02), and interestingly, the highest correlation is measured for [CI](1--0) ($\rho_{L'_{\textrm{[CI](1-0)}}-L_{\rm AGN}}=0.68$ and {\it p}-value = $3.7\times10^{-3}$). If this result were to be confirmed with larger statistics, this could hint of different behaviors for the two gas tracers in the presence of AGN.

\section{The CO and [CI] line datasets: tables and final reduced spectra}\label{sec:appendix_spectra}

This Appendix collects plots and tables describing the CO(1--0), CO(2--1), CO(3--2) and [CI](1--0) spectral line data sets used in this work.

Figures~\ref{fig:spectra1} to \ref{fig:spectra6} report the final reduced CO(1--0), CO(2--1), CO(3--2), and [CI](1--0) line spectra of our sample of (U)LIRGs. All data, including archival ones, were re-reduced and re-analyzed by us in a consistent and uniform way, following the description provided in Section~\ref{sec:observations}. The spectra are presented together with their best-fit multi-component spectral models, computed following Section~\ref{sec:spectralfitting}.

Table~\ref{table:data_available} collects the telescope and project number information for all data used in this work, including the duplicated data not employed in the main analysis (see Appendix~\ref{sec:appendix_duplication_spectra}).

Table~\ref{tab:properties_observations} lists the values of spectral extraction aperture, the observing time, and the RMS noise values for all the spectral line datasets employed in the analysis.

\onecolumn
\begin{tiny}
	\begin{landscape}
		\begin{longtable}{@{} l|cc|cc|cc|cc@{}}
			\caption{Data analyzed on this paper}\label{table:data_available} \\
			\hline
			\hline
			\multirow{2}{*}{Galaxy}              & \multicolumn{2}{|c}{CO(1--0)}   & \multicolumn{2}{|c}{CO(2--1)}   & \multicolumn{2}{|c}{CO(3--2)}   & \multicolumn{2}{|c}{[CI](1--0)} \\ 
			& \multicolumn{1}{|c}{Telescope} &   \multicolumn{1}{c}{Proj ID}      &   \multicolumn{1}{|c}{Telescope}   &   \multicolumn{1}{c}{Proj ID}      &   \multicolumn{1}{|c}{Telescope}   &   \multicolumn{1}{c}{Proj ID}       & \multicolumn{1}{|c}{Telescope} &   \multicolumn{1}{c}{Proj ID} \\
			\hline 
			\endfirsthead
			\multicolumn{9}{c}{\footnotesize Following from previous page}\\
			\hline
			\hline
			\multirow{2}{*}{Galaxy}              & \multicolumn{2}{|c}{CO(1--0)}   & \multicolumn{2}{|c}{CO(2--1)}   & \multicolumn{2}{|c}{CO(3--2)}   & \multicolumn{2}{|c}{[CI](1--0)} \\ 
			& \multicolumn{1}{|c}{Telescope} &   \multicolumn{1}{c}{Proj ID}      &   \multicolumn{1}{|c}{Telescope}   &   \multicolumn{1}{c}{Proj ID}      &   \multicolumn{1}{|c}{Telescope}   &   \multicolumn{1}{c}{Proj ID}       & \multicolumn{1}{|c}{Telescope} &   \multicolumn{1}{c}{Proj ID} \\
			\hline 
			\endhead
			\hline 
			\multicolumn{9}{c}{\footnotesize Follows on next page}\\
			\endfoot
			\hline 
			\endlastfoot
			\hline
			\multirow{1}{*}{IRAS 00188$$-$$0856}     & \multirow{1}{*}{ACA archive}  &  \multirow{1}{*}{2018.1.00503.S} &   \multirow{1}{*}{APEX archive}     &  \multirow{1}{*}{099.F$-$9709}   &   \multirow{1}{*}{ACA archive}     &  \multirow{1}{*}{2018.1.00503.S}   & &  \\*
			\hdashline
			\multirow{2}{*}{IRAS 01003$$-$$2238}     & \multirow{2}{*}{ALMA archive}  &  \multirow{2}{*}{2013.1.00535.S} &   \multirow{2}{*}{APEX archive}     &  \multirow{2}{*}{099.F$-$9709}   &  APEX PI     & 0106.B$-$0674    & & \\*
			&   &   &       &    &  APEX archive   &  090.B$-$0404 & &  \\* 
			\hdashline
			\multirow{2}{*}{IRAS F01572+0009}     & \multirow{2}{*}{ALMA archive}  & \multirow{2}{*}{2013.1.00535.S} &   \multirow{2}{*}{APEX PI}     &  \multirow{2}{*}{0104.B$-$0672}   &   APEX PI     &  0106.B$-$0674   & &  \\ 
			&               &                &               &                & ALMA archive  & 2012.1.00611.S  & &  \\ 
			\hdashline
			\multirow{1}{*}{IRAS 03521+0028}     &   &   &   \multirow{1}{*}{APEX PI}     &  \multirow{1}{*}{0104.B$-$0672}   &   \multirow{1}{*}{APEX PI}     &  \multirow{1}{*}{0106.B$-$0674}    & & \\ 
			\hdashline
			\multirow{1}{*}{IRAS F05024$-$1941}     &   &   &   \multirow{1}{*}{APEX PI}     &  \multirow{1}{*}{0104.B$-$0672}   &   \multirow{1}{*}{APEX PI}     &  \multirow{1}{*}{0106.B$-$0674}    & & \\ 
			\hdashline
			\multirow{2}{*}{IRAS F05189$-$2524}     & \multirow{2}{*}{ALMA archive}  & \multirow{2}{*}{2013.1.00535.S} &   \multirow{2}{*}{APEX PI}     &  \multirow{2}{*}{0104.B$-$0672}   &   APEX PI     &  0106.B$-$0674    & \multirow{2}{*}{APEX PI}  &  \multirow{2}{*}{0104.B$-$0672}\\ 
			&               &                &               &                & ACA archive  & 2012.1.00611.S  & & \\ 
			\hdashline
			\multirow{1}{*}{IRAS 06035$$-$$7102}     & \multirow{1}{*}{ACA archive}  &  \multirow{1}{*}{2018.1.00503.S} &   \multirow{1}{*}{APEX PI}     &  \multirow{1}{*}{0104.B$-$0672}   &  APEX PI     & 0106.B$-$0674    & APEX PI  &  0104.B$-$0672 \\ 
			\hdashline
			\multirow{2}{*}{IRAS 06206$$-$$6315}     &   &   &   \multirow{2}{*}{APEX PI}     &  \multirow{2}{*}{0104.B$-$0672}   &  APEX PI     & 0106.B$-$0674   & &  \\ 
			&   &   &       &    &  APEX archive   &  090.B$-$0404  & & \\ 
			\hdashline
			\multirow{2}{*}{IRAS 07251$-$0248}     &   &   &   ALMA archive     &  2018.1.00699.S   &  \multirow{2}{*}{ APEX archive}     &  \multirow{2}{*}{086.F$-$9321}   &  & \\ 
			&   &   &    APEX archive  &   086.F$-$9321  &   &   & & \\
			\hdashline
			\multirow{1}{*}{IRAS 08311$$-$$2459}     &   &   &   \multirow{1}{*}{APEX PI}     &  \multirow{1}{*}{0104.B$-$0672}   &   \multirow{1}{*}{APEX archive}     &  \multirow{1}{*}{090.B$-$0404}    & & \\ 
			\hdashline
			\multirow{2}{*}{IRAS 09022$-$3615}     &   &   &   \multirow{2}{*}{APEX PI}     &  \multirow{2}{*}{0104.B$-$0672}   &   \multirow{2}{*}{APEX PI}     &  \multirow{2}{*}{0106.B$-$0674}    &  APEX PI  &  0104.B$-$0672 \\ 
			&               &                &               &                &                  &                & ACA archive & 2018.1.00994.S \\ 
			\hdashline
			\multirow{1}{*}{IRAS 10378+1109}     &  &   &   \multirow{1}{*}{APEX PI}     &  \multirow{1}{*}{0104.B$-$0672}   &  APEX archive     & 090.B$-$0404    & & \\ 
			\hdashline
			\multirow{1}{*}{IRAS 11095$$-$$0238}     &   &   &   \multirow{1}{*}{APEX PI}     &  \multirow{1}{*}{0104.B$-$0672}   &   \multirow{1}{*}{APEX archive}     &  \multirow{1}{*}{090.B$-$0404}  & &   \\ 
			\hdashline
			\multirow{1}{*}{IRAS F12072$-$0444}     & \multirow{1}{*}{ALMA archive}  & \multirow{1}{*}{2013.1.00535.S}  &   \multirow{1}{*}{APEX PI}     &  \multirow{1}{*}{0104.B$-$0672}   &   \multirow{1}{*}{APEX archive}     &  \multirow{1}{*}{090.B$-$0404}    & & \\ 
			\hdashline
			\multirow{2}{*}{IRAS F12112+0305}     & \multirow{2}{*}{ALMA archive}  & \multirow{2}{*}{2016.1.00140.S} &   \multirow{2}{*}{APEX PI}     &  \multirow{2}{*}{0104.B$-$0672}   &   APEX PI     &  0106.B$-$0674   & APEX PI  &  0104.B$-$0672    \\ 
			&               &                &               &                & ACA archive  & 2016.2.00042.S & ACA archive & 2018.1.00994.S \\ 
			\hdashline
			\multirow{2}{*}{IRAS 13120$-$5453}     & \multirow{2}{*}{ALMA archive}  & \multirow{2}{*}{2015.1.00287.S} &  APEX archive   &   090.B$-$040  &   APEX archive     &  090.B$-$040   & \multirow{2}{*}{APEX PI}  &  \multirow{2}{*}{0104.B$-$0672} \\ 
			&               &                &    APEX archive   &   092.F$-$9325    & APEX archive  & 092.F$-$9325  & & \\ 
			\hdashline
			\multirow{1}{*}{IRAS F13305$-$1739}     &   &   &   \multirow{1}{*}{APEX PI}     &  \multirow{1}{*}{0104.B$-$0672}   &      &      & & \\ 
			\hdashline
			\multirow{2}{*}{IRAS F13451+1232}     & \multirow{2}{*}{ALMA archive}  & \multirow{2}{*}{2013.1.00180.S} &   \multirow{2}{*}{APEX PI}     &  \multirow{2}{*}{0104.B$-$0672}   &   APEX PI     &  0106.B$-$0674    & & \\ 
			&               &                &               &                & ALMA archive  & 2013.1.00180.S  & & \\ 
			\hdashline
			\multirow{1}{*}{IRAS F14348$-$1447}     &   &   &   \multirow{1}{*}{APEX PI}     &  \multirow{1}{*}{0104.B$-$0672}   &      &      & APEX PI  &  0104.B$-$0672 \\ 
			\hdashline
			\multirow{2}{*}{IRAS F14378$-$3651}     & \multirow{2}{*}{ACA archive}  & \multirow{2}{*}{2018.1.00503.S}  &   APEX archive     &  090.B$-$0404   &      &      & ACA archive & 2018.1.00994.S \\ 
			&               &                &   ACA archive   &   2018.1.00503.S   &   &   & APEX PI  &  0104.B$-$0672 \\ 
			\hdashline
			\multirow{1}{*}{IRAS F15462$-$0450}     & \multirow{1}{*}{ALMA archive}  & \multirow{1}{*}{2015.1.01147.S}  &   \multirow{1}{*}{APEX archive}     &  \multirow{1}{*}{099.F$-$9709}   &      &     & & \\ 
			\hdashline
			\multirow{2}{*}{IRAS 16090$$-$$0139}     & \multirow{2}{*}{ACA archive}  &  \multirow{2}{*}{2018.1.00503.S} &   \multirow{2}{*}{APEX PI}     &  \multirow{2}{*}{0104.B$-$0672}   &  APEX PI     & 0106.B$-$0674    & \multirow{2}{*}{APEX PI}  &  \multirow{2}{*}{0104.B$-$0672} \\ 
			&   &   &       &    &  ACA archive   &  2018.1.00503.S  & & \\ 
			\hdashline
			\multirow{2}{*}{IRAS 17208$-$0014}     & \multirow{2}{*}{ALMA archive}  & \multirow{2}{*}{2016.1.00177.S} &   \multirow{2}{*}{APEX PI}     &  \multirow{2}{*}{0104.B$-$0672}   &   APEX PI     &  0106.B$-$0674   & \multirow{2}{*}{APEX PI}  &  \multirow{2}{*}{0104.B$-$0672} \\ 
			&               &                &               &                & ALMA archive  & 2015.1.00102.S  &  &  \\ 
			\hdashline
			\multirow{1}{*}{IRAS 19254$$-$$7245}     &   &   &   \multirow{1}{*}{APEX archive}     &  \multirow{1}{*}{090.B$-$0404}   &      &      & APEX PI  &  0104.B$-$0672  \\
            \hdashline
		    \multirow{1}{*}{IRAS F19297$-$0406}     &   &   &   \multirow{1}{*}{APEX PI}     &  \multirow{1}{*}{0104.B$-$0672}   &   \multirow{1}{*}{ACA archive}   &  \multirow{1}{*}{2018.1.00503.S}   & & \\ 
			\hdashline
			\multirow{1}{*}{IRAS 19542+1110}     & \multirow{1}{*}{ALMA archive}  & \multirow{1}{*}{2017.1.01235.S}  &   \multirow{1}{*}{APEX PI}     &  \multirow{1}{*}{0104.B$-$0672}   &     &   & APEX PI  &  0104.B$-$0672 \\ 
			\hdashline
			\multirow{1}{*}{IRAS 20087$$-$$0308}     & \multirow{1}{*}{ACA archive}  &  \multirow{1}{*}{2018.1.00503.S}  &   \multirow{1}{*}{APEX PI}     &  \multirow{1}{*}{0104.B$-$0672}   & \multirow{1}{*}{ACA archive}  &  \multirow{1}{*}{2018.1.00503.S}  & & \\ 
			\hdashline
			\multirow{2}{*}{IRAS 20100$$-$$4156}     & \multirow{2}{*}{ALMA archive}  &  \multirow{2}{*}{2013.1.00659.S} &   \multirow{2}{*}{APEX PI}     &  \multirow{2}{*}{0104.B$-$0672}   &  APEX archive     & 090.B$-$0404  & &   \\ 
			&   &   &       &    &  ALMA archive   &  2018.1.00888.S  & & \\ 
			\hdashline
			\multirow{2}{*}{IRAS 20414$$-$$1651}     &    &   & APEX PI     &  0104.B$-$0672   &  APEX PI     &   0106.B$-$0674    & & \\ 
			&   &   &   ALMA archive    &  2018.1.00699.S  &  APEX archive   &  090.B$-$0404  & & \\ 
			\hdashline
			\multirow{2}{*}{IRAS F20551$-$4250}     & \multirow{2}{*}{ACA archive}  & \multirow{2}{*}{2016.2.00006.S}  &   \multirow{2}{*}{APEX PI}     &  \multirow{2}{*}{0104.B$-$0672}   &   APEX PI   &  0106.B$-$0674   & \multirow{2}{*}{ACA archive}  &  \multirow{2}{*}{2018.1.00994.S} \\ 
			&               &                &               &                & APEX archive  & 084.F$-$9306 & &  \\ 
			\hdashline
			\multirow{2}{*}{IRAS F22491$$-$$1808}     &   &   &   \multirow{2}{*}{APEX PI}     &  \multirow{2}{*}{0104.B$-$0672}   &   \multirow{2}{*}{APEX PI}     &  \multirow{2}{*}{0106.B$-$0674}   & APEX PI  &  0104.B$-$0672 \\ 
			&               &                &               &                &                  &                & ACA archive & 2018.1.00994.S \\ 
			\hdashline
			\multirow{1}{*}{IRAS F23060+0505}     &  IRAM PdBI  & UF2A  &   \multirow{1}{*}{APEX PI}     &  \multirow{1}{*}{0104.B$-$0672}   &   \multirow{1}{*}{APEX PI}     &  \multirow{1}{*}{0106.B$-$0674}   & &  \\ 
			\hdashline
            \multirow{2}{*}{IRAS F23128$$-$$5919}     & \multirow{2}{*}{ALMA archive}  & \multirow{2}{*}{2017.1.01398.S}  &    \multirow{2}{*}{APEX PI}     &   \multirow{2}{*}{0104.B$-$0672}   &   APEX archive     &  084.F$-$9306  & APEX PI  &  0104.B$-$0672  \\* 
			&               &                &     &      & ACA archive  & 2016.2.00042.S  & ACA archive & 2018.1.00994.S \\
			\hdashline
			\multirow{2}{*}{IRAS 23230$$-$$6926}     & \multirow{2}{*}{ACA archive}  &  \multirow{2}{*}{2018.1.00503.S} &   \multirow{2}{*}{APEX PI}     &  \multirow{2}{*}{0104.B$-$0672}   &  ACA archive   &  2018.1.00503.S  & & \\*
			&   &   &       &    &  APEX archive     & 090.B$-$0404    & & \\ 
			\hdashline
			\multirow{1}{*}{IRAS 23253$$-$$5415}     &   &   &   \multirow{1}{*}{APEX PI}     &  \multirow{1}{*}{0104.B$-$0672}   &   \multirow{1}{*}{APEX archive}     &  \multirow{1}{*}{090.B$-$0404}    & & \\ 
			\hdashline
            \multirow{1}{*}{IRAS F23389+0300}     &   &   &   \multirow{1}{*}{APEX PI}     &  \multirow{1}{*}{0104.B$-$0672}   &   \multirow{1}{*}{APEX PI}     &  \multirow{1}{*}{0106.B$-$0674}    & & \\ 
			\hline
            \hline
            \multirow{1}{*}{IRAS F00509+1225}     & \multirow{1}{*}{ALMA archive}  & \multirow{1}{*}{2015.1.01147.S} &   \multirow{1}{*}{APEX PI}     &  \multirow{1}{*}{0104.B$-$0672}   &        &     &  APEX PI  &  0104.B$-$0672\\
            \hdashline
			\multirow{2}{*}{PG1126$$-$$041}     &   &   &   ACA archive     &  2017.1.00297.S   &     &   & &  \\ 
			&   &   &   APEX PI     &  0104.B$-$0672   &     &    & & \\ 
			\hdashline
            \multirow{2}{*}{IRAS F12243$-$0036}     & \multirow{2}{*}{ALMA archive}  & \multirow{2}{*}{2016.1.00177.S} &   \multirow{2}{*}{ACA archive}     &  \multirow{2}{*}{2012.1.00377.S}   &   \multirow{2}{*}{APEX archive}     &  \multirow{2}{*}{077.F$-$9300}    & APEX PI  &  0104.B$-$0672   \\ 
			&               &                &               &                &                  &                & ACA archive & 2018.1.00994.S \\ 
			\hdashline
			\multirow{2}{*}{PG2130+099}     &   &   &   ACA archive     &  2017.1.00297.S   &     &    & \multirow{2}{*}{APEX PI}  &  \multirow{2}{*}{0104.B$-$0672} \\ 
			&   &   &   APEX PI     &  0104.B$-$0672   &     &    & & \\ 
			\hline
		\end{longtable}
		
		\tablefoot{The data collected for the analysis performed throughout this paper is a combination of PI and archival data from several telescopes: APEX, ALMA, ACA and PdBI. We report all the data sets used with their respective project ID's. For the cases (line + source) where there is more than one data set available, we report the data set used on top, followed by the rest. The [CI](1--0) line observation for the source PG2130+099 is a non-detection, though we still report it, and in Table \ref{tab:fluxlumino}, we report the upper limit.}
\end{landscape}
\end{tiny}
\twocolumn

\onecolumn
\begin{tiny}
\begin{landscape}
	\begin{longtable}{@{}l|ccc|ccc|ccc|ccc@{}}
		\caption{Description of the observations} \label{tab:properties_observations} \\
		\hline
		\hline 
		\multirow{2}{*}{Galaxy name} & \multicolumn{3}{|c}{CO(1--0)}  & \multicolumn{3}{|c}{CO(2--1)} & \multicolumn{3}{|c}{CO(3--2)} & \multicolumn{3}{|c}{[CI](1--0)}\\ 
		& $d_{\rm aperture}$ [''] & $t_{obs}$ [h] & rms [mJy] & 
		$d_{\rm aperture}$ [''] & $t_{obs}$ [h] & rms [mJy] &  $d_{\rm aperture}$ [''] & $t_{obs}$ [h]  & rms [mJy]  &  $d_{\rm aperture}$ [''] & $t_{obs}$ [h]  & rms [mJy]\\
		(1) & (2) & (3) & (4) & (5) & (6) & (7) & (8) & (9) & (10) & (11) & (12) & (13) \\ 
		\hline 
		\endfirsthead
		\multicolumn{13}{c}{\footnotesize Following from previous page}\\
		\hline
		\hline
		\multirow{2}{*}{Galaxy name} & \multicolumn{3}{|c}{CO(1--0)}  & \multicolumn{3}{|c}{CO(2--1)} & \multicolumn{3}{|c}{CO(3--2)} & \multicolumn{3}{|c}{[CI](1--0)}\\ 
		& $d_{\rm aperture}$ [''] & $t_{obs}$ [h] & rms [mJy] & 
		$d_{\rm aperture}$ [''] & $t_{obs}$ [h] & rms [mJy] &  $d_{\rm aperture}$ [''] & $t_{obs}$ [h]  & rms [mJy]  &  $d_{\rm aperture}$ [''] & $t_{obs}$ [h]  & rms [mJy]\\
		(1) & (2) & (3) & (4) & (5) & (6) & (7) & (8) & (9) & (10) & (11) & (12) & (13) \\ 
		\hline 
		\endhead
		\hline 
		\multicolumn{13}{c}{\footnotesize Follows on next page}\\
		\endfoot
		\hline 
		\endlastfoot
		\hline
		\multirow{1}{*}{IRAS 00188$-$0856}  &  \multirow{1}{*}{30}  &  \multirow{1}{*}{0.65}  &  \multirow{1}{*}{3}  &  \multirow{1}{*}{28}  &  \multirow{1}{*}{1.4}  &  \multirow{1}{*}{32}  &  \multirow{1}{*}{15}   &  \multirow{1}{*}{0.34}   &  \multirow{1}{*}{5}  & & &   \\ 
		\hdashline
		\multirow{2}{*}{IRAS 01003$-$2238}  &  \multirow{2}{*}{5} &  \multirow{2}{*}{0.03} &  \multirow{2}{*}{5}  &  \multirow{2}{*}{28}   &  \multirow{2}{*}{3.3}   &  \multirow{2}{*}{35}   & 20   & 0.84   &  45  & & &   \\ 
		&   &   &    &    &   &   &  20  &  0.49  &  93  & & & \\ 
		\hdashline
		\multirow{2}{*}{IRAS F01572+0009$^{\dagger}$}    & \multirow{2}{*}{7}  & \multirow{2}{*}{0.03}   & \multirow{2}{*}{10}   &  \multirow{2}{*}{34}  &  \multirow{2}{*}{1.3}   &  \multirow{2}{*}{12}     &  20  &  1.6  &  30   & & &  \\ 
		&    &     &     &     &     &    &  8  & 0.54   & 5  & & & \\ 
		\hdashline
		\multirow{1}{*}{IRAS 03521+0028}   &   &   &   &  \multirow{1}{*}{34}  &  \multirow{1}{*}{2.0}  &  \multirow{1}{*}{10}  &  \multirow{1}{*}{20}   &  \multirow{1}{*}{0.95}   &  \multirow{1}{*}{28}  & & &   \\ 
		\hdashline
		\multirow{1}{*}{IRAS F05024$-$1941}  &   &   &   &  \multirow{1}{*}{34} &  \multirow{1}{*}{2.3}  &  \multirow{1}{*}{10}  &  \multirow{1}{*}{20}  &  \multirow{1}{*}{1.24}   &  \multirow{1}{*}{22}  & & &   \\ 
		\hdashline
		\multirow{2}{*}{IRAS F05189$-$2524}   & \multirow{2}{*}{30}  & \multirow{2}{*}{1.5}   & \multirow{2}{*}{4}  &  \multirow{2}{*}{27}  &  \multirow{2}{*}{0.21}  &  \multirow{2}{*}{25}  &  20  &  0.75  &  120 &\multirow{2}{*}{13} & \multirow{2}{*}{0.82} & \multirow{2}{*}{148}  \\ 
		&    &     &    &     &     &    & 12  & 1.86   & 13  & & & \\ 
		\hdashline
		\multirow{1}{*}{IRAS 06035$-$7102}   &  \multirow{1}{*}{41} &  \multirow{1}{*}{0.24} &  \multirow{1}{*}{11}  &  \multirow{1}{*}{27}  &  \multirow{1}{*}{0.19}  &  \multirow{1}{*}{32}   & 20  & 3.3   & 486  & 13 & 0.6 & 158  \\
		\hdashline
		\multirow{2}{*}{IRAS 06206$-$6315}  &   &   &   &  \multirow{2}{*}{27}  &  \multirow{2}{*}{0.6}  &  \multirow{2}{*}{26} & 20  &  0.44 & 59 & & &  \\ 
		&   &   &   &    &    &    &  20 &  0.8 &  45 & & & \\ 
		\hdashline
		\multirow{2}{*}{IRAS 07251$-$0248}    &   &   &   &   5  &   0.69   &   3.3  &  \multirow{2}{*}{20}   &  \multirow{2}{*}{1.11}   &  \multirow{2}{*}{110}    & & & \\ 
		&  &   &    &  28  &  1.22   &  24    &   &   & & & &  \\
		\hdashline
		\multirow{1}{*}{IRAS 08311$-$2459}  &   &   &   &  \multirow{1}{*}{27}   &  \multirow{1}{*}{0.43}   &  \multirow{1}{*}{41}  &  \multirow{1}{*}{20}   &  \multirow{1}{*}{0.74}   &  \multirow{1}{*}{98}   & & &  \\ 
		\hdashline
		\multirow{2}{*}{IRAS 09022$-$3615}    &   &   &    &  \multirow{2}{*}{27}  &  \multirow{2}{*}{0.46}   &  \multirow{2}{*}{39}  &  \multirow{2}{*}{20}  &  \multirow{2}{*}{0.63}   &  \multirow{2}{*}{152}  & 13 & 0.17 & 139  \\ 
		&  &   &    &   &   &   &   &   & & 8 & 0.08 & 46 \\
		\hdashline
		\multirow{1}{*}{IRAS 10378+1109}  &   &   &   &  \multirow{1}{*}{27}  & \multirow{1}{*}{1.66}  &  \multirow{1}{*}{22} & 20  & 1.43   & 42  & & & \\ 
		\hdashline
		\multirow{1}{*}{IRAS 11095$-$0238} &   &   &   &  \multirow{1}{*}{27}  &  \multirow{1}{*}{0.76}  &  \multirow{1}{*}{26}  &  \multirow{1}{*}{20}   &  \multirow{1}{*}{0.37}   &  \multirow{1}{*}{144}   & & &  \\ 
		\hdashline
		\multirow{1}{*}{IRAS F12072$-$0444$^{\dagger}$}   & \multirow{1}{*}{5}  & \multirow{1}{*}{0.03}  & \multirow{1}{*}{7}  &  \multirow{1}{*}{27}   &  \multirow{1}{*}{0.33}   &  \multirow{1}{*}{29}   &  \multirow{1}{*}{20}   &  \multirow{1}{*}{0.86}   &  \multirow{1}{*}{65}   & & &  \\ 
		\hdashline
		\multirow{2}{*}{IRAS F12112+0305}   & \multirow{2}{*}{10}  & \multirow{2}{*}{0.09}  & \multirow{2}{*}{5} &  \multirow{2}{*}{27}   &  \multirow{2}{*}{0.4}   &  \multirow{2}{*}{37}   &  20   &  1.47   &  113  & 13 & 0.23 &  148 \\ 
		&    &     &    &     &     &    & 15  & 2.72  & 42  & 10 & 0.08 & 36 \\ 
		\hdashline
		\multirow{2}{*}{IRAS 13120$-$5453}   & \multirow{2}{*}{17} & \multirow{2}{*}{0.22} & \multirow{2}{*}{12} &   28  &  \multirow{2}{*}{0.35}   &   \multirow{2}{*}{81}   &  20   &  \multirow{2}{*}{0.5}   &  \multirow{2}{*}{130} & \multirow{2}{*}{13} & \multirow{2}{*}{0.17} & \multirow{2}{*}{185}   \\ 
		&        &        &         &   30   &     &     & 20  &   &  & & &  \\ 
		\hdashline
		\multirow{1}{*}{IRAS F13305$-$1739}   &   &   &   &  \multirow{1}{*}{34}  &  \multirow{1}{*}{0.71}  &  \multirow{1}{*}{12}  &    &    &    & & &   \\ 
		\hdashline
		\multirow{2}{*}{IRAS F13451+1232}   & \multirow{2}{*}{12} & \multirow{2}{*}{0.4} & \multirow{2}{*}{4}   &  \multirow{2}{*}{27}  &  \multirow{2}{*}{0.24}  &  \multirow{2}{*}{26}   &  20   &  1.26   &  37  & & &  \\ 
		&       &      &     &        &       &       & 6 & 0.08 & 6 & & &  \\ 
		\hdashline
		\multirow{1}{*}{IRAS F14348$-$1447}   &   &  &    &  \multirow{1}{*}{27}  &  \multirow{1}{*}{0.21}  &  \multirow{1}{*}{32}  &    &    &  & 13 & 0.24 & 292 \\ 
		\hdashline
		\multirow{2}{*}{IRAS F14378$-$3651} & \multirow{2}{*}{30} & \multirow{2}{*}{0.15} & \multirow{2}{*}{8} &  28  &  0.21  &  107  &   &  & & 9 & 0.08 & 15   \\ 
		&       &        &      &   20  &   0.08  &   9   &   &   &  & 13 & 0.08 & 191 \\ 
		\hdashline
		\multirow{1}{*}{IRAS F15462$-$0450}  & \multirow{1}{*}{17}  & \multirow{1}{*}{0.3}  & \multirow{1}{*}{17}  &  \multirow{1}{*}{28}  &  \multirow{1}{*}{1.9}  &  \multirow{1}{*}{20}   &    &    &    & & &  \\ 
		\hdashline
		\multirow{2}{*}{IRAS 16090$-$0139}  &  \multirow{2}{*}{30} &  \multirow{2}{*}{0.15} &  \multirow{2}{*}{6}  &  \multirow{2}{*}{27}  &  \multirow{2}{*}{0.68}  &  \multirow{2}{*}{21}  & 20  & 0.53  & 77 & \multirow{2}{*}{13}& \multirow{2}{*}{1.8}& \multirow{2}{*}{131} \\* 
		&   &   &   &   &    &   &  15  &  0.17  &  13 & & &   \\ 
		\hdashline
	    \multirow{2}{*}{IRAS 17208$-$0014}     & \multirow{2}{*}{16}  & \multirow{2}{*}{0.53} & \multirow{2}{*}{1.35}   &  \multirow{2}{*}{27}  &  \multirow{2}{*}{0.11}  &  \multirow{2}{*}{39} &   20  &   0.34   &   239   & \multirow{2}{*}{13}  & \multirow{2}{*}{0.18} & \multirow{2}{*}{232}  \\ 
		&        &        &        &       &        &       &  11   &  0.27  &    81.5  & & &  \\ 
		\hdashline
		\multirow{1}{*}{IRAS 19254$-$7245}  &   &   &   &  \multirow{1}{*}{28}   &  \multirow{1}{*}{1.2}   &  \multirow{1}{*}{27}  &    &    &   & 13 & 0.41 & 79  \\ 
		\hdashline
		\multirow{1}{*}{IRAS F19297$-$0406}  &   &    &    &  \multirow{1}{*}{27}  &  \multirow{1}{*}{0.42}  &  \multirow{1}{*}{19}   &  \multirow{1}{*}{30}  &  \multirow{1}{*}{0.17}  &  \multirow{1}{*}{19}  & & &  \\ 
		\hdashline
		\multirow{1}{*}{IRAS 19542+1110}  & \multirow{1}{*}{9} & \multirow{1}{*}{0.35} & \multirow{1}{*}{5}  &  \multirow{1}{*}{27}  &  \multirow{1}{*}{0.64}  &  \multirow{1}{*}{34}  &    &   &  & 13 & 0.08 & 138 \\ 
		\hdashline
		\multirow{1}{*}{IRAS 20087$-$0308}  &  \multirow{1}{*}{30}  &  \multirow{1}{*}{0.08}  &  \multirow{1}{*}{9}  &  \multirow{1}{*}{27}   &  \multirow{1}{*}{0.59}   &  \multirow{1}{*}{22}  &  \multirow{1}{*}{20}   &  \multirow{1}{*}{0.17}   &  \multirow{1}{*}{14} & & &  \\ 
		\hdashline
		\multirow{2}{*}{IRAS 20100$-$4156}  &  \multirow{2}{*}{25} &  \multirow{2}{*}{3.5} &  \multirow{2}{*}{1}   &  \multirow{2}{*}{27}   &  \multirow{2}{*}{1.09}   &  \multirow{2}{*}{17}  & 20   & 1.58   & 59  & & &   \\ 
		&   &   &   &   &   &   &  7  &  0.94  &  6 & & &  \\ 
		\hdashline
		\multirow{2}{*}{IRAS 20414$-$1651}     &    &    &    &    27  &  0.98   &   19   &  20   &  1.24  &   70    & & &  \\ 
		&   &   &   &   4  &  0.78   &   122    &  20  &   0.76   &    90 & & &    \\ 
		\hdashline
		\multirow{2}{*}{IRAS F20551$-$4250}   & \multirow{2}{*}{35} & \multirow{2}{*}{0.4} & \multirow{2}{*}{11}  &  \multirow{2}{*}{27}  &  \multirow{2}{*}{0.4}  &  \multirow{2}{*}{42}  &  20  &  0.42  &  150 &\multirow{2}{*}{9} & \multirow{2}{*}{0.08} & \multirow{2}{*}{44}  \\ 
		&        &        &        &        &        &        & 20  & 0.2  & 132  & & & \\
		\hdashline
		\multirow{2}{*}{IRAS F22491$-$1808}   &   &   &   &  \multirow{2}{*}{27}  &  \multirow{2}{*}{0.32}  &  \multirow{2}{*}{30}  &  \multirow{2}{*}{20}   & \multirow{2}{*}{2}   & \multirow{2}{*}{69}  & 13 & 0.53 & 135  \\ 
		&    &     &    &     &     &    &   &      &   & 9 & 0.08 & 29 \\
		\hdashline
		IRAS F23060+0505     &  10 & 3  &  1 &   34     &  1.32   &  16   &   20     &  2.7    &  28 & & &   \\ 
        \hdashline
        \multirow{2}{*}{IRAS F23128$-$5919}  & \multirow{2}{*}{23}  & \multirow{2}{*}{2.27}  & \multirow{2}{*}{2}  & \multirow{2}{*}{27}  & \multirow{2}{*}{0.68}  & \multirow{2}{*}{26}   & 15 & 2.7 & 20 & 13 & 0.16 & 414  \\ 
		&         &         &       &    &    &    &  20   &  0.35   &  102   & 11 & 0.08 &  43  \\
		\hdashline
		\multirow{2}{*}{IRAS 23230$-$6926}     & \multirow{2}{*}{30}  &  \multirow{2}{*}{1.29}   &  \multirow{2}{*}{4}   &   \multirow{2}{*}{27}     &  \multirow{2}{*}{0.73}  &  \multirow{2}{*}{21}    &  20   &  0.58   &  8    & & &    \\ 
		&   &   &   &   &   &   &  15  &  0.6  &  116  & & & \\ 
		\hdashline
		\multirow{1}{*}{IRAS 23253$-$5415}     &   &   &   &   \multirow{1}{*}{27}     &  \multirow{1}{*}{0.83}    &  \multirow{1}{*}{21}   &   \multirow{1}{*}{20}     &  \multirow{1}{*}{1.28}    &  \multirow{1}{*}{75}  & & &  \\ 
		\hdashline
		\multirow{1}{*}{IRAS F23389+0300}   &   &   &    &  \multirow{1}{*}{34}   &  \multirow{1}{*}{2.4}   &  \multirow{1}{*}{11}  &  \multirow{1}{*}{20}   &  \multirow{1}{*}{1.8}   &  \multirow{1}{*}{25}  & & &   \\ 
		\hline
        \hline
        \multirow{1}{*}{IRAS F00509+1225}    & \multirow{1}{*}{23} & \multirow{1}{*}{0.07} & \multirow{1}{*}{26}  &  \multirow{1}{*}{27} &  \multirow{1}{*}{0.29} &  \multirow{1}{*}{42}  &    &    &  & 13 & 0.72 & 650 \\ 
		\hdashline
		\multirow{2}{*}{PG1126$-$041}  &   &   &   &  32  &  2.5  &  2  &   &   &   & & &  \\ 
		&   &   &    &  27  &  0.59  &  25 &   &   &  & & &   \\ 
		\hdashline
        \multirow{2}{*}{IRAS F12243$-$0036}   & \multirow{2}{*}{12}  & \multirow{2}{*}{0.5}  & \multirow{2}{*}{4}   &  \multirow{2}{*}{29}  & \multirow{2}{*}{1.3}  & \multirow{2}{*}{18}   &  \multirow{2}{*}{20}    &  \multirow{2}{*}{0.06}   &  \multirow{2}{*}{562} & 13 & 0.11 & 327 \\
		&    &     &    &     &     &    &   &      &   & 9 & 0.45 & 23 \\
		\hdashline
		\multirow{2}{*}{PG2130+099}  &   &   &    &  30  &  2.06  &  3  &   &   &  & \multirow{2}{*}{13}& \multirow{2}{*}{1.8}& \multirow{2}{*}{70}  \\ 
		&   &   &    &  27  &  2  &  17  &   &   &  & & &   \\ 
		\hline
	\end{longtable}
	
	\tablefoot{For each transition, we report single-dish beam or spectral extraction aperture ($d_{aperture}$), the observing time ($t_{obs}$) (corresponding to the effective value after excluding the discarded sub-scans), and the spectral rms  computed using $\Delta v=50$~\kms velocity channels.\\
    $^{\dagger}$ Sources for which \textit{uv}-tapering was applied for the ALMA CO(1--0) archival data.}
\end{landscape}
\end{tiny}
\twocolumn

\begin{figure*}[tbp]
	\includegraphics[clip=true,trim=0.35cm 0.3cm 0.48cm 0.3cm,width=0.23\textwidth]{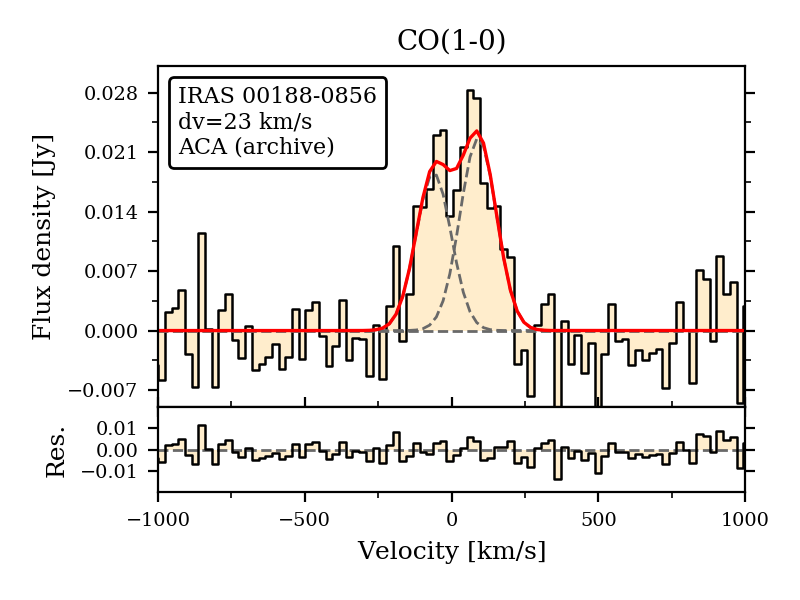}\quad
	\includegraphics[clip=true,trim=0.35cm 0.3cm 0.48cm 0.3cm,width=0.23\textwidth]{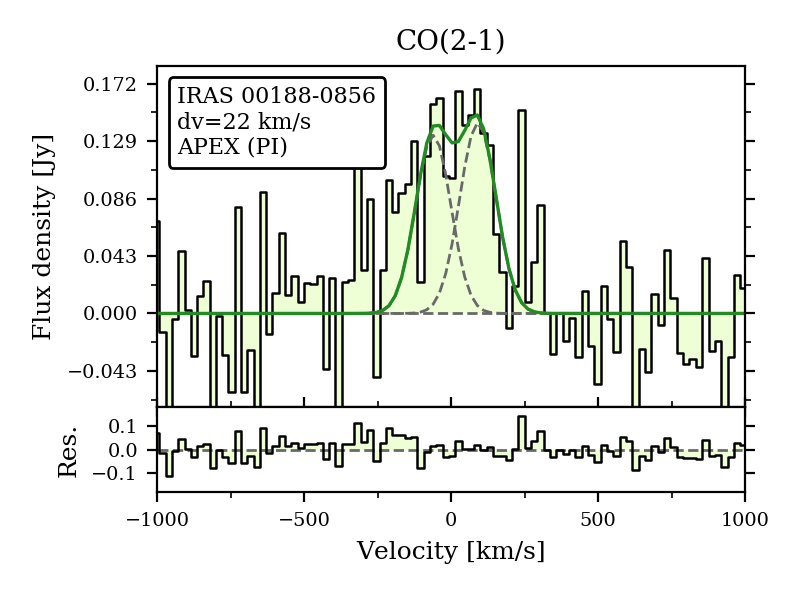}\quad
	\includegraphics[clip=true,trim=0.35cm 0.3cm 0.48cm 0.3cm,width=0.23\textwidth]{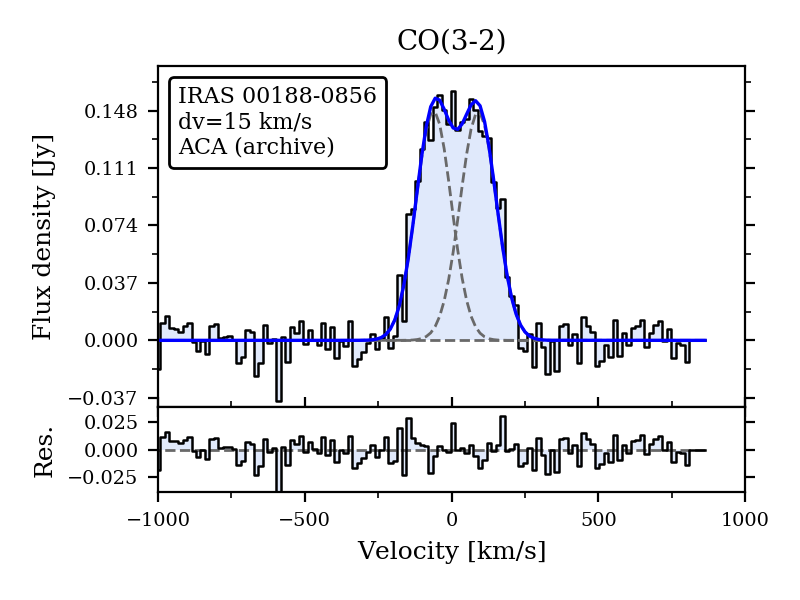}\quad
	\includegraphics[clip=true,trim=0.35cm 0.3cm 0.48cm 0.3cm,width=0.23\textwidth]{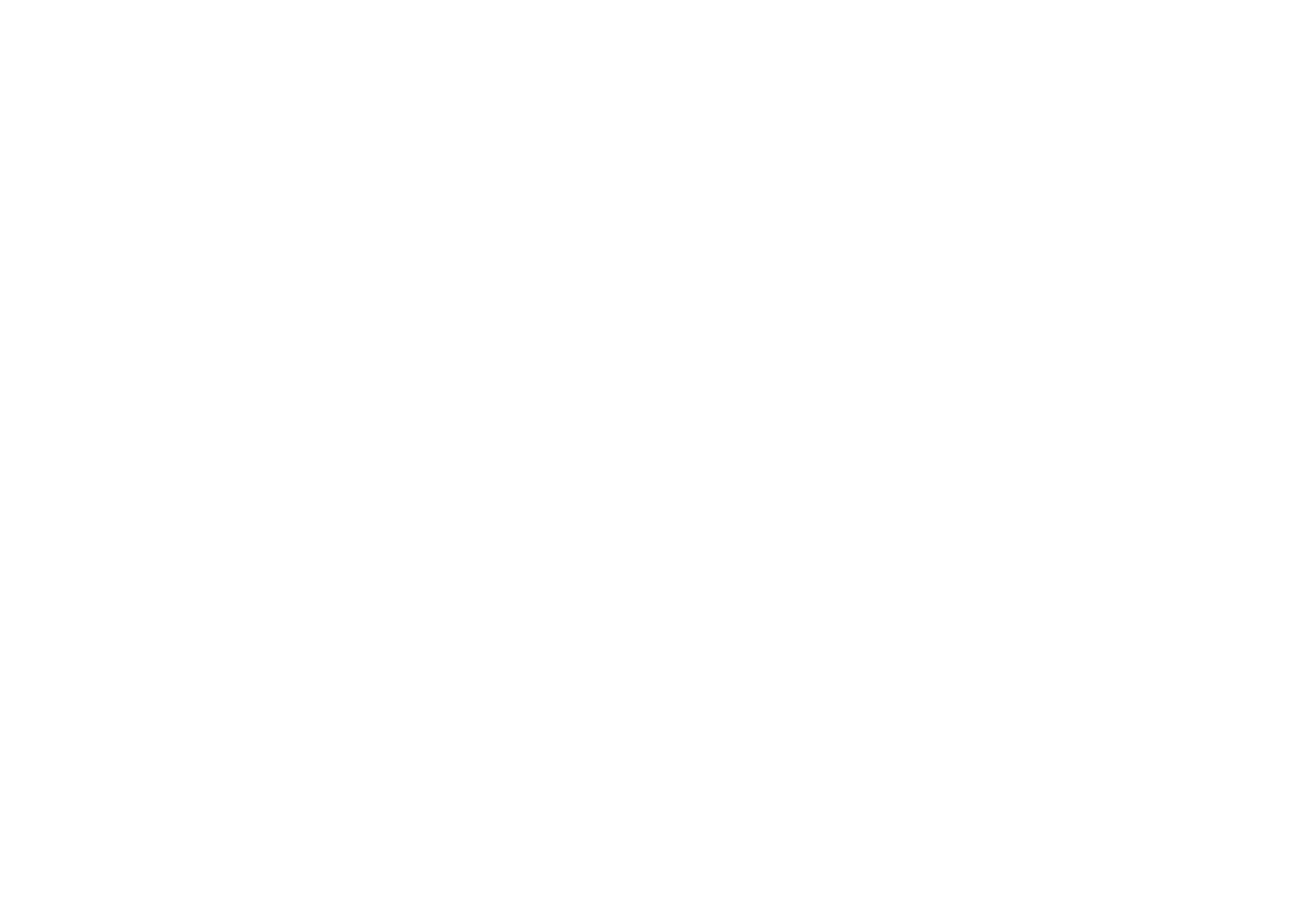}\\  
	\includegraphics[clip=true,trim=0.35cm 0.3cm 0.48cm 0.3cm,width=0.23\textwidth]{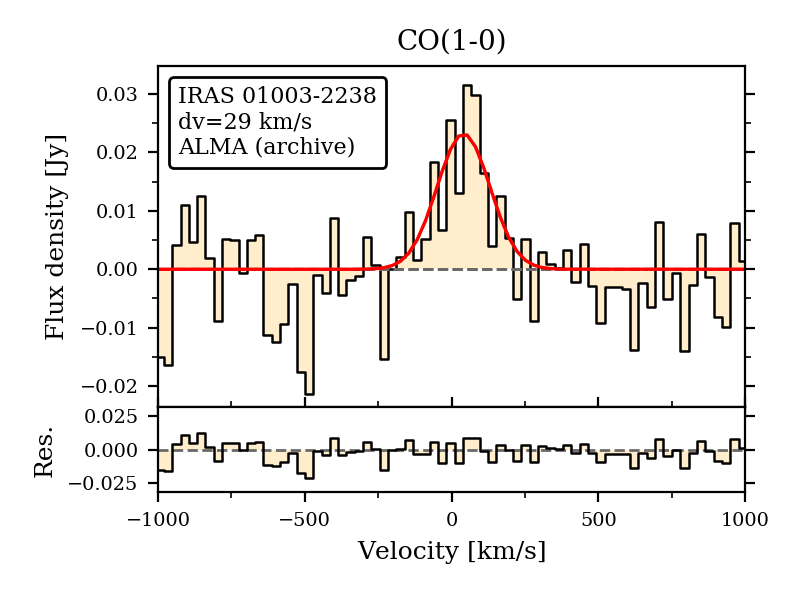}\quad
	\includegraphics[clip=true,trim=0.35cm 0.3cm 0.48cm 0.3cm,width=0.23\textwidth]{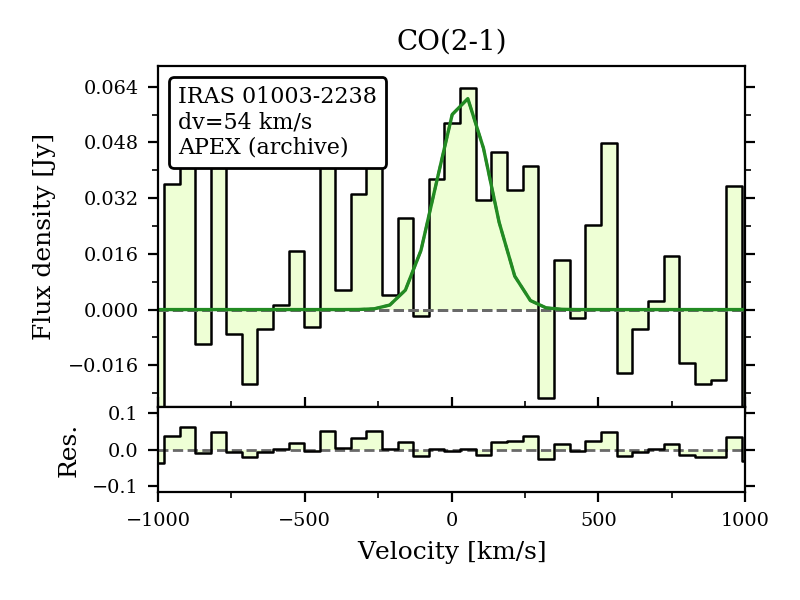}\quad
	\includegraphics[clip=true,trim=0.35cm 0.3cm 0.48cm 0.3cm,width=0.23\textwidth]{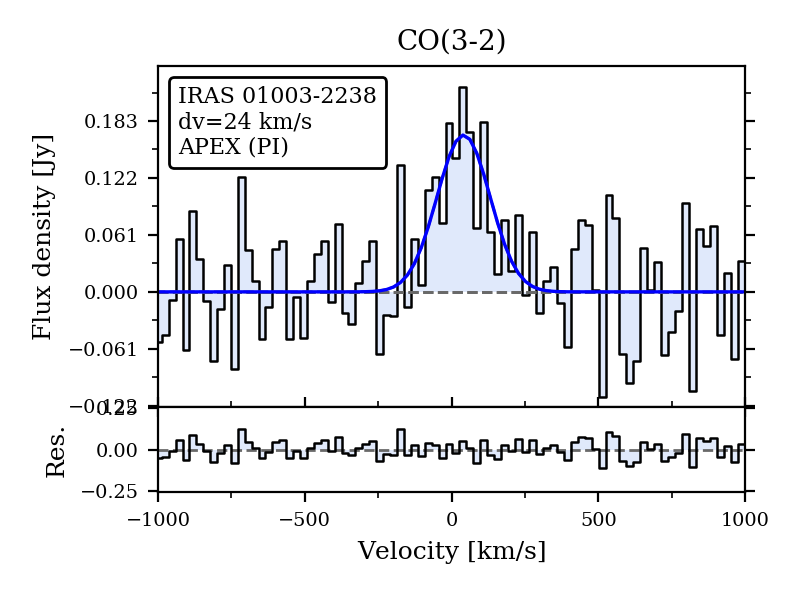}\quad
	\includegraphics[clip=true,trim=0.35cm 0.3cm 0.48cm 0.3cm,width=0.23\textwidth]{figures/empty}\\  
	\includegraphics[clip=true,trim=0.35cm 0.3cm 0.48cm 0.3cm,width=0.23\textwidth]{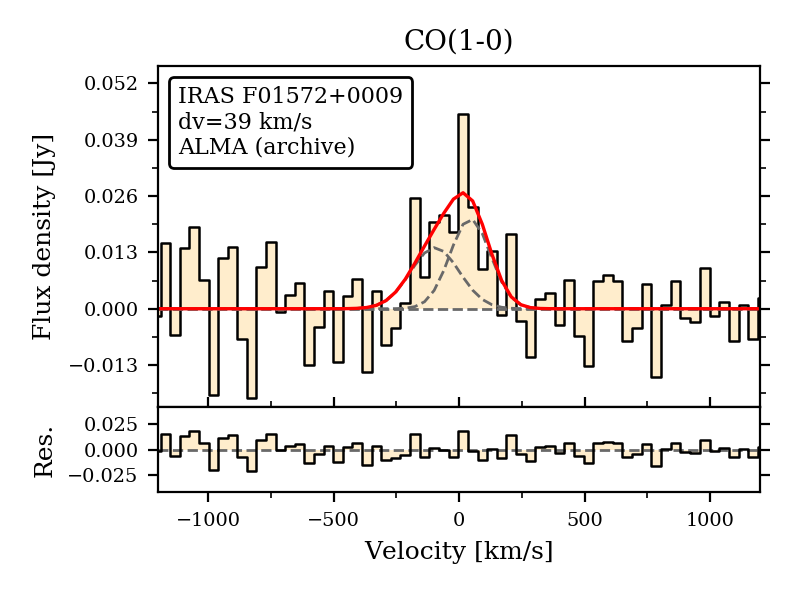}\quad
	\includegraphics[clip=true,trim=0.35cm 0.3cm 0.48cm 0.3cm,width=0.23\textwidth]{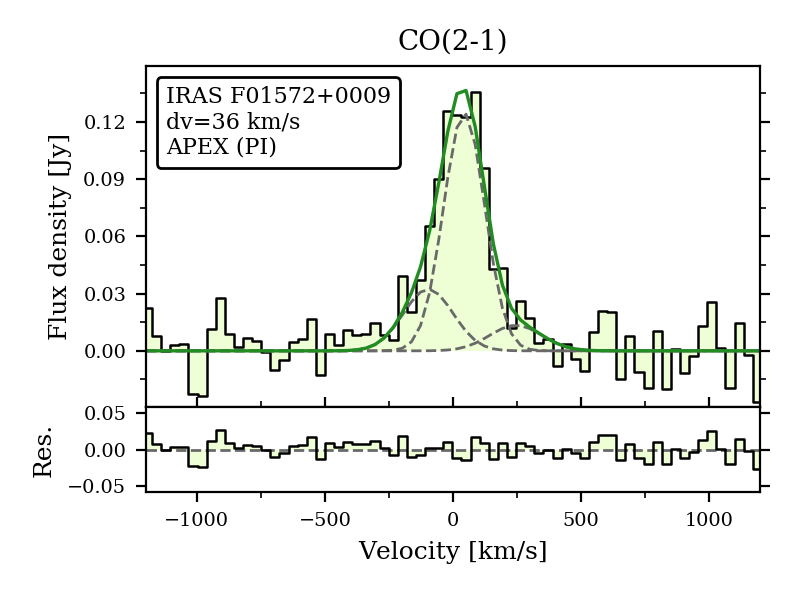}\quad
	\includegraphics[clip=true,trim=0.35cm 0.3cm 0.48cm 0.3cm,width=0.23\textwidth]{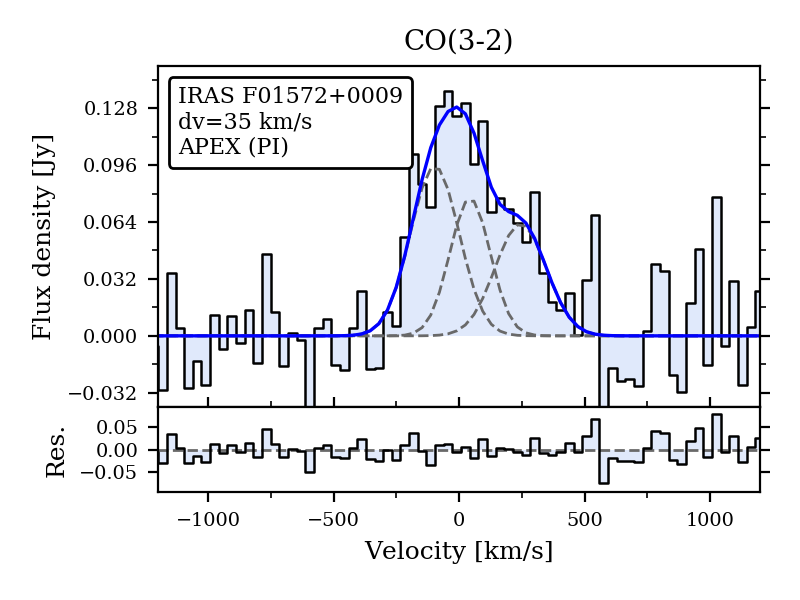}\quad
	\includegraphics[clip=true,trim=0.35cm 0.3cm 0.48cm 0.3cm,width=0.23\textwidth]{figures/empty}\\
	\includegraphics[clip=true,trim=0.35cm 0.3cm 0.48cm 0.3cm,width=0.23\textwidth]{figures/empty}\quad
	\includegraphics[clip=true,trim=0.35cm 0.3cm 0.48cm 0.3cm,width=0.23\textwidth]{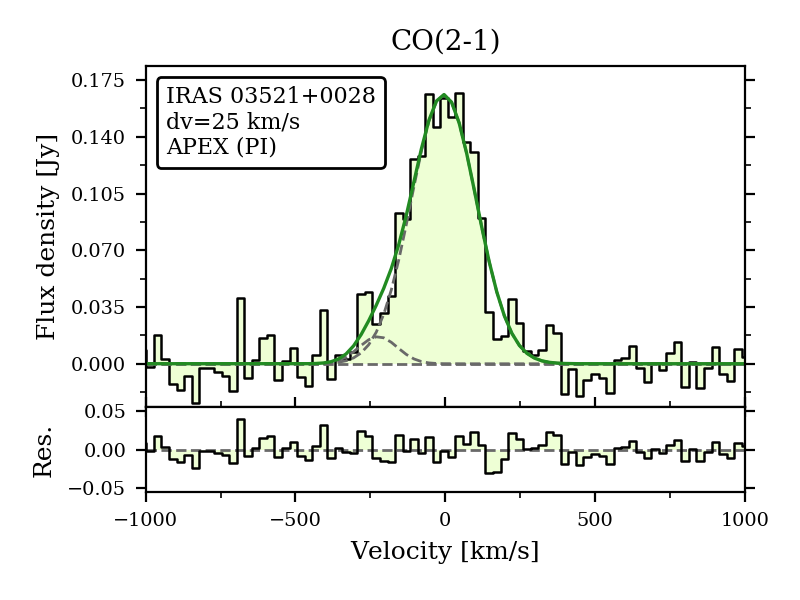}\quad
	\includegraphics[clip=true,trim=0.35cm 0.3cm 0.48cm 0.3cm,width=0.23\textwidth]{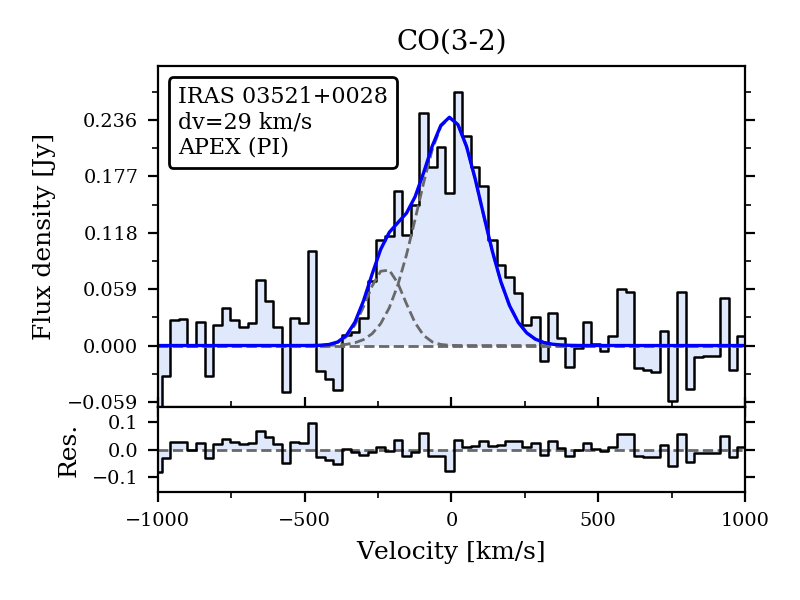}\quad
	\includegraphics[clip=true,trim=0.35cm 0.3cm 0.48cm 0.3cm,width=0.23\textwidth]{figures/empty}\\  
	\includegraphics[clip=true,trim=0.35cm 0.3cm 0.48cm 0.3cm,width=0.23\textwidth]{figures/empty}\quad
	\includegraphics[clip=true,trim=0.35cm 0.3cm 0.48cm 0.3cm,width=0.23\textwidth]{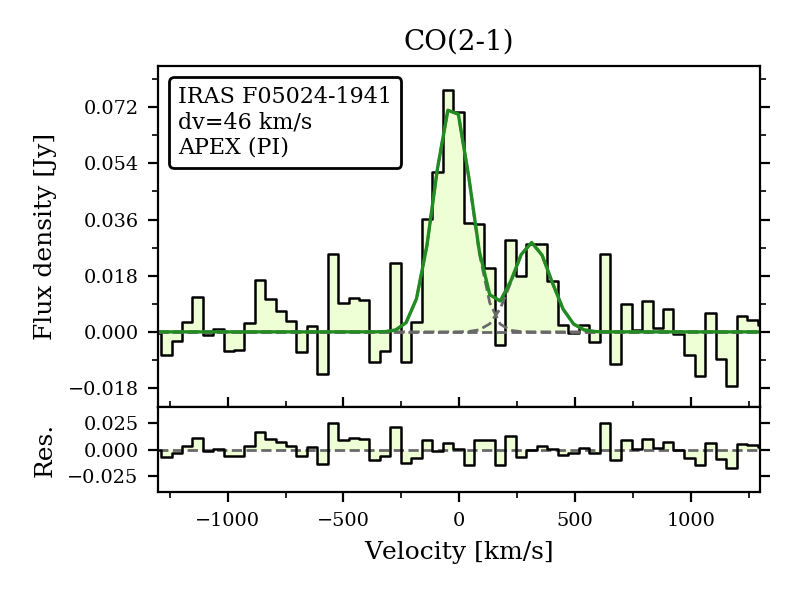}\quad
	\includegraphics[clip=true,trim=0.35cm 0.3cm 0.48cm 0.3cm,width=0.23\textwidth]{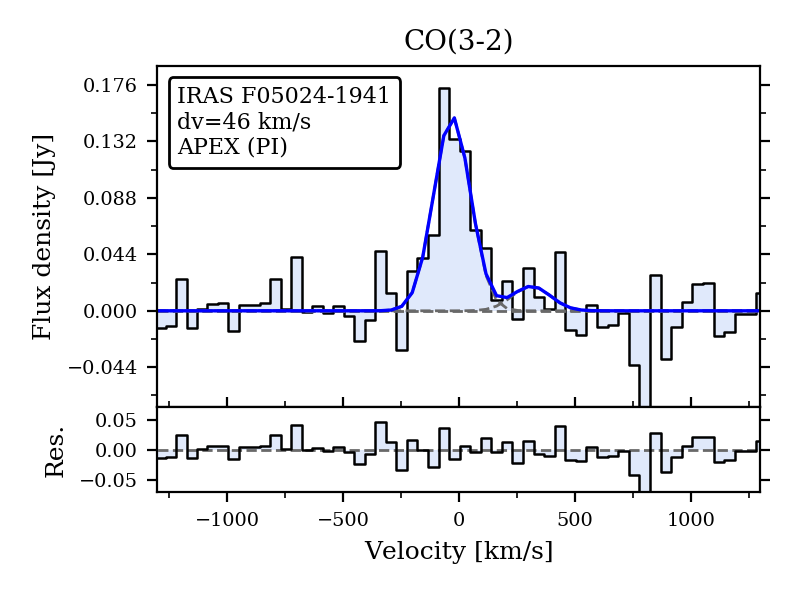}\quad
	\includegraphics[clip=true,trim=0.35cm 0.3cm 0.48cm 0.3cm,width=0.23\textwidth]{figures/empty}\\
	\includegraphics[clip=true,trim=0.35cm 0.3cm 0.48cm 0.3cm,width=0.23\textwidth]{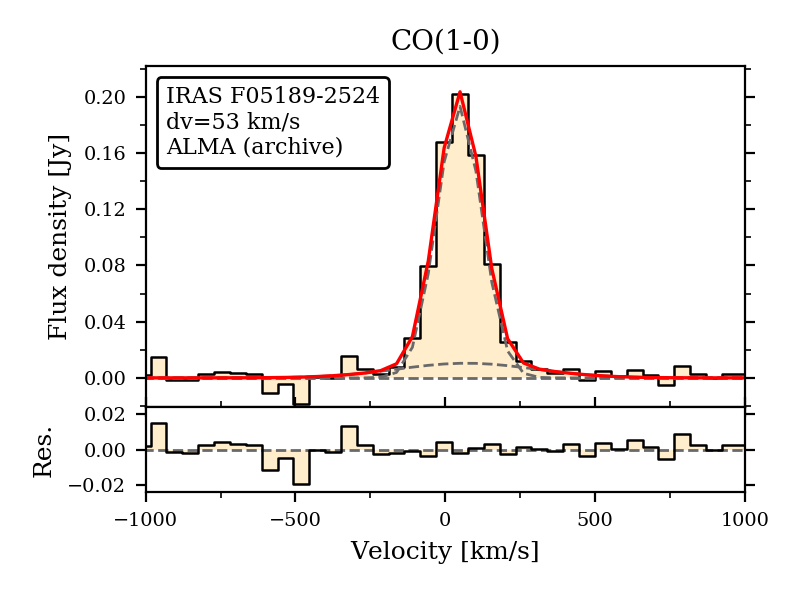}\quad
	\includegraphics[clip=true,trim=0.35cm 0.3cm 0.48cm 0.3cm,width=0.23\textwidth]{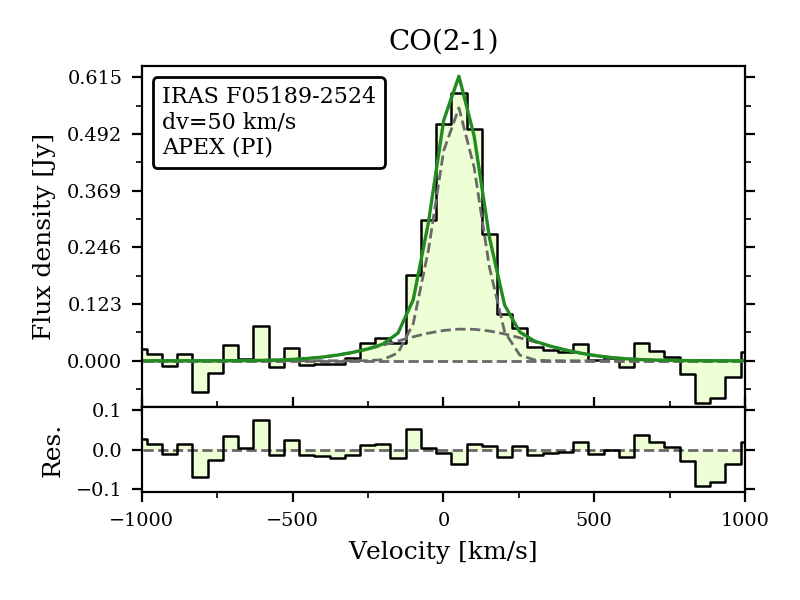}\quad
	\includegraphics[clip=true,trim=0.35cm 0.3cm 0.48cm 0.3cm,width=0.23\textwidth]{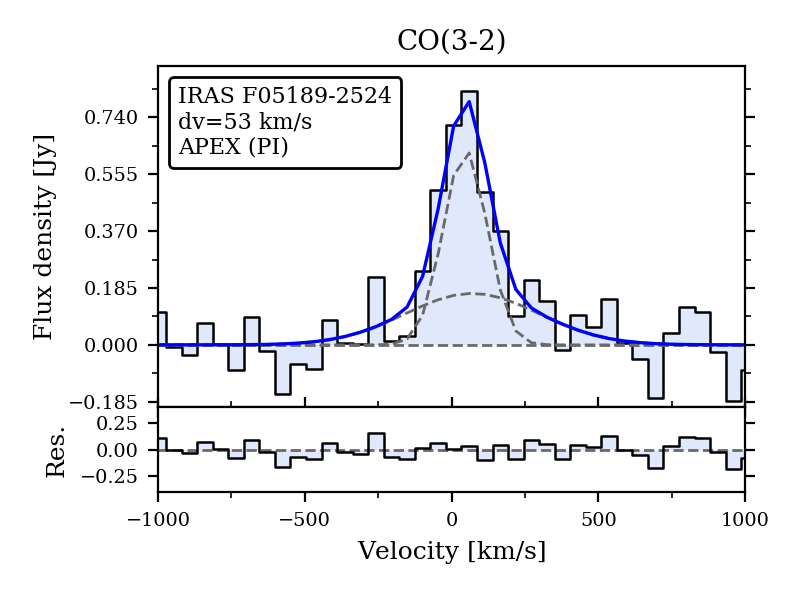}\quad
	\includegraphics[clip=true,trim=0.35cm 0.3cm 0.48cm 0.3cm,width=0.23\textwidth]{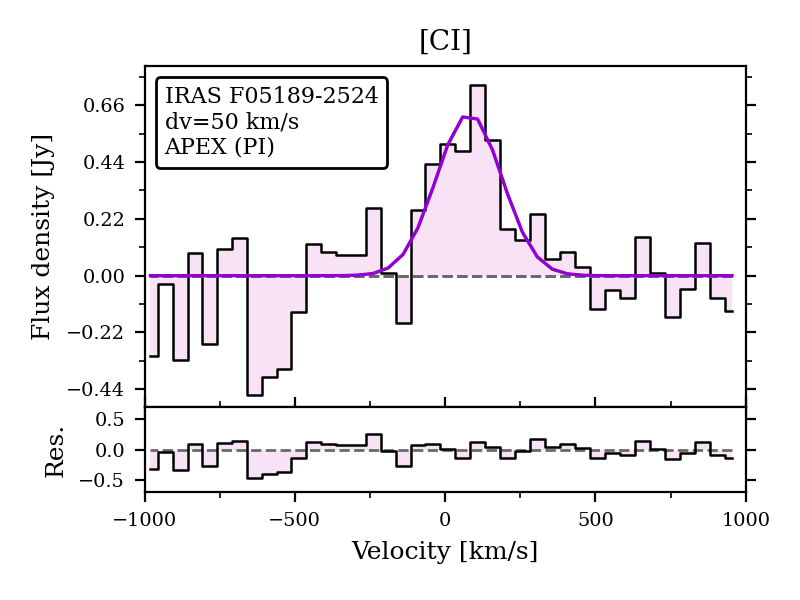}\\
	\includegraphics[clip=true,trim=0.35cm 0.3cm 0.48cm 0.3cm,width=0.23\textwidth]{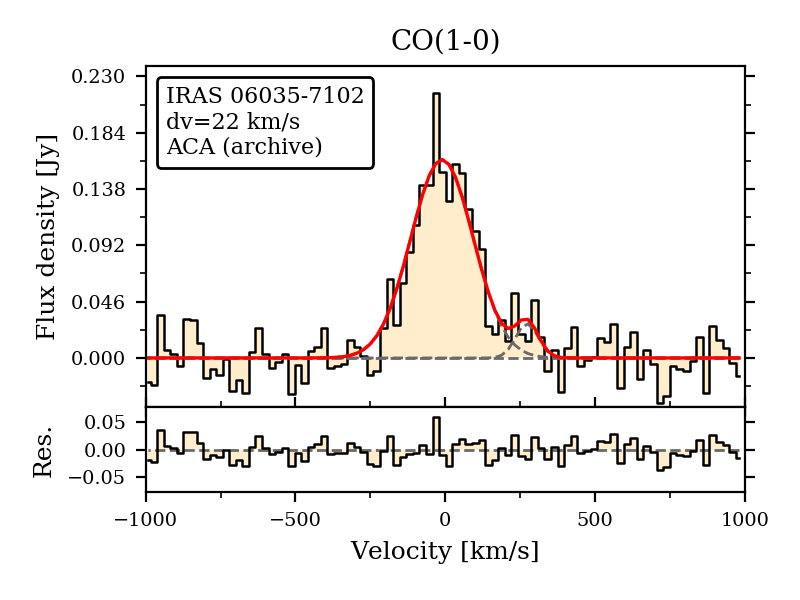}\quad
	\includegraphics[clip=true,trim=0.35cm 0.3cm 0.48cm 0.3cm,width=0.23\textwidth]{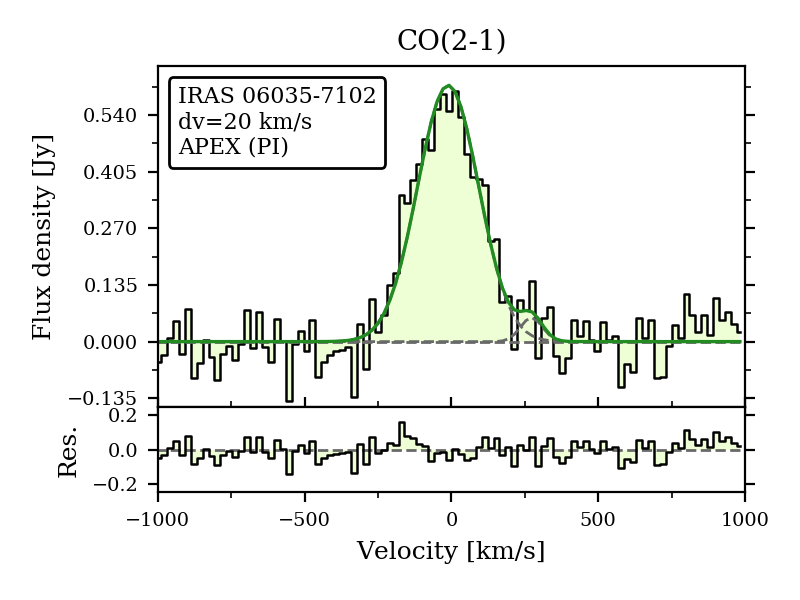}\quad
	\includegraphics[clip=true,trim=0.35cm 0.3cm 0.48cm 0.3cm,width=0.23\textwidth]{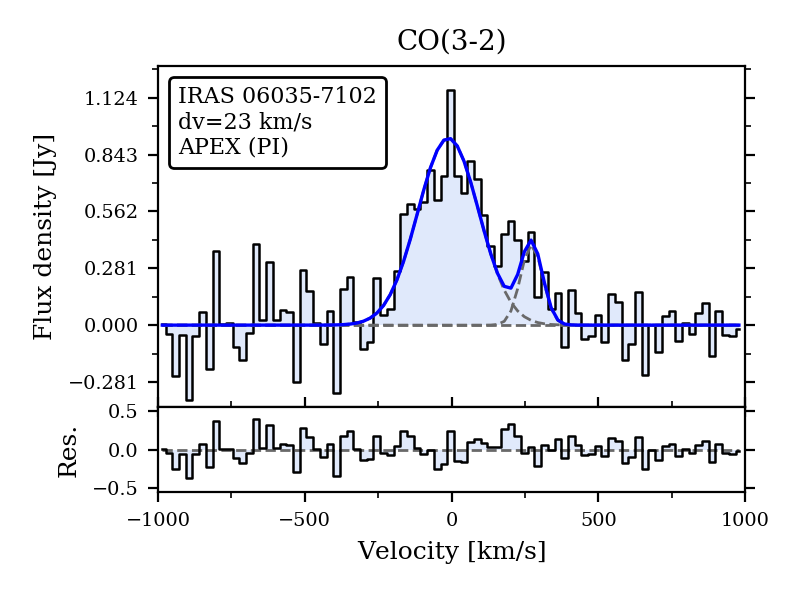}\quad
	\includegraphics[clip=true,trim=0.35cm 0.3cm 0.48cm 0.3cm,width=0.23\textwidth]{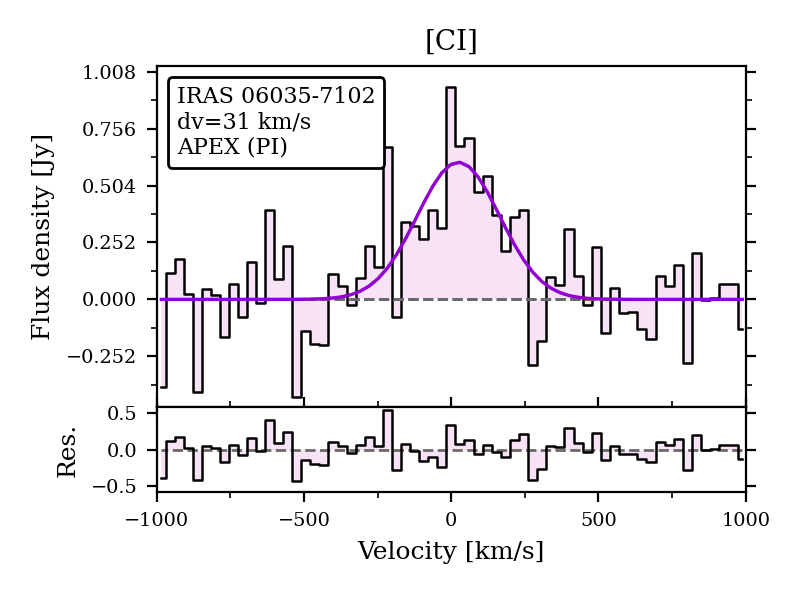}\\ 
	\caption{Integrated, continuum-subtracted CO(1--0), CO(2--1), CO(3--2) and [CI](1--0) spectra, including only the data employed in the analysis. The source name, spectral binning, and telescope are reported on the top-left corner of each plot. This figure shows sources:
	IRAS 00188-0856, IRAS 01003-2238, IRAS F01572+0009, IRAS 03521+0028, IRAS F05024-1941, IRAS F05189-2524, and IRAS 06035-7102.}
 \label{fig:spectra1}
\end{figure*}

\begin{figure*}[tbp]
	\includegraphics[clip=true,trim=0.35cm 0.3cm 0.48cm 0.3cm,width=0.23\textwidth]{figures/empty}\quad
	\includegraphics[clip=true,trim=0.35cm 0.3cm 0.48cm 0.3cm,width=0.23\textwidth]{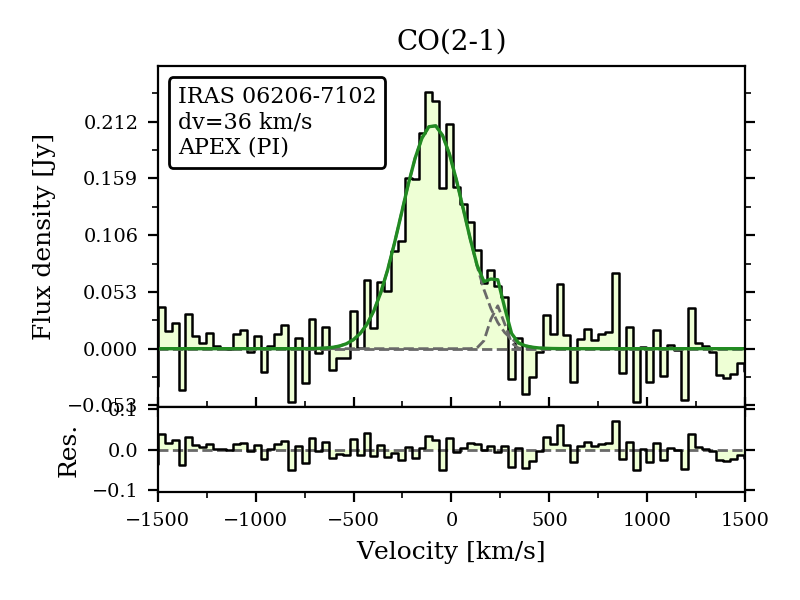}\quad
	\includegraphics[clip=true,trim=0.35cm 0.3cm 0.48cm 0.3cm,width=0.23\textwidth]{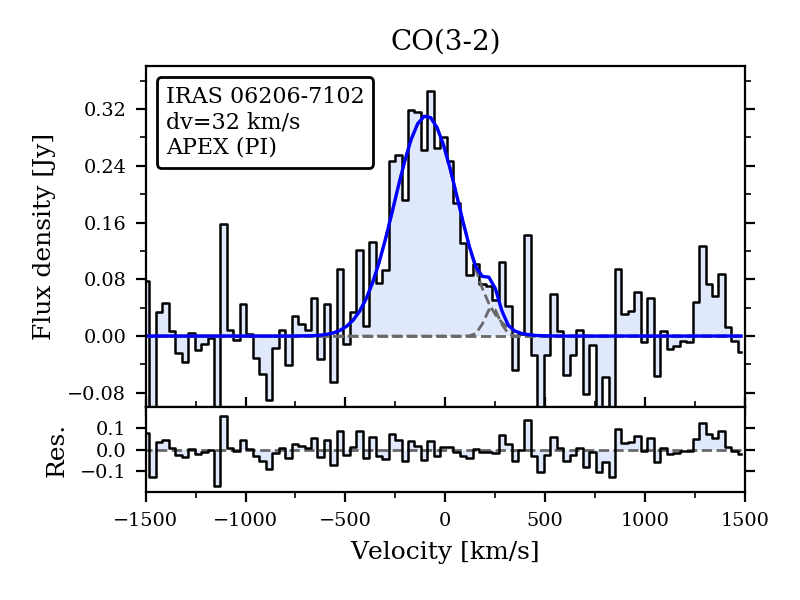}\quad
	\includegraphics[clip=true,trim=0.35cm 0.3cm 0.48cm 0.3cm,width=0.23\textwidth]{figures/empty}\\  
	\includegraphics[clip=true,trim=0.35cm 0.3cm 0.48cm 0.3cm,width=0.23\textwidth]{figures/empty}\quad
	\includegraphics[clip=true,trim=0.35cm 0.3cm 0.48cm 0.3cm,width=0.23\textwidth]{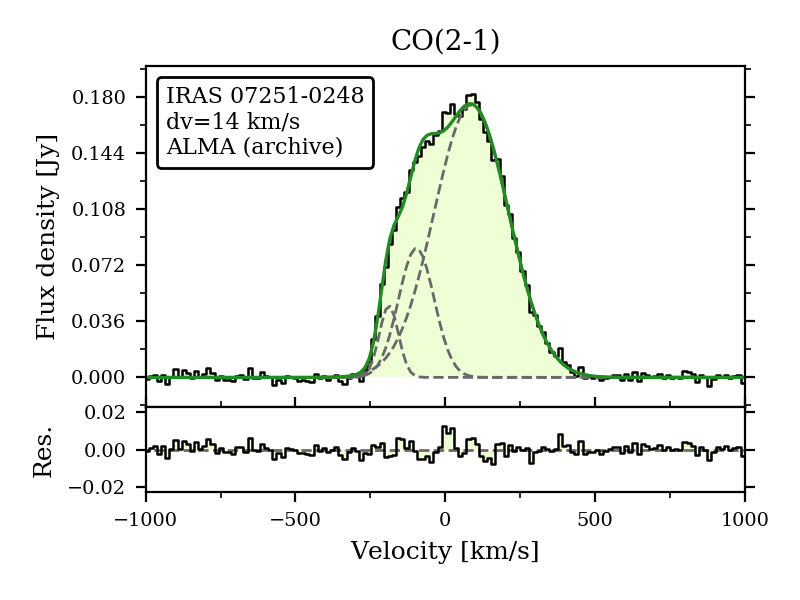}\quad
	\includegraphics[clip=true,trim=0.35cm 0.3cm 0.48cm 0.3cm,width=0.23\textwidth]{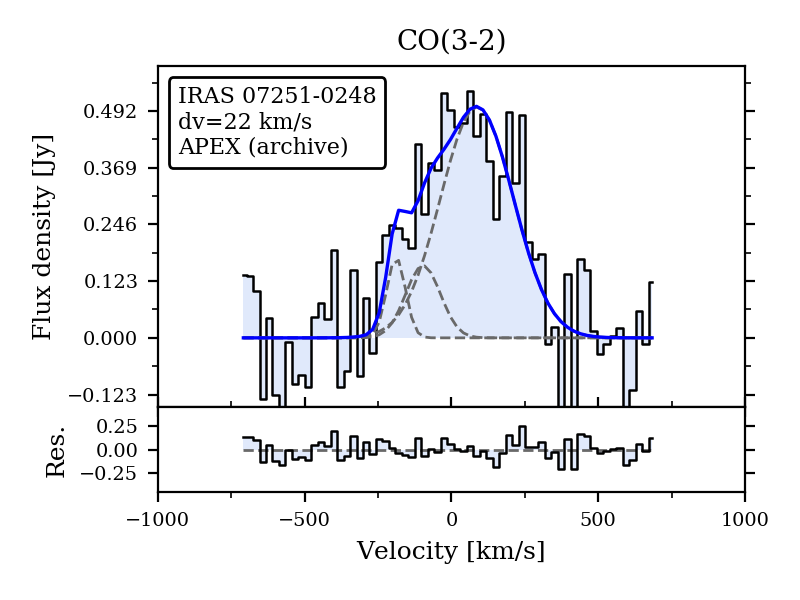}\quad
	\includegraphics[clip=true,trim=0.35cm 0.3cm 0.48cm 0.3cm,width=0.23\textwidth]{figures/empty}\\
	\includegraphics[clip=true,trim=0.35cm 0.3cm 0.48cm 0.3cm,width=0.23\textwidth]{figures/empty}\quad
	\includegraphics[clip=true,trim=0.35cm 0.3cm 0.48cm 0.3cm,width=0.23\textwidth]{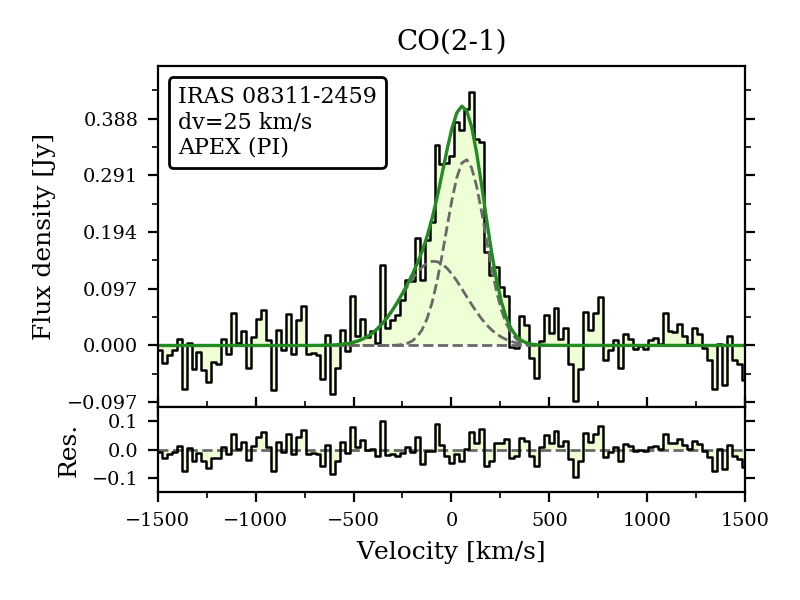}\quad
	\includegraphics[clip=true,trim=0.35cm 0.3cm 0.48cm 0.3cm,width=0.23\textwidth]{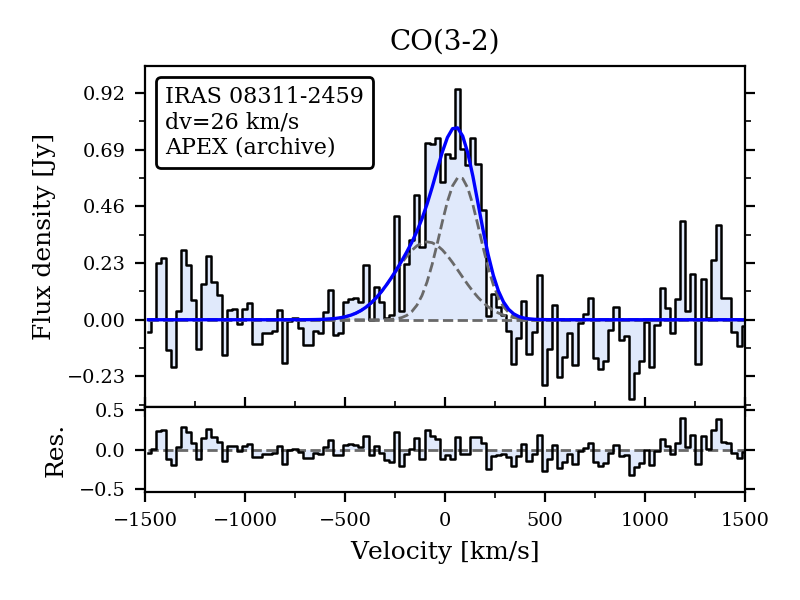}\quad
	\includegraphics[clip=true,trim=0.35cm 0.3cm 0.48cm 0.3cm,width=0.23\textwidth]{figures/empty}\\  
	\includegraphics[clip=true,trim=0.35cm 0.3cm 0.48cm 0.3cm,width=0.23\textwidth]{figures/empty}\quad
	\includegraphics[clip=true,trim=0.35cm 0.3cm 0.48cm 0.3cm,width=0.23\textwidth]{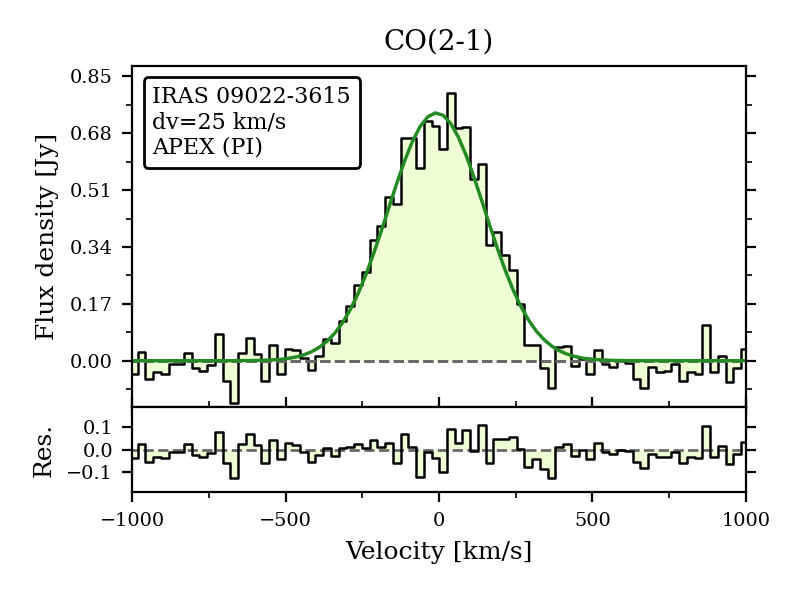}\quad
	\includegraphics[clip=true,trim=0.35cm 0.3cm 0.48cm 0.3cm,width=0.23\textwidth]{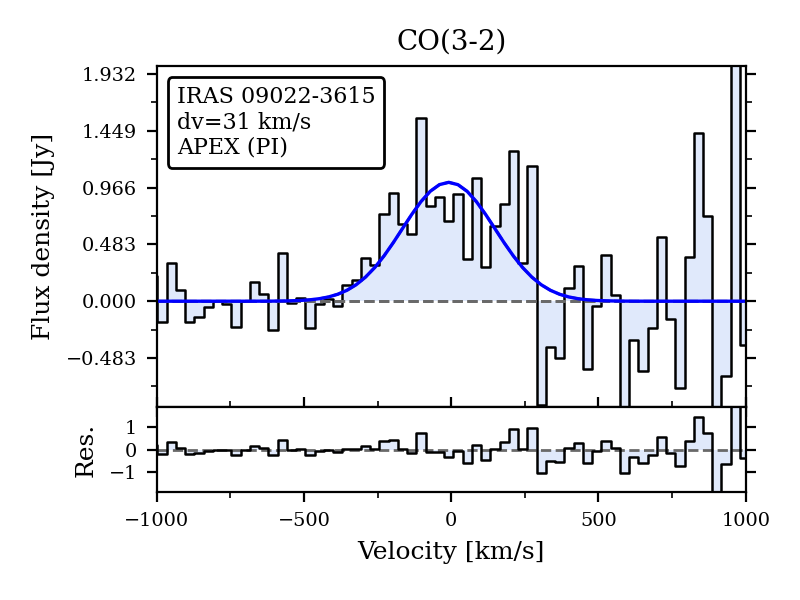}\quad
	\includegraphics[clip=true,trim=0.35cm 0.3cm 0.48cm 0.3cm,width=0.23\textwidth]{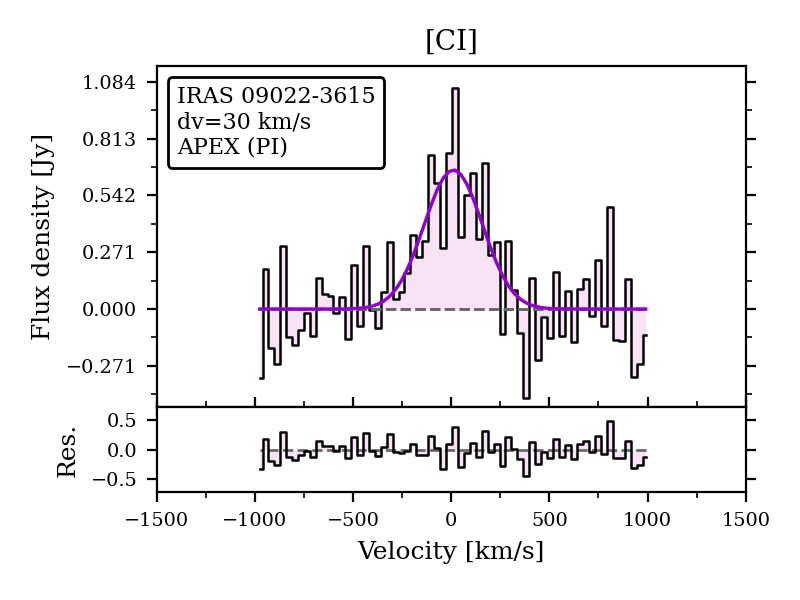}\\  
	\includegraphics[clip=true,trim=0.35cm 0.3cm 0.48cm 0.3cm,width=0.23\textwidth]{figures/empty}\quad
	\includegraphics[clip=true,trim=0.35cm 0.3cm 0.48cm 0.3cm,width=0.23\textwidth]{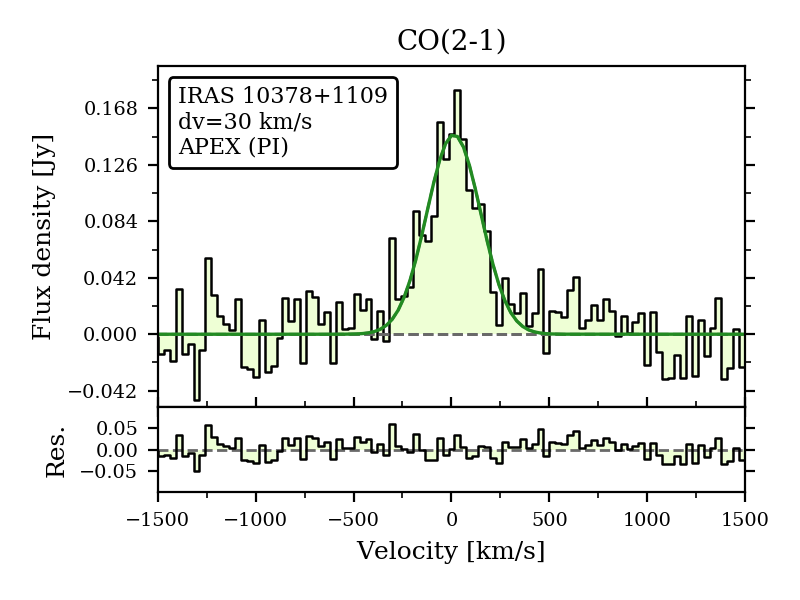}\quad
	\includegraphics[clip=true,trim=0.35cm 0.3cm 0.48cm 0.3cm,width=0.23\textwidth]{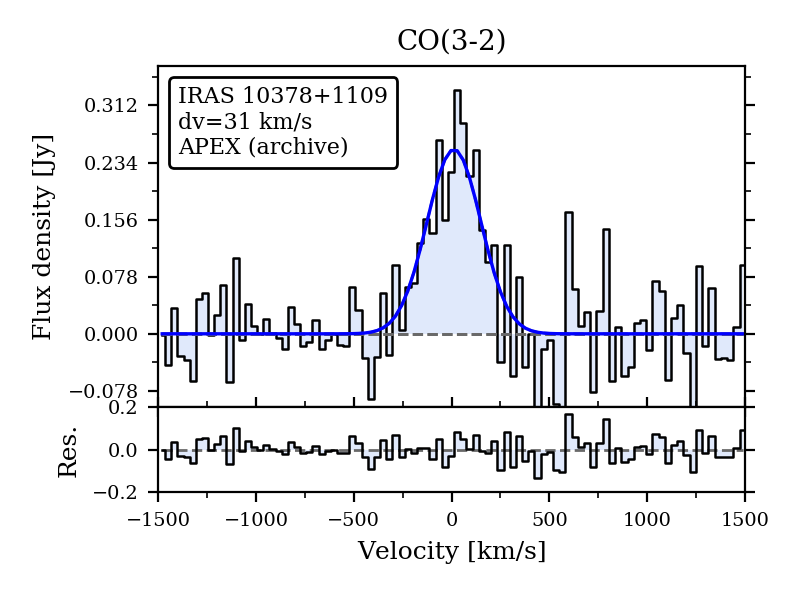}\quad
	\includegraphics[clip=true,trim=0.35cm 0.3cm 0.48cm 0.3cm,width=0.23\textwidth]{figures/empty}\\  
	\includegraphics[clip=true,trim=0.35cm 0.3cm 0.48cm 0.3cm,width=0.23\textwidth]{figures/empty}\quad
	\includegraphics[clip=true,trim=0.35cm 0.3cm 0.48cm 0.3cm,width=0.23\textwidth]{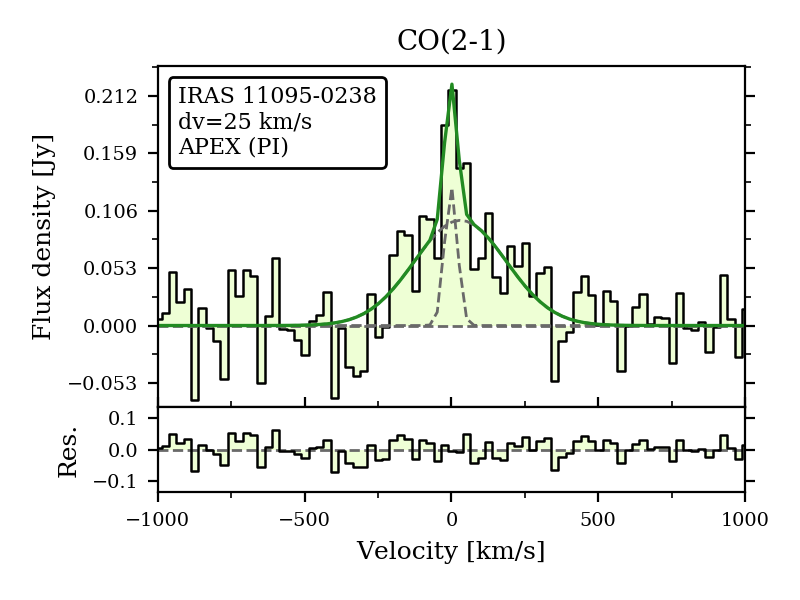}\quad
	\includegraphics[clip=true,trim=0.35cm 0.3cm 0.48cm 0.3cm,width=0.23\textwidth]{figures/empty}\quad
	\includegraphics[clip=true,trim=0.35cm 0.3cm 0.48cm 0.3cm,width=0.23\textwidth]{figures/empty}\\  
	\includegraphics[clip=true,trim=0.35cm 0.3cm 0.48cm 0.3cm,width=0.23\textwidth]{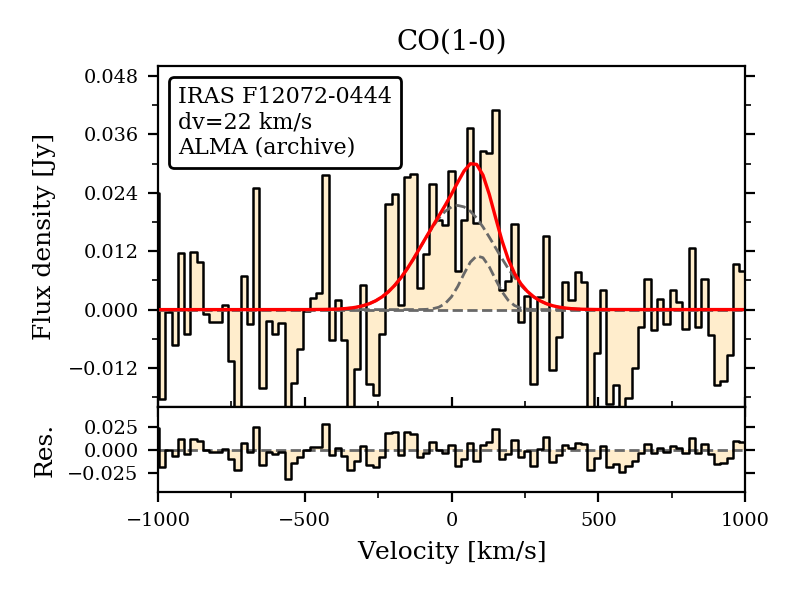}\quad
	\includegraphics[clip=true,trim=0.35cm 0.3cm 0.48cm 0.3cm,width=0.23\textwidth]{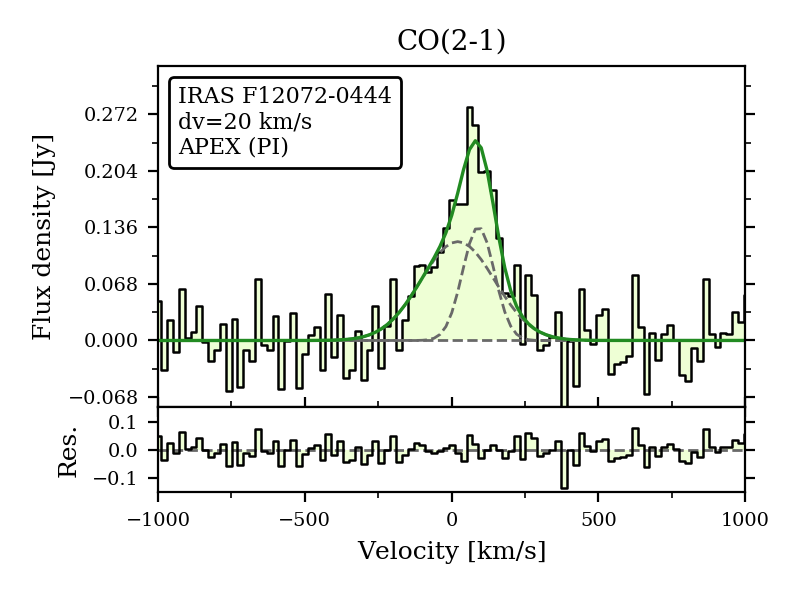}\quad
	\includegraphics[clip=true,trim=0.35cm 0.3cm 0.48cm 0.3cm,width=0.23\textwidth]{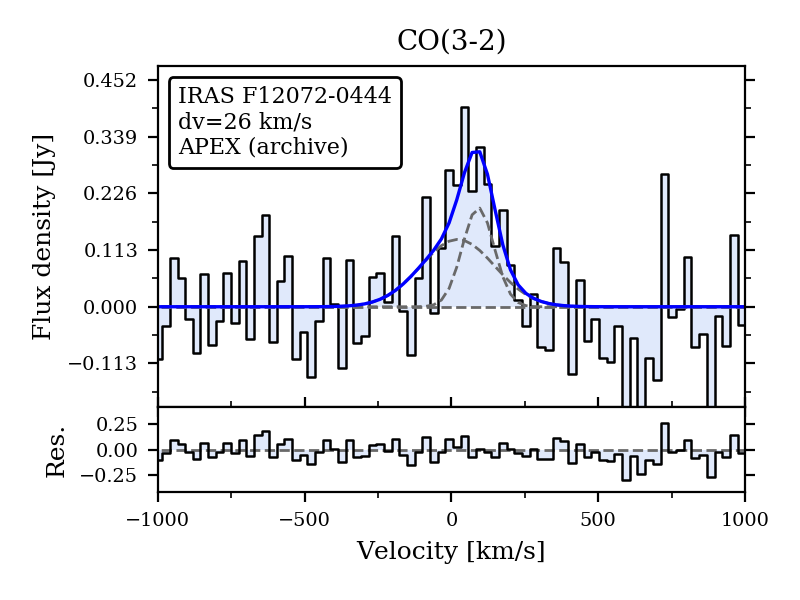}\quad
	\includegraphics[clip=true,trim=0.35cm 0.3cm 0.48cm 0.3cm,width=0.23\textwidth]{figures/empty}\\  
	\caption{Continued from Fig.~\ref{fig:spectra1}. This figure shows sources:
		IRAS 06206-6315, IRAS 07251-0248, IRAS 08311-2459, IRAS 09022-3615, IRAS 10378+1109, IRAS 11095-0238, and IRAS F12074-0444.}\label{fig:spectra2}
\end{figure*}

\begin{figure*}[tbp]
	\includegraphics[clip=true,trim=0.35cm 0.3cm 0.48cm 0.3cm,width=0.23\textwidth]{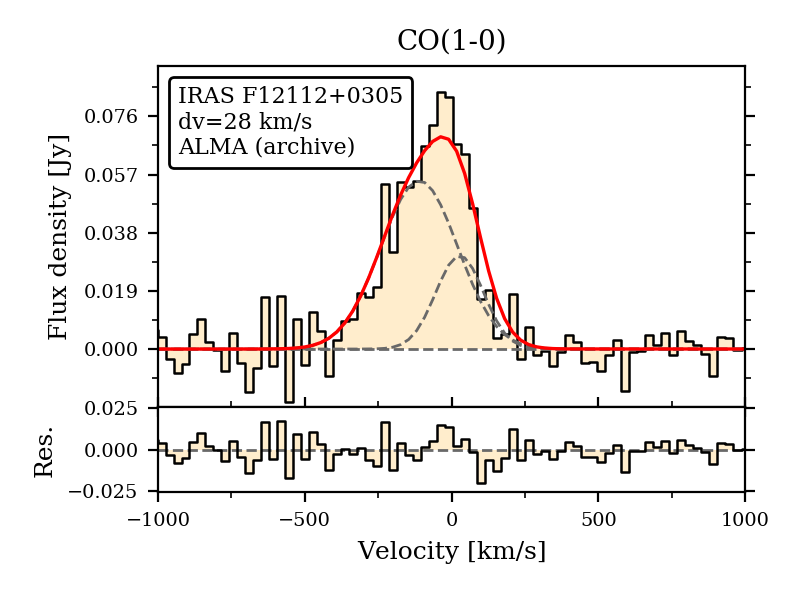}\quad
	\includegraphics[clip=true,trim=0.35cm 0.3cm 0.48cm 0.3cm,width=0.23\textwidth]{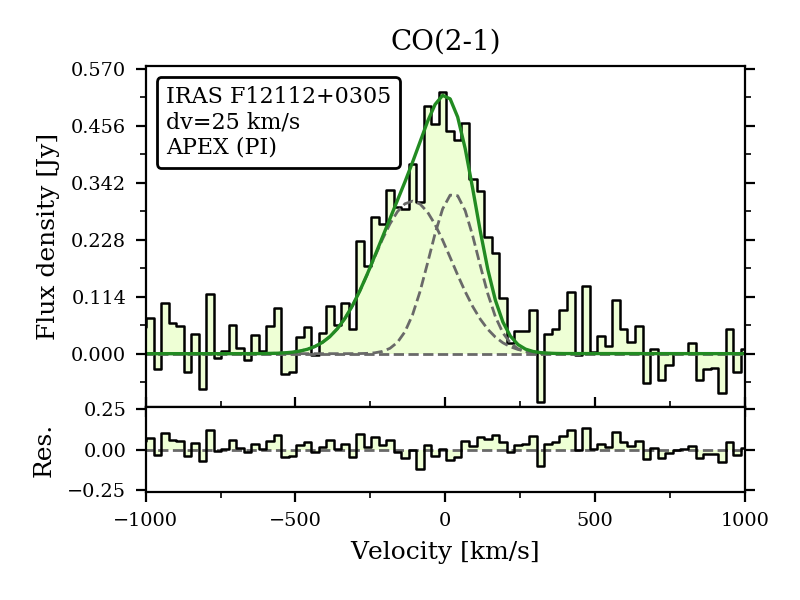}\quad
	\includegraphics[clip=true,trim=0.35cm 0.3cm 0.48cm 0.3cm,width=0.23\textwidth]{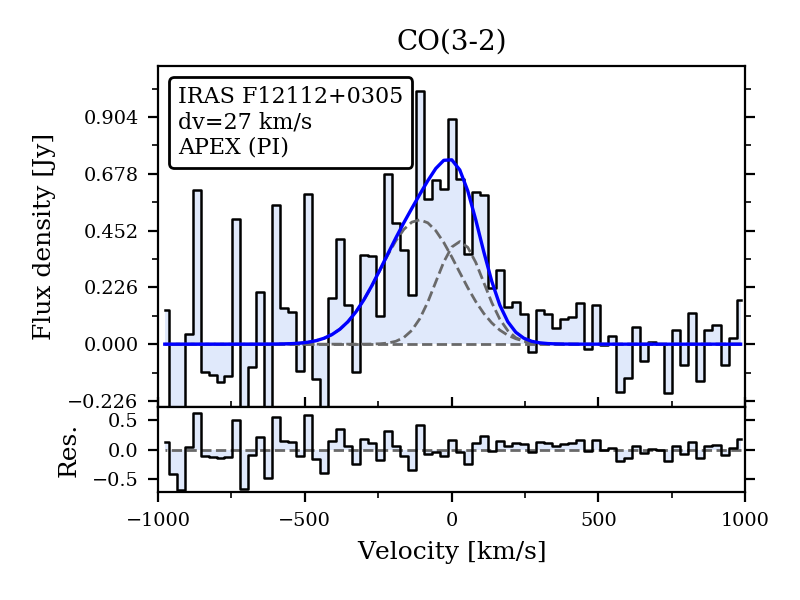}\quad
	\includegraphics[clip=true,trim=0.35cm 0.3cm 0.48cm 0.3cm,width=0.23\textwidth]{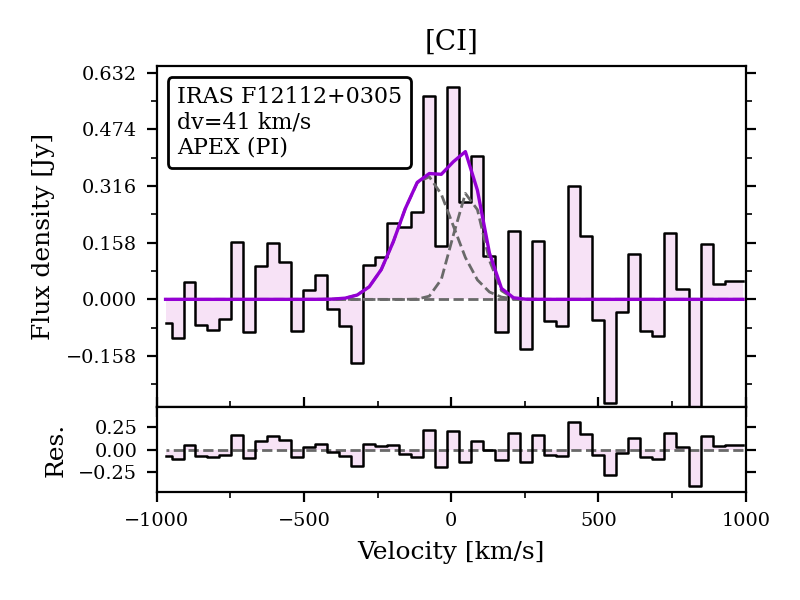}\\ 
	\includegraphics[clip=true,trim=0.35cm 0.3cm 0.48cm 0.3cm,width=0.23\textwidth]{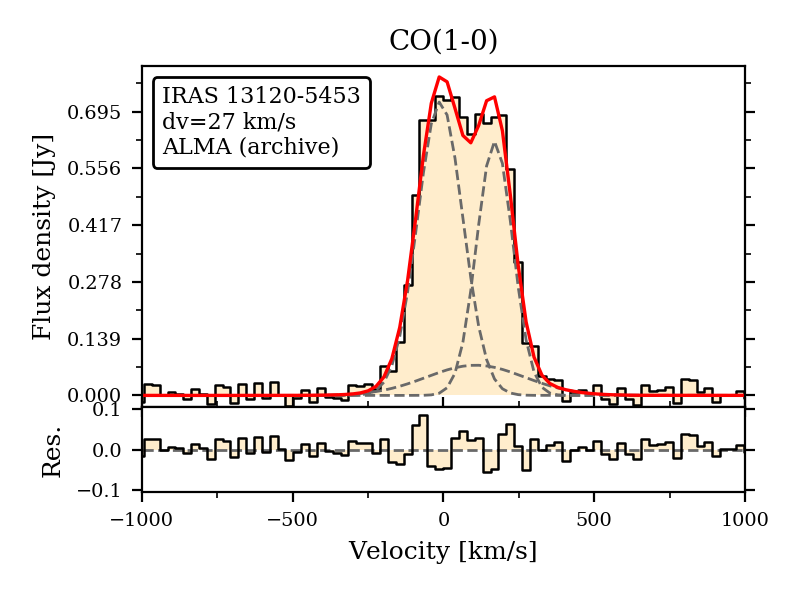}\quad
	\includegraphics[clip=true,trim=0.35cm 0.3cm 0.48cm 0.3cm,width=0.23\textwidth]{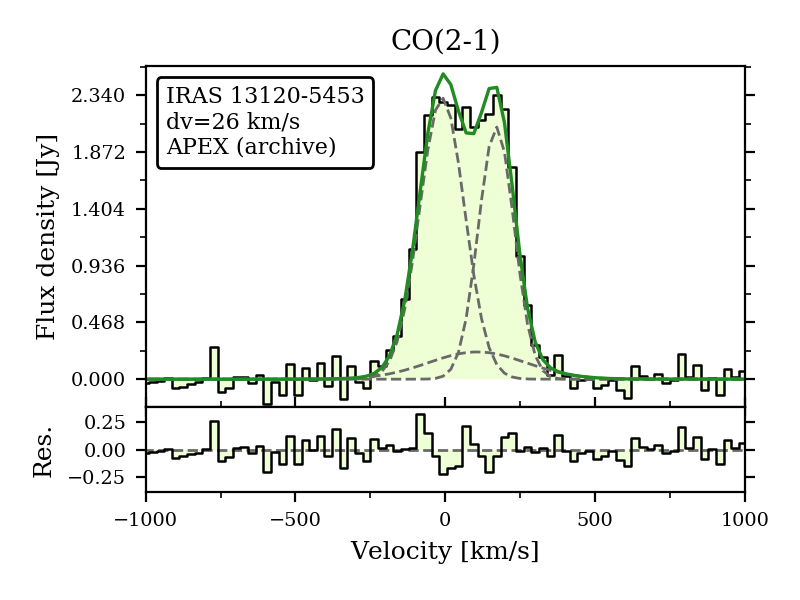}\quad
	\includegraphics[clip=true,trim=0.35cm 0.3cm 0.48cm 0.3cm,width=0.23\textwidth]{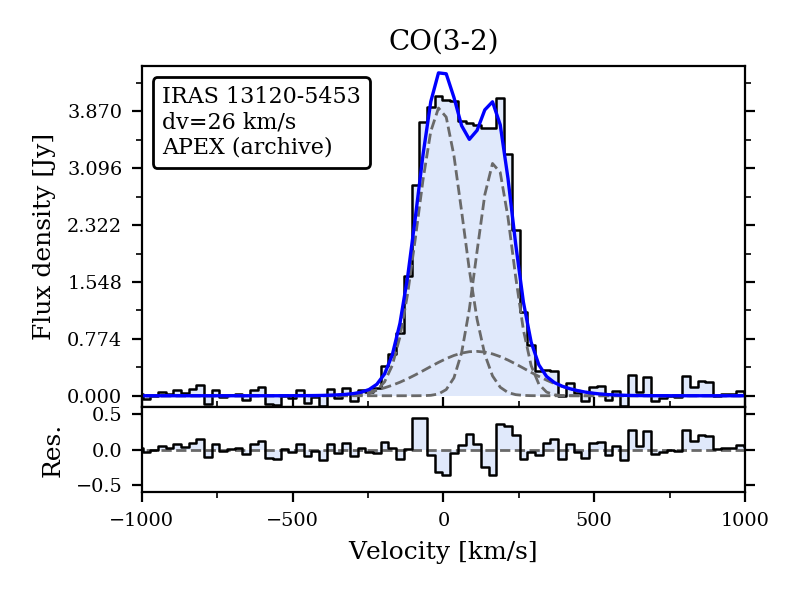}\quad
	\includegraphics[clip=true,trim=0.35cm 0.3cm 0.48cm 0.3cm,width=0.23\textwidth]{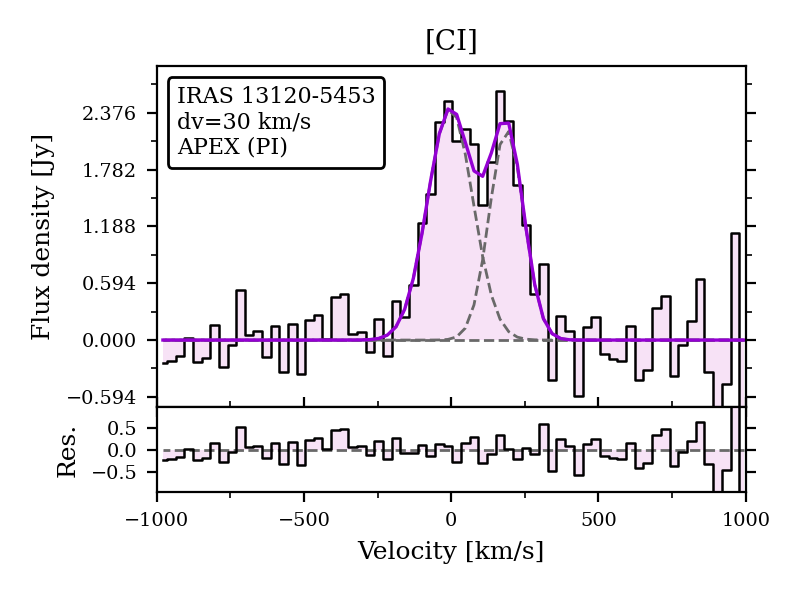}\\ 
	\includegraphics[clip=true,trim=0.35cm 0.3cm 0.48cm 0.3cm,width=0.23\textwidth]{figures/empty}\quad
	\includegraphics[clip=true,trim=0.35cm 0.3cm 0.48cm 0.3cm,width=0.23\textwidth]{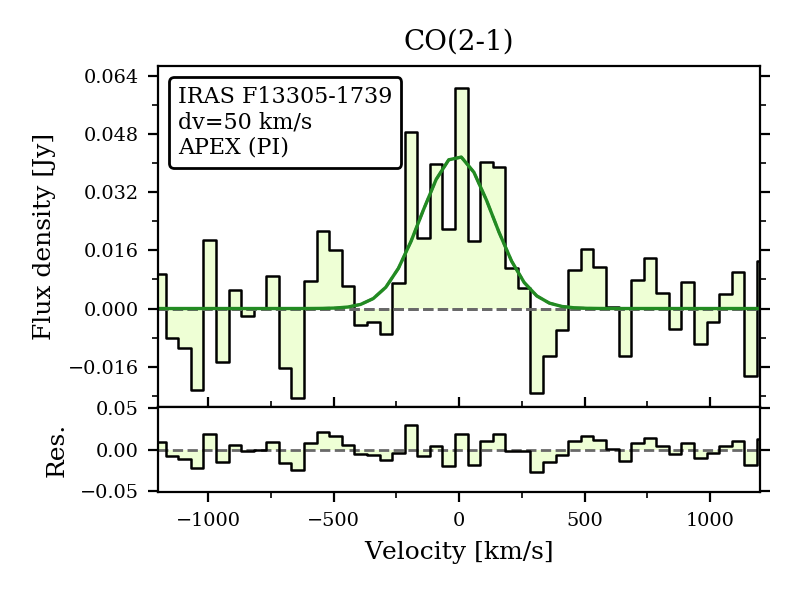}\quad
	\includegraphics[clip=true,trim=0.35cm 0.3cm 0.48cm 0.3cm,width=0.23\textwidth]{figures/empty}\quad
	\includegraphics[clip=true,trim=0.35cm 0.3cm 0.48cm 0.3cm,width=0.23\textwidth]{figures/empty}\\ 
	\includegraphics[clip=true,trim=0.35cm 0.3cm 0.48cm 0.3cm,width=0.23\textwidth]{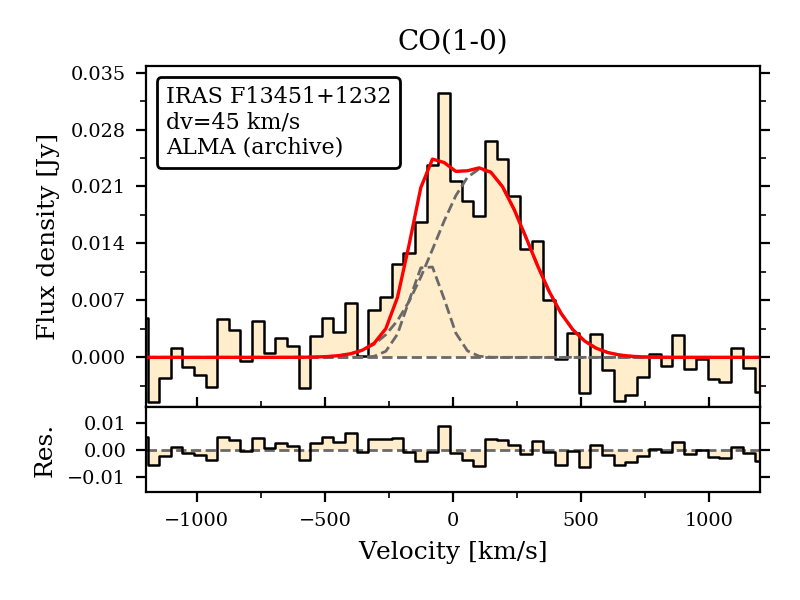}\quad
	\includegraphics[clip=true,trim=0.35cm 0.3cm 0.48cm 0.3cm,width=0.23\textwidth]{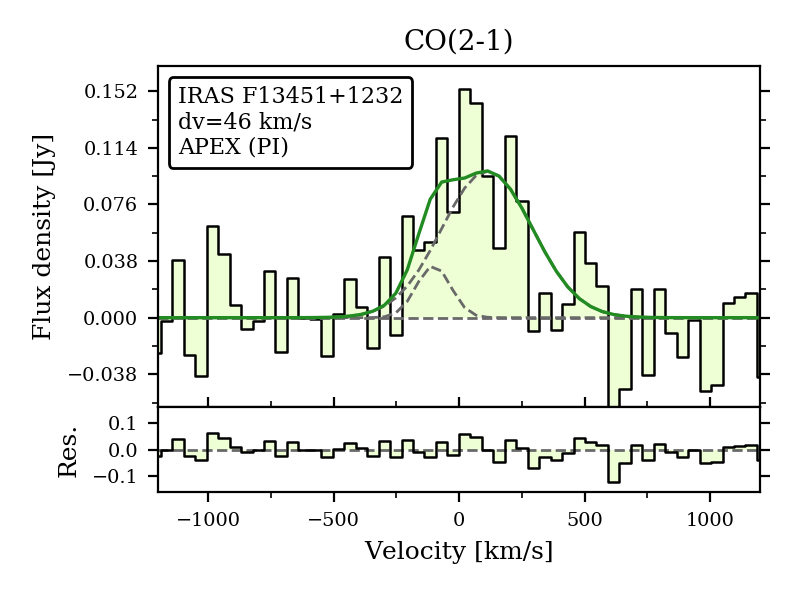}\quad
	\includegraphics[clip=true,trim=0.35cm 0.3cm 0.48cm 0.3cm,width=0.23\textwidth]{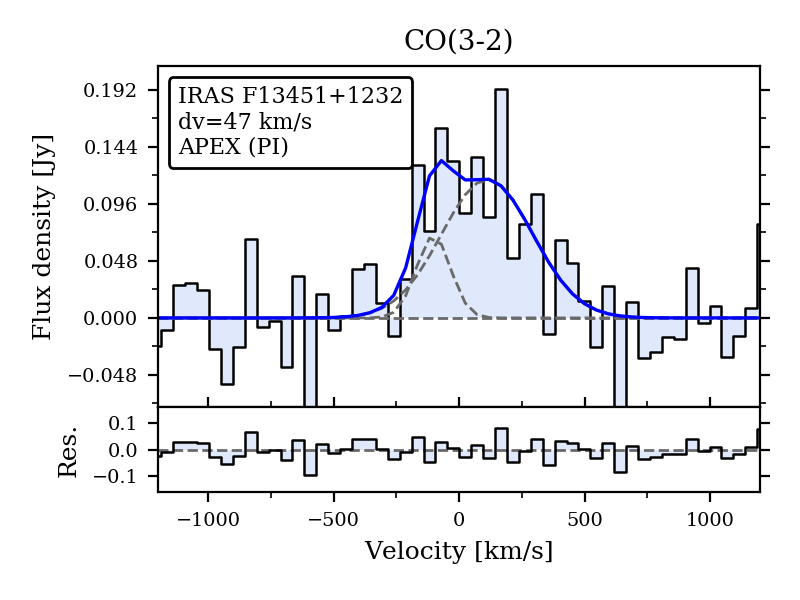}\quad
	\includegraphics[clip=true,trim=0.35cm 0.3cm 0.48cm 0.3cm,width=0.23\textwidth]{figures/empty}\\ 
	\includegraphics[clip=true,trim=0.35cm 0.3cm 0.48cm 0.3cm,width=0.23\textwidth]{figures/empty}\quad
	\includegraphics[clip=true,trim=0.35cm 0.3cm 0.48cm 0.3cm,width=0.23\textwidth]{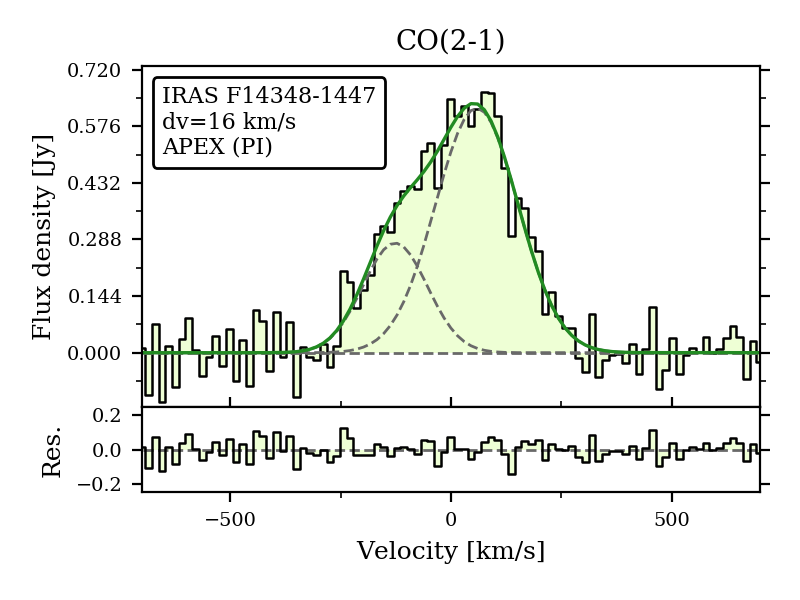}\quad
	\includegraphics[clip=true,trim=0.35cm 0.3cm 0.48cm 0.3cm,width=0.23\textwidth]{figures/empty}\quad
	\includegraphics[clip=true,trim=0.35cm 0.3cm 0.48cm 0.3cm,width=0.23\textwidth]{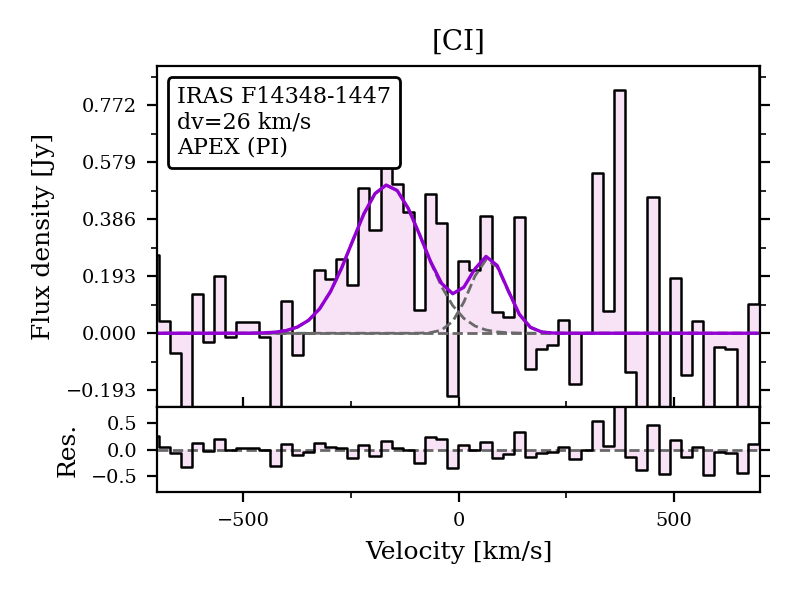}\\ 
	\includegraphics[clip=true,trim=0.35cm 0.3cm 0.48cm 0.3cm,width=0.23\textwidth]{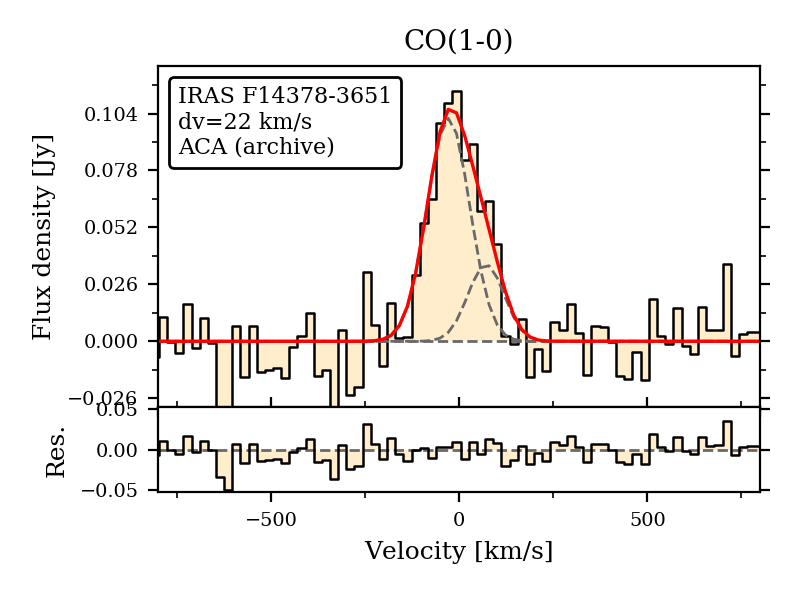}\quad
	\includegraphics[clip=true,trim=0.35cm 0.3cm 0.48cm 0.3cm,width=0.23\textwidth]{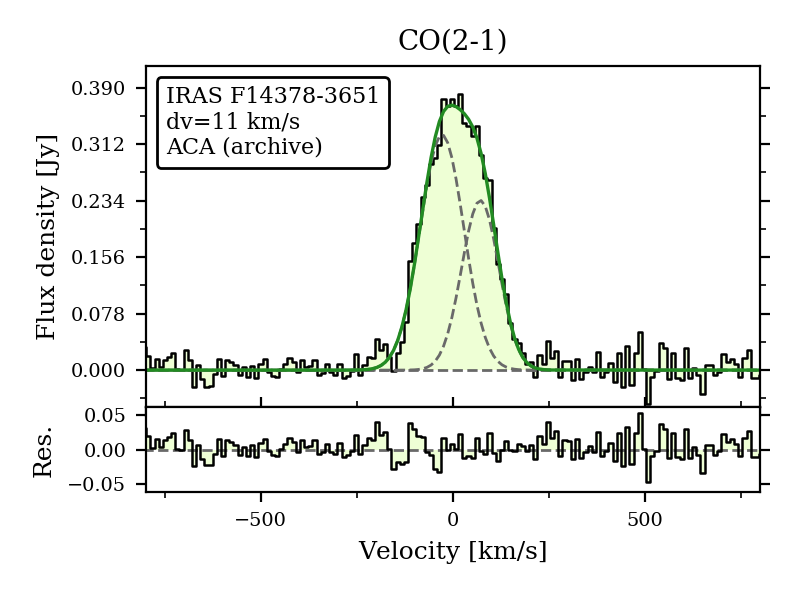}\quad
	\includegraphics[clip=true,trim=0.35cm 0.3cm 0.48cm 0.3cm,width=0.23\textwidth]{figures/empty}\quad
	\includegraphics[clip=true,trim=0.35cm 0.3cm 0.48cm 0.3cm,width=0.23\textwidth]{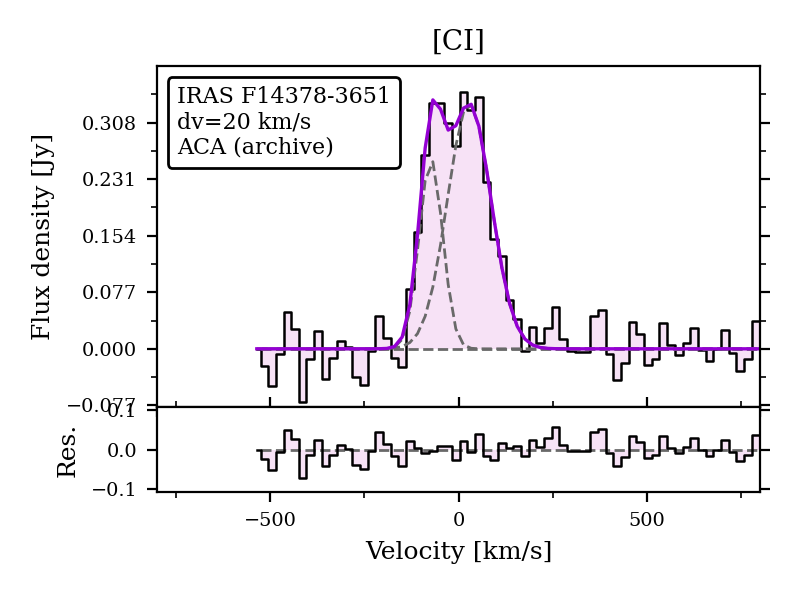}\\ 
	\includegraphics[clip=true,trim=0.35cm 0.3cm 0.48cm 0.3cm,width=0.23\textwidth]{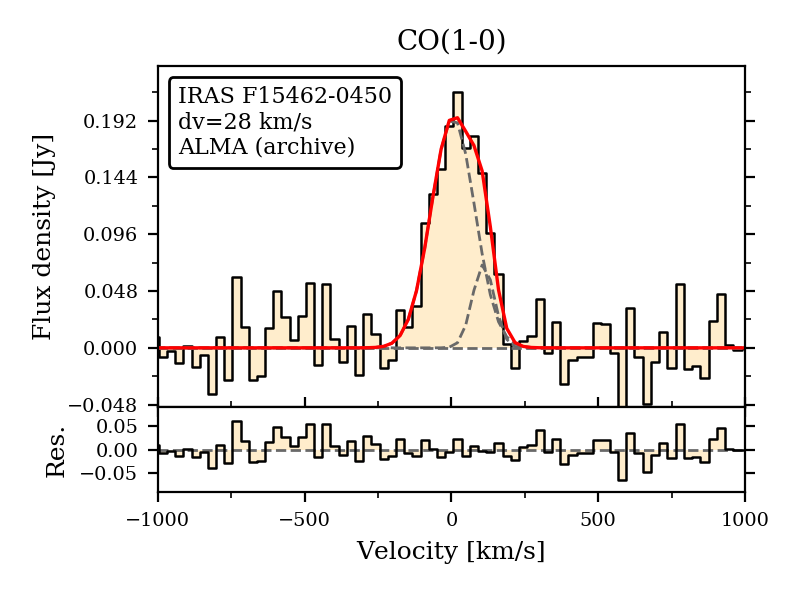}\quad
	\includegraphics[clip=true,trim=0.35cm 0.3cm 0.48cm 0.3cm,width=0.23\textwidth]{figures/empty}\quad
	\includegraphics[clip=true,trim=0.35cm 0.3cm 0.48cm 0.3cm,width=0.23\textwidth]{figures/empty}\quad
	\includegraphics[clip=true,trim=0.35cm 0.3cm 0.48cm 0.3cm,width=0.23\textwidth]{figures/empty}\\  
\caption{Continued from Fig.~\ref{fig:spectra2}. This figure shows sources: IRAS F12112+0305, IRAS 13120-5453, IRAS F13305-1739,  IRAS F13451+1232, IRAS F14348-1447, IRAS F14378-3651, IRAS F15462-0450.}\label{fig:spectra3}
\end{figure*}

\begin{figure*}[tbp]
	\includegraphics[clip=true,trim=0.35cm 0.3cm 0.48cm 0.3cm,width=0.23\textwidth]{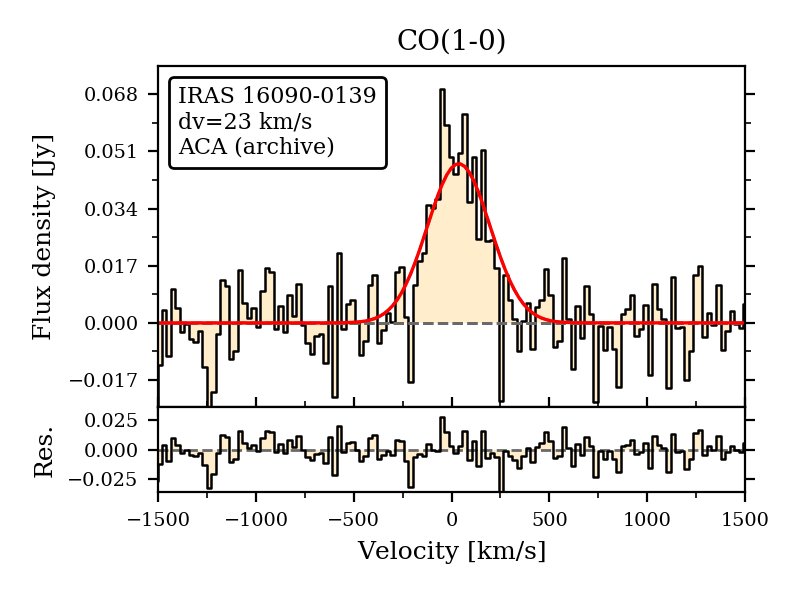}\quad
	\includegraphics[clip=true,trim=0.35cm 0.3cm 0.48cm 0.3cm,width=0.23\textwidth]{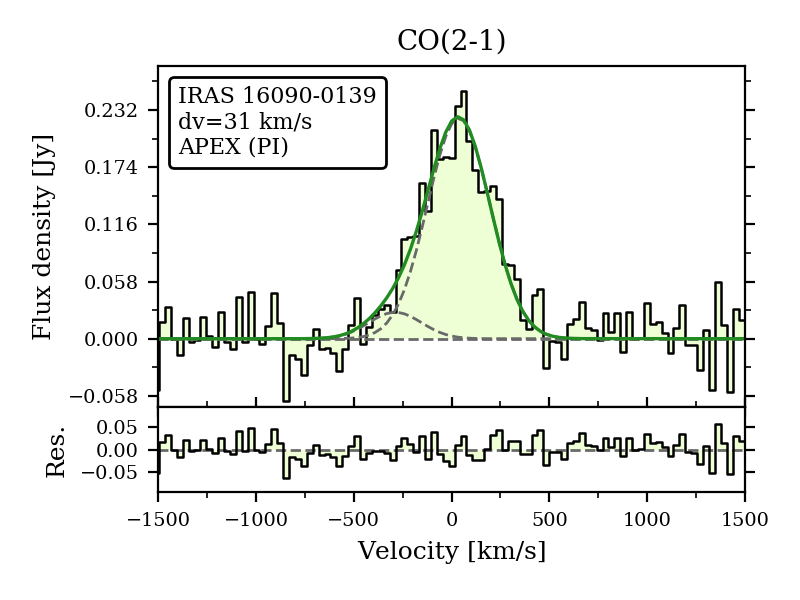}\quad
	\includegraphics[clip=true,trim=0.35cm 0.3cm 0.48cm 0.3cm,width=0.23\textwidth]{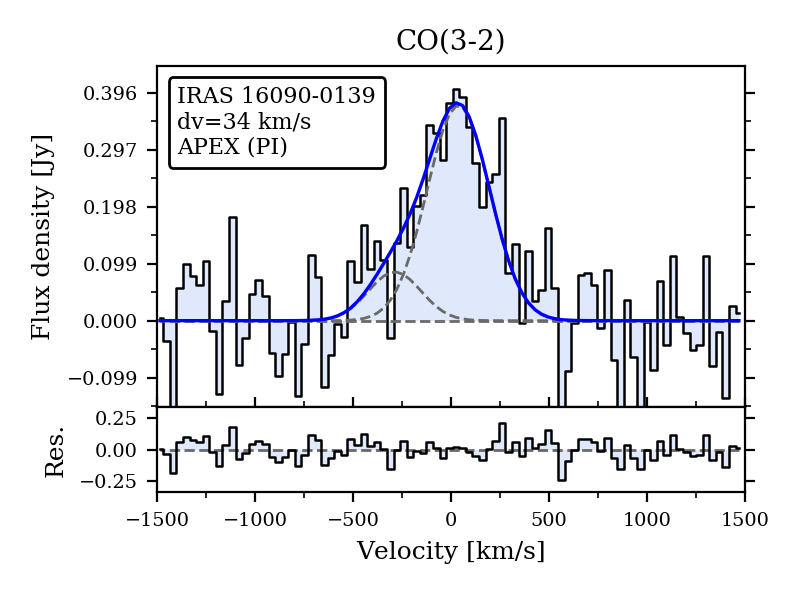}\quad
	\includegraphics[clip=true,trim=0.35cm 0.3cm 0.48cm 0.3cm,width=0.23\textwidth]{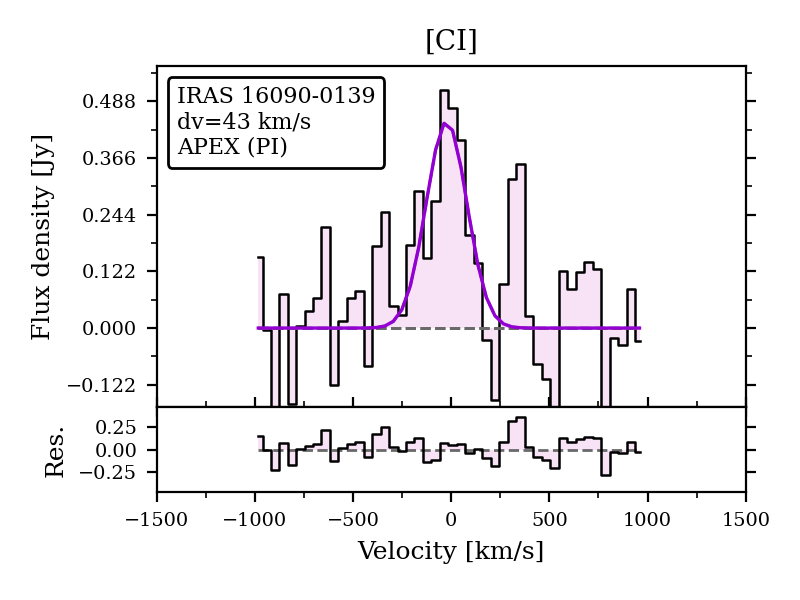}\\  
	\includegraphics[clip=true,trim=0.35cm 0.3cm 0.48cm 0.3cm,width=0.23\textwidth]{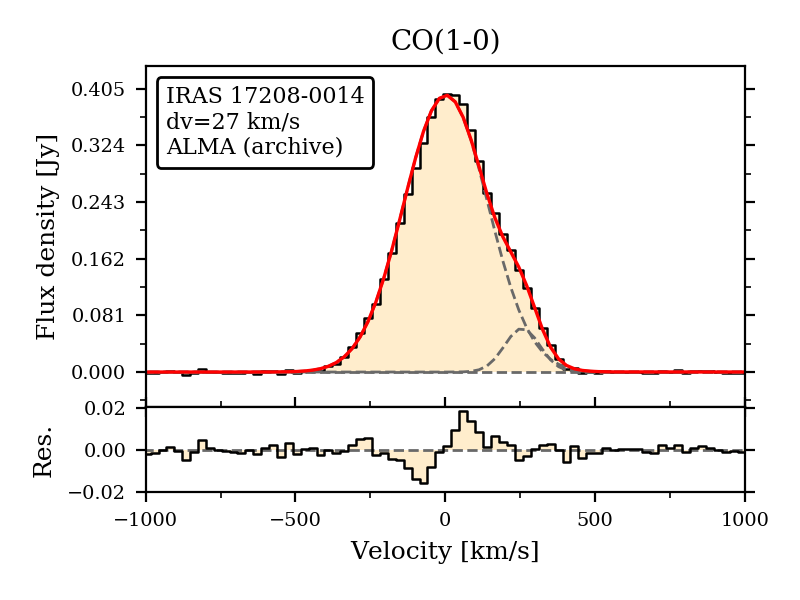}\quad
	\includegraphics[clip=true,trim=0.35cm 0.3cm 0.48cm 0.3cm,width=0.23\textwidth]{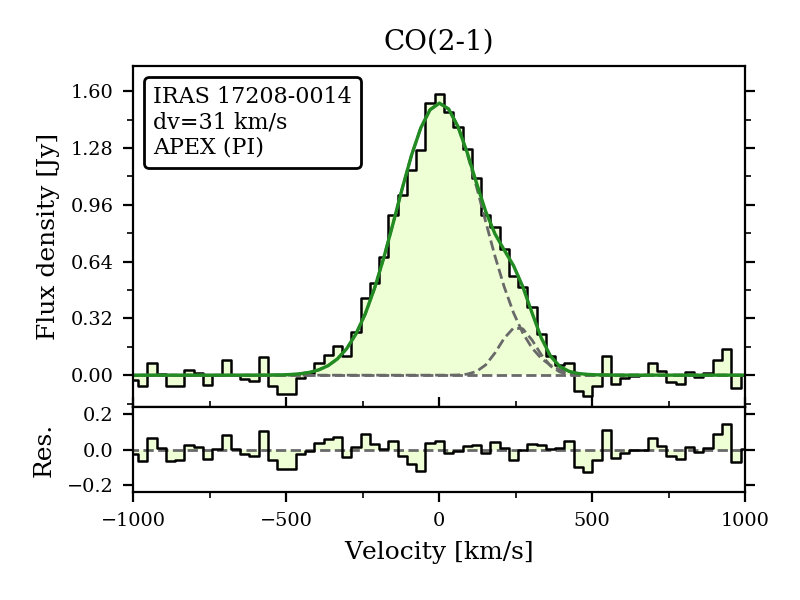}\quad
	\includegraphics[clip=true,trim=0.35cm 0.3cm 0.48cm 0.3cm,width=0.23\textwidth]{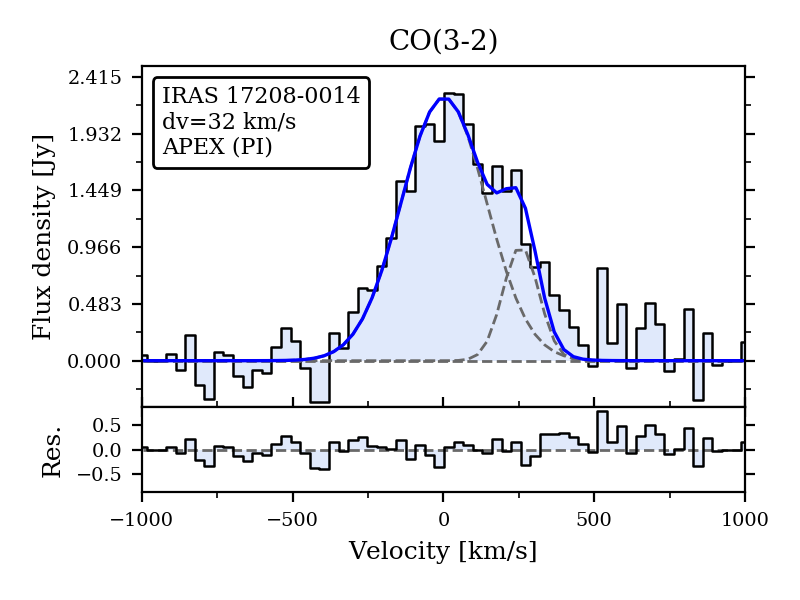}\quad
	\includegraphics[clip=true,trim=0.35cm 0.3cm 0.48cm 0.3cm,width=0.23\textwidth]{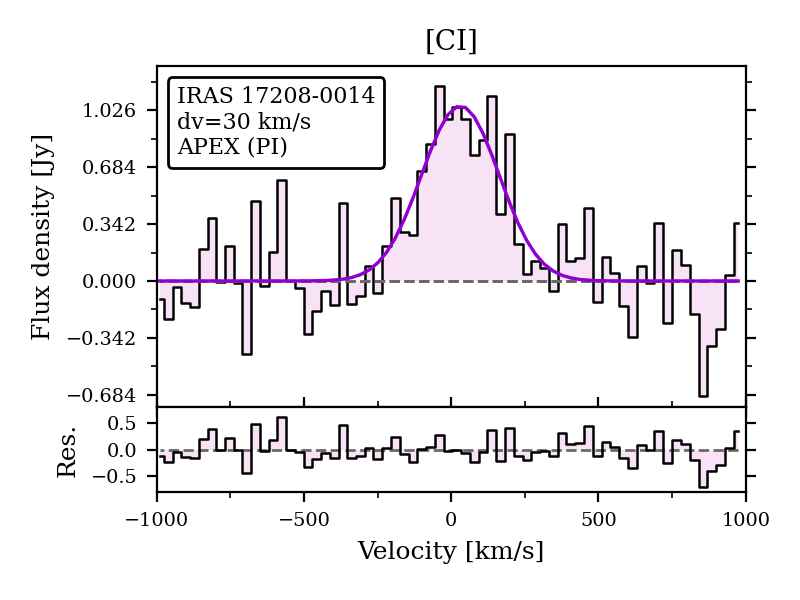}\\
	\includegraphics[clip=true,trim=0.35cm 0.3cm 0.48cm 0.3cm,width=0.23\textwidth]{figures/empty}\quad
	\includegraphics[clip=true,trim=0.35cm 0.3cm 0.48cm 0.3cm,width=0.23\textwidth]{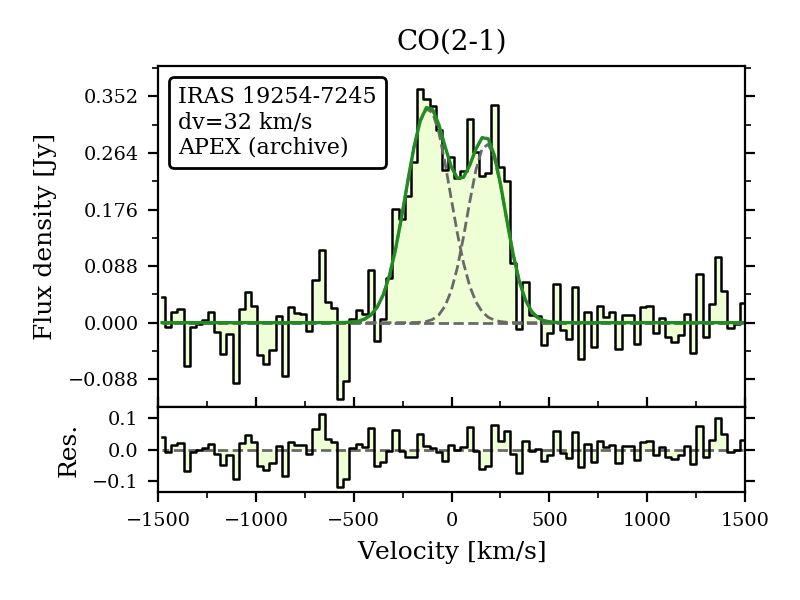}\quad
	\includegraphics[clip=true,trim=0.35cm 0.3cm 0.48cm 0.3cm,width=0.23\textwidth]{figures/empty}\quad
	\includegraphics[clip=true,trim=0.35cm 0.3cm 0.48cm 0.3cm,width=0.23\textwidth]{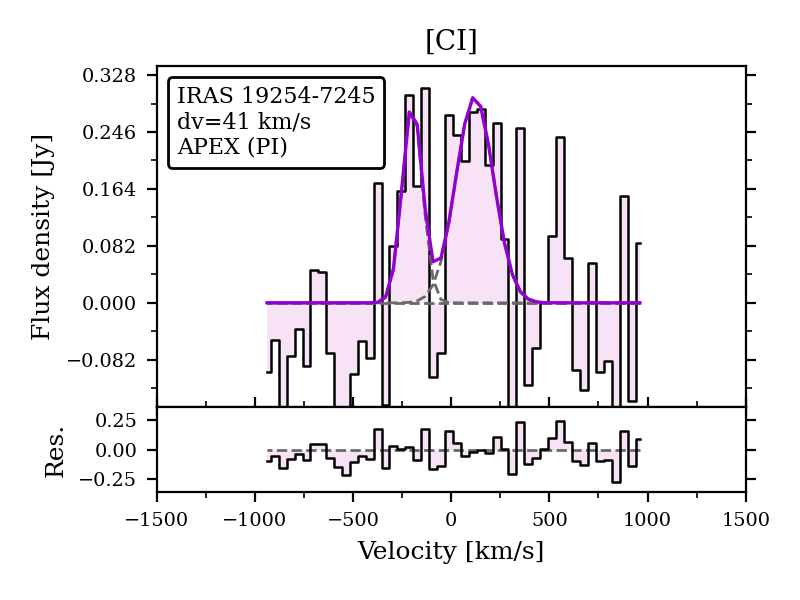}\\  
	\includegraphics[clip=true,trim=0.35cm 0.3cm 0.48cm 0.3cm,width=0.23\textwidth]{figures/empty}\quad
	\includegraphics[clip=true,trim=0.35cm 0.3cm 0.48cm 0.3cm,width=0.23\textwidth]{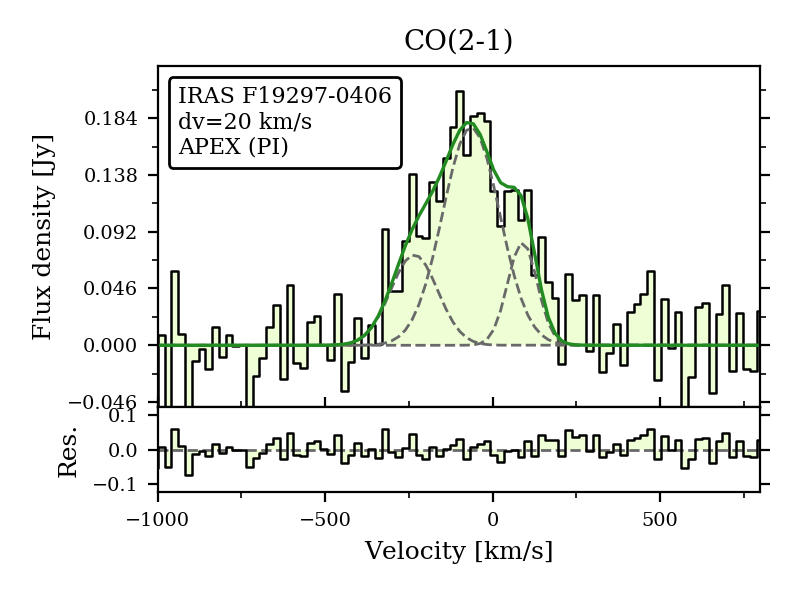}\quad
	\includegraphics[clip=true,trim=0.35cm 0.3cm 0.48cm 0.3cm,width=0.23\textwidth]{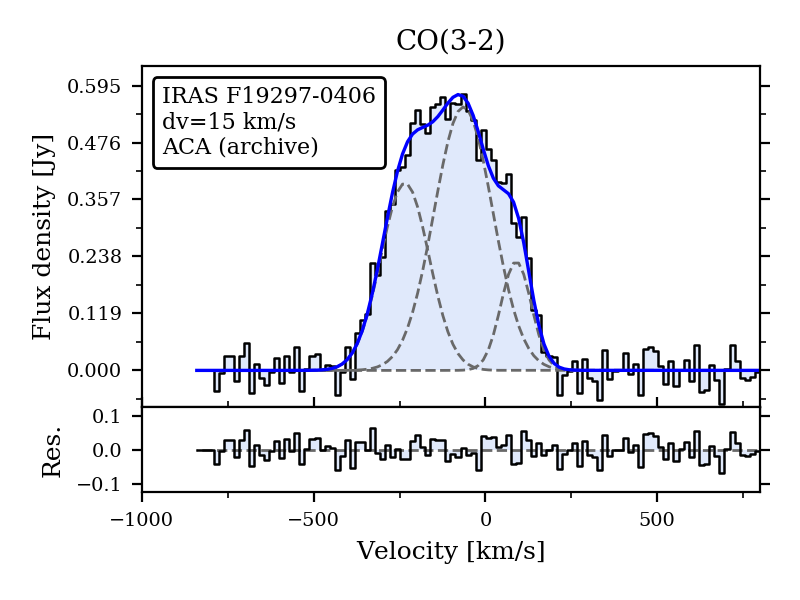}\quad
	\includegraphics[clip=true,trim=0.35cm 0.3cm 0.48cm 0.3cm,width=0.23\textwidth]{figures/empty}\\  
	\includegraphics[clip=true,trim=0.35cm 0.3cm 0.48cm 0.3cm,width=0.23\textwidth]{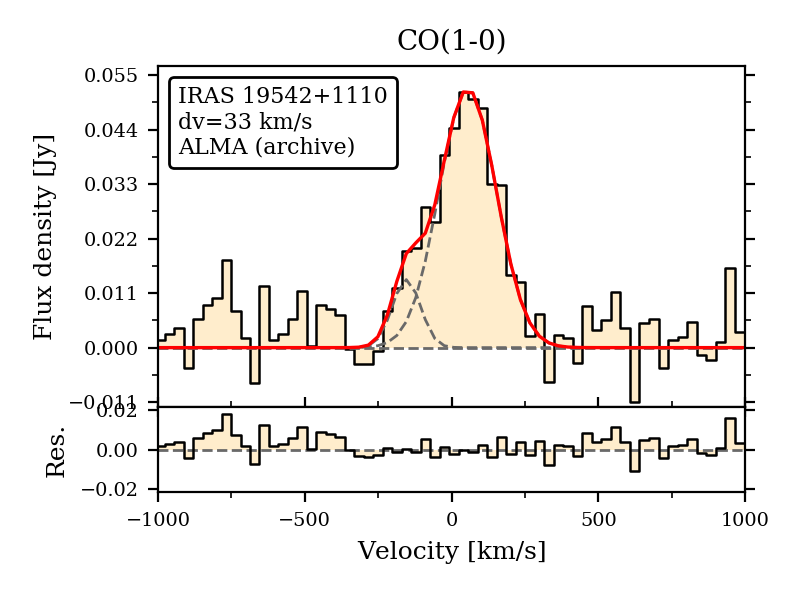}\quad
	\includegraphics[clip=true,trim=0.35cm 0.3cm 0.48cm 0.3cm,width=0.23\textwidth]{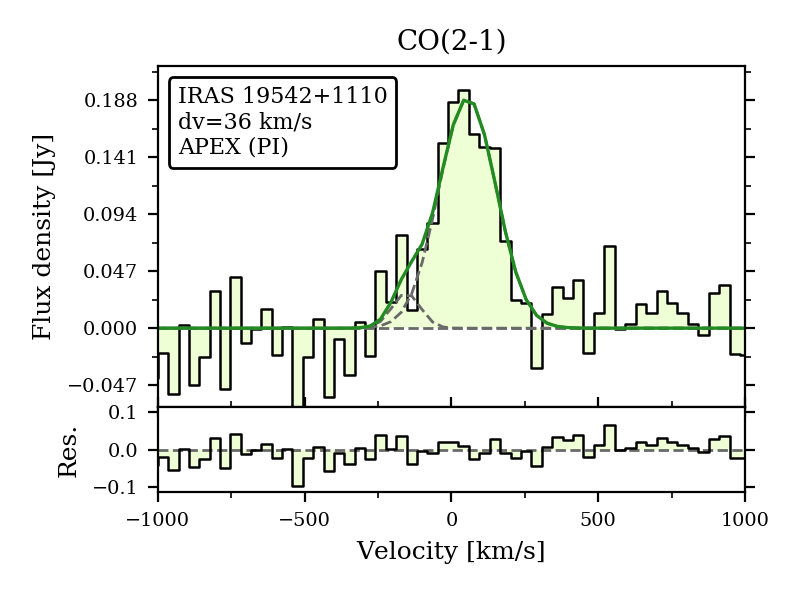}\quad
	\includegraphics[clip=true,trim=0.35cm 0.3cm 0.48cm 0.3cm,width=0.23\textwidth]{figures/empty}\quad
	\includegraphics[clip=true,trim=0.35cm 0.3cm 0.48cm 0.3cm,width=0.23\textwidth]{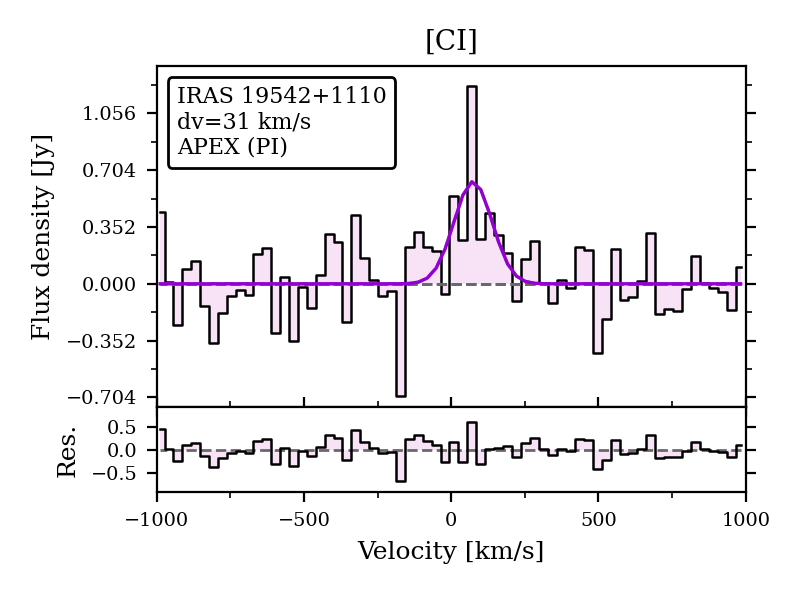}\\  
	\includegraphics[clip=true,trim=0.35cm 0.3cm 0.48cm 0.3cm,width=0.23\textwidth]{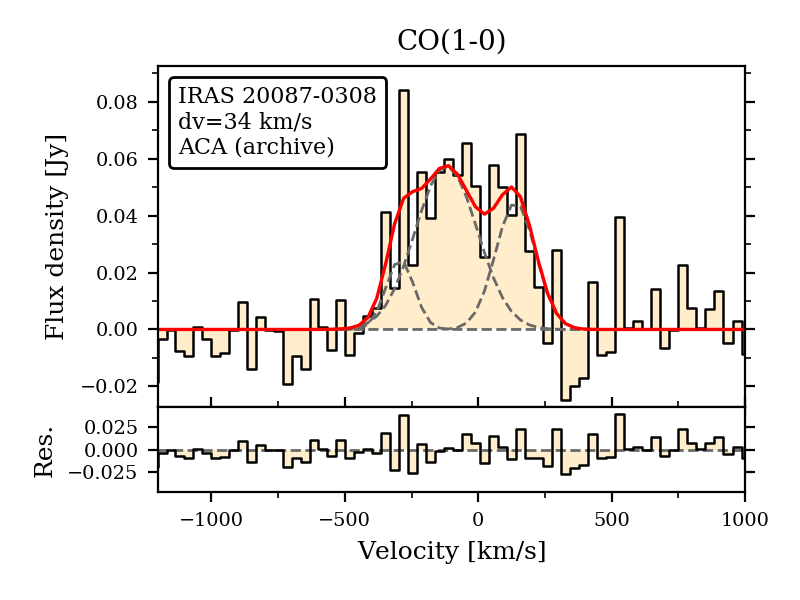}\quad
	\includegraphics[clip=true,trim=0.35cm 0.3cm 0.48cm 0.3cm,width=0.23\textwidth]{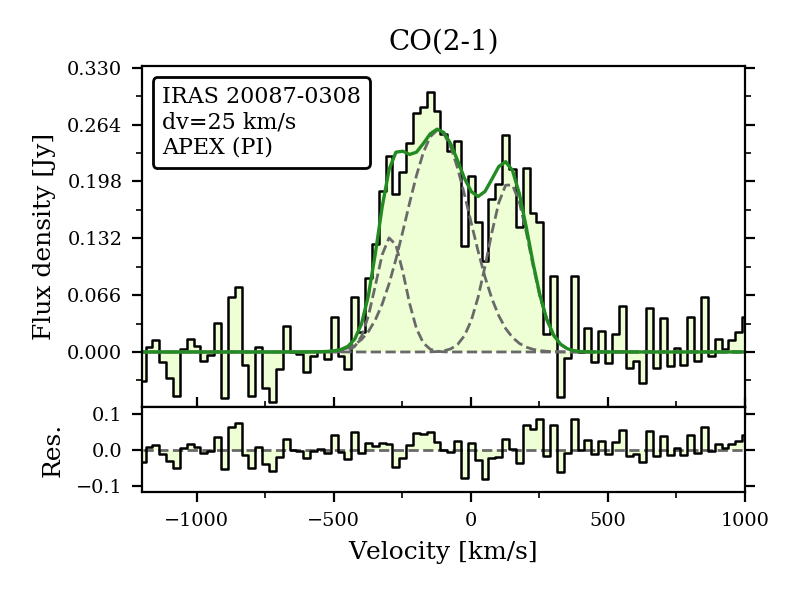}\quad
	\includegraphics[clip=true,trim=0.35cm 0.3cm 0.48cm 0.3cm,width=0.23\textwidth]{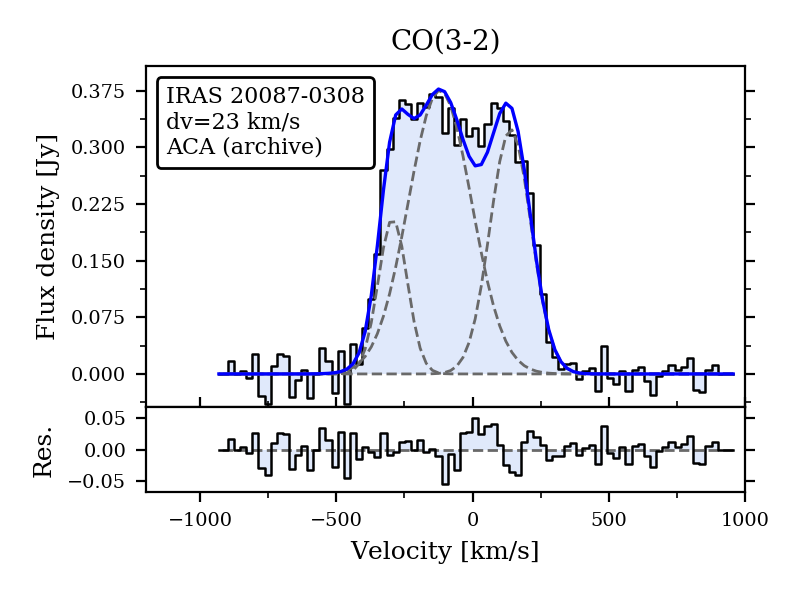}\quad
	\includegraphics[clip=true,trim=0.35cm 0.3cm 0.48cm 0.3cm,width=0.23\textwidth]{figures/empty}\\ 
	\includegraphics[clip=true,trim=0.35cm 0.3cm 0.48cm 0.3cm,width=0.23\textwidth]{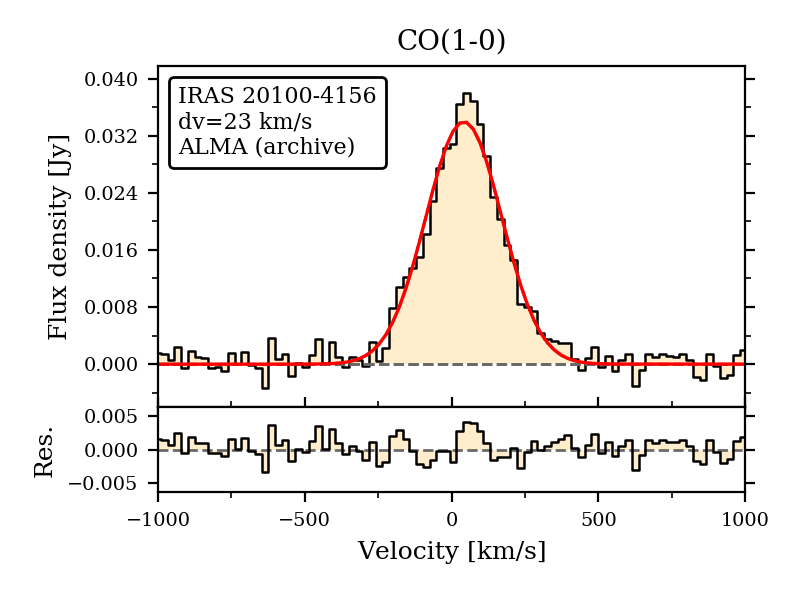}\quad
	\includegraphics[clip=true,trim=0.35cm 0.3cm 0.48cm 0.3cm,width=0.23\textwidth]{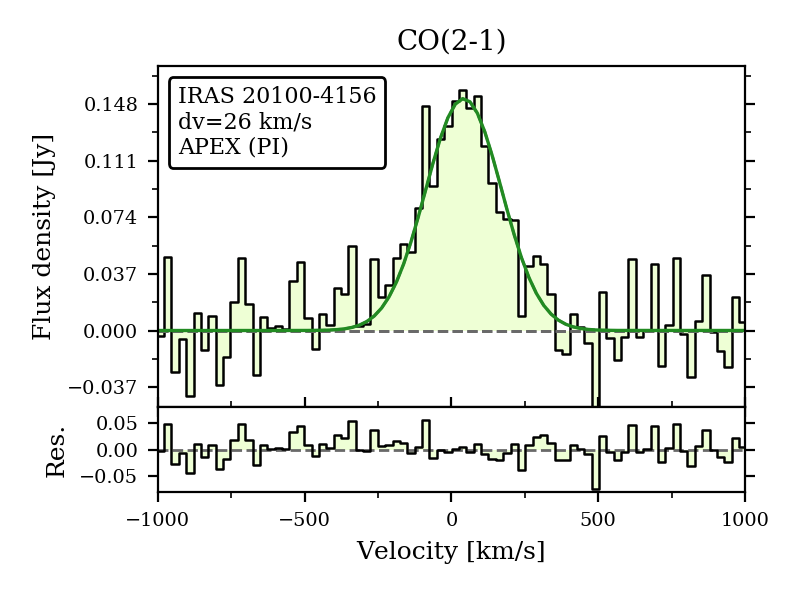}\quad
	\includegraphics[clip=true,trim=0.35cm 0.3cm 0.48cm 0.3cm,width=0.23\textwidth]{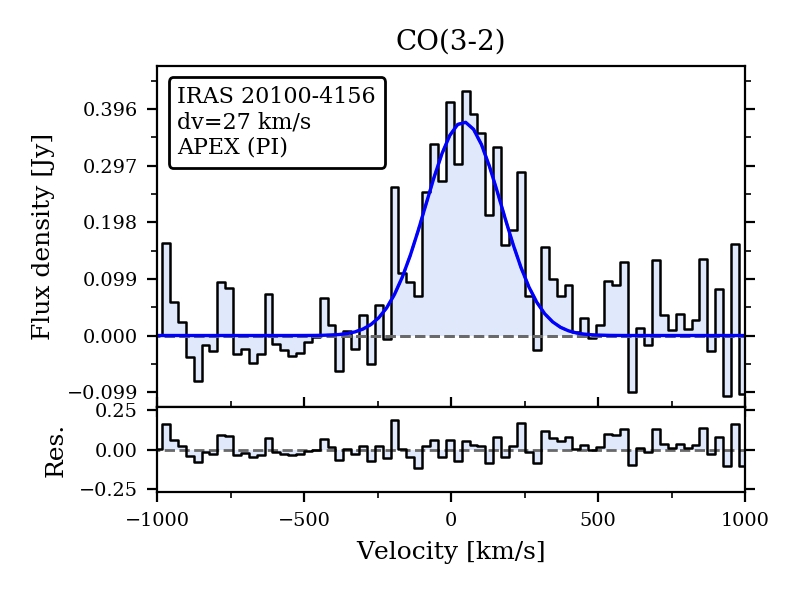}\quad
	\includegraphics[clip=true,trim=0.35cm 0.3cm 0.48cm 0.3cm,width=0.23\textwidth]{figures/empty}\\  
\caption{Continued from Fig.~\ref{fig:spectra3}. This figure shows sources: IRAS 16090-0139, IRAS 17208-0014, IRAS 19254-7245, IRAS F19297-0406, IRAS 19542+1110, IRAS 20087-0308, IRAS 20100-4156.}\label{fig:spectra4} 
\end{figure*}

\begin{figure*}[tbp]
	\includegraphics[clip=true,trim=0.35cm 0.3cm 0.48cm 0.3cm,width=0.23\textwidth]{figures/empty}\quad
	\includegraphics[clip=true,trim=0.35cm 0.3cm 0.48cm 0.3cm,width=0.23\textwidth]{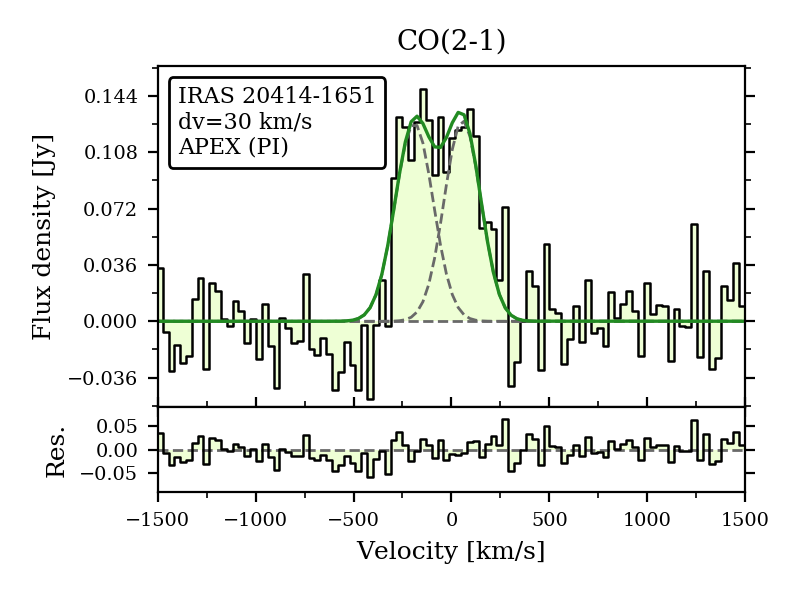}\quad
	\includegraphics[clip=true,trim=0.35cm 0.3cm 0.48cm 0.3cm,width=0.23\textwidth]{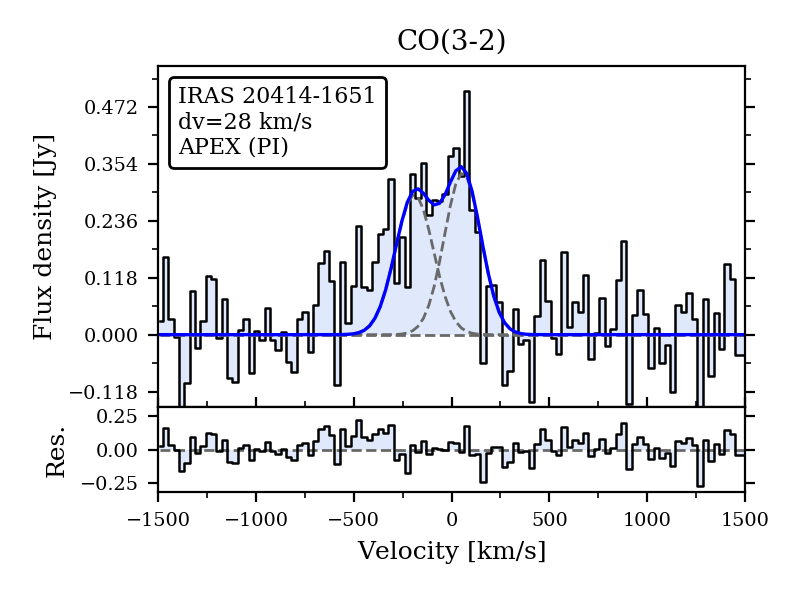}\quad
	\includegraphics[clip=true,trim=0.35cm 0.3cm 0.48cm 0.3cm,width=0.23\textwidth]{figures/empty}\\  
	\includegraphics[clip=true,trim=0.35cm 0.3cm 0.48cm 0.3cm,width=0.23\textwidth]{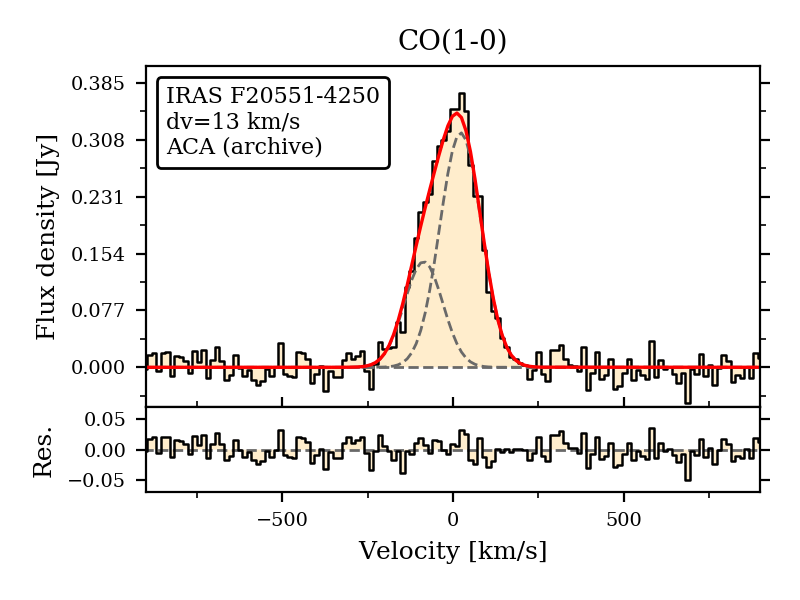}\quad
	\includegraphics[clip=true,trim=0.35cm 0.3cm 0.48cm 0.3cm,width=0.23\textwidth]{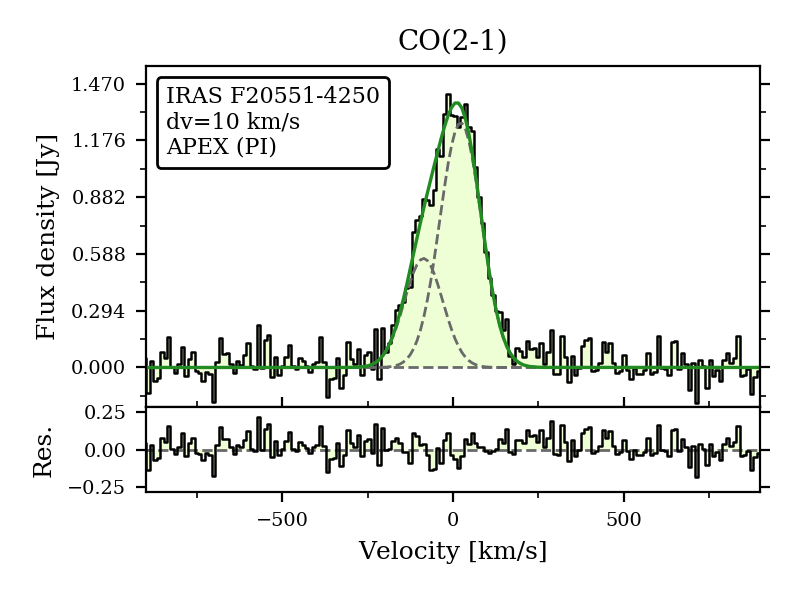}\quad
	\includegraphics[clip=true,trim=0.35cm 0.3cm 0.48cm 0.3cm,width=0.23\textwidth]{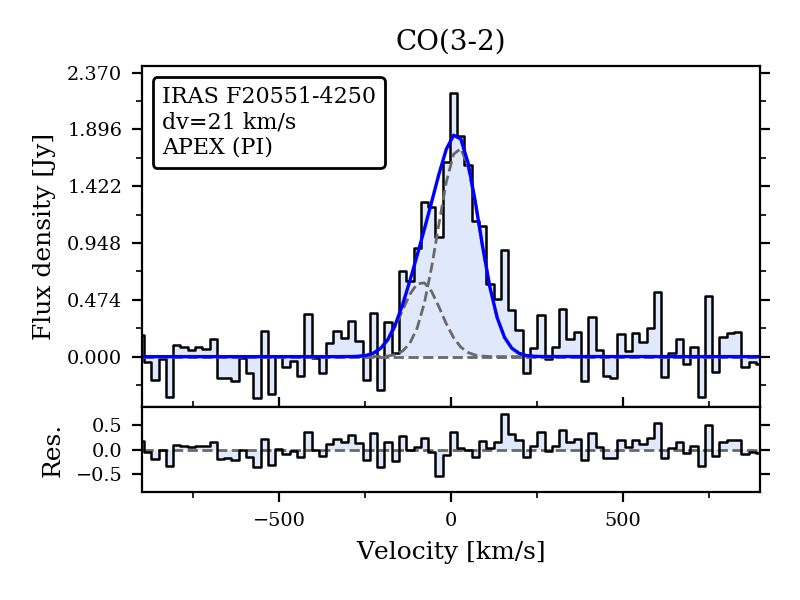}\quad
	\includegraphics[clip=true,trim=0.35cm 0.3cm 0.48cm 0.3cm,width=0.23\textwidth]{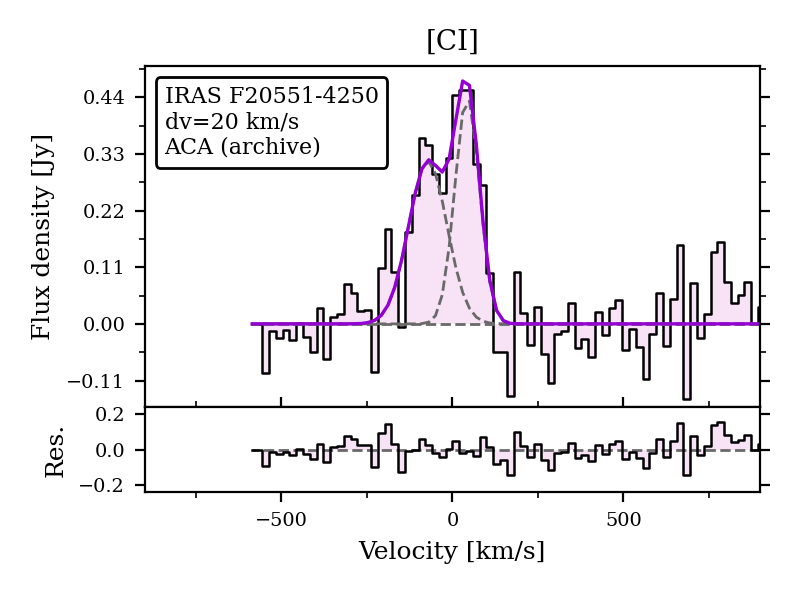}\\  
	\includegraphics[clip=true,trim=0.35cm 0.3cm 0.48cm 0.3cm,width=0.23\textwidth]{figures/empty}\quad
	\includegraphics[clip=true,trim=0.35cm 0.3cm 0.48cm 0.3cm,width=0.23\textwidth]{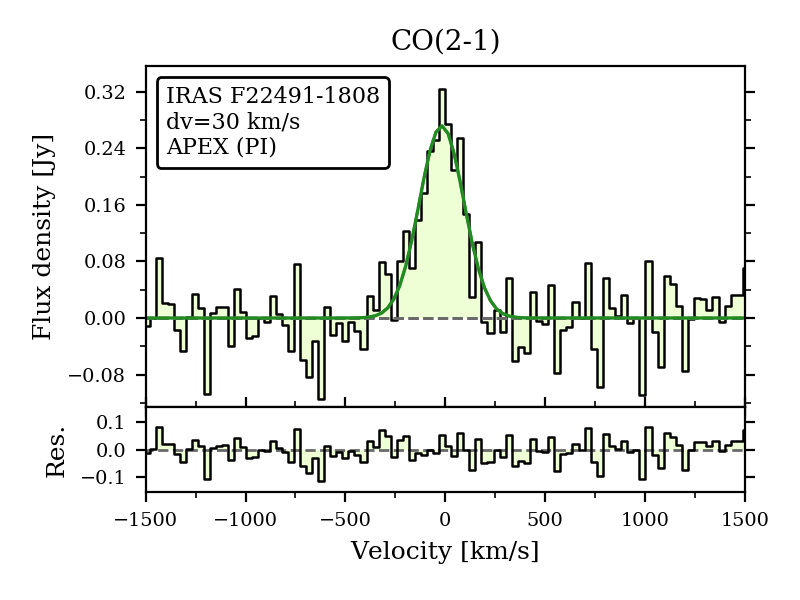}\quad
	\includegraphics[clip=true,trim=0.35cm 0.3cm 0.48cm 0.3cm,width=0.23\textwidth]{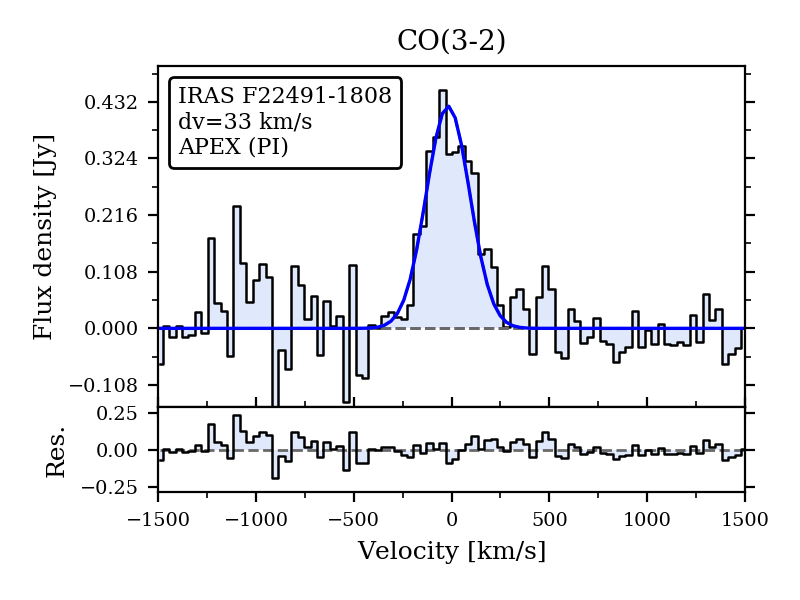}\quad
	\includegraphics[clip=true,trim=0.35cm 0.3cm 0.48cm 0.3cm,width=0.23\textwidth]{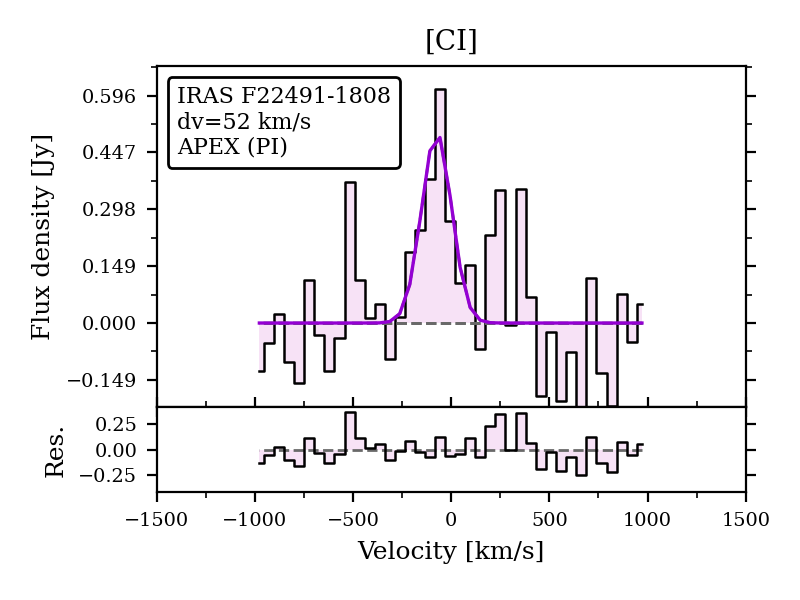}\\  
	\includegraphics[clip=true,trim=0.35cm 0.3cm 0.48cm 0.3cm,width=0.23\textwidth]{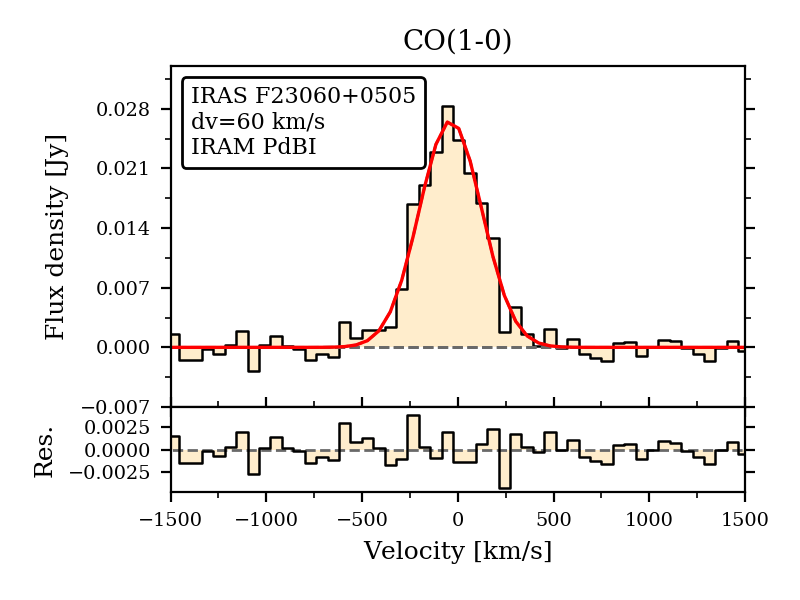}\quad
	\includegraphics[clip=true,trim=0.35cm 0.3cm 0.48cm 0.3cm,width=0.23\textwidth]{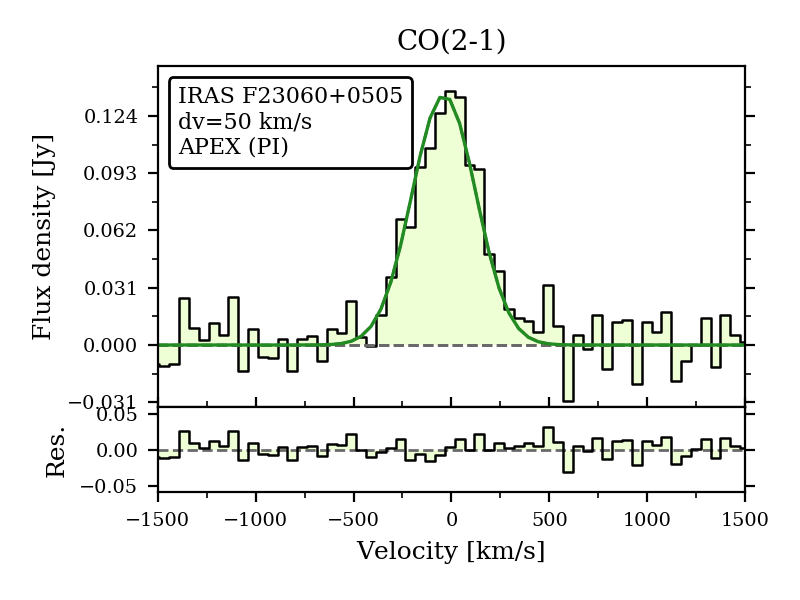}\quad
	\includegraphics[clip=true,trim=0.35cm 0.3cm 0.48cm 0.3cm,width=0.23\textwidth]{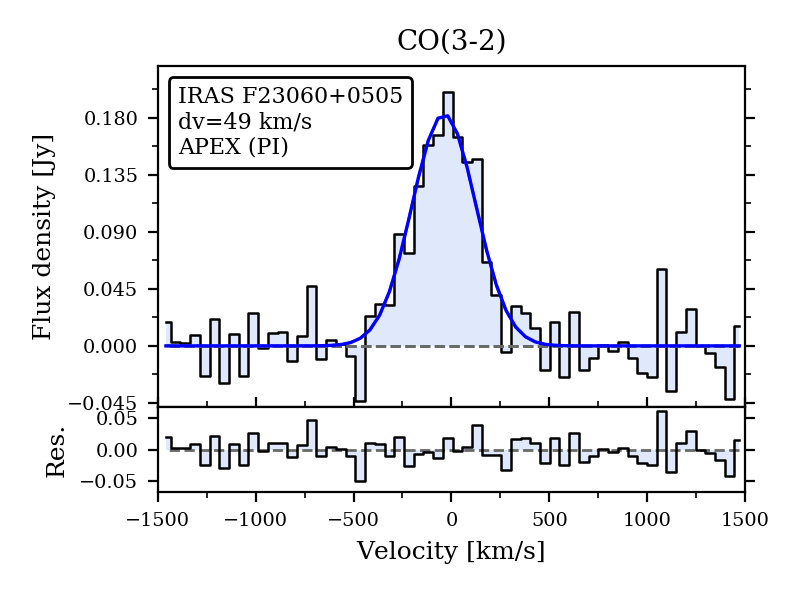}\quad
	\includegraphics[clip=true,trim=0.35cm 0.3cm 0.48cm 0.3cm,width=0.23\textwidth]{figures/empty}\\  
	\includegraphics[clip=true,trim=0.35cm 0.3cm 0.48cm 0.3cm,width=0.23\textwidth]{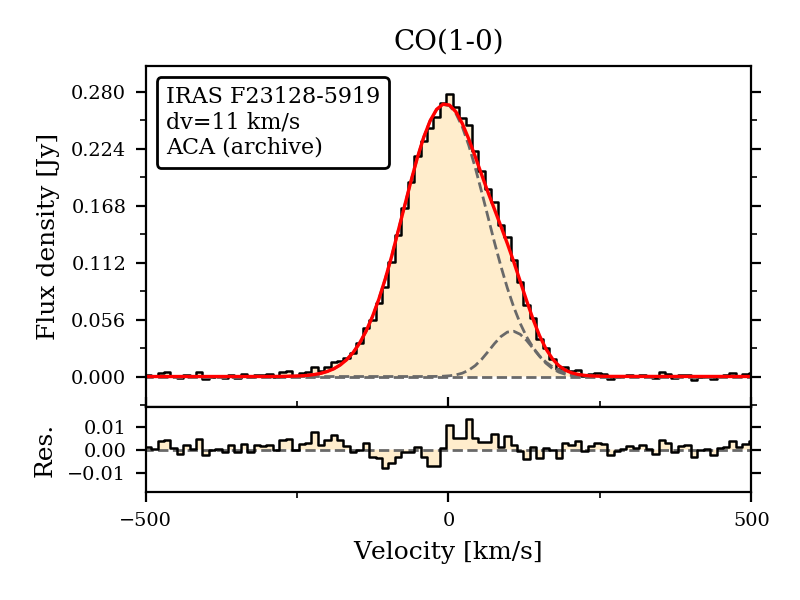}\quad
	\includegraphics[clip=true,trim=0.35cm 0.3cm 0.48cm 0.3cm,width=0.23\textwidth]{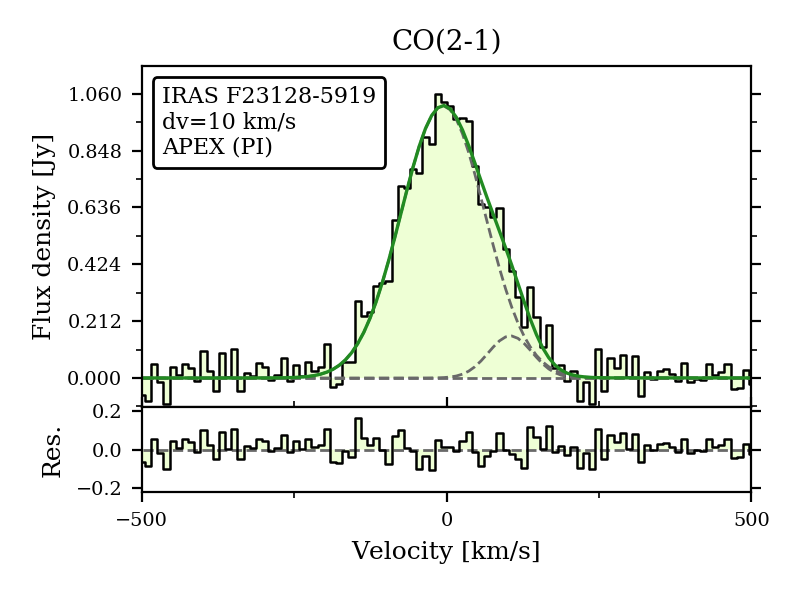}\quad
	\includegraphics[clip=true,trim=0.35cm 0.3cm 0.48cm 0.3cm,width=0.23\textwidth]{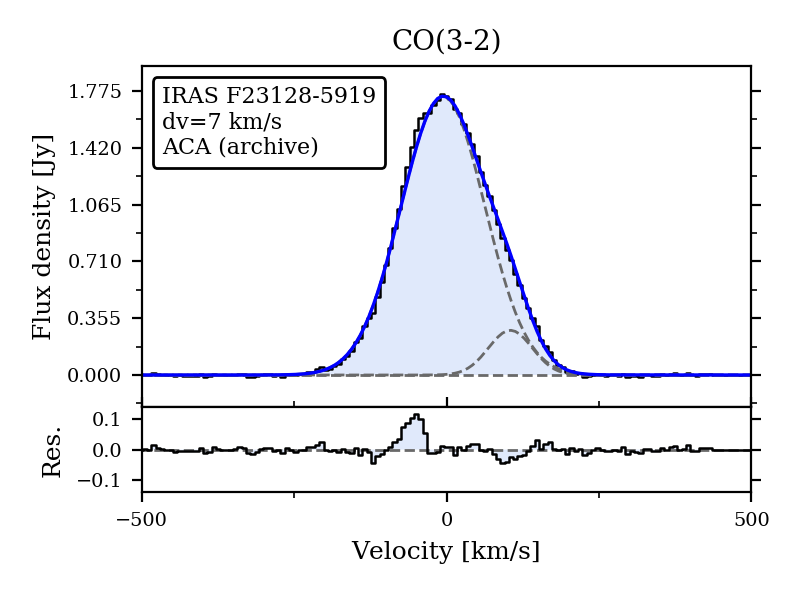}\quad
	\includegraphics[clip=true,trim=0.35cm 0.3cm 0.48cm 0.3cm,width=0.23\textwidth]{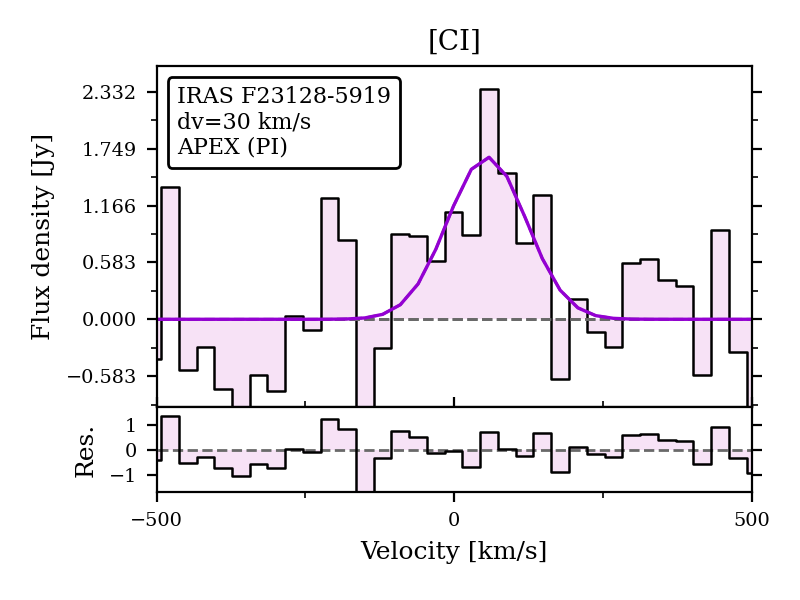}\\  
	\includegraphics[clip=true,trim=0.35cm 0.3cm 0.48cm 0.3cm,width=0.23\textwidth]{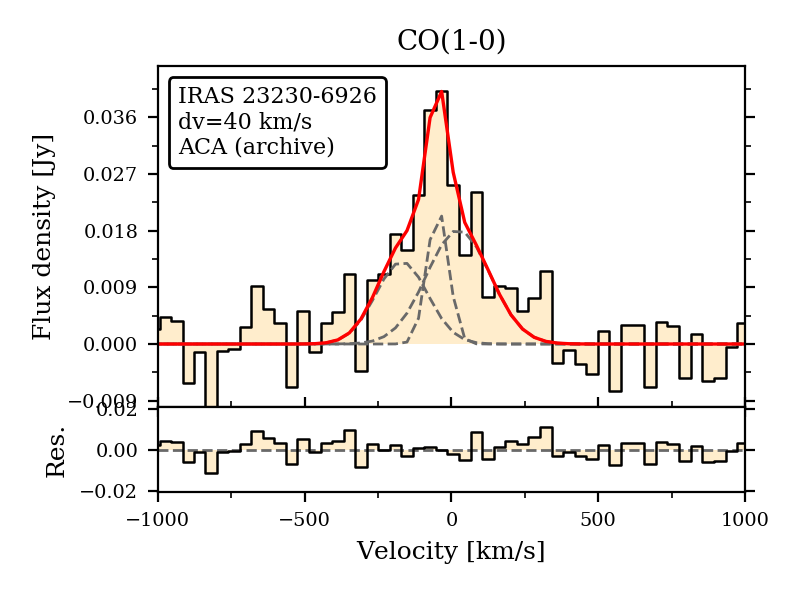}\quad
	\includegraphics[clip=true,trim=0.35cm 0.3cm 0.48cm 0.3cm,width=0.23\textwidth]{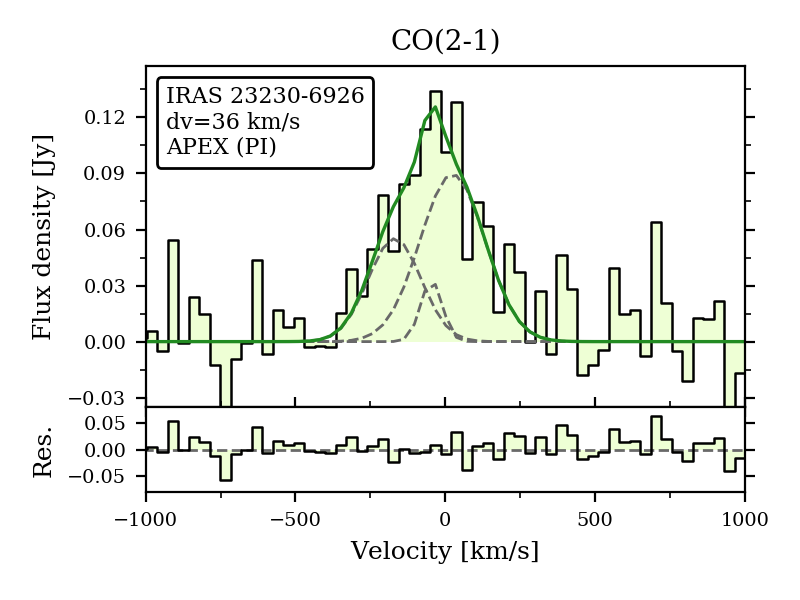}\quad
	\includegraphics[clip=true,trim=0.35cm 0.3cm 0.48cm 0.3cm,width=0.23\textwidth]{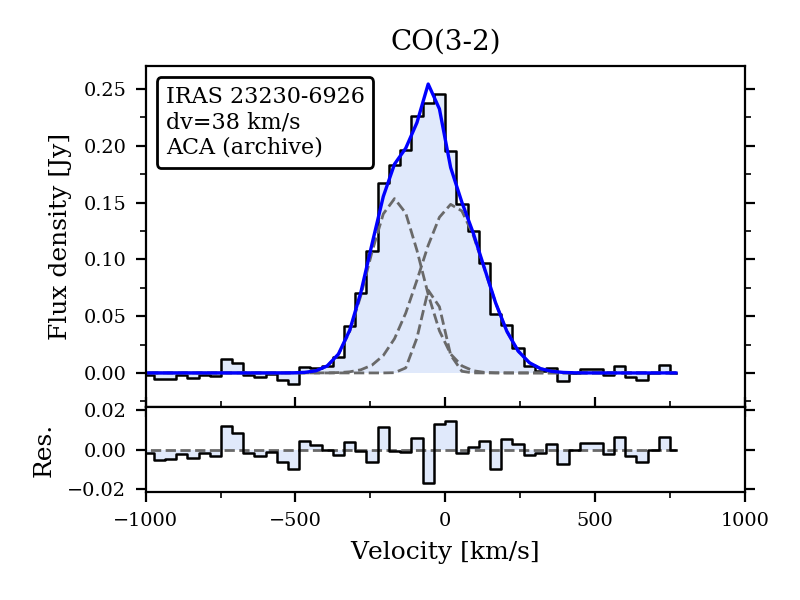}\quad
	\includegraphics[clip=true,trim=0.35cm 0.3cm 0.48cm 0.3cm,width=0.23\textwidth]{figures/empty}\\  
	\includegraphics[clip=true,trim=0.35cm 0.3cm 0.48cm 0.3cm,width=0.23\textwidth]{figures/empty}\quad
	\includegraphics[clip=true,trim=0.35cm 0.3cm 0.48cm 0.3cm,width=0.23\textwidth]{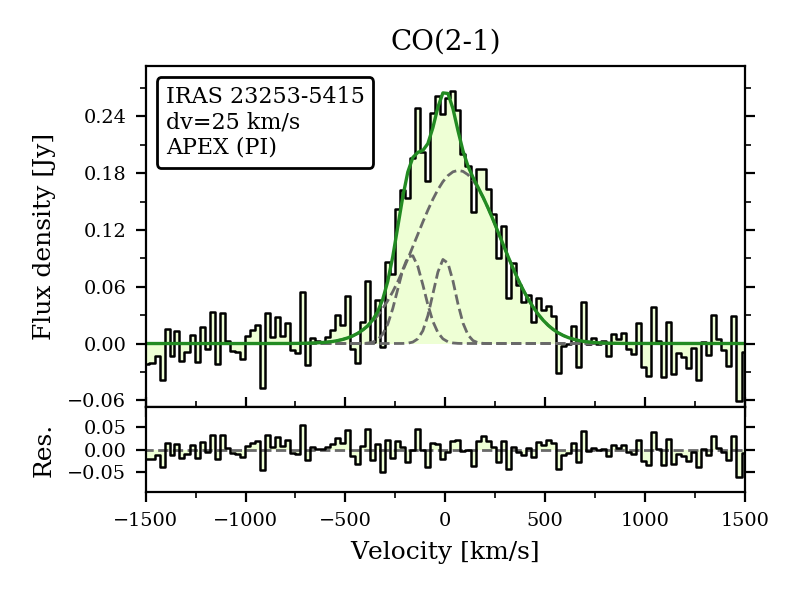}\quad
	\includegraphics[clip=true,trim=0.35cm 0.3cm 0.48cm 0.3cm,width=0.23\textwidth]{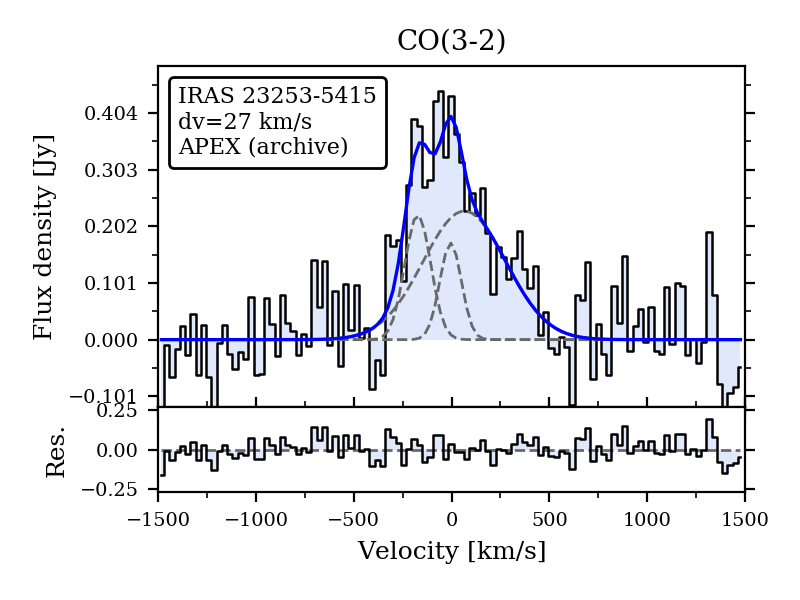}\quad
	\includegraphics[clip=true,trim=0.35cm 0.3cm 0.48cm 0.3cm,width=0.23\textwidth]{figures/empty}\\  
\caption{Continued from Fig.~\ref{fig:spectra4}. This figure shows sources: IRAS 20414-1651, IRAS F20551-4250, IRAS F22491-1808, IRAS F23060+0505, IRAS F23128-5919, IRAS 23230-6926, IRAS 23253-5415.}\label{fig:spectra5} 
\end{figure*}

\begin{figure*}[tbp]
	\includegraphics[clip=true,trim=0.35cm 0.3cm 0.48cm 0.3cm,width=0.23\textwidth]{figures/empty}\quad
	\includegraphics[clip=true,trim=0.35cm 0.3cm 0.48cm 0.3cm,width=0.23\textwidth]{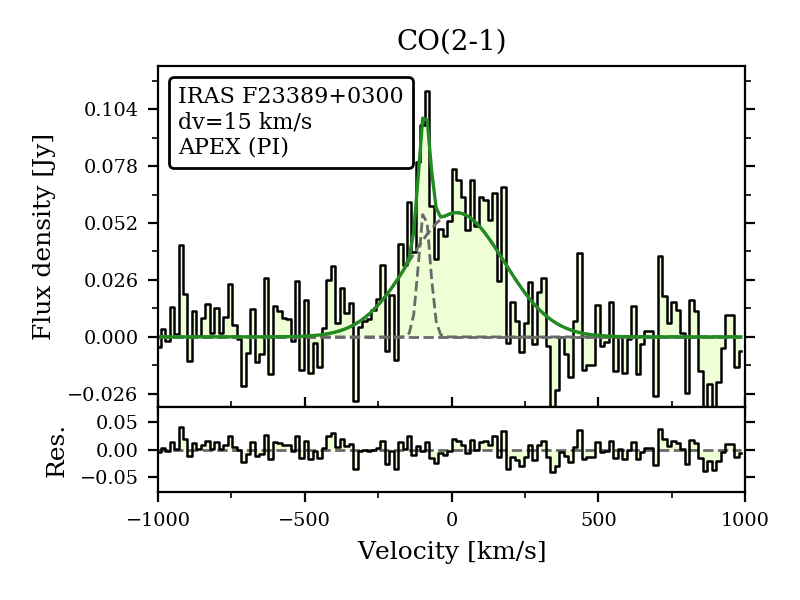}\quad
	\includegraphics[clip=true,trim=0.35cm 0.3cm 0.48cm 0.3cm,width=0.23\textwidth]{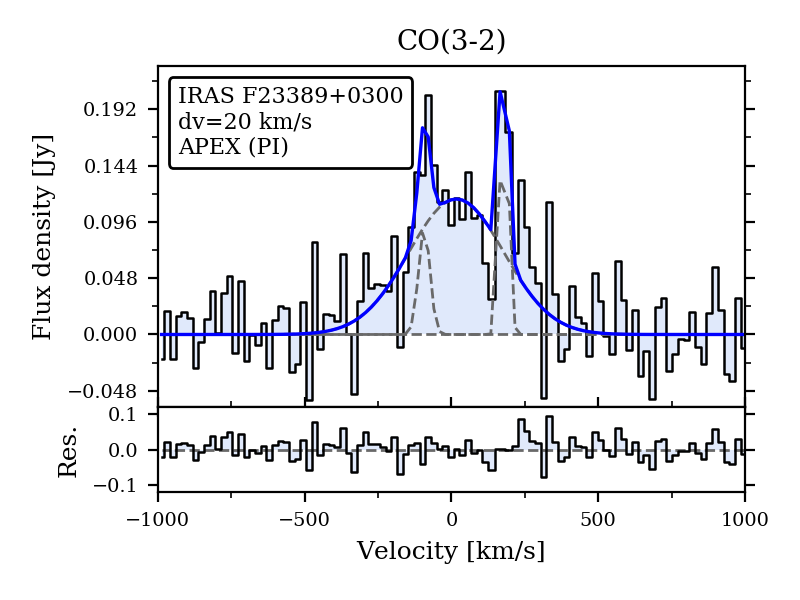}\quad
	\includegraphics[clip=true,trim=0.35cm 0.3cm 0.48cm 0.3cm,width=0.23\textwidth]{figures/empty}\\  
	\includegraphics[clip=true,trim=0.35cm 0.3cm 0.48cm 0.3cm,width=0.23\textwidth]{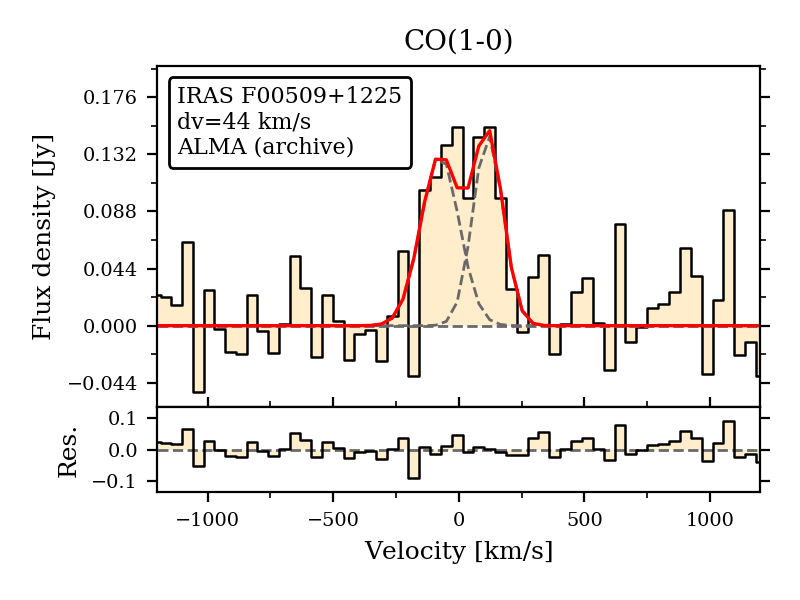}\quad
	\includegraphics[clip=true,trim=0.35cm 0.3cm 0.48cm 0.3cm,width=0.23\textwidth]{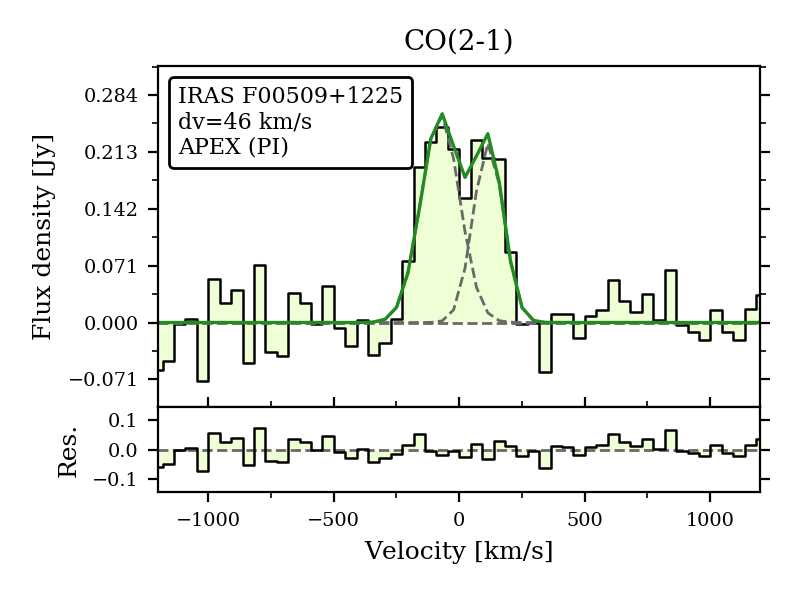}\quad
	\includegraphics[clip=true,trim=0.35cm 0.3cm 0.48cm 0.3cm,width=0.23\textwidth]{figures/empty}\quad
	\includegraphics[clip=true,trim=0.35cm 0.3cm 0.48cm 0.3cm,width=0.23\textwidth]{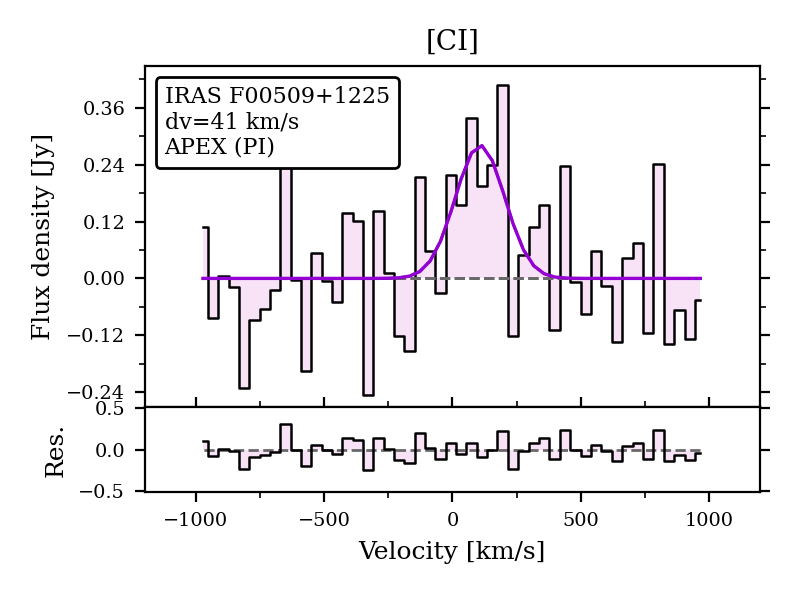}\\
	\includegraphics[clip=true,trim=0.35cm 0.3cm 0.48cm 0.3cm,width=0.23\textwidth]{figures/empty}\quad
	\includegraphics[clip=true,trim=0.35cm 0.3cm 0.48cm 0.3cm,width=0.23\textwidth]{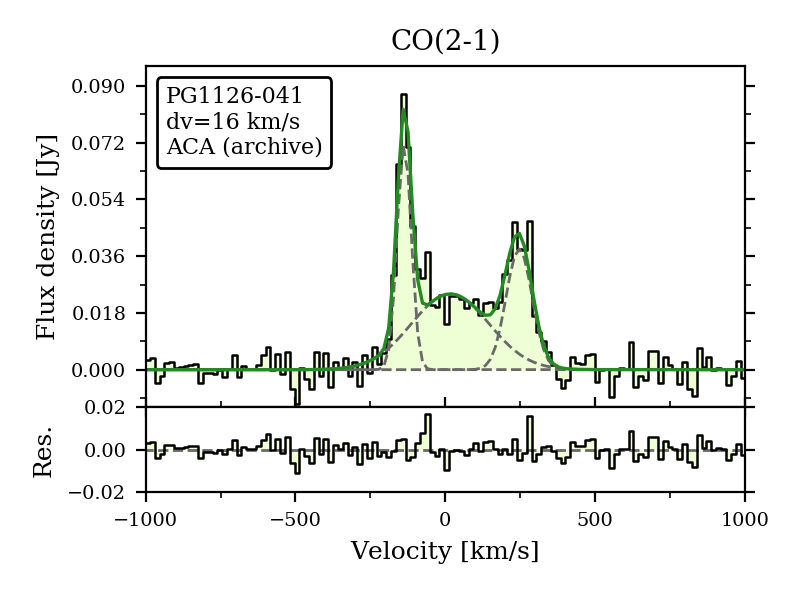}\quad
	\includegraphics[clip=true,trim=0.35cm 0.3cm 0.48cm 0.3cm,width=0.23\textwidth]{figures/empty}\quad
	\includegraphics[clip=true,trim=0.35cm 0.3cm 0.48cm 0.3cm,width=0.23\textwidth]{figures/empty}\\  
	\includegraphics[clip=true,trim=0.35cm 0.3cm 0.48cm 0.3cm,width=0.23\textwidth]{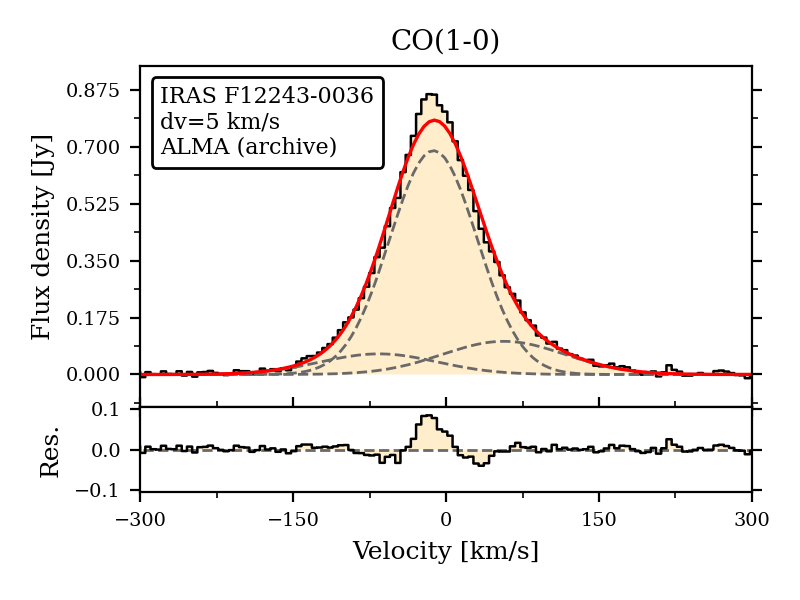}\quad
	\includegraphics[clip=true,trim=0.35cm 0.3cm 0.48cm 0.3cm,width=0.23\textwidth]{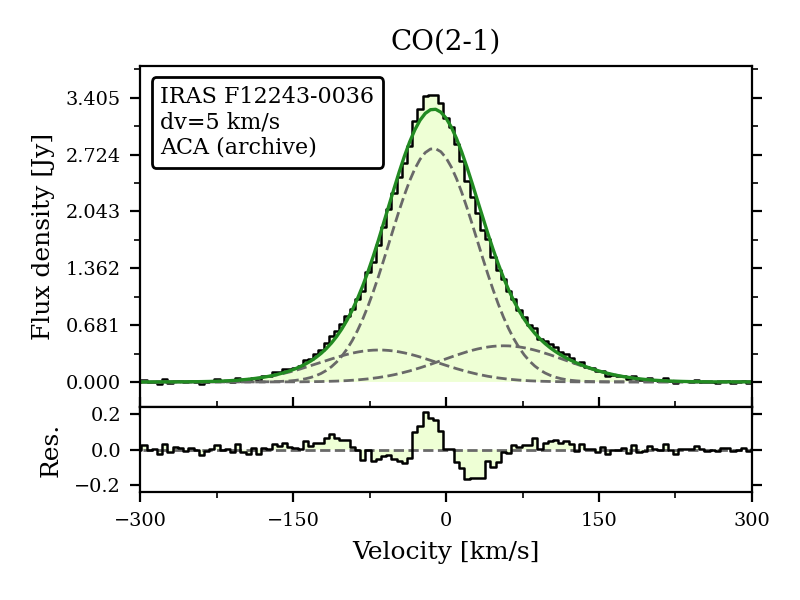}\quad
	\includegraphics[clip=true,trim=0.35cm 0.3cm 0.48cm 0.3cm,width=0.23\textwidth]{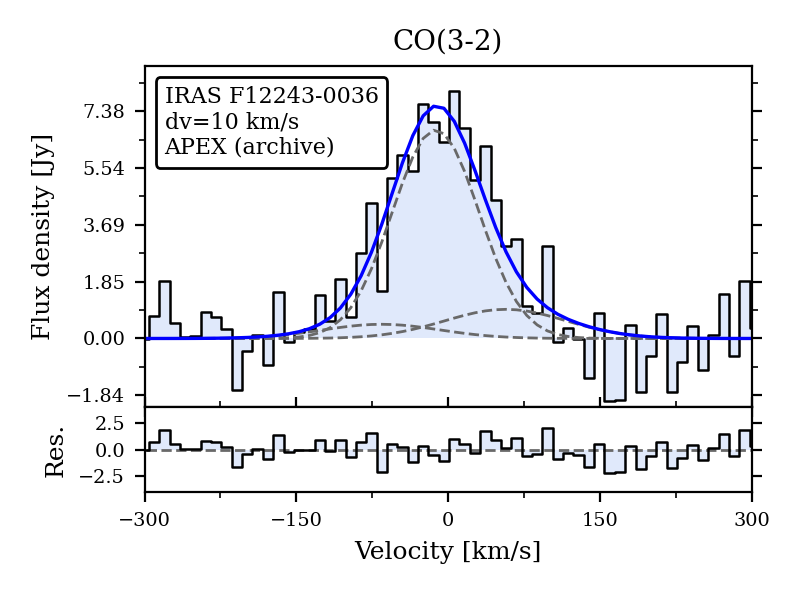}\quad
	\includegraphics[clip=true,trim=0.35cm 0.3cm 0.48cm 0.3cm,width=0.23\textwidth]{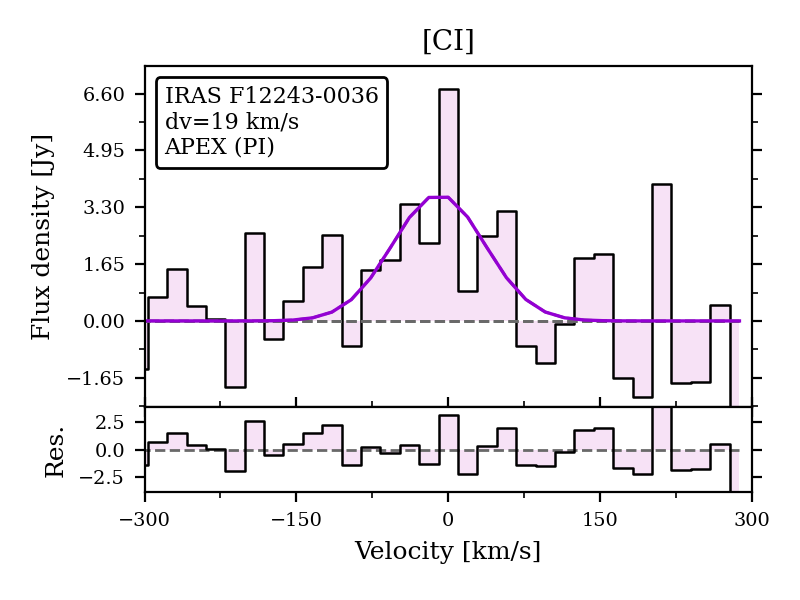}\\
	\includegraphics[clip=true,trim=0.35cm 0.3cm 0.48cm 0.3cm,width=0.23\textwidth]{figures/empty}\quad
	\includegraphics[clip=true,trim=0.35cm 0.3cm 0.48cm 0.3cm,width=0.23\textwidth]{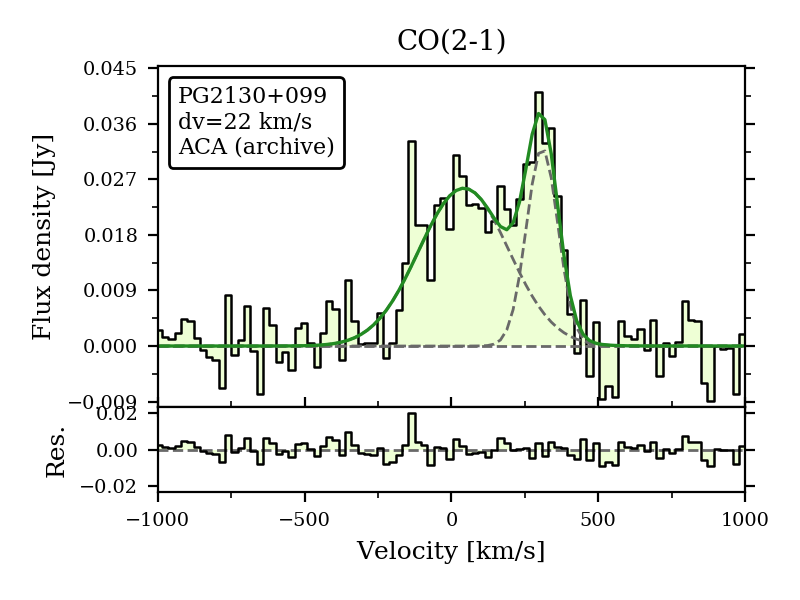}\quad
	\includegraphics[clip=true,trim=0.35cm 0.3cm 0.48cm 0.3cm,width=0.23\textwidth]{figures/empty}\quad
	\includegraphics[clip=true,trim=0.35cm 0.3cm 0.48cm 0.3cm,width=0.23\textwidth]{figures/empty}\\  
\caption{Continued from Fig.~\ref{fig:spectra5}. This figure shows sources: IRAS F23389+0300, IRAS F00509+1225, PG1126-041, IRAS F12243-0036 and PG2130+099.}\label{fig:spectra6} 
\end{figure*}

\section{Duplicated data sets not employed in main analysis}\label{sec:appendix_duplication_spectra}

\begin{figure*}[tbp]
	\includegraphics[clip=true,trim=0.35cm 0.3cm 0.48cm 0.3cm,width=0.23\textwidth]{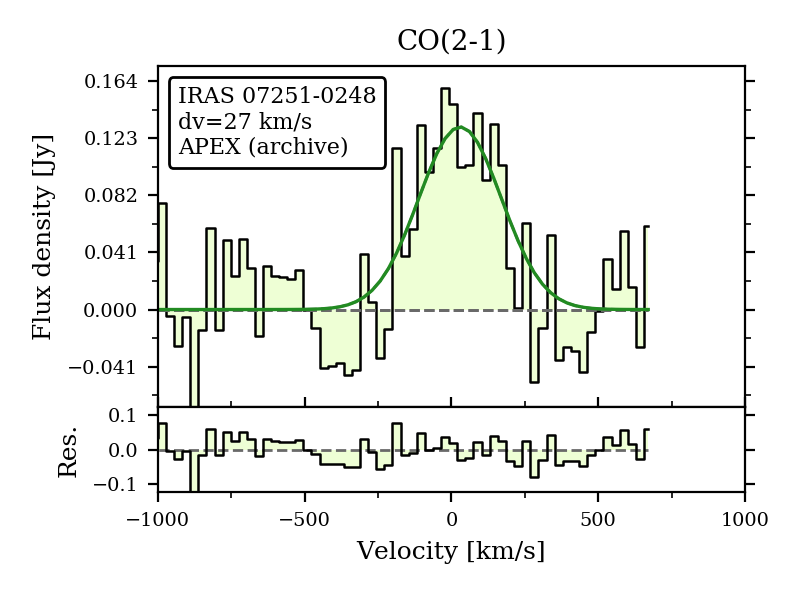}\quad
	\includegraphics[clip=true,trim=0.35cm 0.3cm 0.48cm 0.3cm,width=0.23\textwidth]{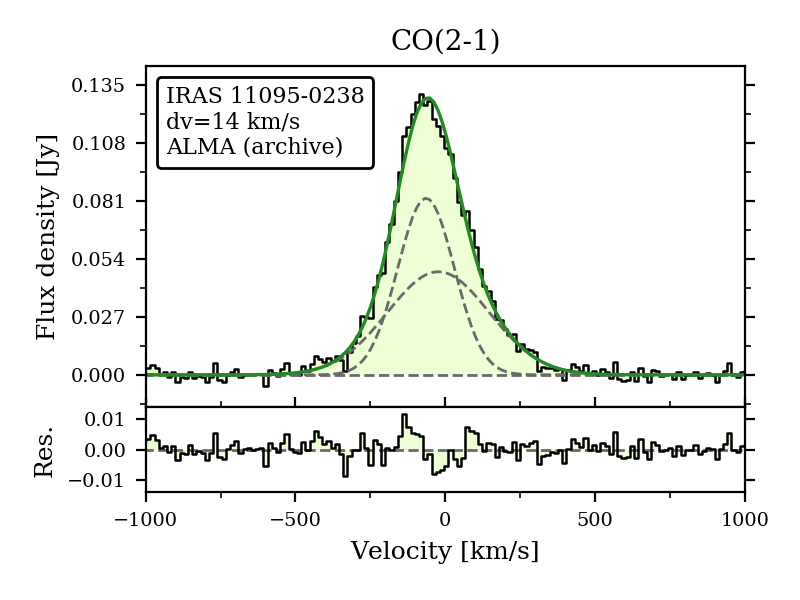}\quad
	\includegraphics[clip=true,trim=0.35cm 0.3cm 0.48cm 0.3cm,width=0.23\textwidth]{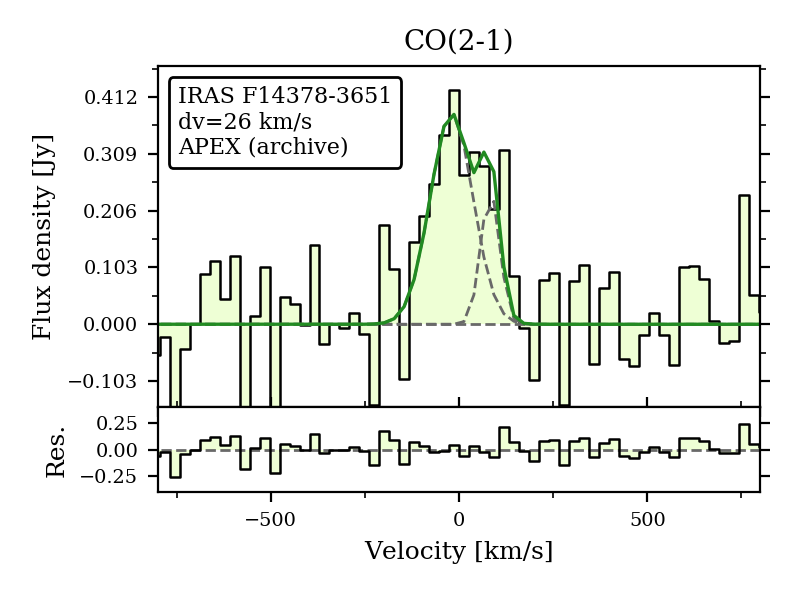}\quad
    \includegraphics[clip=true,trim=0.35cm 0.3cm 0.48cm 0.3cm,width=0.23\textwidth]{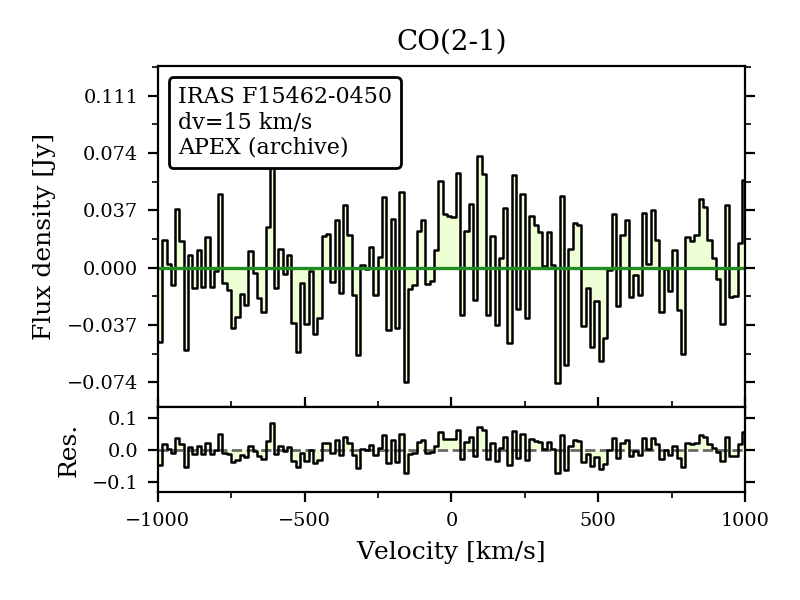}\\
    \includegraphics[clip=true,trim=0.35cm 0.3cm 0.48cm 0.3cm,width=0.23\textwidth]{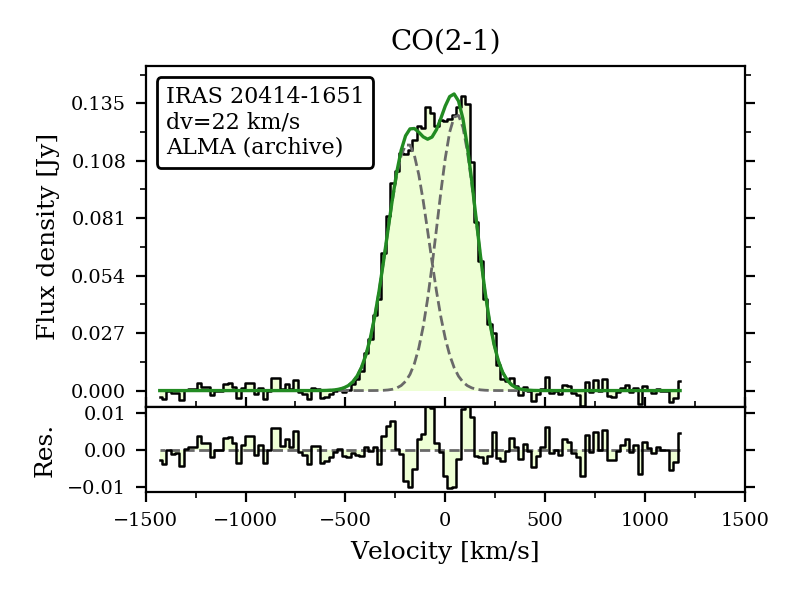}\quad
	\includegraphics[clip=true,trim=0.35cm 0.3cm 0.48cm 0.3cm,width=0.23\textwidth]{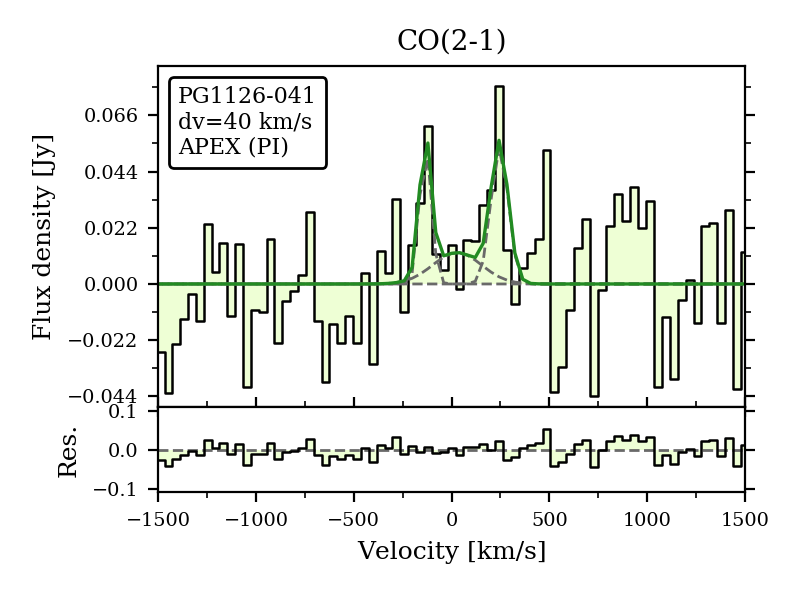}\quad
	\includegraphics[clip=true,trim=0.35cm 0.3cm 0.48cm 0.3cm,width=0.23\textwidth]{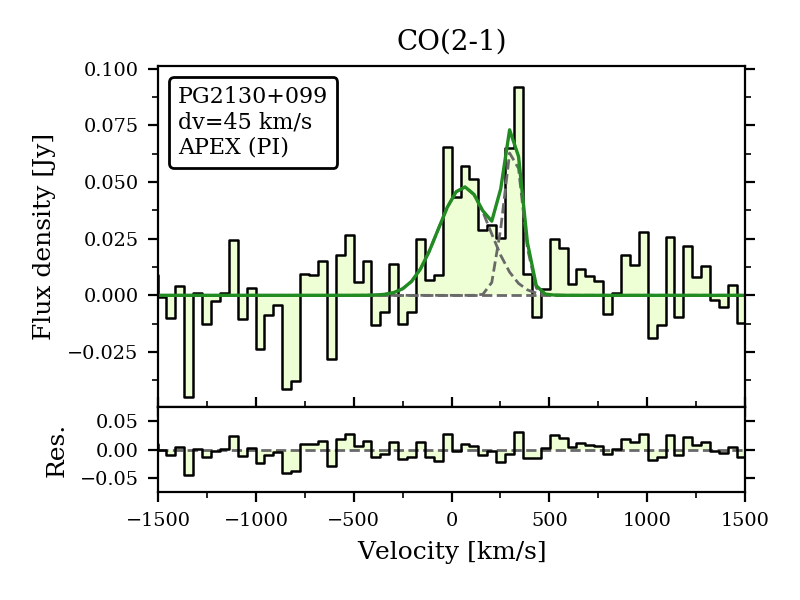}\quad
	\includegraphics[clip=true,trim=0.35cm 0.3cm 0.48cm 0.3cm,width=0.23\textwidth]{figures/empty}\\
	\caption{Duplicated CO(2--1) spectra not used in the analysis.}\label{fig:spectra_dupli_co21}
\end{figure*}

\begin{figure*}[tbp]
	\includegraphics[clip=true,trim=0.35cm 0.3cm 0.48cm 0.3cm,width=0.23\textwidth]{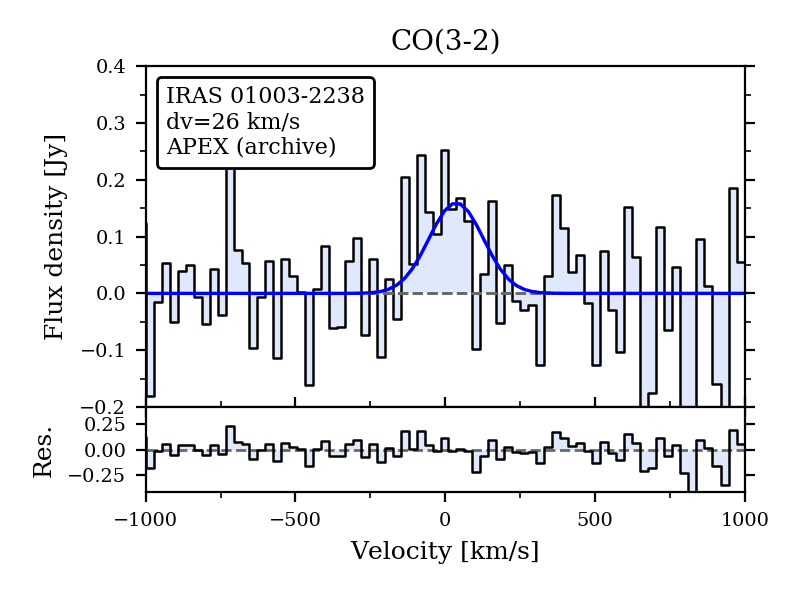}\quad
	\includegraphics[clip=true,trim=0.35cm 0.3cm 0.48cm 0.3cm,width=0.23\textwidth]{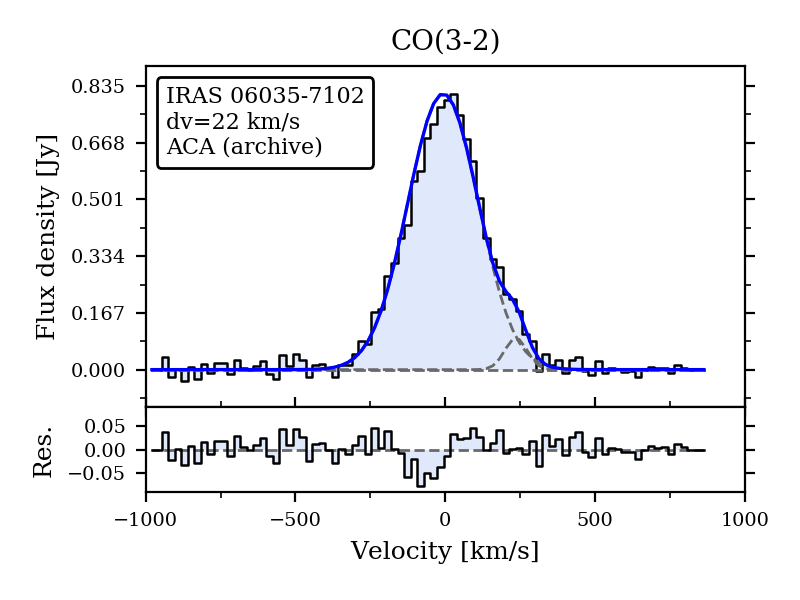}\quad
	\includegraphics[clip=true,trim=0.35cm 0.3cm 0.48cm 0.3cm,width=0.23\textwidth]{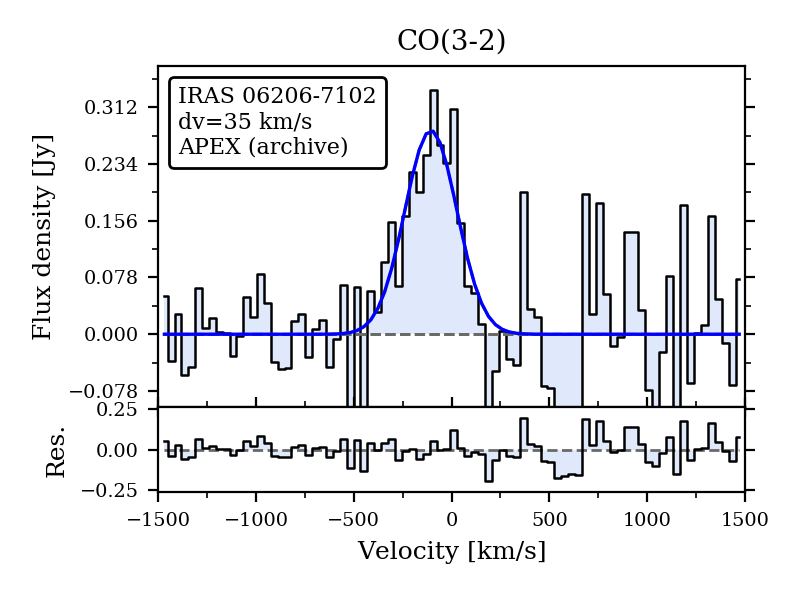}\quad
	\includegraphics[clip=true,trim=0.35cm 0.3cm 0.48cm 0.3cm,width=0.23\textwidth]{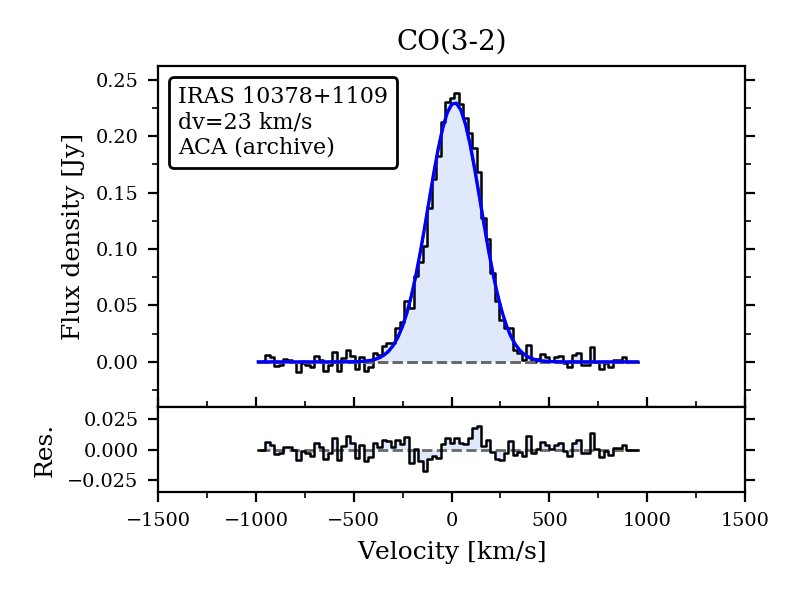}\\  
	\includegraphics[clip=true,trim=0.35cm 0.3cm 0.48cm 0.3cm,width=0.23\textwidth]{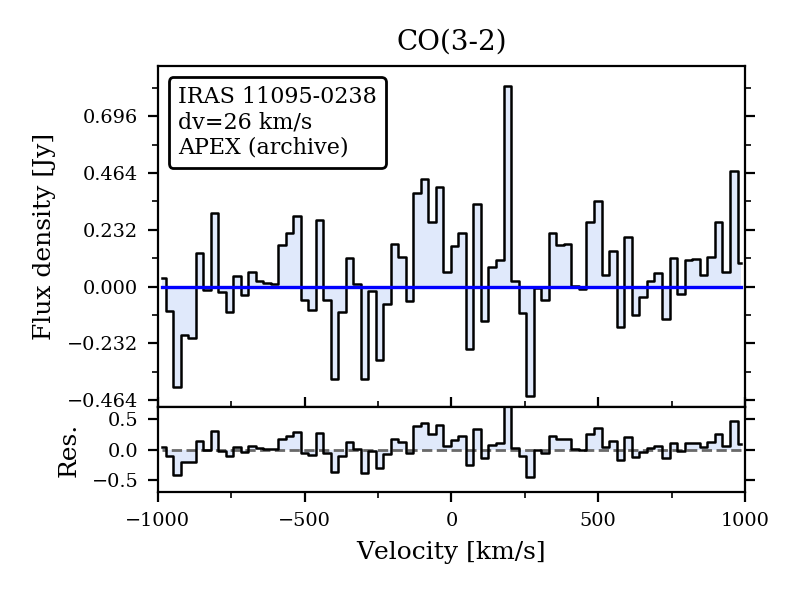}\quad
	\includegraphics[clip=true,trim=0.35cm 0.3cm 0.48cm 0.3cm,width=0.23\textwidth]{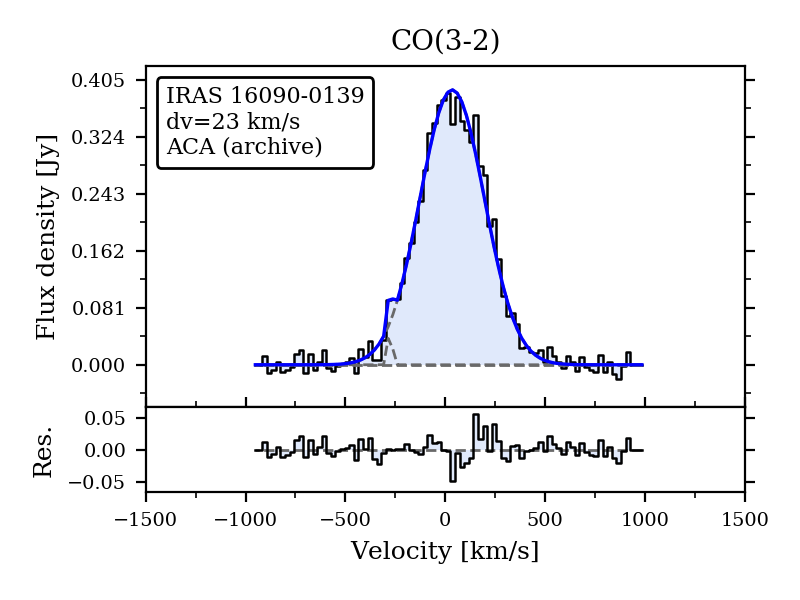}\quad
	\includegraphics[clip=true,trim=0.35cm 0.3cm 0.48cm 0.3cm,width=0.23\textwidth]{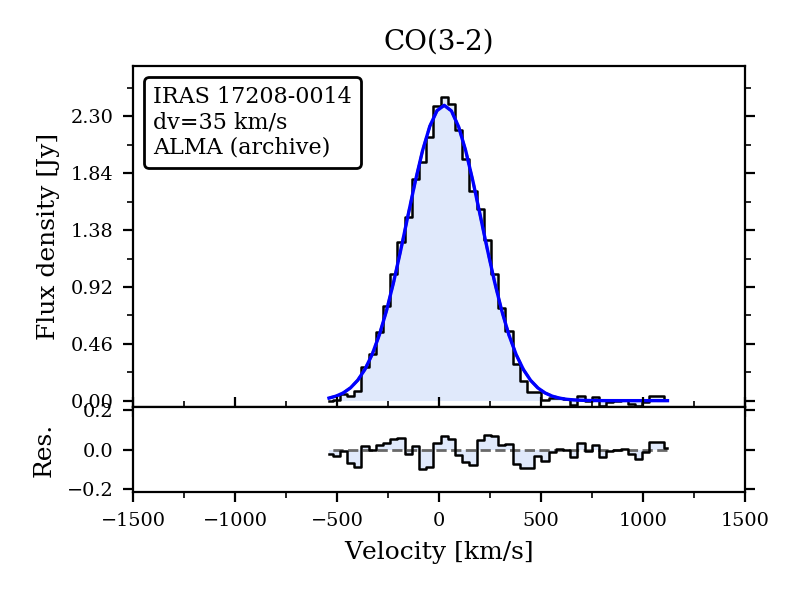}\quad
	\includegraphics[clip=true,trim=0.35cm 0.3cm 0.48cm 0.3cm,width=0.23\textwidth]{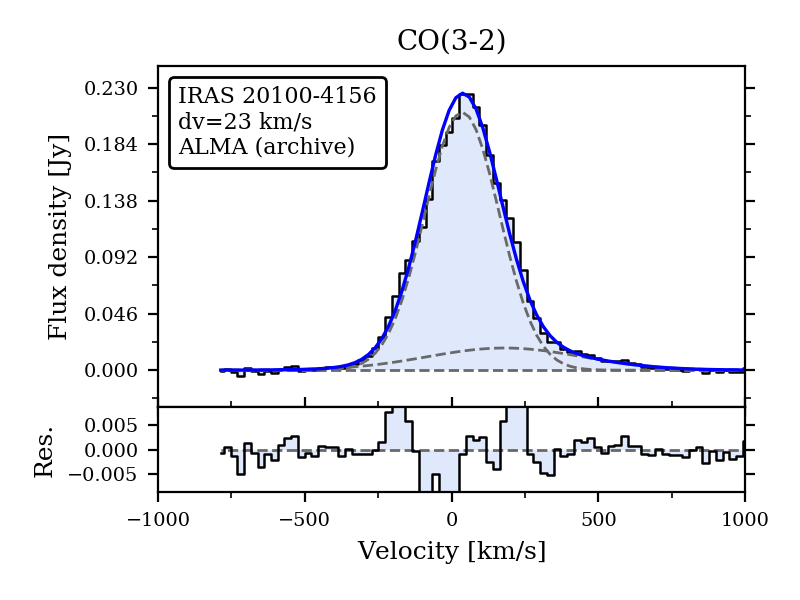}\\
	\includegraphics[clip=true,trim=0.35cm 0.3cm 0.48cm 0.3cm,width=0.23\textwidth]{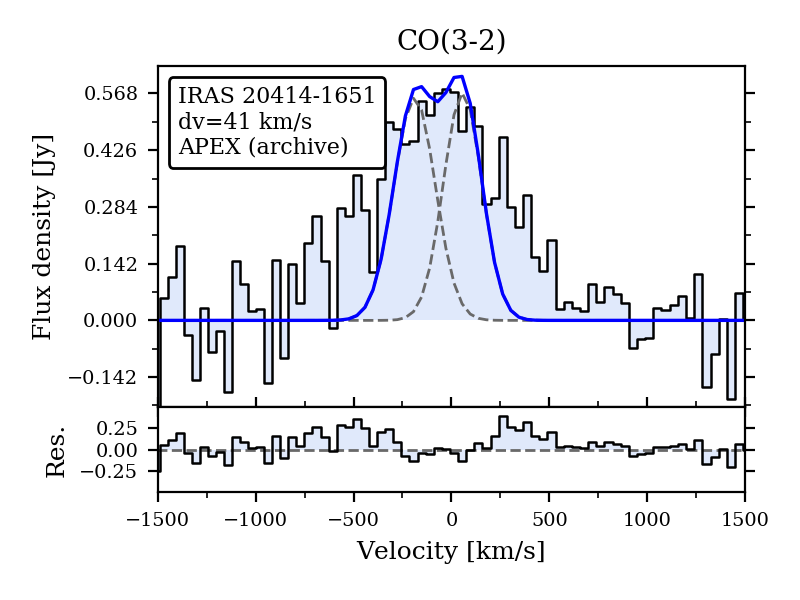}\quad
	\includegraphics[clip=true,trim=0.35cm 0.3cm 0.48cm 0.3cm,width=0.23\textwidth]{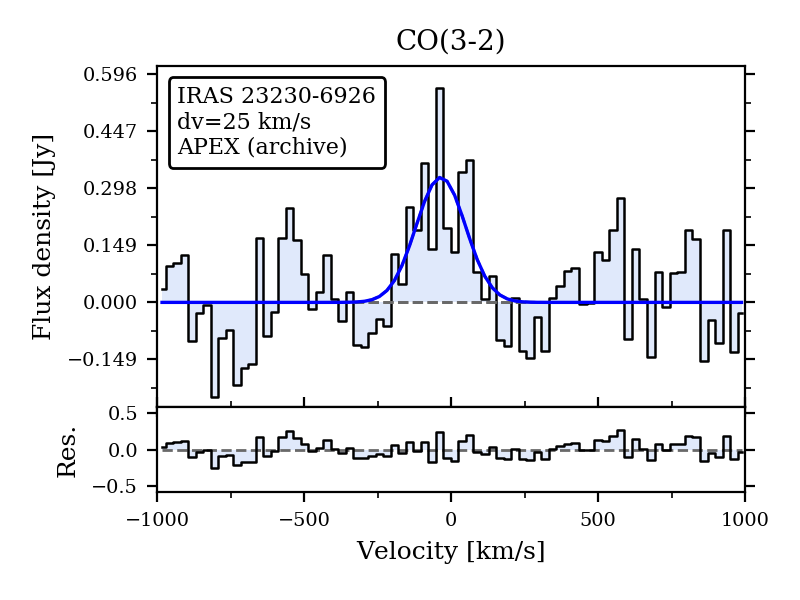}\quad
	\includegraphics[clip=true,trim=0.35cm 0.3cm 0.48cm 0.3cm,width=0.23\textwidth]{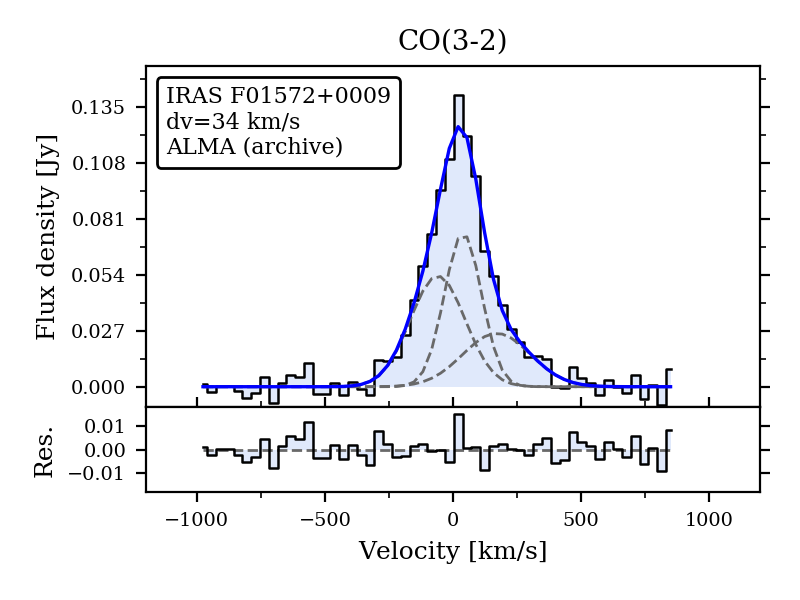}\quad
	\includegraphics[clip=true,trim=0.35cm 0.3cm 0.48cm 0.3cm,width=0.23\textwidth]{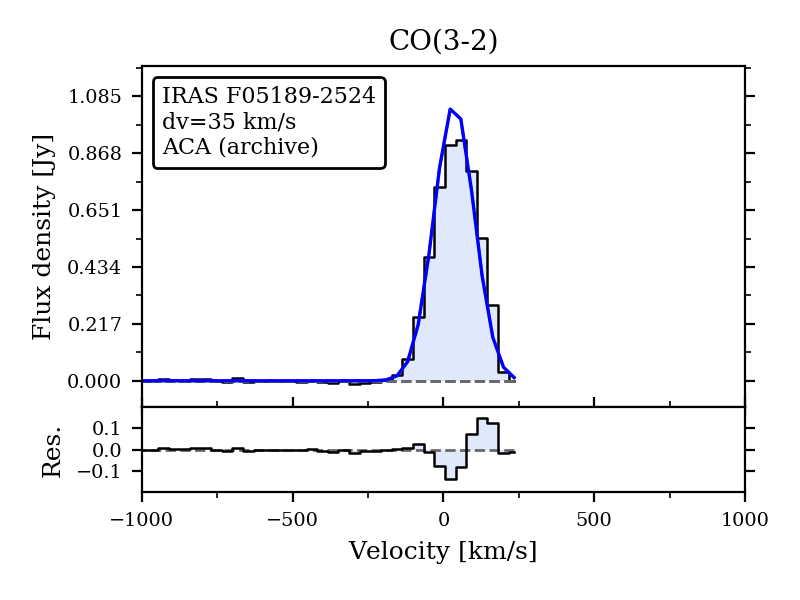}\\
	\includegraphics[clip=true,trim=0.35cm 0.3cm 0.48cm 0.3cm,width=0.23\textwidth]{figures/20414_CO32_dv41_APEXarch.png}\quad
	\includegraphics[clip=true,trim=0.35cm 0.3cm 0.48cm 0.3cm,width=0.23\textwidth]{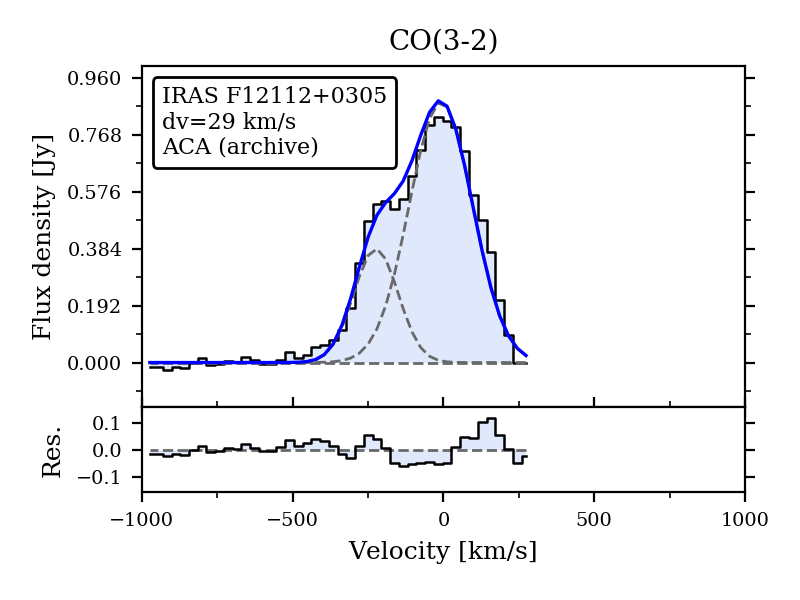}\quad
	\includegraphics[clip=true,trim=0.35cm 0.3cm 0.48cm 0.3cm,width=0.23\textwidth]{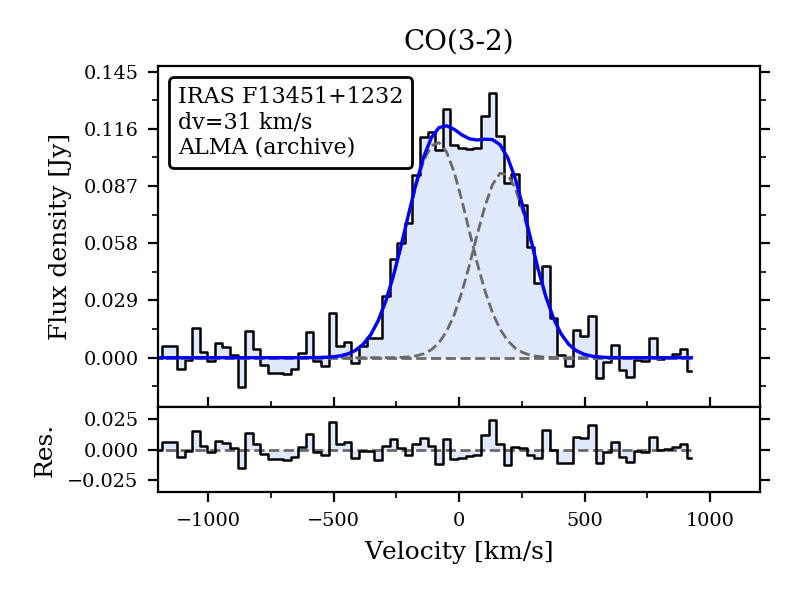}\quad
	\includegraphics[clip=true,trim=0.35cm 0.3cm 0.48cm 0.3cm,width=0.23\textwidth]{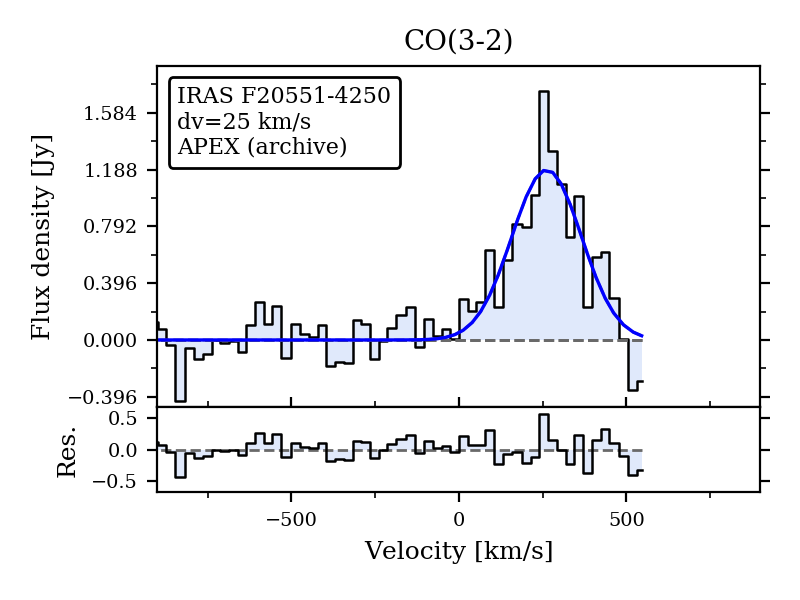}\\
	\includegraphics[clip=true,trim=0.35cm 0.3cm 0.48cm 0.3cm,width=0.23\textwidth]{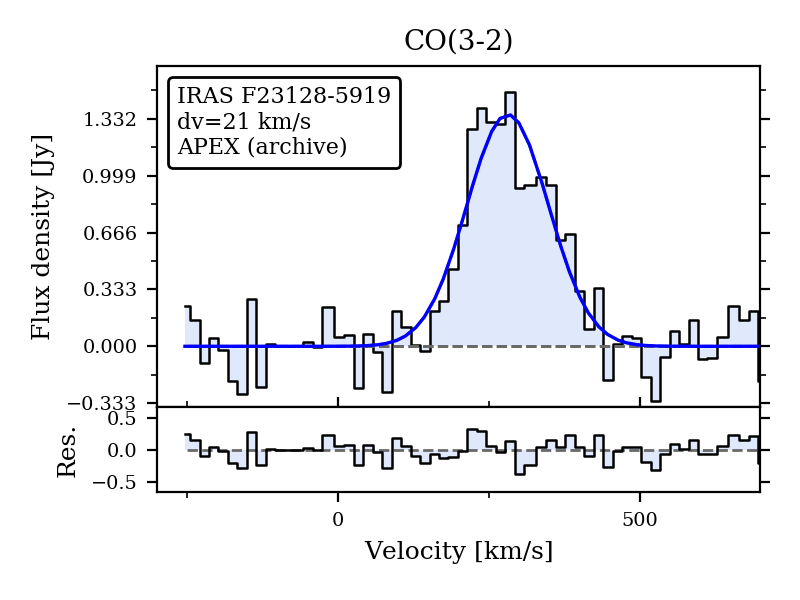}\quad
	\includegraphics[clip=true,trim=0.35cm 0.3cm 0.48cm 0.3cm,width=0.23\textwidth]{figures/empty}\quad
	\includegraphics[clip=true,trim=0.35cm 0.3cm 0.48cm 0.3cm,width=0.23\textwidth]{figures/empty}\quad
	\includegraphics[clip=true,trim=0.35cm 0.3cm 0.48cm 0.3cm,width=0.23\textwidth]{figures/empty}\\
\caption{Duplicated CO(3--2) spectra not used in the analysis.}\label{fig:spectra_dupli_co32}
\end{figure*}

\begin{figure*}[tbp]
	\includegraphics[clip=true,trim=0.35cm 0.3cm 0.48cm 0.3cm,width=0.23\textwidth]{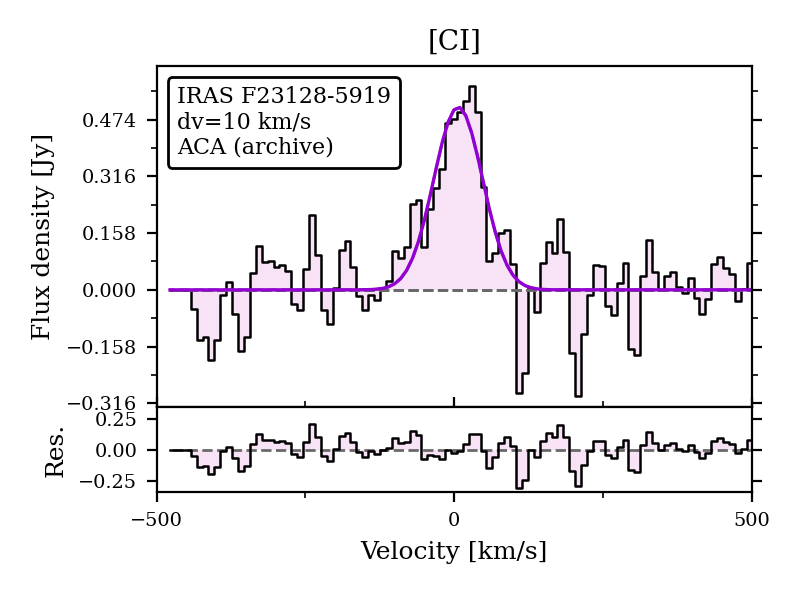}\quad
	\includegraphics[clip=true,trim=0.35cm 0.3cm 0.48cm 0.3cm,width=0.23\textwidth]{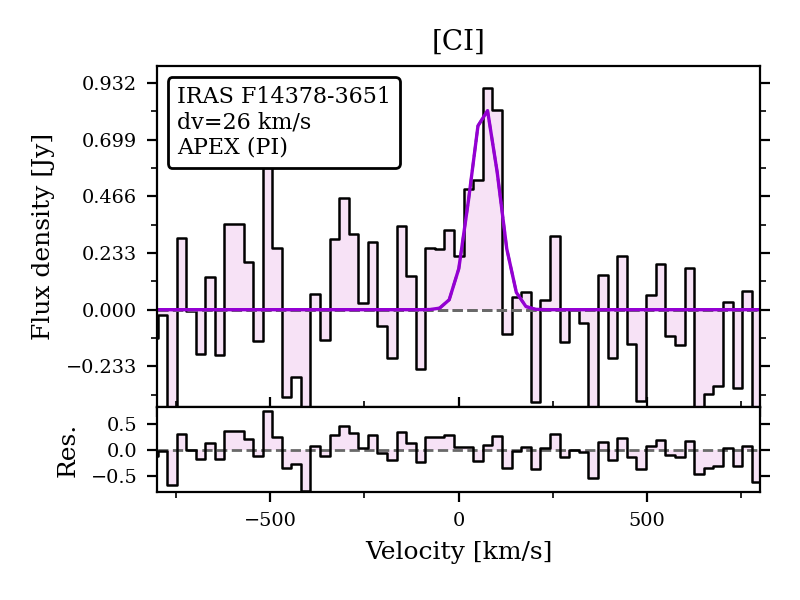}\quad
	\includegraphics[clip=true,trim=0.35cm 0.3cm 0.48cm 0.3cm,width=0.23\textwidth]{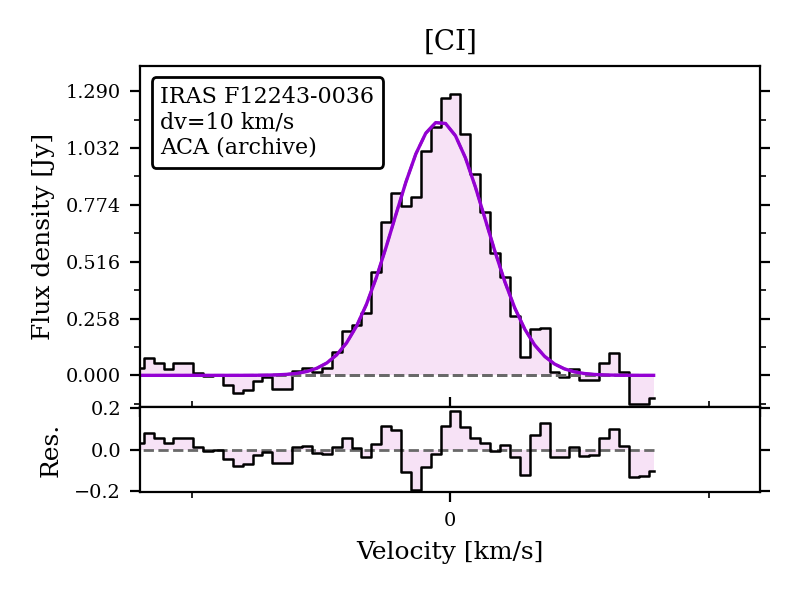}\quad
	\includegraphics[clip=true,trim=0.35cm 0.3cm 0.48cm 0.3cm,width=0.23\textwidth]{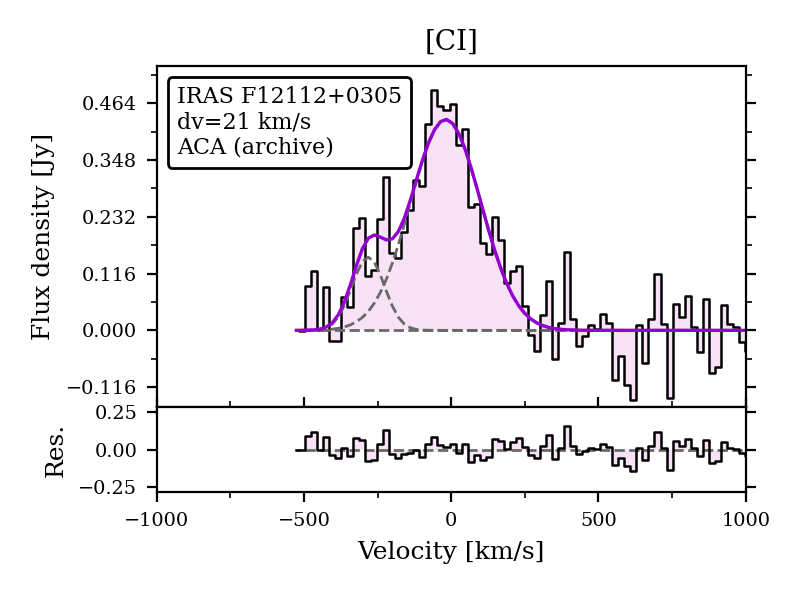}\\
	\includegraphics[clip=true,trim=0.35cm 0.3cm 0.48cm 0.3cm,width=0.23\textwidth]{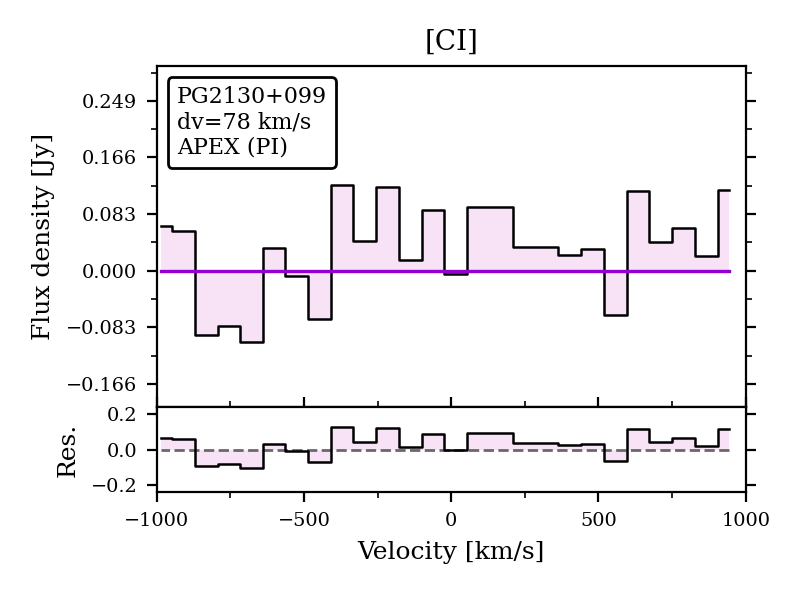}\quad
	\includegraphics[clip=true,trim=0.35cm 0.3cm 0.48cm 0.3cm,width=0.23\textwidth]{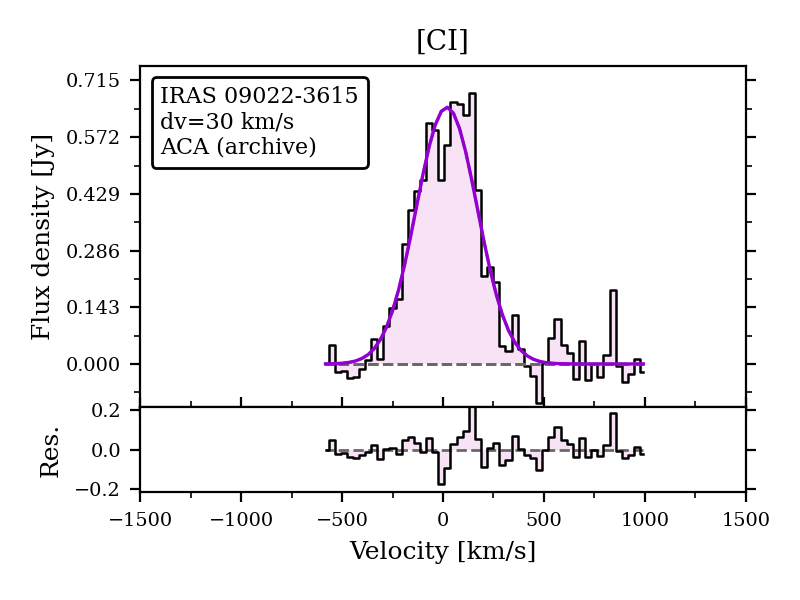}\\
	\caption{Duplicated [CI](1--0) spectra not used in the analysis.}\label{fig:spectra_dupli_ci10}
\end{figure*}

Here we report additional spectra reduced for the sources in our sample, which were not included in the main analysis because of the availability of better quality and/or higher priority data, as explained in Section~\ref{sec:datareduction}. Figures~\ref{fig:spectra_dupli_co21},~\ref{fig:spectra_dupli_co32}, and ~\ref{fig:spectra_dupli_ci10} show the CO(2--1), CO(3--2), and [CI](1--0) duplicated spectra, respectively. 

Additional details concerning individual sources are described below.

\subsubsection*{IRAS F15462-0450}
The APEX archival CO(2--1) dataset for the source IRAS F15462-0450 was discarded from the analysis because an inspection of the header signals a suspected pointing issue during the observation. This may be the cause of significant flux loss that led to a non detection. The ALMA archival CO(1--0) data for this source show a total flux of $43\pm 16$ [Jy~\kms], which would lead to detectable CO(2--1) transition by APEX, which supports our hypothesis of pointing issues. We checked the literature for previous CO(2--1) observations of this source and found an IRAM 30m telescope dataset from 2008 published by \cite{Xia+12}, where they computed a total integrated CO(2--1) flux of $21.04\pm 1.58$ [Jy \kms] for IRAS F15462-0450. 

\subsubsection*{IRAS F12112+0305}
The ALMA CO(1--0) line observations of IRAS F12112+0305 show a low flux value of $34.6\pm1.1$ [Jy \kms], hinting at a significant missing flux. We proceeded to check the literature for previous CO(1--0) observations of this particular source, and found the work by \cite{Chung+09} reporting a Five College Radio Astronomy Observatory (FCRAO) 14 m telescope observation carried out between 2007 and 2008, with a total CO(1--0) line flux of $64.3 \pm 21.04$ [Jy \kms].

\section{Tables listing line fluxes, luminosities, and luminosity ratios}\label{sec:appendix_tables}

Table~\ref{tab:fluxlumino} lists the integrated line fluxes and luminosities, obtained through the spectral line fitting, computed using the relations in Section~\ref{sec:linelumrat}.

\begin{tiny}
	\onecolumn
	\begin{longtable}{@{}lcccccccc@{}}
		\caption{Total CO fluxes and luminosities}\label{tab:fluxlumino} \\
		\hline
		\hline 
		\multirow{3}{*}{Galaxy name} & \multicolumn{2}{c}{CO(1--0)}  & \multicolumn{2}{c}{CO(2--1)} & \multicolumn{2}{c}{CO(3--2)} & \multicolumn{2}{c}{[CI](1--0)} \\ 
		& $S_{\textrm{CO}}~dv$ & $L^{\prime}_{\textrm{CO}}\times 10^{-9}$  & $S_{\textrm{CO}}~dv$ & $L^{\prime}_{\textrm{CO}}\times 10^{-9}$ &
		$S_{\textrm{CO}}~dv$ & $L^{\prime}_{\textrm{CO}}\times 10^{-9}$ & $S_{\textrm{[CI]}}~dv$ & $L^{\prime}_{\textrm{[CI]}}\times 10^{-9}$\\
		& [Jy~km~s$^{-1}$] & [K~km~s$^{-1}$~pc$^2$]  & [Jy~km~s$^{-1}$] & [K~km~s$^{-1}$~pc$^2$] & [Jy~km~s$^{-1}$] & [K~km~s$^{-1}$~pc$^2$] & [Jy~km~s$^{-1}$] & [K~km~s$^{-1}$~pc$^2$] \\ 
		(1) & (2) & (3) & (4) & (5) & (6) & (7) & (8) & (9) \\ 
		\hline 
		\endfirsthead
		\multicolumn{9}{c}{\footnotesize Following from previous page}\\
		\hline
		\multirow{3}{*}{Galaxy name} & \multicolumn{2}{c}{CO(1--0)}  & \multicolumn{2}{c}{CO(2--1)} & \multicolumn{2}{c}{CO(3--2)} & \multicolumn{2}{c}{[CI](1--0)} \\ 
		& $S_{\textrm{CO}}~dv$ & $L^{\prime}_{\textrm{CO}}\times 10^{-9}$  & $S_{\textrm{CO}}~dv$ & $L^{\prime}_{\textrm{CO}}\times 10^{-9}$ &
		$S_{\textrm{CO}}~dv$ & $L^{\prime}_{\textrm{CO}}\times 10^{-9}$ & $S_{\textrm{[CI]}}~dv$ & $L^{\prime}_{\textrm{[CI]}}\times 10^{-9}$\\
		& [Jy~km~s$^{-1}$] & [K~km~s$^{-1}$~pc$^2$]  & [Jy~km~s$^{-1}$] & [K~km~s$^{-1}$~pc$^2$] & [Jy~km~s$^{-1}$] & [K~km~s$^{-1}$~pc$^2$] & [Jy~km~s$^{-1}$] & [K~km~s$^{-1}$~pc$^2$] \\ 
		(1) & (2) & (3) & (4) & (5) & (6) & (7) & (8) & (9) \\ 
		\hline 
		\endhead
		\hline 
		\multicolumn{9}{c}{\footnotesize Follows on next page}\\
		\endfoot
		\hline 
		\endlastfoot
		\hline
        IRAS 00188-0856 & 6.4 $\pm$ 0.5 & 5.4 $\pm$ 0.4 & 46.6 $\pm$ 4.4 & 9.8 $\pm$ 0.9 & 45.4 $\pm$ 1.2 & 4.2 $\pm$ 0.1 &               &               \\ 
        IRAS 01003-2238 & 3.5 $\pm$ 0.5 & 2.5 $\pm$ 0.4 & 9.6 $\pm$ 2.5 & 1.7 $\pm$ 0.4 & 21.2 $\pm$ 3.8 & 1.7 $\pm$ 0.3 &               &               \\ 
        IRAS F01572+0009 & 5.9 $\pm$ 0.8 & 8.1 $\pm$ 1.2 & 32.7 $\pm$ 2.2 & 11.2 $\pm$ 0.8 & 49.7 $\pm$ 3.9 & 7.5 $\pm$ 0.6 &               &               \\ 
        IRAS 03521+0028 &               &               & 47.8 $\pm$ 1.8 & 14.1 $\pm$ 0.5 & 73.1 $\pm$ 5.6 & 9.6 $\pm$ 0.7 &               &               \\ 
        IRAS F05024-1941 &               &               & 22.0 $\pm$ 1.6 & 10.7 $\pm$ 0.8 & 35.5 $\pm$ 2.6 & 7.7 $\pm$ 0.6 &               &               \\ 
        IRAS F05189-2524 & 42.0 $\pm$ 1.3 & 3.7 $\pm$ 0.1 & 133.6 $\pm$ 5.8 & 3.0 $\pm$ 0.1 & 167.0 $\pm$ 15.5 & 1.6 $\pm$ 0.2 & 134.4 $\pm$ 17.1 & 0.7 $\pm$ 0.1 \\ 
        IRAS 06035-7102 & 40.3 $\pm$ 1.9 & 12.7 $\pm$ 0.6 & 154.4 $\pm$ 6.7 & 12.1 $\pm$ 0.5 & 326.5 $\pm$ 14.8 & 11.4 $\pm$ 0.5 & 224.9 $\pm$ 17.4 & 3.9 $\pm$ 0.3 \\ 
        IRAS 06206-6315 &               &               & 81.6 $\pm$ 4.9 & 8.7 $\pm$ 0.5 & 115.0 $\pm$ 7.0 & 5.5 $\pm$ 0.3 &               &               \\ 
        IRAS 07251-0248 &               &               & 71.3 $\pm$ 1.0 & 6.8 $\pm$ 0.1 & 145.5 $\pm$ 12.4 & 6.2 $\pm$ 0.5 &               &               \\ 
        IRAS 08311-2459 &               &               & 136.7 $\pm$ 5.3 & 17.3 $\pm$ 0.7 & 247.4 $\pm$ 16.5 & 13.9 $\pm$ 0.9 &               &               \\ 
        IRAS 09022-3615 &               &               & 284.4 $\pm$ 7.9 & 12.5 $\pm$ 0.3 & 368.2 $\pm$ 40.3 & 7.2 $\pm$ 0.8 & 168.0 $\pm$ 16.3 & 1.6 $\pm$ 0.2 \\ 
        IRAS 10378+1109 &               &               & 46.7 $\pm$ 3.1 & 11.1 $\pm$ 0.7 & 89.0 $\pm$ 6.8 & 9.4 $\pm$ 0.7 &               &               \\ 
        IRAS 11095-0238 &               &               & 43.4 $\pm$ 2.9 & 6.2 $\pm$ 0.4 &      761.5$^{*}$     &   48.4$^{*}$      &               &               \\ 
        IRAS F12072-0444 & 6.6 $\pm$ 0.8 & 5.6 $\pm$ 0.7 & 49.3 $\pm$ 3.4 & 10.3 $\pm$ 0.7 & 45.2 $\pm$ 7.8 & 4.2 $\pm$ 0.7 &               &               \\ 
        IRAS F12112+0305 & 34.6 $\pm$ 1.1 & 9.2 $\pm$ 0.3 & 183.8 $\pm$ 9.0 & 12.2 $\pm$ 0.6 & 306.7 $\pm$ 19.6 & 9.0 $\pm$ 0.6 & 88.0 $\pm$ 12.2 & 1.3 $\pm$ 0.2 \\ 
        IRAS 13120-5453 & 263.8 $\pm$ 4.6 & 12.2 $\pm$ 0.2 & 838.3 $\pm$ 28.6 & 9.7 $\pm$ 0.3 & 1477.2 $\pm$ 55.2 & 7.6 $\pm$ 0.3 & 794.2 $\pm$ 42.8 & 2.0 $\pm$ 0.1 \\ 
        IRAS F13305-1739 &               &               & 11.4 $\pm$ 1.3 & 3.2 $\pm$ 0.4 &               &               &               &               \\ 
        IRAS F13451+1232 & 11.8 $\pm$ 0.6 & 8.8 $\pm$ 0.5 & 29.2 $\pm$ 4.2 & 5.5 $\pm$ 0.8 & 80.1 $\pm$ 0.2 & 6.7 $\pm$ 0.02 &               &               \\ 
        IRAS F14348-1447 &               &               & 201.0 $\pm$ 6.4 & 17.3 $\pm$ 0.5 &               &               & 115.4 $\pm$ 16.7 & 2.2 $\pm$ 0.3 \\ 
        IRAS F14378-3651 & 19.4 $\pm$ 1.0 & 4.4 $\pm$ 0.2 & 78.0 $\pm$ 1.6 & 4.5 $\pm$ 0.1 &               &               & 68.5 $\pm$ 3.1 & 0.9 $\pm$ 0.01 \\ 
        IRAS F15462-0450 & 47.0 $\pm$ 2.8 & 23.8 $\pm$ 1.4 &               &               &               &               &               &               \\ 
        IRAS 16090-0139 & 17.0 $\pm$ 0.9 & 15.5 $\pm$ 0.8 & 97.5 $\pm$ 3.9 & 22.1 $\pm$ 0.9 & 148.3 $\pm$ 12.9 & 15.0 $\pm$ 1.3 & 123.8 $\pm$ 13.6 & 6.2 $\pm$ 0.7 \\ 
        IRAS 17208-0014 & 150.9 $\pm$ 2.0 & 13.5 $\pm$ 0.2 & 583.0 $\pm$ 14.6 & 13.1 $\pm$ 0.3 & 1041.2 $\pm$ 36.6 & 10.4 $\pm$ 0.4 & 339.1 $\pm$ 26.0 & 1.7 $\pm$ 0.1 \\ 
        IRAS 19254-7245 &               &               & 170.0 $\pm$ 6.4 & 7.9 $\pm$ 0.3 &               &               & 98.7 $\pm$ 13.5 & 1.0 $\pm$ 0.1 \\ 
        IRAS F19297-0406 &               &               & 57.5 $\pm$ 3.0 & 5.3 $\pm$ 0.3 & 214.7 $\pm$ 4.6 & 8.7 $\pm$ 0.2 &               &               \\ 
        IRAS 19542+1110 & 16.5 $\pm$ 0.8 & 3.2 $\pm$ 0.15 & 48.8 $\pm$ 3.6 & 2.3 $\pm$ 0.2 &               &               & 93.9 $\pm$ 14.0 & 1.0 $\pm$ 0.1 \\ 
        IRAS 20087-0308 & 30.3 $\pm$ 1.6 & 17.0 $\pm$ 0.9 & 141.8 $\pm$ 4.6 & 19.9 $\pm$ 0.6 & 202.1 $\pm$ 4.2 & 12.6 $\pm$ 0.3 &               &               \\ 
        IRAS 20100-4156 & 11.3 $\pm$ 0.2 & 9.7 $\pm$ 0.2 & 54.7 $\pm$ 3.0 & 11.7 $\pm$ 0.6 & 133.5 $\pm$ 8.4 & 12.7 $\pm$ 0.8 &               &               \\ 
        IRAS 20414-1651 &               &               & 63.7 $\pm$ 3.1 & 6.0 $\pm$ 0.3 & 163.0 $\pm$ 11.1 & 6.9 $\pm$ 0.5 &               &               \\ 
        IRAS F20551-4250 & 67.1 $\pm$ 1.5 & 6.1 $\pm$ 0.13 & 292.1 $\pm$ 6.5 & 6.6 $\pm$ 0.15 & 389.0 $\pm$ 19.2 & 3.9 $\pm$ 0.2 & 90.8 $\pm$ 5.2 & 0.5 $\pm$ 0.03 \\ 
        IRAS F22491-1808 &               &               & 77.9 $\pm$ 4.6 & 5.9 $\pm$ 0.3 & 133.8 $\pm$ 7.4 & 4.5 $\pm$ 0.2 & 106.2 $\pm$ 16.0 & 1.8 $\pm$ 0.3 \\ 
        IRAS F23060+0505 & 10.7 $\pm$ 0.3 & 16.5 $\pm$ 0.5 & 54.2 $\pm$ 3.1 & 20.9 $\pm$ 1.2 & 70.3 $\pm$ 3.6 & 12.1 $\pm$ 0.6 &               &               \\ 
        IRAS F23128-5919 & 51.2 $\pm$ 0.7 & 5.0 $\pm$ 0.1 & 191.9 $\pm$ 4.4 & 4.7 $\pm$ 0.1 & 329.5 $\pm$ 5.4 & 3.6 $\pm$ 0.1 & 233.3 $\pm$ 43.0 & 1.2 $\pm$ 0.2 \\ 
        IRAS 23230-6926 & 8.3 $\pm$ 0.7 & 4.8 $\pm$ 0.4 & 44.1 $\pm$ 3.3 & 6.3 $\pm$ 0.5 & 83.5 $\pm$ 2.4 & 5.3 $\pm$ 0.2 &               &               \\ 
        IRAS 23253-5415 &               &               & 119.6 $\pm$ 3.9 & 25.7 $\pm$ 0.8 & 160.1 $\pm$ 10.5 & 15.3 $\pm$ 1.0 &               &               \\ 
        IRAS F23389+0300 &               &               & 24.1 $\pm$ 1.3 & 6.5 $\pm$ 0.4 & 52.4 $\pm$ 3.1 & 6.3 $\pm$ 0.4 &               &               \\ 
        \hline
        IRAS F00509+1225 & 47.4 $\pm$ 4.7 & 8.7 $\pm$ 0.9 & 80.6 $\pm$ 6.1 & 3.7 $\pm$ 0.3 &               &               & 67.7 $\pm$ 16.4 & 0.7 $\pm$ 0.2 \\ 
        PG1126-041 &               &               & 16.7 $\pm$ 0.4 & 0.7 $\pm$ 0.02 &               &               &               &               \\ 
        IRAS F12243-0036 & 101.6 $\pm$ 1.0 & 0.2 $\pm$ 0.003 & 428.2 $\pm$ 5.6 & 0.3 $\pm$ 0.003 & 862.5 $\pm$ 56.3 & 0.2 $\pm$ 0.02 & 520.4 $\pm$ 61.5 & 0.1 $\pm$ 0.001 \\ 
        PG2130+099 &               &               & 22.5 $\pm$ 1.8 & 1.1 $\pm$ 0.1 &               &               &       72.4$^{*}$       &       0.78$^{*}$        \\ 
		\hline
	\end{longtable}
	
	\tablefoot{$^{*}$We compute the $3\sigma$ upper limit on the total integrated line flux, following $\int S_{\rm line,dv} < 3\sigma_{\rm rms, channel} \sqrt{\delta v_{\rm channel}}\Delta v_{\rm line}$, where $\sigma_{\rm rms, channel}$ is the rms noise per spectral channel, $\delta v_{\rm channel}$ is the channel width, and $\Delta v_{\rm line}$ is the expected line width (assumed to be equal to the CO(2--1) line width). }
	
	\twocolumn
\end{tiny}

\section{Additional relations explored}

\begin{figure*}[tbp]
	\includegraphics[width=.31\textwidth]{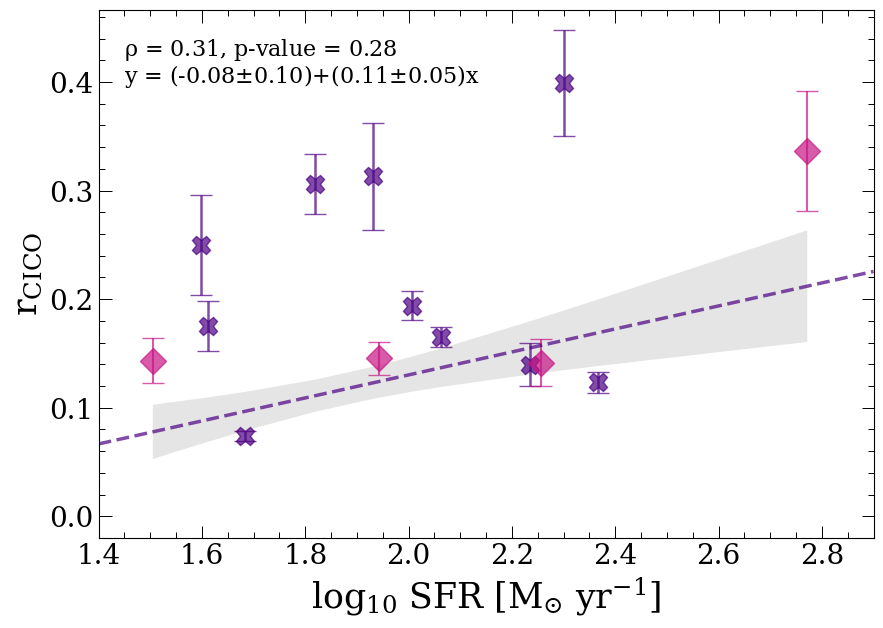}\quad
	\includegraphics[width=.31\textwidth]{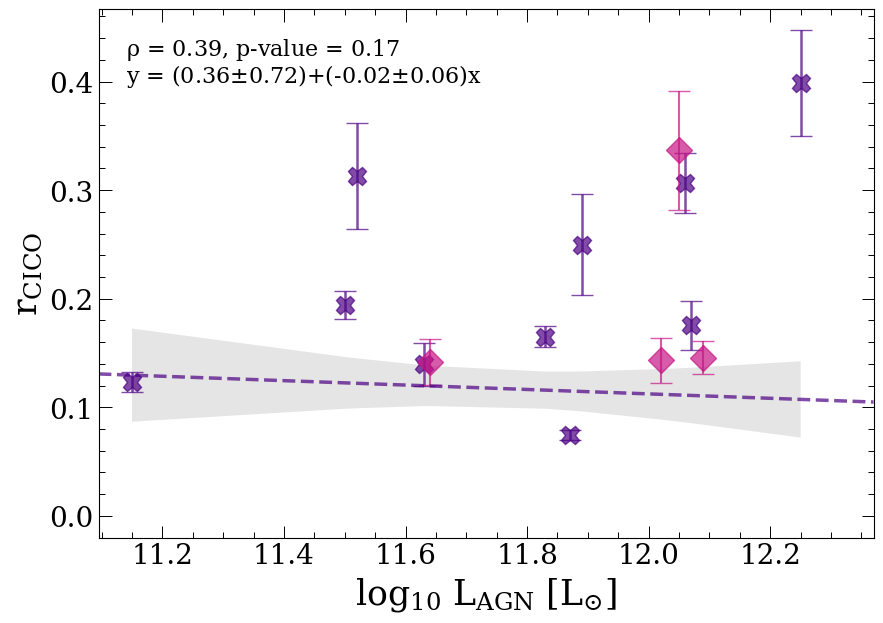}\quad
	\includegraphics[width=.31\textwidth]{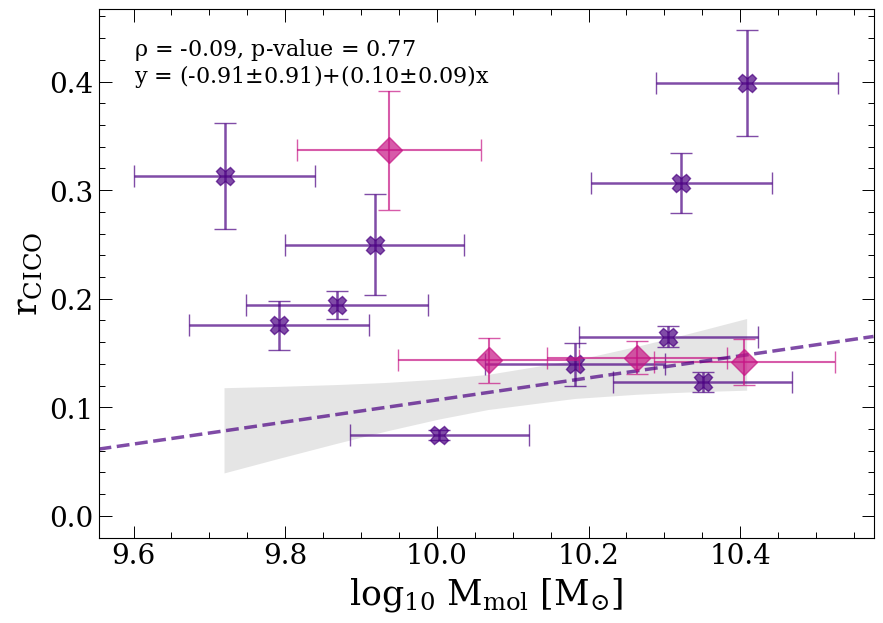}\\
	\includegraphics[width=.31\textwidth]{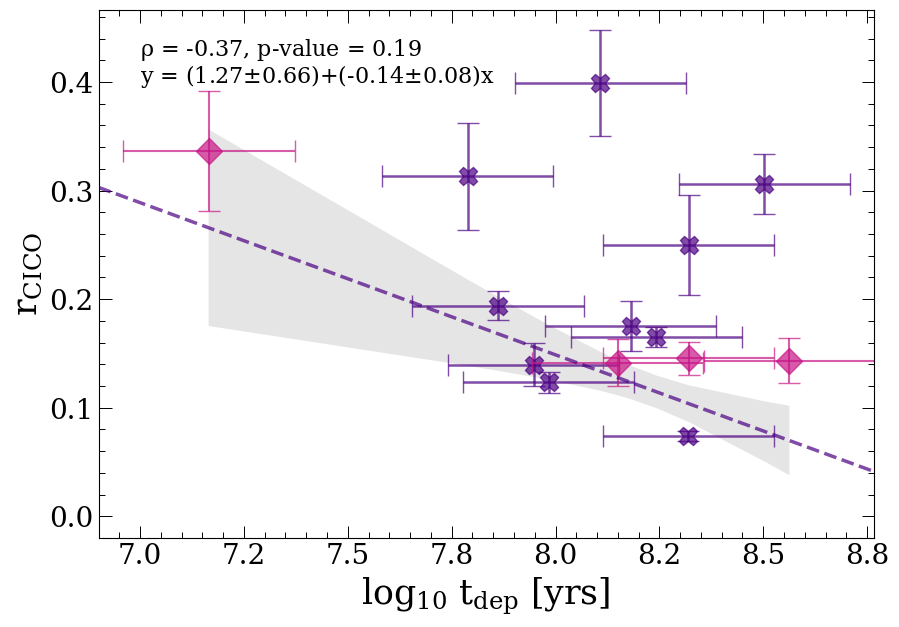}\quad
	\includegraphics[width=.31\textwidth]{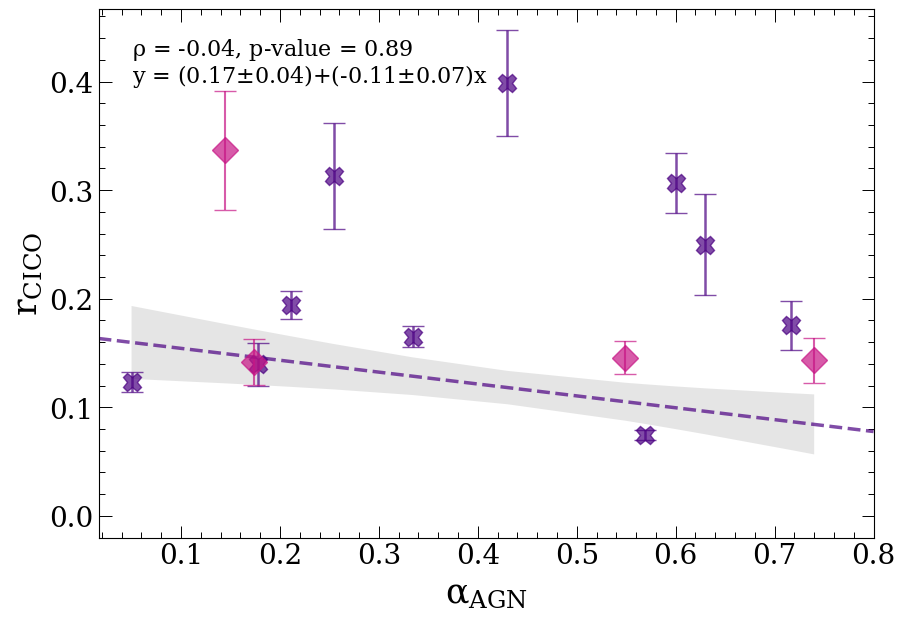}\\
	\caption{$r_{\textrm{CICO}}$ plotted as a function of different galaxy properties, namely: SFR, $L_{\mathrm{AGN}}$, $M_{\mathrm{mol}}$, $\rm \tau_{dep}$, and $\alpha_{\rm AGN}$. Symbols and notation as in Figure~\ref{fig:rCICO_LIR} for our sample of ULIRGs.}
	\label{fig:rCICO_properties}
\end{figure*}

In this appendix, we show the relations between the $r_{\rm CICO}$ and different galaxy properties: SFR, $L_{\rm AGN}$, $M_{\rm mol}$, $\rm \tau_{dep}$, and $\alpha_{\rm AGN}$ (Figure \ref{fig:rCICO_properties}). In the main body of the text we plot the relation with $L_{\rm IR}$ (see Fig. \ref{fig:rCICO_LIR}). We do not retrieve any significant relations for any of the quantities studied here, as can be seen from the Pearson correlation coefficients and \textit{p}-values reported in each graph.

\end{appendix}

\end{document}